\documentclass[utf8]{aa}
\usepackage[varg]{txfonts}
\usepackage[utf8]{inputenc}
\usepackage{chngcntr} 
\usepackage{longtable}
\usepackage{pdflscape}
\bibpunct{(}{)}{;}{a}{}{,} 

\newcommand{\sybf}{}
\newcommand{\revbf}{}
\newcommand{\lebf}{}

\begin{document}

\title{Reanalysis of nearby open clusters using Gaia DR1/TGAS and HSOY\thanks{Tables C.1 to C.24 are only available in electronic form at the CDS via anonymous ftp to cdsarc.u-strasbg.fr (130.79.128.5) or via http://cdsweb.u-strasbg.fr/cgi-bin/qcat?J/A+A/}} 

\author{Steffi X. Yen\inst{1}
  \and Sabine Reffert\inst{1} 
     \and Elena Schilbach\inst{1,2}
     \and Siegfried Röser\inst{1,2}
     \and Nina V. Kharchenko\inst{1,3}
     \and Anatoly E. Piskunov\inst{1,4}} 

\offprints{S. X. Yen, \email{syen@lsw.uni-heidelberg.de}}
 
 \institute{Zentrum für Astronomie der Universität Heidelberg, Landessternwarte, Königstuhl 12, 69117 Heidelberg, Germany
        \and Zentrum für Astronomie der Universität Heidelberg, Astronomisches Rechen-Institut, Mönchhofstrasse 12-14, 69120 Heidelberg, Germany
        \and Main Astronomical Observatory, 27 Academica Zabolotnogo Str., 03143 Kiev, Ukraine
        \and Institute of Astronomy of the Russian Acad. Sci., 48 Pyatnitskaya Str., 109017 Moscow, Russia}

\date{Received 6 September 2017 / Accepted 18 January 2018}

\abstract
{Open clusters have long been used to gain insights into the structure, composition, and evolution of the Galaxy. With the large amount of stellar data available for many clusters in the Gaia era, new techniques must be developed for analyzing open clusters, as visual inspection of cluster color-magnitude diagrams is no longer feasible. An automatic tool will be required to analyze large samples of open clusters.}
   {We seek to develop an automatic isochrone-fitting procedure to \sybf{consistently} determine cluster membership and the fundamental cluster parameters.}
   {\sybf{Our cluster characterization pipeline first \lebf{determined} cluster membership with precise astrometry, primarily from TGAS and HSOY. With initial cluster members established, isochrones \lebf{were} fitted, using a $\chi^2$ minimization, to the cluster photometry \lebf{in order} to determine cluster mean distances, ages, and reddening. Cluster membership was also refined based on the stellar photometry. We used multiband photometry, which includes ASCC-2.5 $BV$, 2MASS $JHK_s$, Gaia $G$ band.}}
   {We present parameter estimates for all 24 clusters closer than 333 pc as determined by the \lebf{Catalogue} of Open Cluster Data and the Milky Way Star Clusters catalog. We find that our parameters are consistent to those in the Milky Way Star Clusters catalog.}
   {We demonstrate that it is feasible to develop an automated pipeline that determines cluster parameters and membership reliably. After additional modifications, our pipeline will be able to use Gaia DR2 as input, leading to better cluster memberships and more accurate cluster parameters for a much larger number of clusters.}


\maketitle

\section{Introduction}
\sybf{Open clusters are the keys to unlocking the mysteries of stellar evolution, and the structure and chemical evolution of our Galactic disk. In order to study the Milky Way disk with open clusters, a large amount of stellar data and a homogeneous set of cluster parameters (age, distance, and reddening) are required.}

\sybf{Compilations of open clusters and their parameters from the literature have been assembled, for example \citet{1981csca.book.....R} and \citet{1982A&A...109..213L}. Information for these clusters come from individual studies, which use a variety of methods to derive parameters, resulting in a heterogeneous parameter set.} 
\sybf{In general, cluster membership relies on the analysis of positions and kinematics: proper motions and radial velocities. Classical methods vary from using visible groupings of stars to calculating probabilities based on proper motions (e.g., \citealt{1958AJ.....63..387V}, \citealt{1971A&A....14..226S}, \citealt{1985A&A...150..298C}). Photometry can also be used to distinguish member and field stars of a cluster (e.g., \citealt{1971A&A....11..359V}, \citealt{1972A&AS....7..355M}, \citealt{1983A&AS...51..235B}). Nowadays, rigorous mathematical and statistical approaches are developed to derive memberships using kinematics and photometry (e.g., \citealt{2013ApJ...762...88M}, \citealt{2014A&A...563A..45S}). Cluster distances can be determined directly from trigonometric parallax measurements. However, the most common approach to deriving cluster distance, which subsequently also provides estimates of age and reddening, uses photometric data by  fitting theoretical isochrones to the cluster color-magnitude diagram (CMD).}

\sybf{The wealth of information from large photometric surveys has led to the creation of homogeneous cluster catalogs, where the methods used to derive membership and cluster parameters combine the techniques aforementioned with some optimization of the results by eye. The first catalogs deriving homogenous cluster parameters} relied on optical photometry, including the works by \citet{1971A&AS....4..241B}, \citet{1982ApJS...49..425J}, \citet{1998AJ....116.2423P}, \citet{1999A&AS..135....5C}, \citet{2002A&A...389..871D}, and \citet{2005A&A...440..403K, 2005A&A...438.1163K}. Infrared photometry from the 2MASS \citep{2006AJ....131.1163S} survey then allowed for the discovery of many new open clusters (e.g., \citealt{2001A&A...376..434D}, \citealt{2003A&A...397..177B}, \citealt{2003A&A...400..533D}, \citealt{2008A&A...486..771K}, \citealt{2010AstL...36...75G}, \citealt{2011AcA....61..231B}). The Milky Way Star Cluster (MWSC) catalog \citep{2013A&A...558A..53K} is the largest-to-date census on the star cluster populations within the Milky Way. It is nearly complete to $\sim$1.8 kpc from the Sun and contains 2808 open clusters.

\sybf{While these catalogs have analyzed hundreds or thousands of open clusters in a homogeneous way, large discrepancies exist between the final cluster parameters reported by these catalogs, as illustrated by \citet{2015A&A...582A..19N}.} Some of the limitations of these catalogs include: small number of cluster members, low accuracy in proper motion, \sybf{precision and/or accuracy of photometry}, or use of different isochrones. It is clear that a comprehensive assessment requires accurate photometric and kinematic data at least; spectroscopic data would also be a great benefit. 

Taking advantage of the large amount of stellar data available (photometry from large all-sky surveys and refined astrometry from space missions), we have developed an automated pipeline to \sybf{consistently} determine cluster membership and fit the fundamental cluster parameters: distance $d$, reddening $E(B - V)$, and age log~$t$, where $t$ is in years. Due to the limited size \sybf{and precision} of the TGAS catalog \revbf{\citep{2016A&A...595A...2G}}, we performed our analysis on 24 nearby open clusters. These clusters are generally well-studied and are located within 333 pc as given in both MWSC and the \lebf{Catalogue} of Open Cluster Data (COCD, \citealt{2005A&A...440..403K,2005A&A...438.1163K}). \sybf{As the typical error of TGAS parallaxes is 0.3 mas, studying clusters with parallaxes greater than 3 mas allows us to use accurate stellar parallaxes, with errors less than 10\%.} The names and MWSC identifiers of the clusters are provided in Table 1. 

This pipeline was developed to ascertain the possibility of an automated isochrone fitting routine that reliably determines cluster membership and parameters. We note that the techniques described in this paper are specifically designed for working with the current data available and its limitations. 

In Section 2, we describe our input data. The details of our cluster characterization pipeline is described in Section 3. In Section 4, we discuss our results and compare them with the literature. Finally, a summary of our \sybf{technique and first results are} given in Section 5.

\section{Data}
The basis of our data set is largely rooted in the cluster \sybf{field} star lists of the MWSC catalog. \sybf{The selection area around each cluster has a radius of $r_{\textrm{a}}$ = $r_{\textrm{cl}}$ + 0.3$^{\circ}$, where $r_{\textrm{cl}}$ is taken from the literature \citep{2012A&A...543A.156K}.} The primary stellar data for MWSC was compiled from the PPMXL \citep{2010AJ....139.2440R} and 2MASS \citep{2006AJ....131.1163S} all-sky catalogs. The positions from the PPMXL catalog are used and supplemented with Hipparcos \citep{2007A&A...474..653V}, in order to recover any missing bright stars. 

Building upon MWSC, we \sybf{used $B$ and $V$} photometry from the All-Sky Compiled Catalogue of 2.5 million stars (ASCC-2.5, \citealt{2001KFNT...17..409K}), \sybf{which are based mainly on Hipparcos and Tycho-2 \citep{2000A&A...355L..27H}}.
We also incorporated 2MASS $JHK_s$ and Gaia DR1 \sybf{\citep{2016A&A...595A...2G}} $G$ band photometry, for a total of up to six bands for each star. \revbf{The median uncertainties in $G$ magnitudes range from the mmag level to 0.03~mag \citep{2017A&A...599A..32V, 2017A&A...600A..51E}. In order to also account for systematics in the $G$ magnitudes of the brighter stars, we have adopted a conservative error of 0.03~mag for all stars.} 

Furthermore, we included precise stellar astrometric data from TGAS \citep{2016A&A...595A...2G} and HSOY \citep{2017A&A...600L...4A}. TGAS supplies proper motions and parallaxes for roughly 2 million Tycho-2 stars. HSOY, which combines positions from Gaia DR1 and data from PPMXL, gives proper motions for 583 million stars. Hipparcos proper motions were also used to recover any missing bright stars. By combining six-band photometric measurements, proper motions, and parallaxes for the stars, we are able to better constrain and determine cluster membership and parameters.

For this work, we used the Padova isochrone set: PARSEC version 1.2S \citep{2012MNRAS.427..127B} in the Johnson $BV$, 2MASS $JHK_s$, and Gaia $G$ photometric systems with $Z$ = $Z_\odot$ = 0.0152 \citep{2009A&A...498..877C,2011SoPh..268..255C}. \sybf{Most open cluster studies based on isochrone fitting assume solar metallicity for simplicity because cluster metallicities are known for very few open clusters. In the updated catalog by \citet{2002A&A...389..871D}, this parameter is available for roughly 13\% of the $\sim$2000 clusters. Of our 24 clusters, only 14 have metallicities listed in the MWSC. The mean of these metallicities is $-$0.09 dex, which is close to solar metallicity. Furthermore, considering these values also have some error, it is reasonable to assume solar metallicity for all clusters. Nevertheless, for clusters with highly nonsolar metallicities, this will introduce a small bias to the derived parameters.} Our isochrone set spans the age range 6.6 $\leq$ $\log t$ $\leq$ 10.1, at step sizes $\Delta\log$ $t$ = 0.01.

We have also constructed the corresponding zero-age main sequence (ZAMS) for this isochrone set using the evolutionary tracks and ``ptcri" file \footnote{http://people.sissa.it/$\sim$sbressan/parsec.html}. The age at which the ZAMS occurs for each mass in the isochrone set can be determined by matching the \sybf{MS\_BEG} point for a given mass in the ptcri file to the age at the \sybf{MS\_BEG} point in the evolutionary track for that mass. With a list of ages for each mass, the ZAMS $B$, $V$, $J$, $H$, $K_s$, and $G$ magnitudes can then be determined by finding the matching age and mass combination in the isochrone set. 

\section{Pipeline methodology}
Our automated cluster characterization pipeline \sybf{consistently} determines cluster membership and fits the fundamental cluster parameters: distance, reddening, and age.
\sybf{The pipeline follows a sequence of procedures with two main segments: (1) membership determination and (2) isochrone fitting and membership refinement. Cluster membership is first determined} through iterative proper motion, parallax, and photometric selections. After \sybf{the initial} membership selection, isochrones are fitted to the photometric observations of cluster members \sybf{to determine cluster parameters and membership is further refined by removing highly discordant stars. This segment is iterated until membership and cluster parameters are consistent.}

The pipeline relies on a $\chi^2$ minimization to fit the ZAMS and isochrones to the cluster photometry. The minimization method used is the Levenberg-Marquardt as provided by LMFIT \citep{newville_2014_11813}. We chose to use a least squares method because it is a reliable and commonly used fitting approach. In future versions of our pipeline we may implement a maximum likelihood method as \citet{2010A&A...516A...2M}, \citet{2012A&A...539A.125D}, and \citet{2014A&A...564A..49P} have used, should it turn out to be more robust. For now, the least squares fitting has produced reliable results for all nearby clusters\sybf{, as quantified in Sect. {\ref{results}}}.

\subsection{Proper motion and parallax selections}
The proper motion selection routine first computes the weighted mean cluster TGAS proper motion using TGAS proper motions of the most probable cluster members from the MWSC. In the MWSC, the most probable members are defined as stars with a combined probability of kinematic or proper motion, photometric, and spatial components, greater than 0.61 \citep{2013A&A...558A..53K}.
The weighted mean cluster TGAS proper motion in RA and DEC are denoted by $\bar{\mu}_{\alpha^*,\mathrm{T}}$ and $\bar{\mu}_{\delta,\mathrm{T}}$ respectively, where $\alpha^{*} = \alpha \cdot \cos\delta$.

Considering that the median standard proper motion uncertainty of TGAS is 1.2 \sybf{mas yr$^{-1}$} \citep{2016A&A...595A...4L}, all stars within 2 \sybf{mas yr$^{-1}$} of the weighted mean cluster TGAS proper motion are selected as candidate members. For 10\% of TGAS stars, the standard proper motion error is more than 2.7 \sybf{mas yr$^{-1}$} \citep{2016A&A...595A...4L}, which means there are potential cluster members with larger proper motion errors that fall outside the 2 \sybf{mas yr$^{-1}$} radius. To recover these stars, a factor of 2.5 is applied to the first radius, and so stars within 5 \sybf{mas yr$^{-1}$} are also considered cluster candidates if their 3$\sigma$ proper motion error ellipse is consistent with the weighted mean cluster proper motion.

\sybf{While the proper motion precision varies across the sky \citep{2016A&A...595A...4L}, the limits employed here are general and optimal for our cluster sample. Using a smaller value for the inner selection radius would lead to missing members because although some clusters in our sample have very small proper motion errors, roughly 0.2 mas yr$^{-1}$, there are systematics to consider on top of the formal error. Furthermore, there likely exists some internal dispersion in the proper motions for cluster members, as they are not expected to have exactly the same proper motions. On the other hand, the inner selection circle cannot be made too large, otherwise too many nonmembers would be included, especially for clusters with larger proper motion errors. The inner selection radius corresponds to the largest median proper motion error present. Taking these considerations into account, an inner radius of 2 \revbf{mas yr$^{-1}$} is a good compromise for all clusters, regardless of their sky position.}

This first \sybf{selection} allows the MWSC cluster members to be identified. But in order to account for systematic offsets between the data sets, the procedure is iterated. With new cluster membership, the weighted mean TGAS proper motion is recomputed and the selection process is repeated. This routine is iterated, on average, once or twice until the cluster membership is unchanged. Illustrations of the result of this procedure are shown in Fig.~\ref{fig1} (for Blanco~1) and in Appendix B (for all clusters).

\begin{figure}[h]
\centering
\resizebox{\hsize}{!}{\includegraphics[width=9cm]{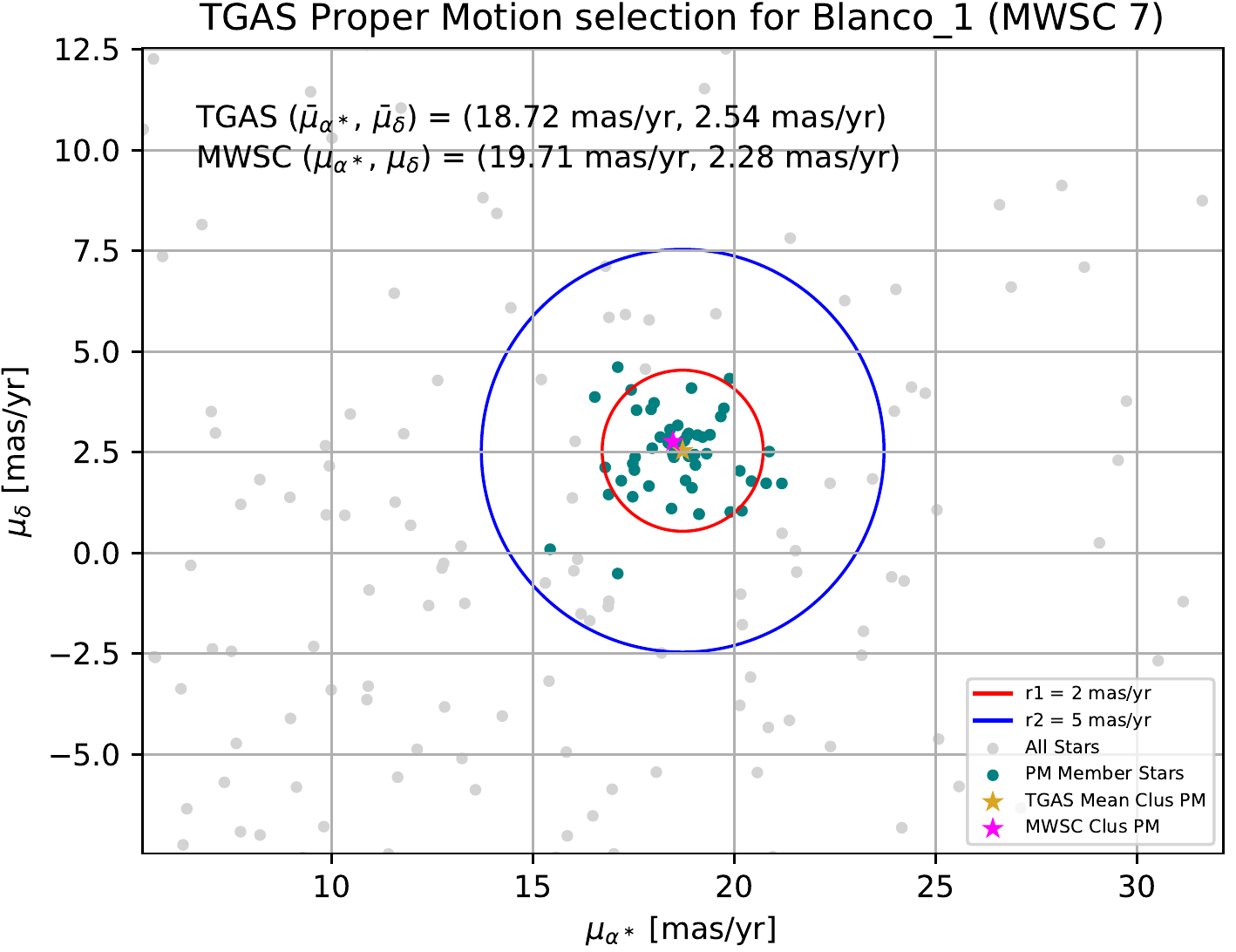}}

\caption{Result of the TGAS proper motion selection for Blanco~1. The final Blanco~1 TGAS proper motion is (18.72 \sybf{mas yr$^{-1}$}, 2.54 \sybf{mas yr$^{-1}$}). The teal points represent the proper motion members, where all stars within the 2 \sybf{mas yr$^{-1}$} radius (red circle) of the mean cluster proper motion are selected and the stars within 5 \sybf{mas yr$^{-1}$} (blue circle) are only selected if their 3$\sigma$ errors are consistent with the mean cluster proper motion.}
 \label{fig1}
\end{figure}

\begin{figure}
\centering
\resizebox{\hsize}{!}{\includegraphics[width=9cm]{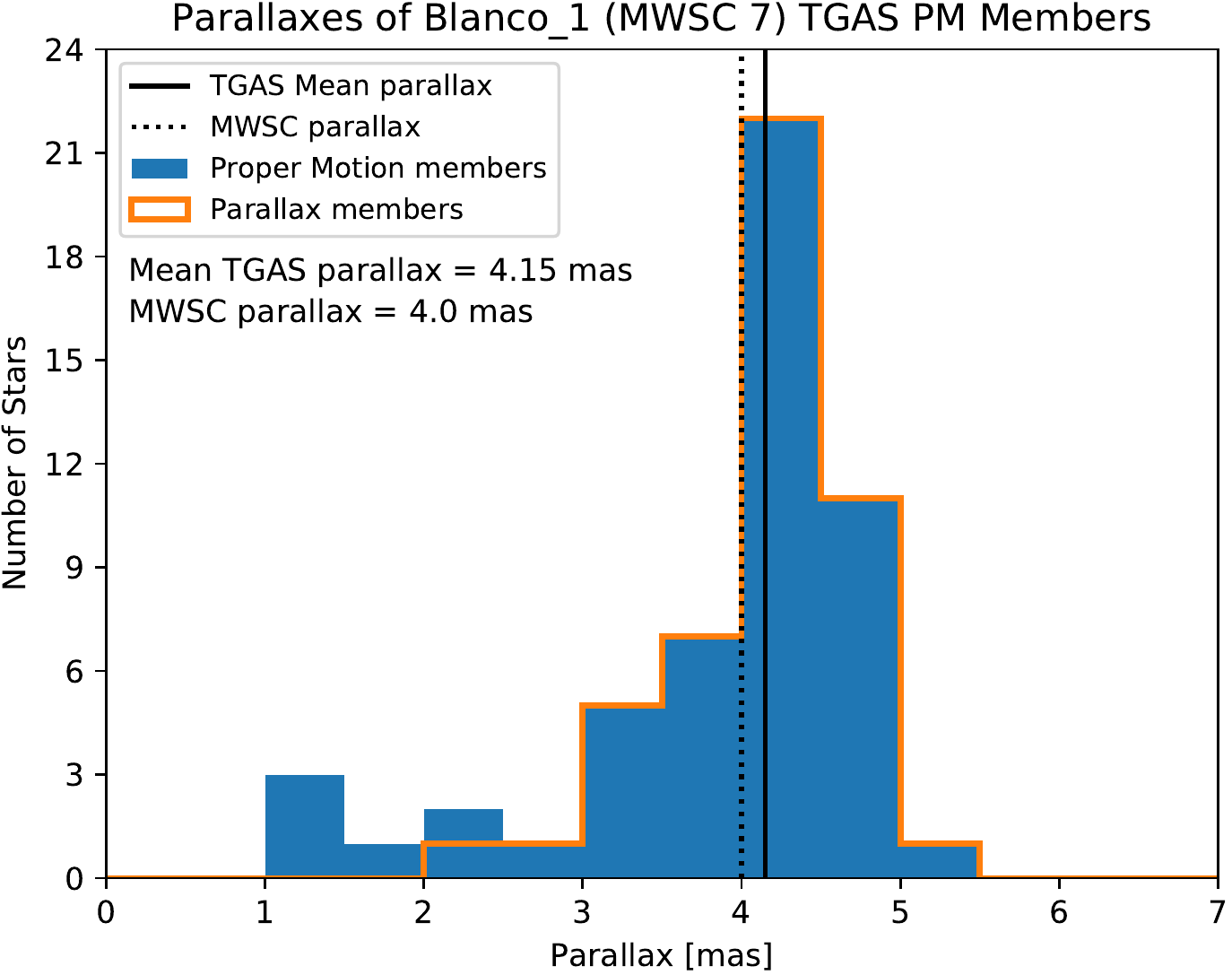}}

\caption{Result of the TGAS parallax selection for Blanco~1 proper motion-selected stars. The final weighted mean TGAS parallax for Blanco~1 is 4.15 mas. The orange outline illustrates the stars with 3$\sigma$ errors consistent with the mean parallax; these stars are the TGAS astrometrically-selected candidates of Blanco~1.}
 \label{fig2}
\end{figure}

Next, a parallax selection is performed on the TGAS proper-motion candidates to further refine cluster membership. \sybf{As an initial estimate, the} cluster parallax is first computed from the cluster's MWSC distance. Stars are considered cluster candidates if the 3$\sigma$ parallax error of the star lies within the cluster parallax. Since a systematic difference in parallaxes may exist between MWSC and TGAS, we compute the weighted mean cluster parallax, $\bar{\varpi}_{\mathrm{T}}$, for the initial parallax-selected stars. Again, the 3$\sigma$ parallax error of the stars are compared to the $\bar{\varpi}_{\mathrm{T}}$ to determine membership. The parallax selection is also iterated until cluster membership no longer changes.

In the cases where there is no clear over density in the cluster parallaxes, or \sybf{if there is a notable peak at the parallax of field stars, so} the resulting $\bar{\varpi}_{\mathrm{T}}$ differs by more than 3$\sigma$ from the MWSC parallax, the parallax iteration is canceled. Instead, stars are considered cluster candidates if their 3$\sigma$ parallax error is consistent with the MWSC parallax. \sybf{\citet{2017A&A...606L...8K} have shown that MWSC and TGAS parallaxes are compatible within 2 kpc from the Sun.} The final selected cluster parallaxes are shown in Fig.~\ref{fig2} for Blanco~1 and illustrate that some accepted TGAS proper motion candidates, from the previous selection, have been eliminated. TGAS proper motion and parallax diagrams for all clusters can be found in Appendix B. \sybf{We note that the parallaxes obtained at this stage are solely used as initial estimates. \lebf{Final} cluster distances are computed from isochrone fitting, independent of these trigonometric parallaxes.}

After defining the TGAS astrometric candidates, stars with HSOY proper motions are considered for membership. If a star in our data set has both TGAS and HSOY proper motions, their membership relies only on the more precise TGAS data. Moreover, stars that did not meet the TGAS selection criteria are no longer be considered for membership even if they may have consistent HSOY proper motions. For Gaia $G$ $\leq$ $\sim$16 mag, the mean standard error in HSOY proper motions is about 2 \sybf{mas yr$^{-1}$} for declinations greater than $-30^{\circ}$ and 3 \sybf{mas yr$^{-1}$} for declinations less than $-30^{\circ}$ \citep{2017A&A...600L...4A}. Considering these errors, the same proper motion selection cuts as used for TGAS are applied.
As the HSOY catalog is much larger than TGAS and because there are no clear cluster overdensities in HSOY proper motion space, the HSOY selection is not iterated.

Our selection considers the majority of the HSOY cluster candidates; the missing HSOY candidates are preferentially the fainter stars with $G$ $>$ 16 mag, which have standard HSOY proper motion errors larger than 3 \sybf{mas yr$^{-1}$}. At this stage we are not concerned with completeness, that is, obtaining all possible cluster members, because we do not derive the cluster mass function. We have focused on obtaining true cluster members, so the cluster parameters are accurately determined. 

Lastly, we recover the bright stars missing from the cluster, that is, with no TGAS or HSOY proper motions, by using Hipparcos proper motions.
These bright stars are crucial for accurate determination of the cluster age. The median proper motion error for stars with broad-band Hipparcos magnitudes brighter than 9 mag is about 0.88 \sybf{mas yr$^{-1}$} \citep{2007A&A...474..653V}. Taking these errors into account, the same proper motion cuts as used for TGAS are applied. The Hipparcos proper motion selection is not iterated because there are very few stars with Hipparcos proper motions.

\subsection{Photometric selection}
The second procedure in the pipeline is a photometric membership selection. Before cluster membership is further refined, any highly evolved stars, such as red giant branch (RGB) stars or supergiants, are identified and temporarily removed. The highly evolved members are temporarily removed because we perform initial fitting, to determine rough estimates for distance and reddening, with the ZAMS (see Sect. 3.3), and since these stars have evolved off the main sequence, they should not be considered in this initial fit. To select the highly evolved stars, the ZAMS is placed in the $V$ vs. $B-V$ CMD at the cluster's MWSC $E(B-V)$ and $d$ values with an offset of $\Delta V$ = $-$5 mag, which nicely divides the cluster main sequence and the upper right corner of the cluster CMD, where giant stars are located. Stars are identified as evolved if they have $V$ magnitudes brighter than the offset ZAMS. This usually removes all potentially highly evolved stars; repeating a similar procedure in the infrared CMDs \sybf{does} not yield additional stars to remove.

It is important to note that TGAS parallaxes are currently only used for membership selection and not in the pipeline for distance determination because the measurements are available for very few stars. This will certainly change in the next version of the pipeline, after the release of Gaia DR2. Nevertheless, this general isochrone fitting method, which determines cluster distance independent of using TGAS parallaxes (except for membership determination), will still be relevant to use after Gaia DR2; in particular, for the analysis of very distant clusters, which will have poor parallax measurements in Gaia DR2.

Since the HSOY catalog does not provide parallaxes for the stars, a given cluster might end up with hundreds to thousands of candidate members after the HSOY proper motion selection, many of which are field stars and obvious nonmembers. To eliminate the majority of these nonmembers, the ZAMS is placed in the $J$ vs. $G-J$ CMD using the cluster's MWSC $E(B-V)$ and $d$ values with an offset of $\Delta J$ = +1.0 mag, removing candidates with $J$ magnitudes greater than this offset and $G-J$ > 0.5 mag. Candidates with $J$ magnitudes greater than this offset and $G-J$ < 0.5 mag are kept because these are bright stars near the main sequence turn-off and are essential to proper age determination. The ZAMS is also offset by $\Delta (G-J)$ = $+$0.5 mag and candidates with $(G-J)$ magnitudes greater than this offset are removed. This process is repeated in the $K_s$ vs. $J-K_s$ CMD with a $\Delta K_s$ = +1.0 mag and $\Delta (J-K_s)$ = +0.2 mag offset of the ZAMS to remove additional obvious nonmembers, but taking care to keep the stars above the $\Delta K_s$ offset if $J-K_s$ < 0.5 mag. These thresholds were determined after varying the values and looking at the \sybf{CMDs of our 24-cluster sample}. The cuts needed to be such that clear cluster members are kept, while the majority of background stars are rejected. These photometric selections might remove some evolved cluster members, but as aforementioned, we are not concerned with completeness at this stage.

\subsection{Initial cluster parameters}\label{subsec:icp}
In order to efficiently use the Levenberg-Marquardt fitting method, initial guesses for the fitted parameters must be supplied. In order to obtain reliable initial cluster $E(B-V)$ and \sybf{distance} $d$ estimates, the ZAMS is fit to the photometric observations of the astrometrically- and photometrically-selected candidate stars. $E(B-V)$ and $d$ are inserted as parameters by adjusting the ZAMS photometry. For each member star, residuals are calculated for all available measured photometric bands and the corresponding ZAMS magnitudes for all possible masses. \sybf{The residual is defined as difference between the photometric measurement and ZAMS model for a given photometric band, from which the $\chi^2$ contribution for each star is derived.} The mass, and subsequently the ZAMS photometry, yielding the minimum $\chi^2$ contribution is then matched to the star. This method allows a direct comparison between measured and model magnitudes and ensures that the star has a mass consistent in all photometric bands.

\sybf{We also consider the presence of unresolved binaries, which contribute to a spread in the cluster main sequence toward brighter magnitudes. For two stars of equal masses, the offset is 0.75 mag, but for stars of unequal masses, this offset is less (e.g., \citet{1996AJ....112..628F} found binaries with mass ratios less than 0.5 are within 0.1 mag of the main sequence). The number and types of binaries in our clusters are unknown.} 
After testing a range of values, we decided to use a mean offset of $-$0.1 mag, \sybf{as given by \citet{1996AJ....112..628F}}, for all passbands, \sybf{ensuring that the residuals are minimized in such a way that the derived $E(B-V)$ and $d$ fit between the sequences of single and multiple stars.} In the future, we plan to determine the effect of binaries and this binary offset quantity in more detail by taking observed binary fractions and characteristics statistically into account, which will most likely lead to different offsets in the various photometric bands.

After the ZAMS fit, offsets of $\Delta J$ = +0.5 mag and $\Delta (G-J)$ = +0.25 mag are applied to the fitted ZAMS in the $J$ vs. $G-J$ CMD to further remove nonmember stars. Again, stars above the $\Delta J$ offset with $G-J$ < 0.5 mag are kept. The ZAMS is refitted and stars are removed until the cluster membership no longer changes. However, if cluster membership is below 25 stars, the ZAMS fitting process is not iterated and no additional stars are removed; otherwise the cluster would contain too few members to properly fit. The final $E(B-V)$ and $d$ values obtained from the ZAMS fit are then used as the starting values in the isochrone fitting.

\subsection{Isochrone fitting}
With highly probable cluster members determined and initial estimates for the cluster $E(B - V)$ and $d$, isochrones at a step of 0.1 dex are fitted to the cluster photometry. \sybf{Lower and upper limits for $E(B-V)$ are 0 and 0.5 mag. For $d$, the limits are $d\pm$20 pc. These limits are not too restrictive and allow the parameter space around the initial estimates to be explored, \revbf{in case} there is a bias from the ZAMS fitting.} 

\sybf{As mentioned above, a given star can have up to six bands of photometric data; the bright Hipparcos stars, not in TGAS, can have up to five passbands, from ASCC-2.5 and 2MASS, available.} The process of matching a stellar mass, and subsequently the isochrone model magnitudes, to the available photometry for a cluster member is the same as that described in Sect.~\ref{subsec:icp} for the ZAMS fitting. By fitting the photometric measurements, instead of colors and magnitudes (in the CMD), \sybf{the precision of each measurement is accounted for.}

Here, the highly evolved members of the cluster are included and down-weighted (by increasing their error bars by a factor of 10). We down-weight the highly evolved stars for two reasons:
\begin{enumerate}
\item The isochrones are not very well-determined in the late phases of stellar evolution, that is, after the main sequence turn-off.
\item Since the RGB stars are very bright, they have very small photometric errors, which greatly affects the isochrone fitting.
\end{enumerate}
Giving less weight to these stars allows them to still be considered when fitting for the cluster parameters, but prevents them from completely dominating the fitting.

The isochrone yielding the minimum reduced $\chi^2$ is selected and the stars with the largest $\chi^2$ contribution are removed. This process is repeated, starting with the ZAMS fit, until a minimum reduced $\chi^2$ $<$ 8 is achieved. In some cases where this minimum could not be achieved - likely due to many bright Hipparcos stars with very small photometric errors in $B$ and $V$ - a reduced $\chi^2$ $<$ 14 was adopted. After this reduced $\chi^2$ criteria is met, all isochrones at $\Delta \log t$ = 0.01 are then fitted to the photometry of the final cluster members, in order to fine-tune the cluster's age. The isochrone yielding the minimum reduced $\chi^2$ gives the cluster's final age, $E(B - V)$, and $d$.

This reduced $\chi^2$ threshold allows obvious nonmembers to be rejected, while keeping the evident cluster members, \sybf{and was selected after experimenting with many different values for our cluster sample.} Imposing the typical reduced $\chi^2$ = 1 would simply remove too many cluster members, \sybf{but the \lebf{best} value to use cannot be determined because of multiple unquantifiable effects, including systematics in the theoretical models, photometric errors in the Gaia $G$ band calibration, and unresolved binaries}. We do not expect a perfect reduced $\chi^2$ agreement because the isochrone models do not perfectly match the data at all stellar masses and passbands, along small mass ranges of the isochrones. This seems to be evident in the low-mass end of the isochrones in the $K_s$ vs. $J-K_s$ CMD. Additionally, the Gaia $G$ isochrones used are based on the before launch $G$ calibrations \citep{2010A&A...523A..48J}, thus small deviations are expected, and in fact, do exist, when compared to $G$ photometry \sybf{\citep{2016A&A...595A...7C}}.

\begin{table*}[h]
\caption{Derived parameters for 24 clusters}              
\label{table:results}      
\centering                                      
\begin{tabular}{lrrrrrccrrrr}
\hline\hline 
Name & MWSC & $\alpha_{\textrm{T}}$ [h:m:s] & $\bar{\mu}_{\alpha^*,\textrm{T}}$ & $\bar{\mu}_{\delta,\textrm{T}}$ & $\bar{\varpi}$$_{\textrm{T}}$ & $\log t$ [yrs] & $E(B-V)$ & $d$ & $N$$_{\textrm{T}}$ & $N$$_{\textrm{H}}$ &$N$$_{\textrm{Hip}}$\\
 & & $\delta_{\textrm{T}}$ [d:m:s] & $\sigma_{\mu\alpha^*,\textrm{T}}$ & $\sigma_{\mu\delta,\textrm{T}}$ & $\sigma_{\varpi,\textrm{T}}$ & $\sigma_{\log t}$ & $\sigma_{E(B-V)}$ & $\sigma_d$ & & & \\
& & &  \small{[mas yr$^{-1}$]} & \small{[mas yr$^{-1}$]} & \small{[mas]} & \small{[dex]} & \small{[mag]} & \small{[pc]} & & & \\
\hline
Blanco~1 & 7 & 00:04:13.47 & 18.65 & 2.63 & 4.11 & 8.16 & 0.007 & 251.6 & 48 & 237 & 5 \\
 &  & -29:55:40.26 & 0.08 & 0.08 & 0.10 & $_{-0.20}^{+0.59}$ & 0.001 & 0.4 &  &  &  \\

Platais 2 & 109 & 01:11:41.91 & 14.99 & -9.99 & 4.80 & 8.65 & 0.041 & 180.0 & 2 & 3 & 2 \\
 &  & 32:03:34.40 & 0.91 & 0.66 & 0.71 & $_{-0.27}^{+0.14}$ & 0.007 & 1.6 &  &  &  \\

$\alpha$ Per & 274 & 03:25:49.74 & 22.80 & -25.29 & 5.56 & 7.80 & 0.109 & 167.7 & 84 & 79 & 7 \\
 &  & 49:08:21.03 & 0.05 & 0.04 & 0.06 & $_{-0.25}^{+0.05}$ & 0.001 & 0.3 &  &  &  \\

Alessi 13 & 278 & 03:24:19.92 & 37.08 & -4.30 & 9.62 & 8.75 & 0.027 & 97.8 & 9 & 9 & 0 \\
 &  & -35:49:26.86 & 0.11 & 0.12 & 0.16 & $_{-0.05}^{+0.05}$ & 0.004 & 0.7 &  &  &  \\

Pleiades & 305 & 03:46:16.73 & 19.92 & -45.20 & 7.38 & 8.15 & 0.010 & 126.3 & 91 & 280 & 11 \\
 &  & 24:13:27.06 & 0.05 & 0.04 & 0.06 & $_{-0.15}^{+0.08}$ & 0.001 & 0.2 &  &  &  \\

Platais 3 & 395 & 04:39:37.44 & 3.83 & -20.37 & 5.22 & 8.92 & 0.000 & 176.4 & 14 & 27 & 0 \\
 &  & 71:16:03.03 & 0.15 & 0.17 & 0.20 & $_{-0.50}^{+0.18}$ & 0.006 & 0.6 &  &  &  \\

Platais 4 & 467 & 05:06:55.86 & 1.99 & -7.21 & 2.90 & 8.31 & 0.198 & 296.9 & 7 & 123 & 0 \\
 &  & 22:36:15.54 & 0.21 & 0.20 & 0.28 & $_{-0.52}^{+0.09}$ & 0.003 & 0.7 &  &  &  \\

Collinder 65 & 540 & 05:26:34.75 & -0.17 & -5.37 & 2.77 & 8.02 & 0.031 & 375.8 & 18 & 1222 & 2 \\
 &  & 15:43:17.21 & 0.18 & 0.18 & 0.24 & $_{-0.23}^{+0.44}$ & 0.001 & 0.3 &  &  &  \\

NGC 2232 & 871 & 06:27:50.50 & -4.62 & -1.80 & 3.56 & 8.02 & 0.000 & 356.6 & 8 & 218 & 4 \\
 &  & -04:47:30.02 & 0.19 & 0.19 & 0.24 & $_{-0.28}^{+0.13}$ & 0.034 & 0.7 &  &  &  \\

Alessi 3 & 1157 & 07:16:08.68 & -9.74 & 12.13 & 3.81 & 8.90 & 0.035 & 261.5 & 20 & 14 & 0 \\
 &  & -46:33:31.12 & 0.14 & 0.12 & 0.15 & $_{-0.10}^{+0.09}$ & 0.004 & 1.1 &  &  &  \\

NGC 2451A & 1308 & 07:42:32.56 & -21.21 & 15.42 & 5.27 & 8.17 & 0.014 & 196.6 & 24 & 77 & 7 \\
 &  & -38:19:44.51 & 0.13 & 0.13 & 0.15 & $_{-0.31}^{+0.22}$ & 0.002 & 0.6 &  &  &  \\

Praesepe & 1527 & 08:39:54.62 & -36.03 & -12.86 & 5.39 & 8.90 & 0.010 & 183.0 & 56 & 319 & 2 \\
 &  & 19:36:05.43 & 0.07 & 0.06 & 0.08 & $_{-0.18}^{+0.12}$ & 0.001 & 0.2 &  &  &  \\

IC 2391 & 1529 & 08:40:28.73 & -24.51 & 23.28 & 6.74 & 7.91 & 0.057 & 158.5 & 24 & 18 & 4 \\
 &  & -53:10:02.84 & 0.10 & 0.09 & 0.12 & $_{-0.43}^{+0.39}$ & 0.003 & 0.8 &  &  &  \\

Platais 8 & 1629 & 09:06:44.79 & -15.83 & 14.73 & 7.45 & 7.90 & 0.024 & 143.3 & 12 & 25 & 1 \\
 &  & -58:59:11.66 & 0.12 & 0.12 & 0.16 & $_{-0.09}^{+0.26}$ & 0.003 & 0.7 &  &  &  \\

Platais 9 & 1639 & 09:10:33.81 & -24.62 & 12.91 & 5.89 & 8.09 & 0.005 & 190.7 & 9 & 63 & 4 \\
 &  & -43:53:06.96 & 0.21 & 0.20 & 0.24 & $_{-0.19}^{+0.45}$ & 0.003 & 0.8 &  &  &  \\

IC 2602 & 1841 & 10:42:28.05 & -17.63 & 10.57 & 6.79 & 8.00 & 0.004 & 149.0 & 32 & 99 & 8 \\
 &  & -64:14:37.49 & 0.08 & 0.07 & 0.10 & $_{-0.26}^{+0.05}$ & 0.002 & 0.4 &  &  &  \\

Coma Ber & 2020 & 12:24:23.72 & -12.22 & -9.01 & 11.55 & 8.75 & 0.053 & 85.6 & 33 & 10 & 7 \\
 &  & 25:57:23.45 & 0.07 & 0.07 & 0.09 & $_{-0.15}^{+0.18}$ & 0.001 & 0.1 &  &  &  \\

Platais 10 & 2150 & 13:41:50.68 & -30.51 & -10.52 & 4.02 & 8.29 & 0.093 & 231.0 & 8 & 43 & 0 \\
 &  & -59:07:48.14 & 0.14 & 0.14 & 0.19 & $_{-0.18}^{+0.12}$ & 0.003 & 0.9 &  &  &  \\

Alessi 9 & 2670 & 17:44:59.78 & 9.71 & -8.81 & 4.94 & 8.42 & 0.091 & 224.9 & 11 & 68 & 1 \\
 &  & -47:02:35.25 & 0.20 & 0.21 & 0.26 & $_{-0.32}^{+0.13}$ & 0.002 & 0.6 &  &  &  \\

Collinder 350 & 2700 & 17:48:14.26 & -5.28 & -0.13 & 2.69 & 9.00 & 0.167 & 298.3 & 10 & 165 & 0 \\
 &  & 01:20:25.42 & 0.19 & 0.19 & 0.24 & $_{-0.22}^{+0.14}$ & 0.003 & 0.7 &  &  &  \\

NGC 6475 & 2739 & 17:53:29.50 & 3.15 & -5.51 & 3.45 & 8.29 & 0.156 & 300.8 & 49 & 1428 & 8 \\
 &  & -34:39:22.33 & 0.09 & 0.09 & 0.11 & $_{-0.31}^{+0.18}$ & 0.001 & 0.3 &  &  &  \\

Ruprecht 147 & 3078 & 19:16:18.33 & -1.04 & -26.92 & 3.53 & 8.86 & 0.059 & 265.1 & 18 & 43 & 1 \\
 &  & -16:17:23.19 & 0.18 & 0.19 & 0.23 & $_{-0.65}^{+0.12}$ & 0.003 & 0.9 &  &  &  \\

NGC 7092 & 3521 & 21:32:05.91 & -7.54 & -20.13 & 3.33 & 8.70 & 0.010 & 290.6 & 26 & 24 & 0 \\
 &  & 48:27:02.73 & 0.10 & 0.10 & 0.12 & $_{-0.23}^{+0.06}$ & 0.002 & 0.9 &  &  &  \\

ASCC 123 & 3654 & 22:41:36.00 & 11.67 & -1.09 & 4.38 & 8.10 & 0.097 & 243.5 & 12 & 37 & 4 \\
 &  & 54:09:56.43 & 0.15 & 0.13 & 0.17 & $_{-0.39}^{+0.40}$ & 0.003 & 0.9 &  &  &  \\
\hline
\end{tabular}
\end{table*}

\begin{figure*}[h]
\centering
\includegraphics[width=18cm]{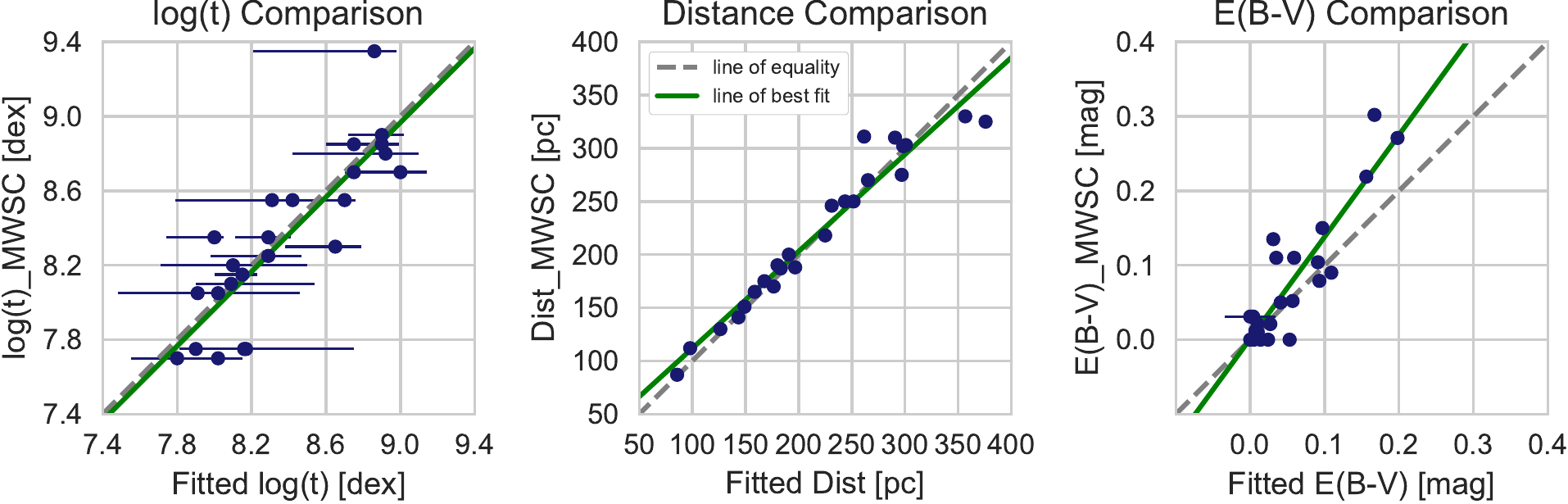}
\caption{Comparison of 24 fitted cluster parameter results from this work to those determined in MWSC. \sybf{Horizontal e}rror bars, from this work, are plotted for \sybf{all three comparisons, but for the fitted distance and fitted $E(B-V)$, the errors are} generally smaller than the point marker used. The dashed gray line indicates the line of equality and the green line shows the best fit. \revbf{Except for the $E(B-V)$}, the agreement between MWSC and newly fitted pipeline values is rather good.}
 \label{figCompResults}
\end{figure*}

\begin{figure*}[h]
\centering
\resizebox{\hsize}{!}{\includegraphics[width=9cm]{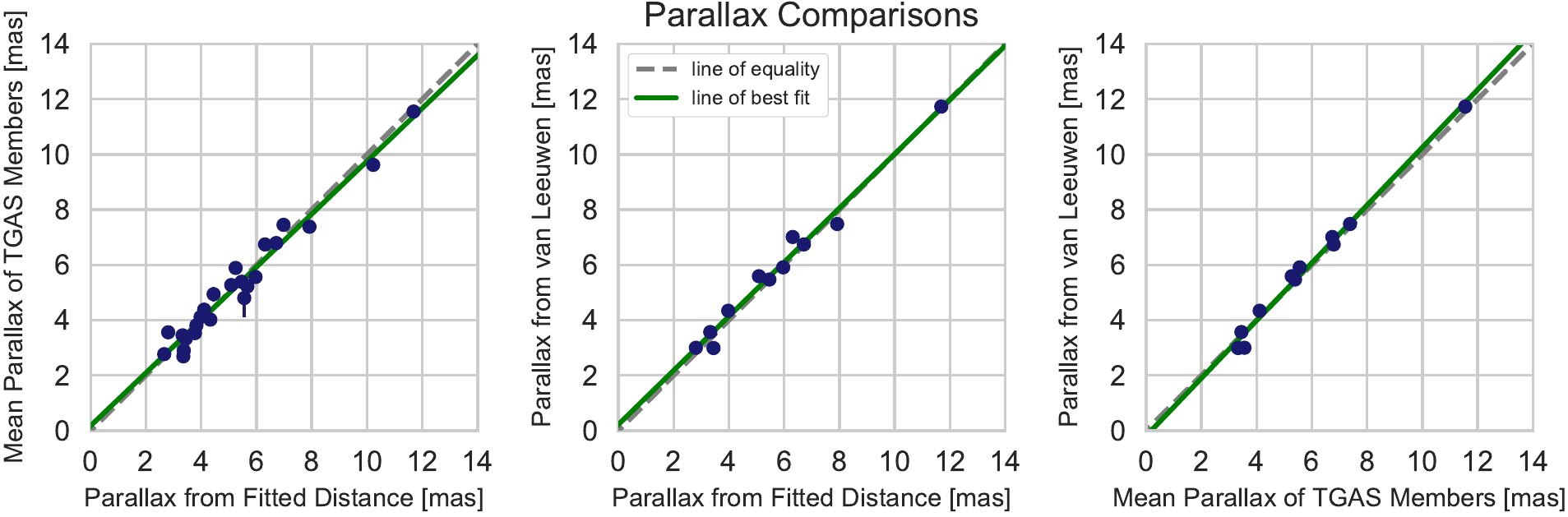}}
\caption{Left panel: Comparison of the weighted mean parallax of TGAS cluster members to parallaxes computed from the fitted distances for 24 clusters.
The cluster with the largest error bar is Platais~2. Middle panel: Comparison of 11 cluster parallaxes from \citet{2017A&A...601A..19G} to the parallaxes computed from the fitted distances. Right panel: Comparison of cluster parallaxes from \citet{2017A&A...601A..19G} to the weighted mean parallax of TGAS cluster members for 11 clusters. In all plots, the error bars are generally smaller than the marker used. The dashed gray line indicates the line of equality and the green line \lebf{represents} the best fit.}
\label{figCompPlxs}
\end{figure*}

\section{Results}\label{results}
We analyzed 24 nearby open clusters with our automated pipeline. The cluster parameter results, $\log$ $t$ [years], \sybf{$\sigma_{\log t}$,} $E(B-V)$, $\sigma_{E(B-V)}$, $d$, and $\sigma_{d}$, returned from the pipeline are listed for each cluster in Table~\ref{table:results}. \sybf{The errors in $E(B-V)$ and $d$ are the formal 1$\sigma$ errors from the $\chi^2$ fit \citep{newville_2014_11813}, which do not fully capture the real errors in these values. The real errors, including the systematics described in Sect. 3.4, are most likely larger.
The error in $\log t$ is estimated from the plateau in the reduced $\chi^2$ distribution of all possible ages at the cluster's $E(B-V)$ and $d$. The large range of errors for $\log t$ shows that} the age is difficult to constrain. The smallest error in age is expected for clusters which harbor turn-off stars, but if turn-off stars are not present in any given cluster, then the age will be relatively uncertain.
Cluster CMDs showing the final cluster membership and fitted isochrone are given in Appendix~\ref{sec.AppCMDs}. Three CMDs are provided for each cluster: $V$ vs. $B-V$, $J$ vs. $G-J$, and $K_s$ vs. $J-K_s$. \sybf{Tables with cluster membership and relevant stellar data (positions, proper motions, parallaxes, and photometry) are also given in Appendix C.
The tables also include stellar masses for each star, as determined by our isochrone fit. This data is only available electronically at the CDS online archive\footnote{via anonymous ftp to cdsarc.u-strasbg.fr (130.79.128.5) or via http://cdsweb.u-strasbg.fr/cgi-bin/qcat?J/A+A/}.}

The contents of Table~\ref{table:results} also include mean cluster TGAS positions in RA and DEC, $\alpha_{\textrm{T}}$ and $\delta_{\textrm{T}}$ respectively, weighted mean cluster TGAS proper motions in RA and DEC, $\bar{\mu}_{\alpha^*,\textrm{T}}$ and $\bar{\mu}_{\delta,\textrm{T}}$ respectively, and parallaxes, $\bar{\varpi}$$_{\textrm{T}}$, from the cluster's final TGAS members. The errors in cluster proper motion, $\sigma_{\mu\alpha^*}$,$_{\textrm{T}}$ and $\sigma_{\mu\delta,\textrm{T}}$, and parallax, $\sigma_{\varpi,\textrm{T}}$, provided are the formal errors of the weighted mean.
The systematic error in parallax from Gaia is 0.3 mas \citep{2016A&A...595A...4L}. This systematic error does not average out and still needs to be considered in addition to the formal errors mentioned. The proper motion and parallax values quoted in the table may differ from those in the figures of Appendix B, in which the values are given for membership before photometric selection.
Lastly, the final numbers of TGAS, HSOY, and Hipparcos stars, $N$$_{\textrm{T}}$, $N$$_{\textrm{H}}$, and $N$$_{\textrm{Hip}}$  determined to be cluster members are provided in the last three columns of Table~\ref{table:results}.

\subsection{Comparison with MWSC}
\sybf{A comparison between our derived parameters to the MWSC values is shown in Fig.~\ref{figCompResults}.
As expected, there is a wide range in the determined ages. The median age difference is 0.13 dex and the dispersion in age is 0.22 dex. The ages of 15 clusters are within 1$\sigma$ and 5 clusters are within 2$\sigma$. The clusters with the largest discrepancies in age are Ruprecht~147, NGC~2451A, Blanco~1, and Platais~2 with differences of 0.49, 0.42, 0.41, and 0.35 dex respectively.}

\sybf{In the case of Platais~2, the difference could very well be due to the fact that the cluster does not really exist. There is no over density in the TGAS proper motion diagram and none of the few proper motion-selected members have a parallax consistent with the MWSC value. Furthermore, the small number of proper motion-selected members do not exhibit a very similar parallax. Another cluster in our sample, whose existence we also doubt, is Platais~4. The TGAS proper motion diagram of Platais~4 also does not show a strong over density and the peak of its TGAS parallax distribution is at 1 mas, which is mostly from many background \lebf{and/or} field stars, as the MWSC parallax is at 3.6 mas. Only four of its proper motion-selected members have TGAS parallaxes around 3.6 mas. If strong over densities are not observed in the TGAS proper motion and parallax domains, it is a possibility that the cluster is not real. The existence of these clusters will be clarified by the stellar parallaxes from Gaia DR2.}

For Ruprecht~147, the age discrepancy clearly results from the addition of three early-type stars. These stars have TGAS proper motions and parallaxes consistent with the corresponding mean cluster values and thus, were considered cluster members by our pipeline. 
But for the age determined in the MWSC, these stars were probably considered blue stragglers and excluded from the isochrone fitting.
\sybf{Manually rejecting these stars from our fitting procedure yields $\log t =$ 9.8 dex, which is marginally more consistent with the MWSC age.}

\sybf{For the distance parameter, we find that our derived results are consistent with those listed in MWSC, as the middle panel of Fig.~\ref{figCompResults} shows most clusters accumulating along the line of equality. The median difference in distance is 6.7 pc, with a dispersion of 17.7 pc. The dispersion in distances is much larger than the median formal distance errors. As mentioned above, the formal distance errors are smaller than expected because they do not include systematic errors. The clusters with the largest differences in distance, of 50.8, 49.5, and 26.6 pc, are Collinder~65, Alessi~3, and NGC~2232 respectively. Converting our distances to distance moduli, ($m_{K_s} - M_{K_s}$), we investigated whether a correlation exists between the difference in distance moduli and the difference in $E(B-V)$ of our determination and the MWSC. No correlation between the two was found.}

\sybf{Comparison of the derived and MWSC $E(B-V)$ values in the right panel of Fig.~\ref{figCompResults} illustrates that the derived values from this work are generally smaller than those found by the MWSC. The median difference and dispersion in $E(B-V)$ are 0.017 mag and 0.04 mag respectively. Due to the small formal errors derived, only 8 clusters are compatible within 3$\sigma$. The clusters with the largest discrepancies in $E(B-V)$ are Collinder~350, Collinder~65, and Alessi~3 with differences of 0.135, 0.101, and 0.075 mag respectively. Differences in $E(B-V)$ result in only small shifts of the isochrone, as seen from the two isochrones plotted in the cluster CMDs in Appendix A.}

Overall, we find good agreement between MWSC values and our fitted cluster parameters, \sybf{as the median deviations in $\log t$, $d$, and $E(B-V)$ are roughly 0.24\%, 2\%, and 18\%.} A linear best fit to the observed differences is almost indistinguishable from the line of equality (Fig. ~\ref{figCompResults}), especially for the distances.

\subsection{Parallax comparison}
Some discrepancies do exist between our mean parallaxes and those computed from our fitted distances, as illustrated in the left plot of Fig. ~\ref{figCompPlxs}. The median difference in parallaxes is \sybf{0.36 mas, with a dispersion of 0.42 mas.} The largest discrepancies, with differences of \sybf{0.76 mas, are Platais~2 and NGC~2232}. Others include \sybf{Collinder~350 and Platais~9, with parallax differences of 0.66 mas and 0.65 mas} respectively. For most of these clusters, there is no clear peak in the TGAS parallax distributions at the MWSC parallax. Currently, TGAS parallaxes are not used to aid in the distance determination from isochrone fitting, so deviations are expected between the two values. \sybf{We also checked to see if the discrepancy in parallaxes could have resulted from our assumption of solar metallicity for all clusters. For the 14 clusters with metallicity values, a plot of the differences between trigonometric parallaxes and photometric parallaxes as a function of metallicity did not reveal a strong correlation. Thus, the assumption of solar metallicity has little or no effect on the distances determined. Overall, the mean trigonometric TGAS parallaxes and parallaxes from the fitted distance are compatible. Comparison of these parallaxes show 21\% agree within 1$\sigma$, 63\% agree within 2$\sigma$, and 80\% agree within 3$\sigma$.}

Our result for the Pleiades is also somewhat inconsistent. Computing the weighted mean parallax from the final TGAS members yields a parallax of \sybf{7.38$\pm$0.06} mas, which is consistent with a parallax of 7.48$\pm$0.03 mas \sybf{\citep{2017A&A...601A..19G}}. However, the parallax computed from our fitted distance is rather high, at about \sybf{$7.92\substack{ +0.01 \\ -0.02}$}~mas. The difference in parallaxes could be due to the small binary offset we used. \sybf{\citet{2008ApJ...678..431C} found an usually large fraction of binaries in the Pleiades, which implies a larger binary offset would be required for our analysis of the Pleiades.} 
As aforementioned, the effect of binaries in clusters will be determined more statistically in the future, and will also enforce consistency between trigonometric parallax and fitted photometric distance once Gaia DR2 data is available.

The parallaxes for 11 of these 24 clusters were also recently investigated by \sybf{\citet{2017A&A...601A..19G}}. The clusters in common are \sybf{listed in Table~\ref{table:compVL}, as well as the number of TGAS members found by each study, denoted by $N_{\textrm{Y18}}$ (this study) and $N_{\textrm{vL17}}$, and the number of TGAS members in common, $N_{\textrm{com}}$. For nearly all clusters, there is very good overlap in TGAS membership. On average, \citet{2017A&A...601A..19G} has found more members per cluster. \revbf{This is mainly due to the significantly larger sky areas they considered for the clusters.
Additionally, their membership determination relied solely on astrometry, while our analysis further refines TGAS membership based on stellar photometry.}

The parallaxes of these clusters are compared to the parallaxes derived from the fitted distance in the middle plot of Fig. \ref{figCompPlxs}. The median difference in parallaxes is 0.25 mas and the dispersion is 0.34 mas. \sybf{Due to the small formal errors on the isochrone-fitted distances, only four clusters are compatible within 3$\sigma$.}

Since the analysis of \citet{2017A&A...601A..19G} focuses on TGAS astrometric data, we also compare those parallaxes with the weighted mean parallax of our TGAS cluster members. In this case, the median parallax difference is 0.23 mas with a dispersion of 0.27 mas. Overall, the TGAS parallaxes of these 11 clusters are in \sybf{great agreement; 18\% agree within 1$\sigma$, 82\% within 2$\sigma$, and 91\% within 3$\sigma$.}

\begin{table}[h]
\caption{Number of TGAS stars in common with \citealt{2017A&A...601A..19G} for 11 clusters}              
\label{table:compVL}      
\centering                                      
\begin{tabular}{lrrr}
\hline\hline 
Name & $N_{\textrm{Y18}}$ & $N_{\textrm{vL17}}$ & $N_{\textrm{com}}$ \\
\hline
Blanco~1 & 48 & 44 & 36 \\
$\alpha$ Per & 84 & 116 & 66 \\
Pleiades & 91 & 154 & 85 \\
NGC 2232 & 8 & 31 & 4 \\
NGC 2451A & 24 & 37 & 19 \\
Praesepe & 56 & 79 & 46 \\
IC 2391 & 24 & 43 & 21 \\
IC 2602 & 32 & 66 & 32 \\
Coma Ber & 33 & 49 & 30 \\
NGC 6475 & 49 & 78 & 36 \\
NGC 7092 & 26 & 23 & 14 \\
\hline
\end{tabular}
\end{table}

\section{Summary and conclusions}
In the Gaia era, it is paramount to develop new analysis techniques and to combine archived data with newly available data, in order to gain new insights about open clusters and subsequently, the Galactic disk. As aforementioned, efforts have already been made by several groups, including \citet{2010A&A...516A...2M}, \citet{2012A&A...539A.125D}, and \citet{2014A&A...564A..49P}. We have also taken a step toward this effort and developed an automatic isochrone-fitting procedure that determines both cluster membership and cluster parameters. Using precise proper motions and parallaxes predominately from TGAS and HSOY \sybf{for initial membership determination} and six-band photometry \sybf{for parameter determination and membership refinement}, we have returned cluster parameters and cluster parallaxes that are similar to those found by \citet{2013A&A...558A..53K} and \citet{2017A&A...601A..19G} respectively, for the 24 closest open clusters in MWSC.

Our pipeline was developed specifically to work with the quality and limitations of the data set used and \sybf{currently} only applies to nearby clusters. We are continuing development of the pipeline to make it more generalized and applicable to more distant clusters. We are also exploring the use of a maximum likelihood method, instead of least squares for our isochrone fitting procedure. We will continue to refine the astrometric and photometric selection criteria and adapt it for use with Gaia DR2. \sybf{A major revision will be to couple the trigonometric parallaxes and parallaxes derived from photometric distances, as the precision of Gaia parallaxes will provide a stronger constraint on cluster distance than any other method. Nevertheless, our current technique, which determines cluster distance independent of TGAS parallaxes, will still be valuable for the analysis of very distant clusters, which will have poor parallaxes in Gaia DR2.}

\sybf{On the bright side, more precise proper motions and parallaxes from Gaia DR2 will allow for more straightforward membership selection.} The present challenge involves developing a single algorithm, \sybf{to obtain accurate estimates of cluster parameters, that can be applied to} all clusters, near and far. A tool such as this is necessary to assemble a large, homogeneous catalog of open clusters and their parameters, \sybf{which can then be used to investigate the structure, dynamics, and evolution of the Milky Way.}

\begin{acknowledgements}
\noindent This work was supported by Sonderforschungsbereich SFB 881 ``The Milky Way System" (subproject B5) of the German Research Foundation (DFG) and the Russian Federation of Basic Research project number 16- 52-12027. We thank the anonymous referee for her/his extensive comments, which have greatly improved this paper. This work has made use of data from the European Space Agency (ESA) mission {\it Gaia} (\url{https://www.cosmos.esa.int/gaia}), processed by the {\it Gaia} Data Processing and Analysis Consortium (DPAC, \url{https://www.cosmos.esa.int/web/gaia/dpac/consortium}). Funding
for the DPAC has been provided by national institutions, in particular the institutions participating in the {\it Gaia} Multilateral Agreement.
This publication has also made use of data products from the Two Micron All Sky Survey, which is a joint project of the University of Massachusetts and the Infrared Processing and Analysis Center/California Institute of Technology, funded by the National Aeronautics and Space Administration and the National Science Foundation. Furthermore, this research has made use of the VizieR catalog access tool, CDS, Strasbourg, France. The original description of the VizieR service was published in A\&AS 143, 23.
\end{acknowledgements}

\bibliographystyle{aa.bst} 
\bibliography{SXyen_et_al_2018.bib} 

\begin{appendix}
\counterwithin{figure}{section}

\section{Cluster color-magnitude diagrams}\label{sec.AppCMDs}
\begin{figure*}[!h]
\centering
\includegraphics[width=6cm]{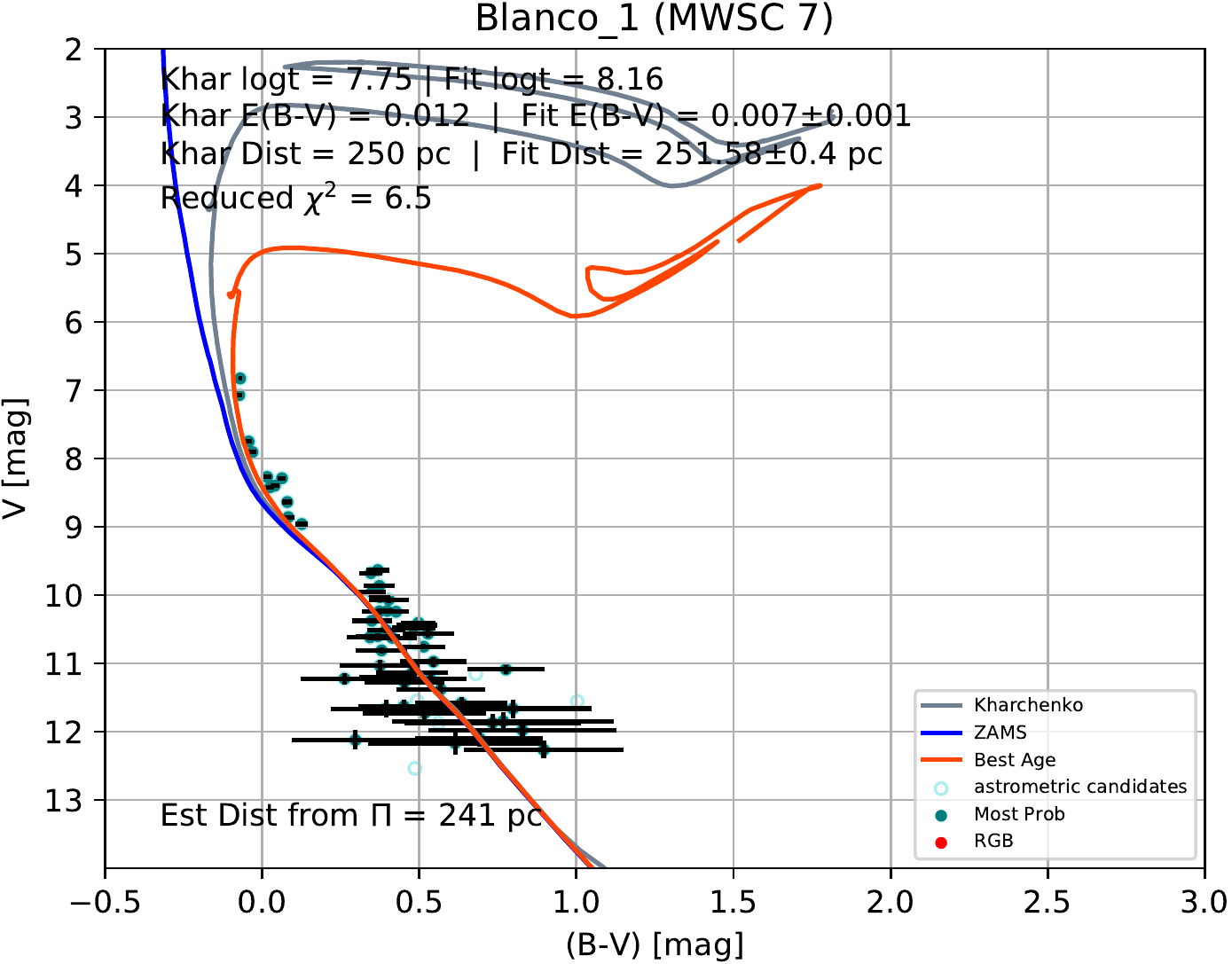}
\includegraphics[width=6cm]{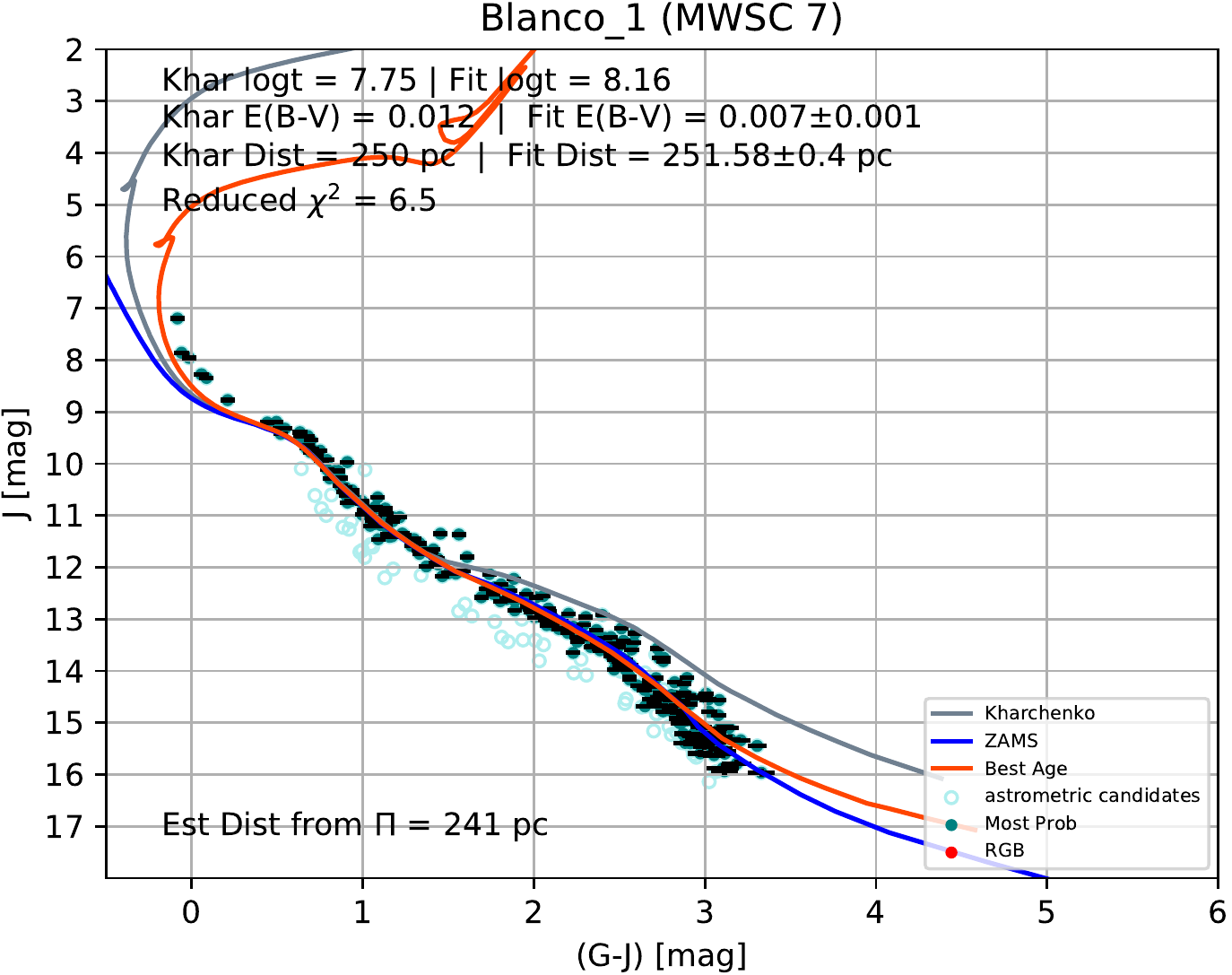}
\includegraphics[width=6cm]{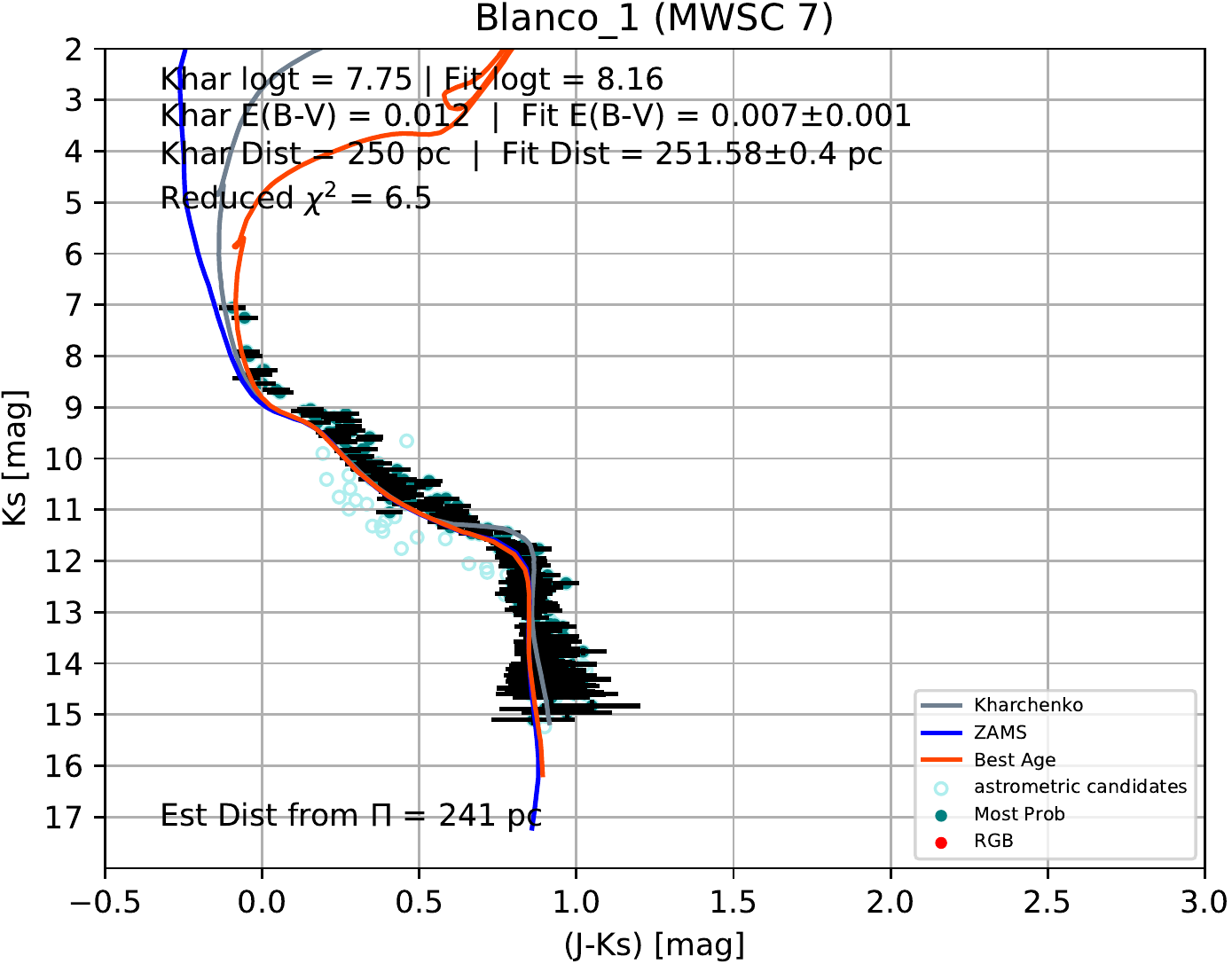}\\
\includegraphics[width=6cm]{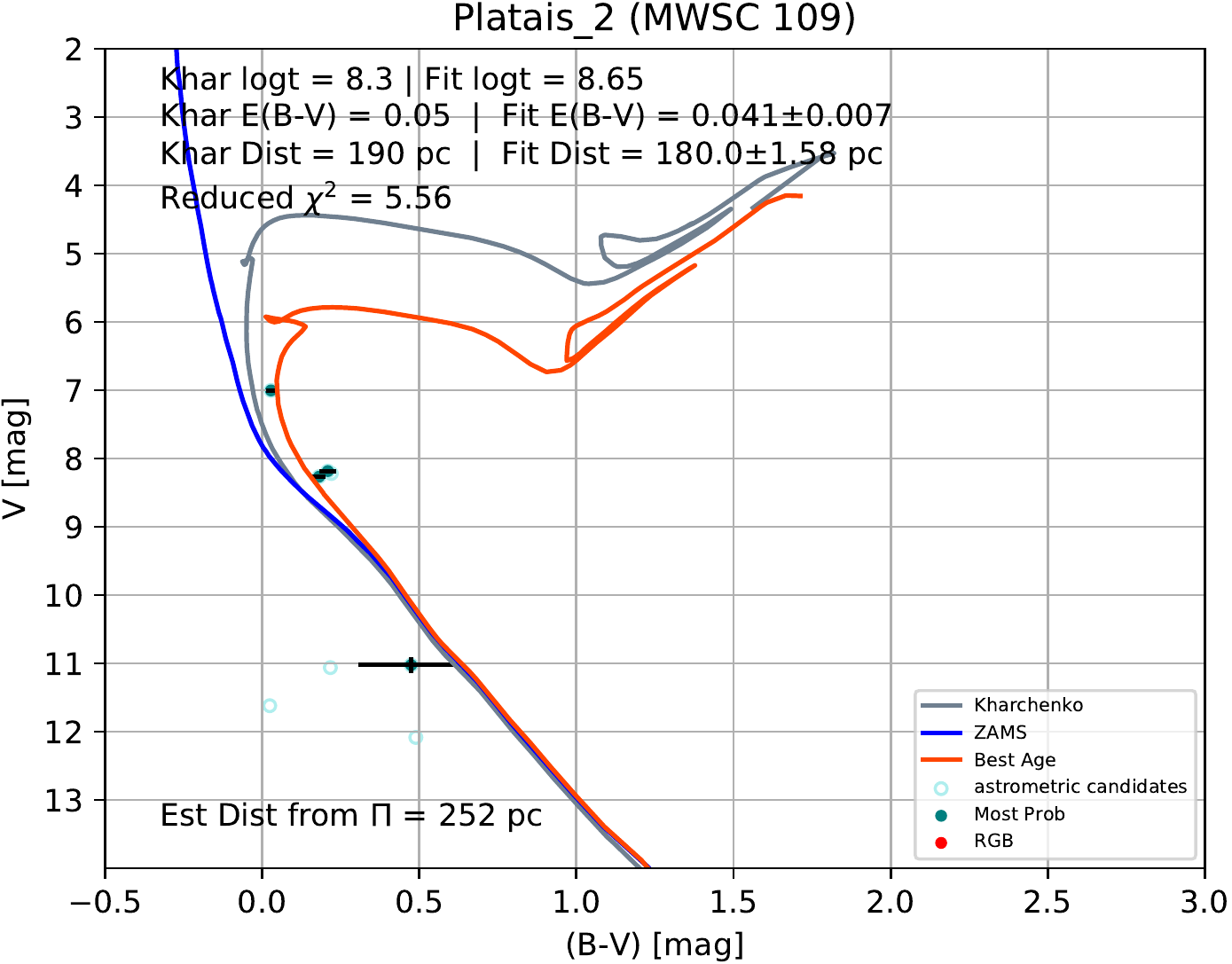}
\includegraphics[width=6cm]{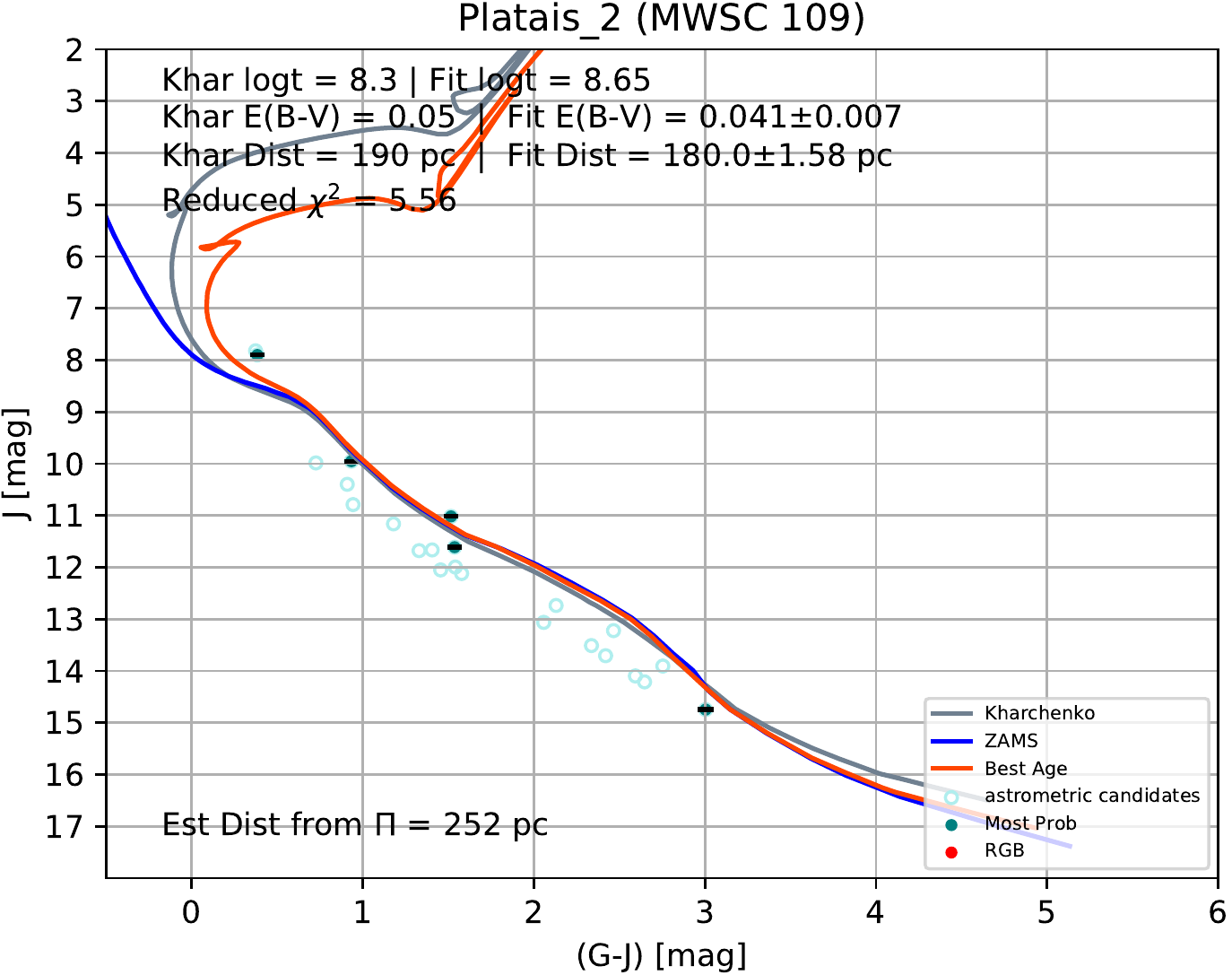}
\includegraphics[width=6cm]{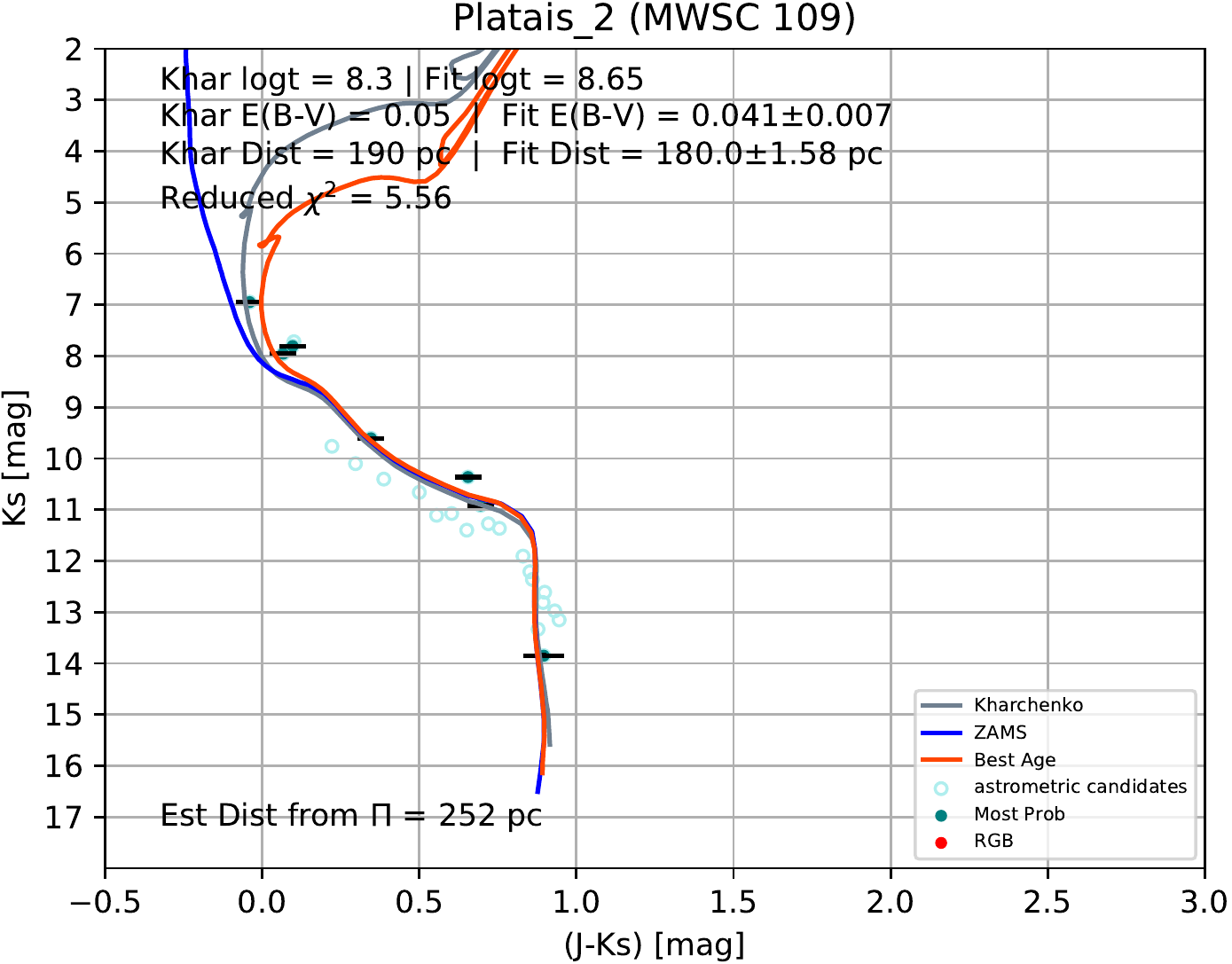}\\
\includegraphics[width=6cm]{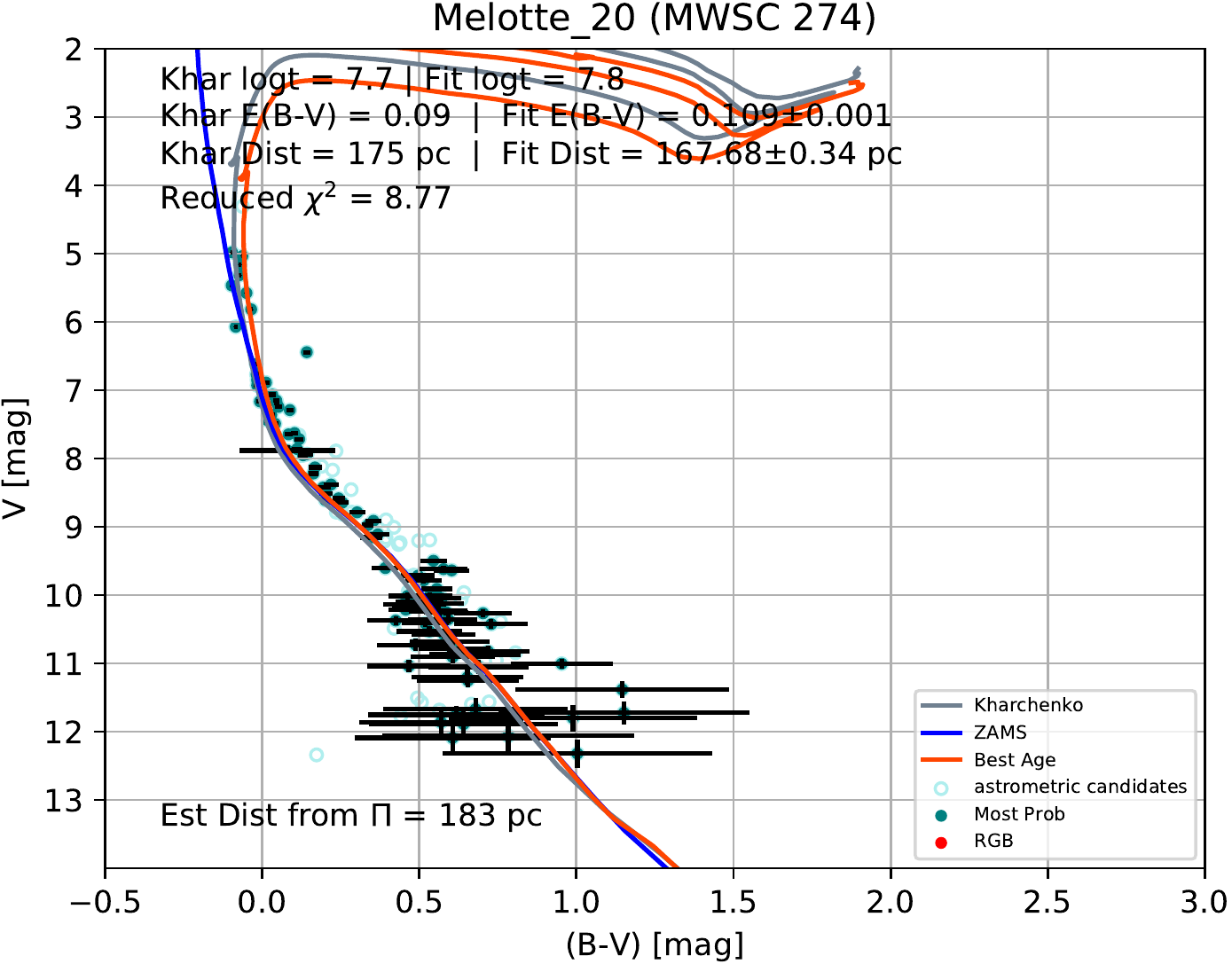}
\includegraphics[width=6cm]{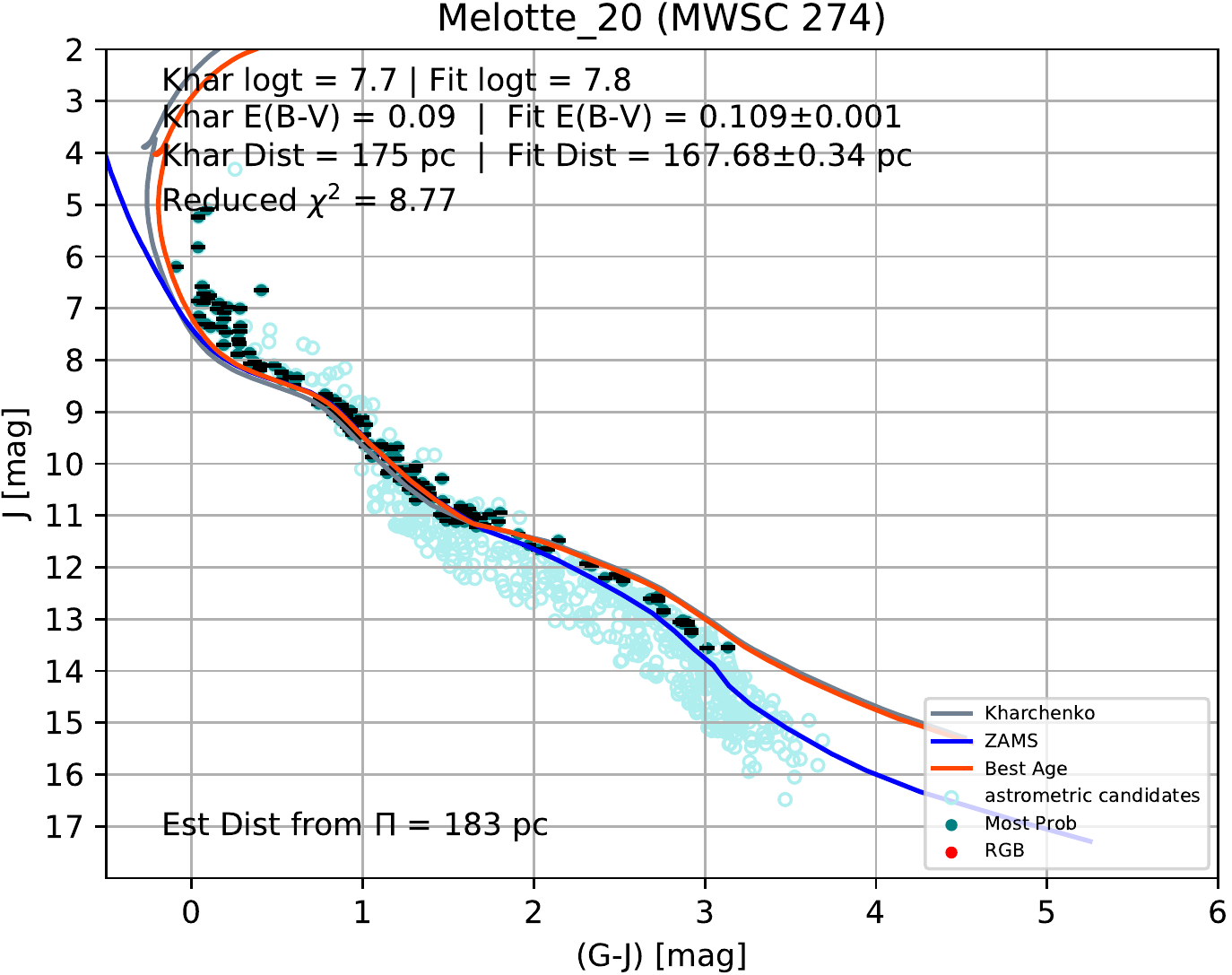}
\includegraphics[width=6cm]{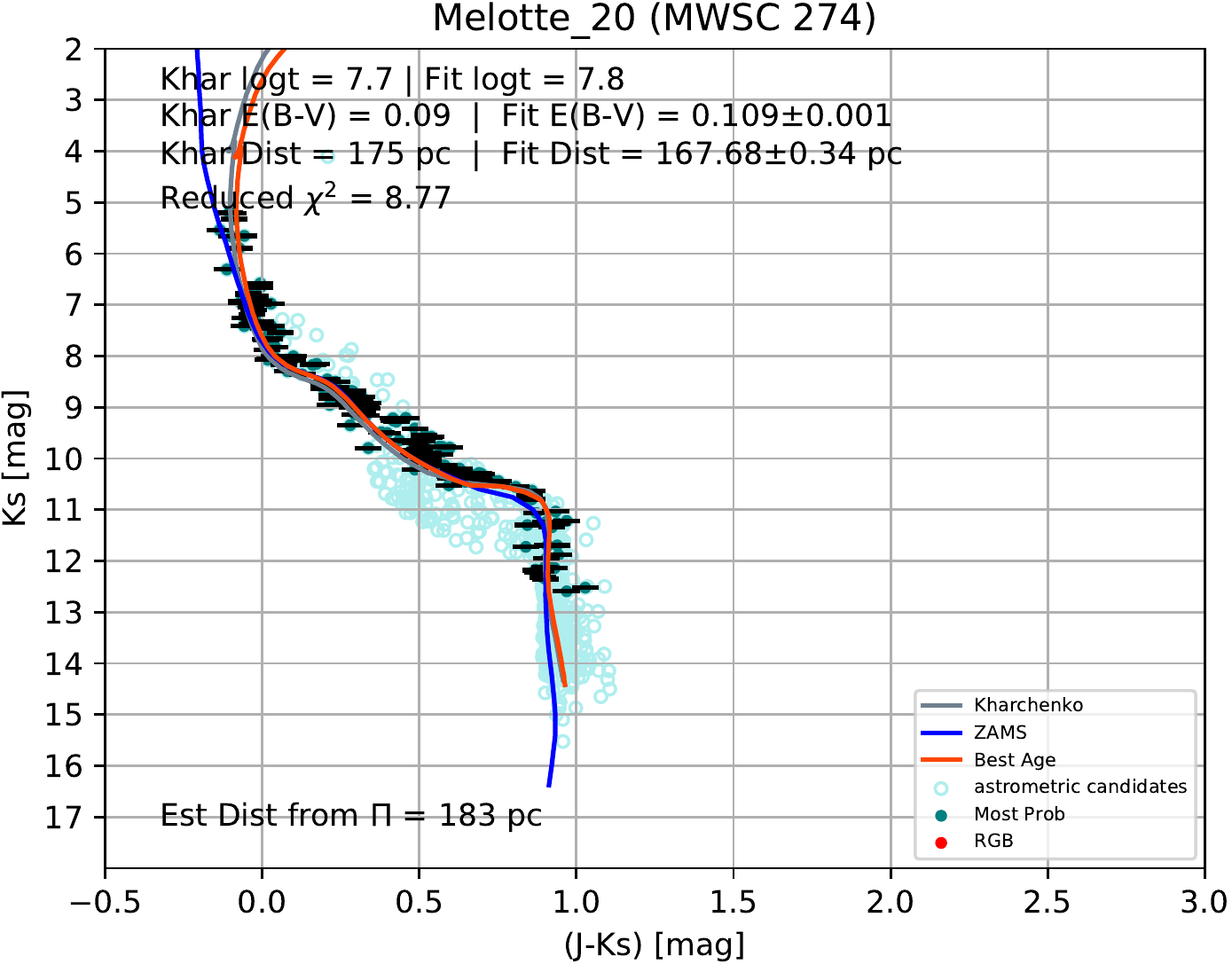}\\
\includegraphics[width=6cm]{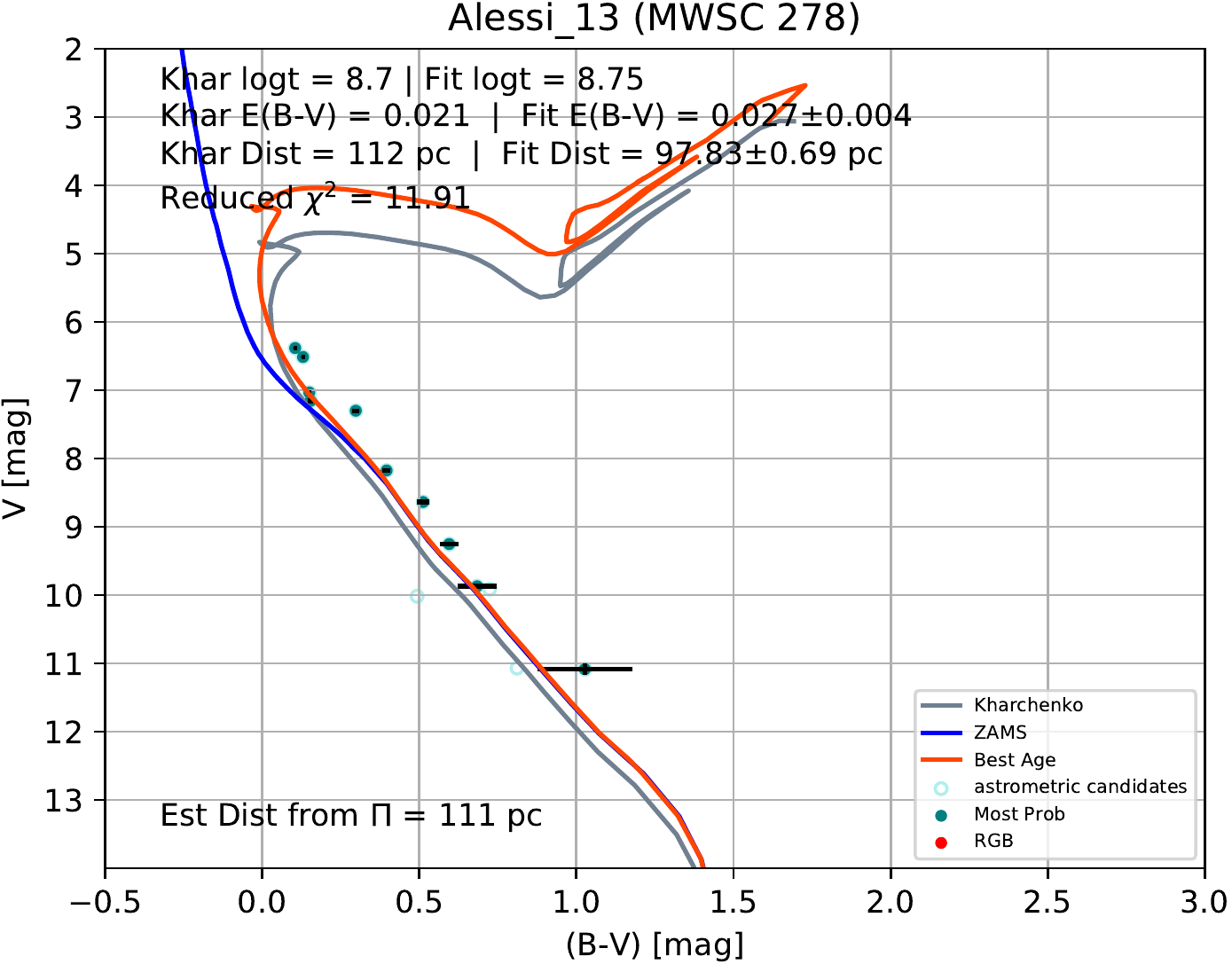}
\includegraphics[width=6cm]{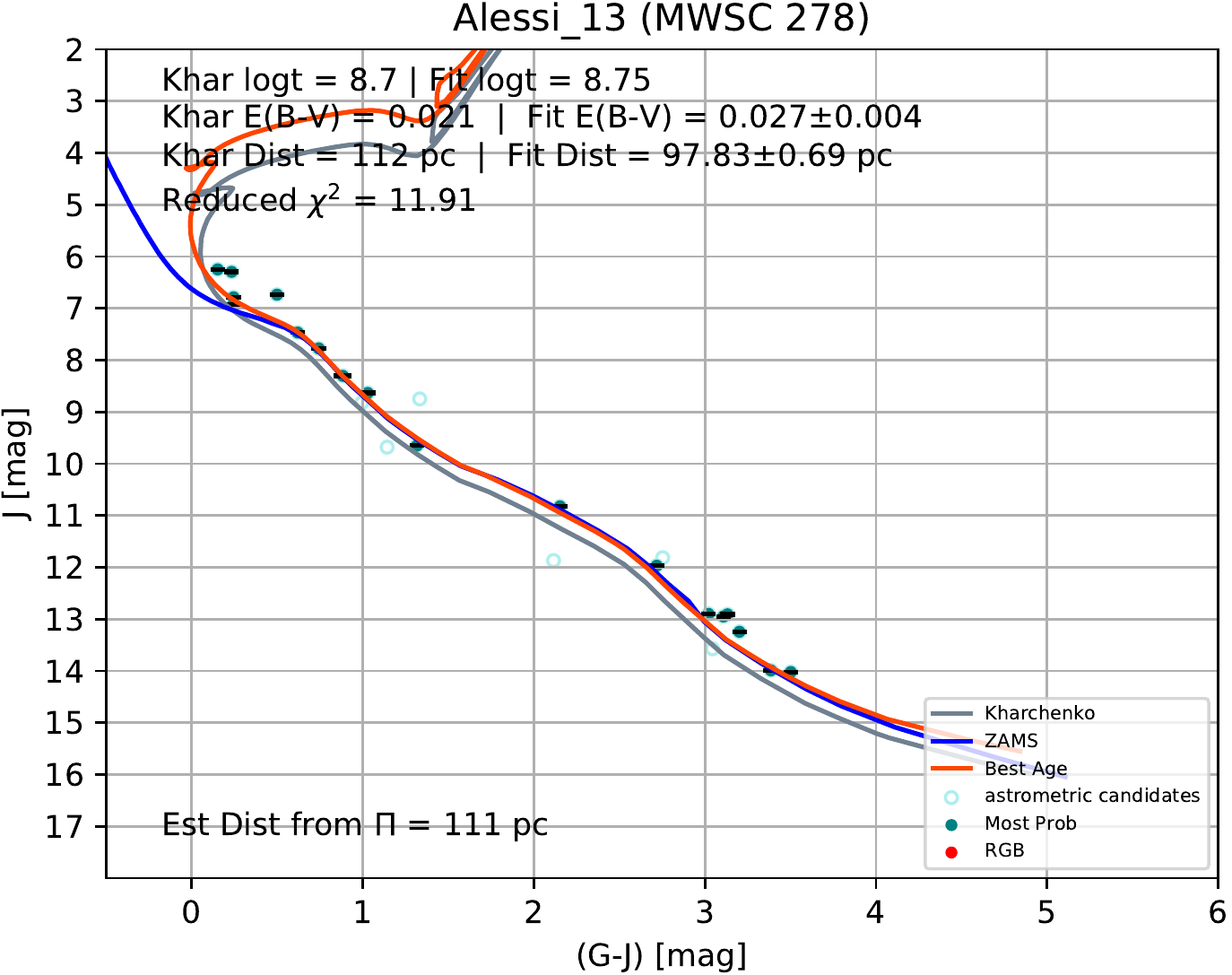}
\includegraphics[width=6cm]{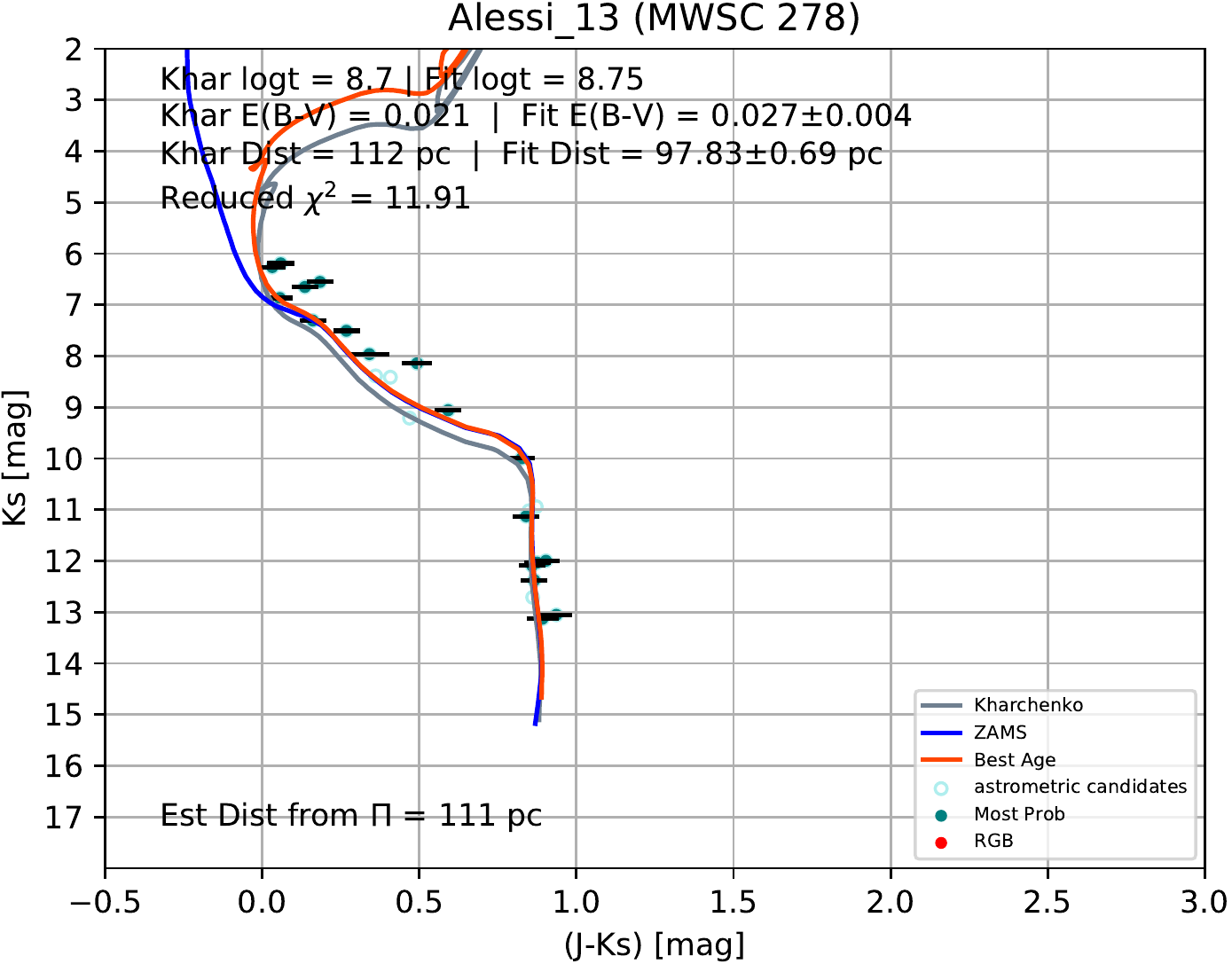}\\
\caption{Color-magnitude diagrams for clusters, from top to bottom: Blanco~1, Platais~2, $\alpha$~Per (Melotte~20), and Alessi~13. From left to right: their respective $V$ vs. $(B-V)$, $J$ vs. $(G-J)$, and $K_s$ vs. $(J-K_s)$ CMDs. The cluster members determined from the pipeline are given by teal circles with their corresponding magnitude and color error bars. The cluster astrometric candidates that were later rejected as cluster members are shown by light blue open circles. RGB stars, if any, are indicated by red circles. The red isochrone is the pipeline selected age, plotted with the fitted cluster $E(B-V)$ and $d$. This isochrone is plotted without the binary offset. The gray isochrone shows the age, $E(B-V)$, and $d$ as determined by \citet{2013A&A...558A..53K}. The blue line is the ZAMS plotted with the fitted cluster $E(B-V)$ and $d$.}
\label{figa1}
\end{figure*}

\newpage
\begin{figure*}
\centering
\includegraphics[width=6cm]{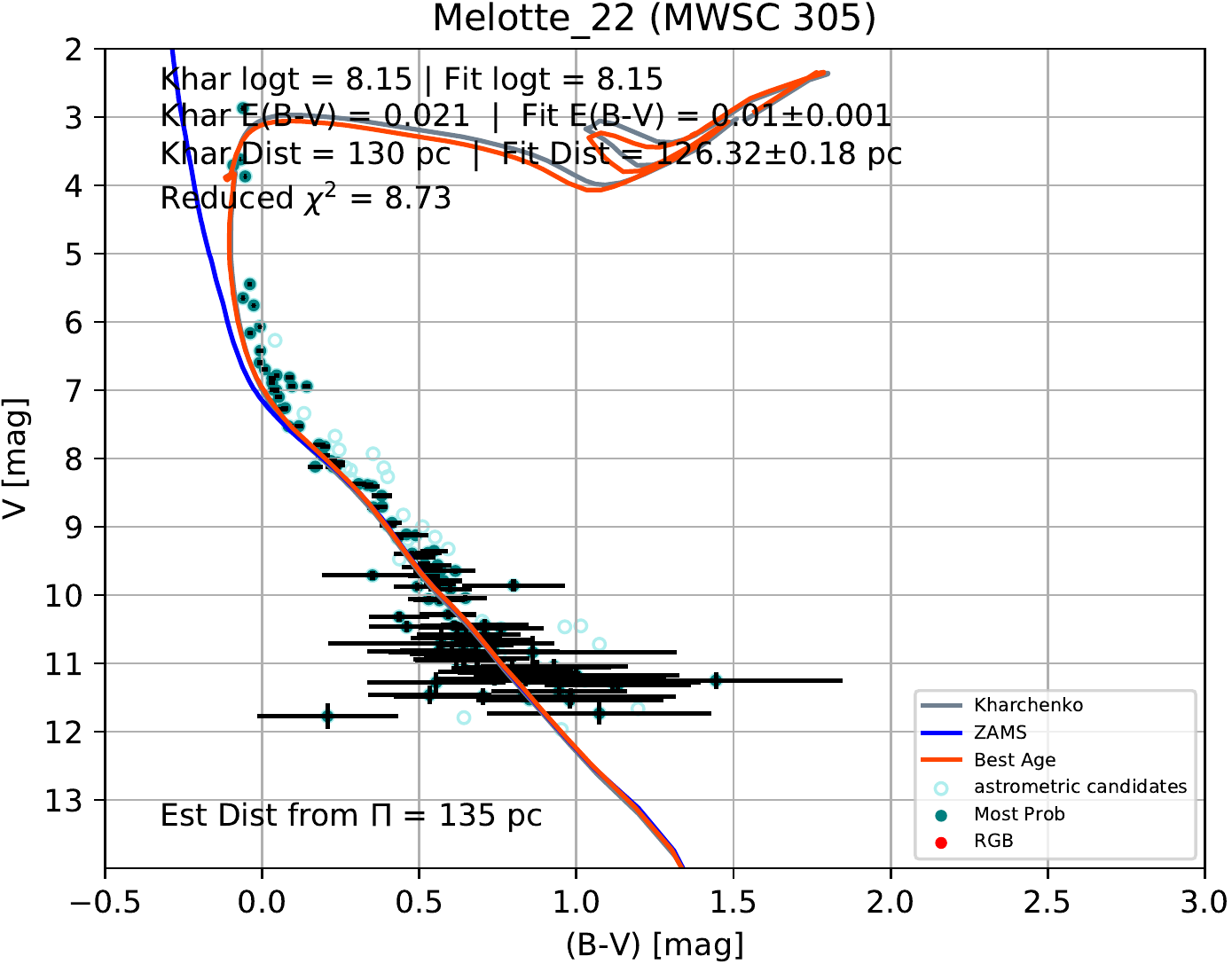}
\includegraphics[width=6cm]{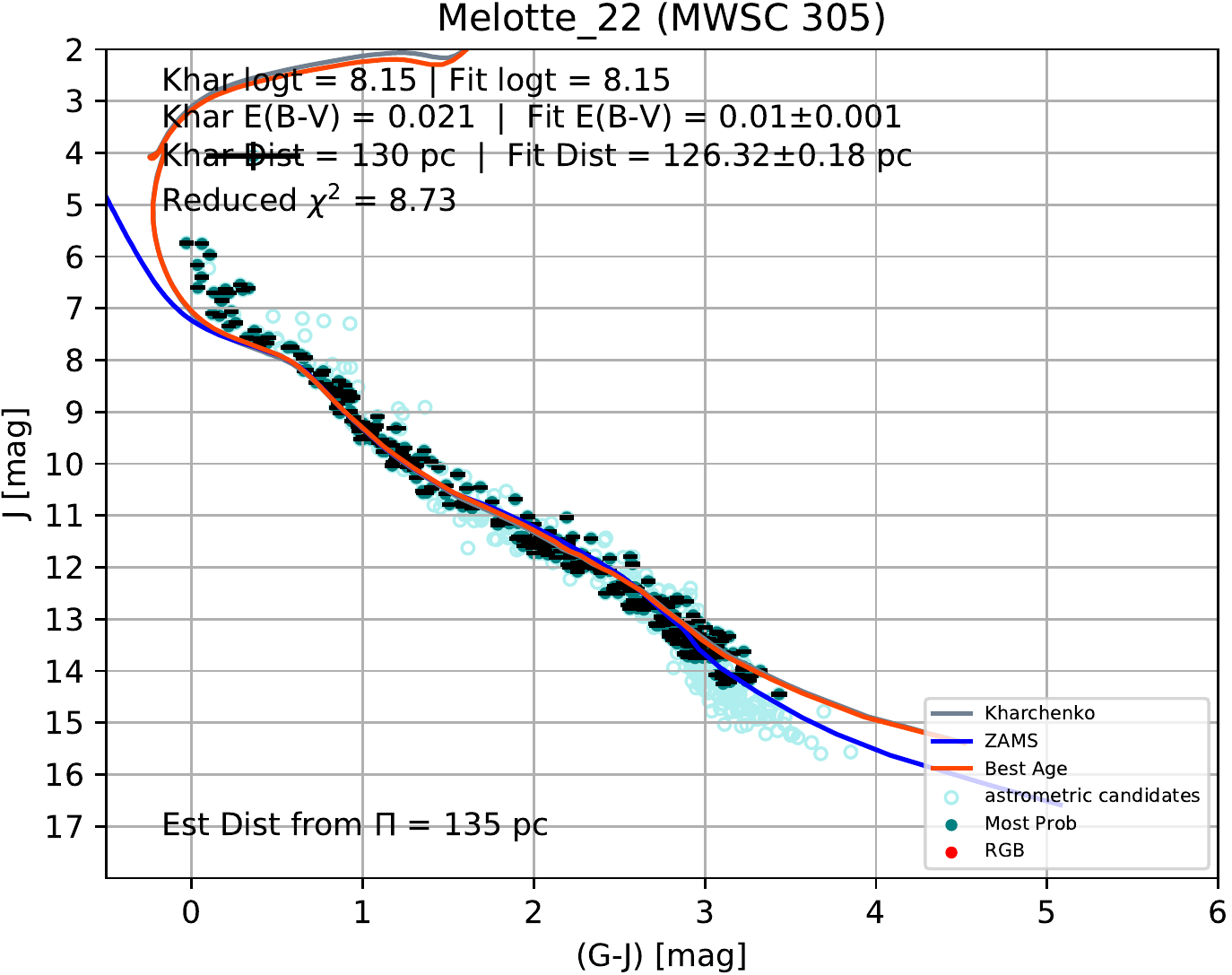}
\includegraphics[width=6cm]{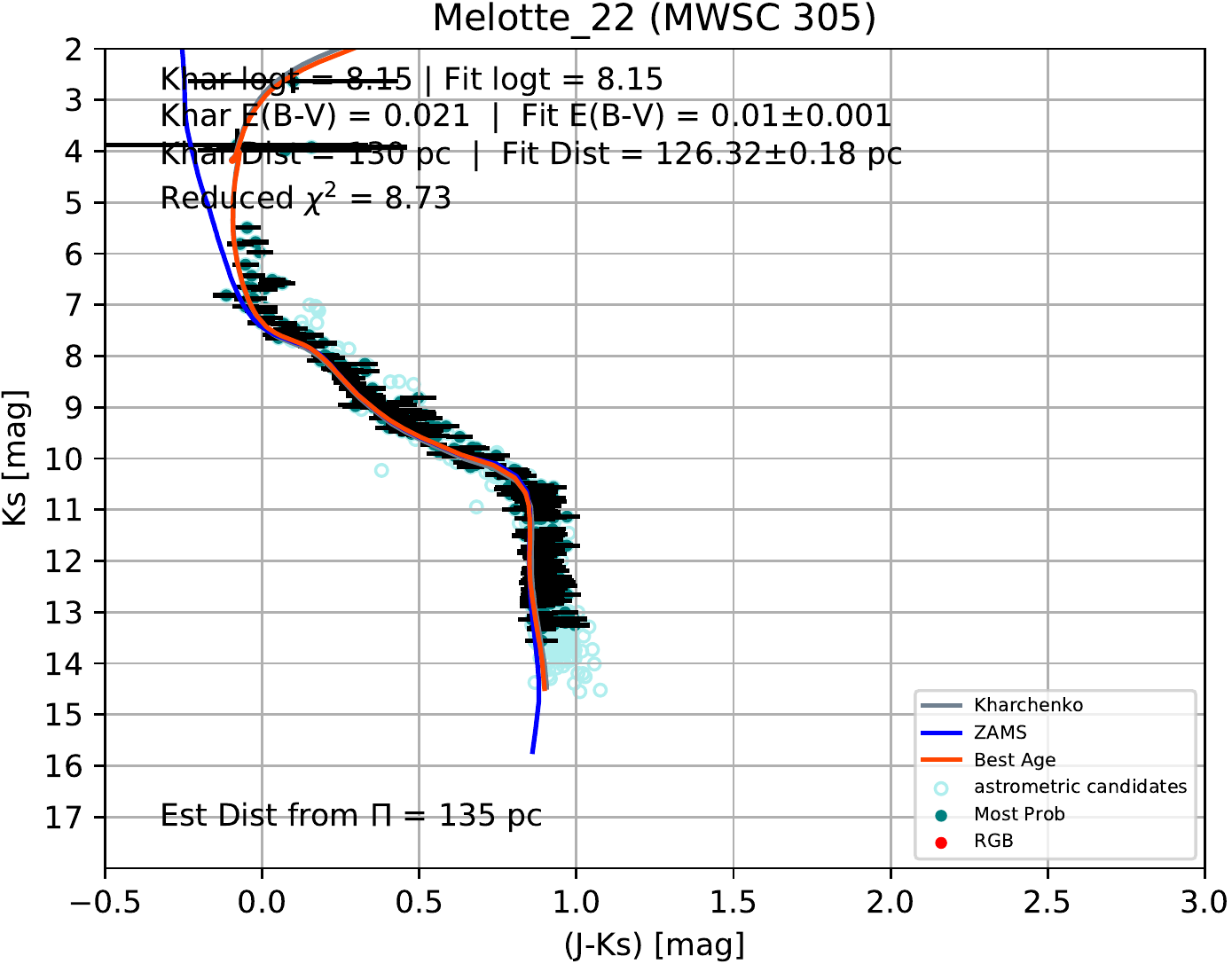}\\
\includegraphics[width=6cm]{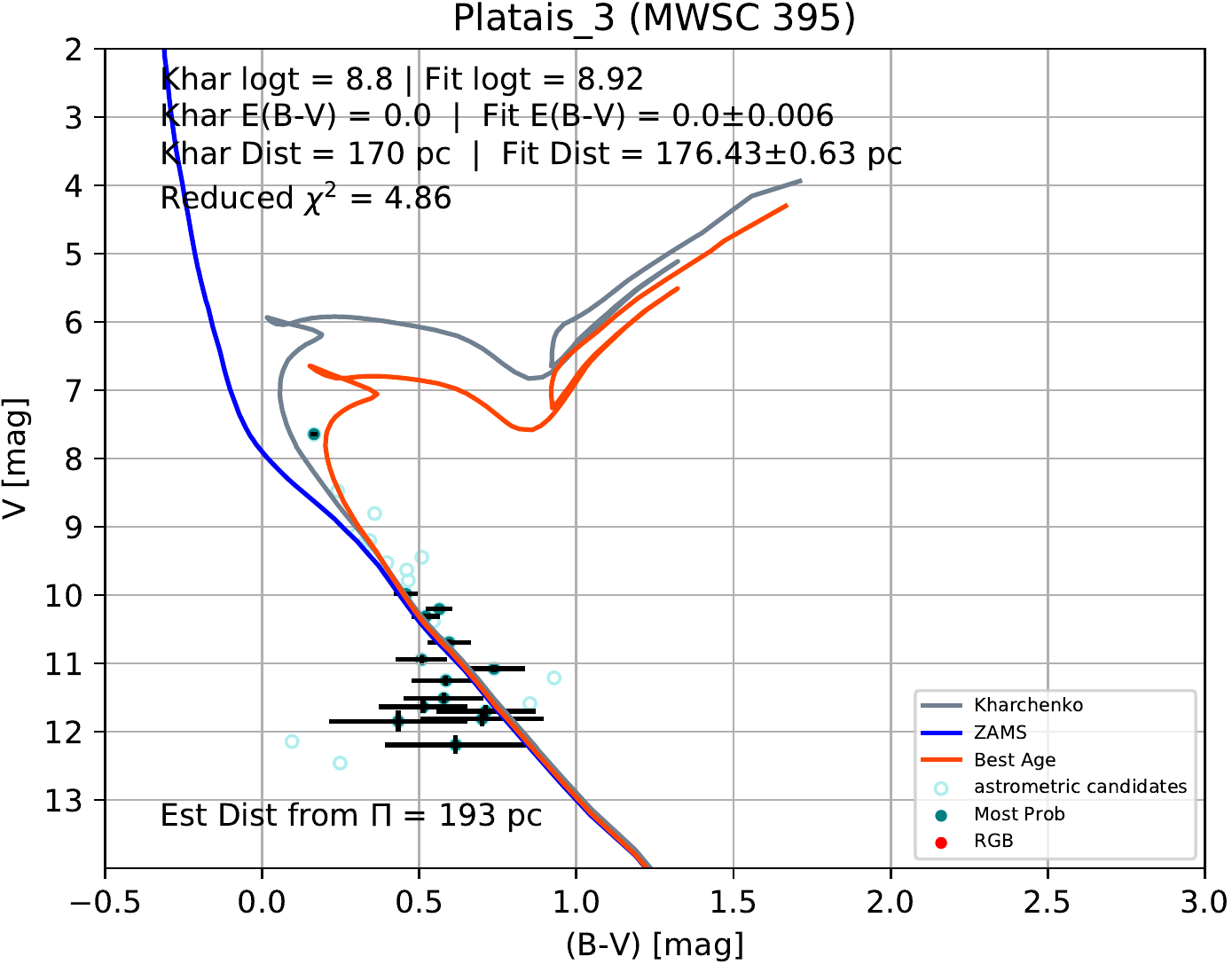}
\includegraphics[width=6cm]{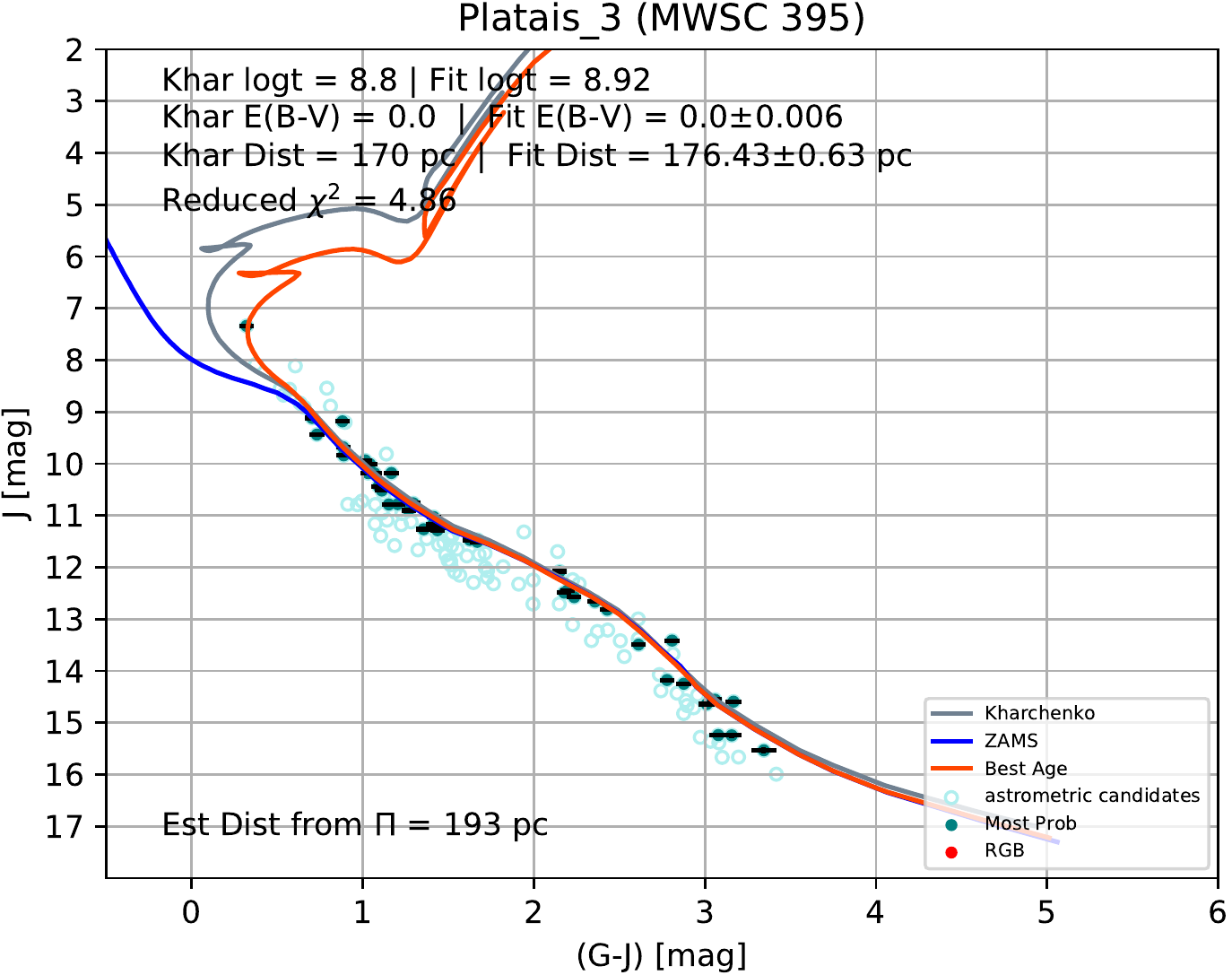}
\includegraphics[width=6cm]{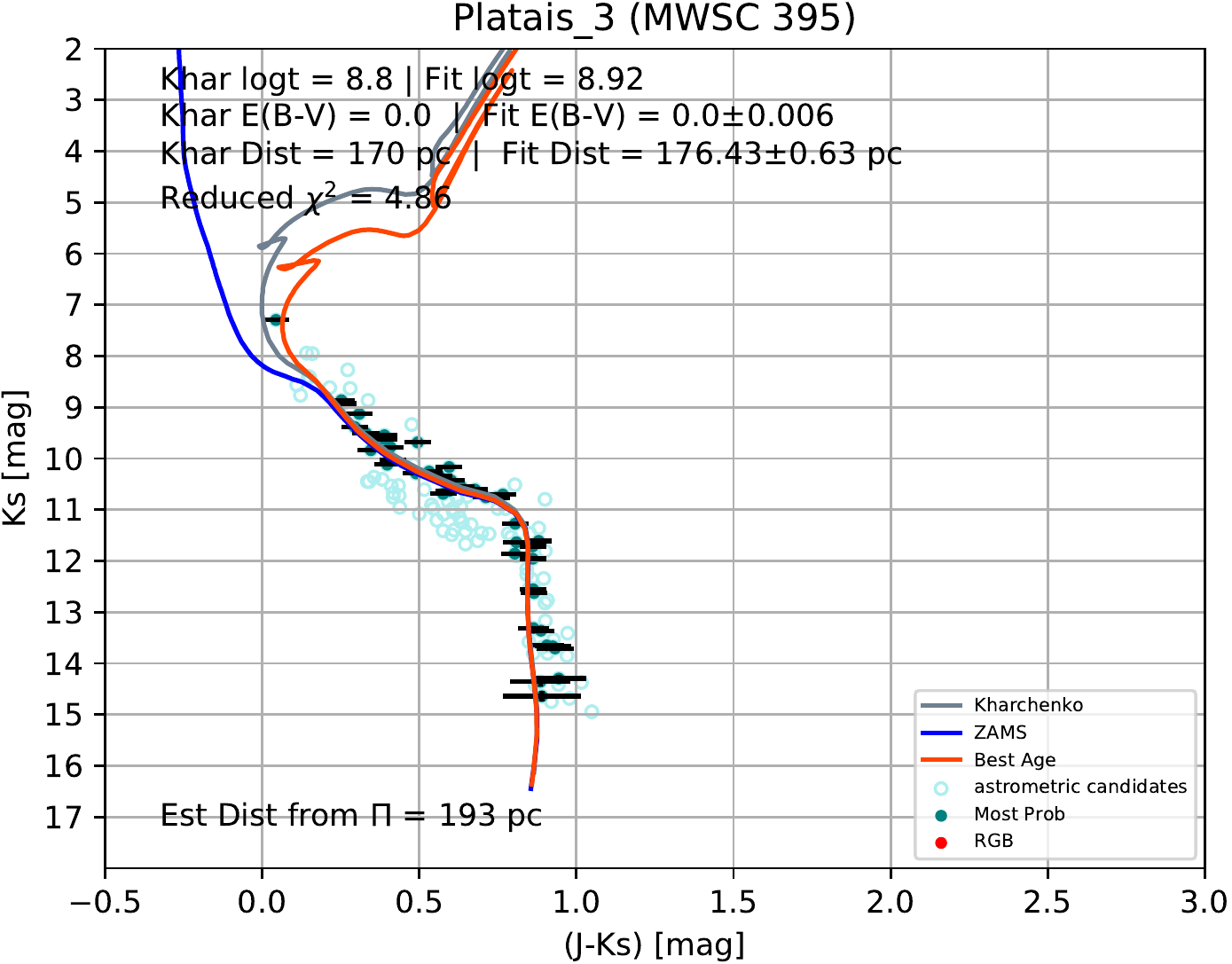}\\
\includegraphics[width=6cm]{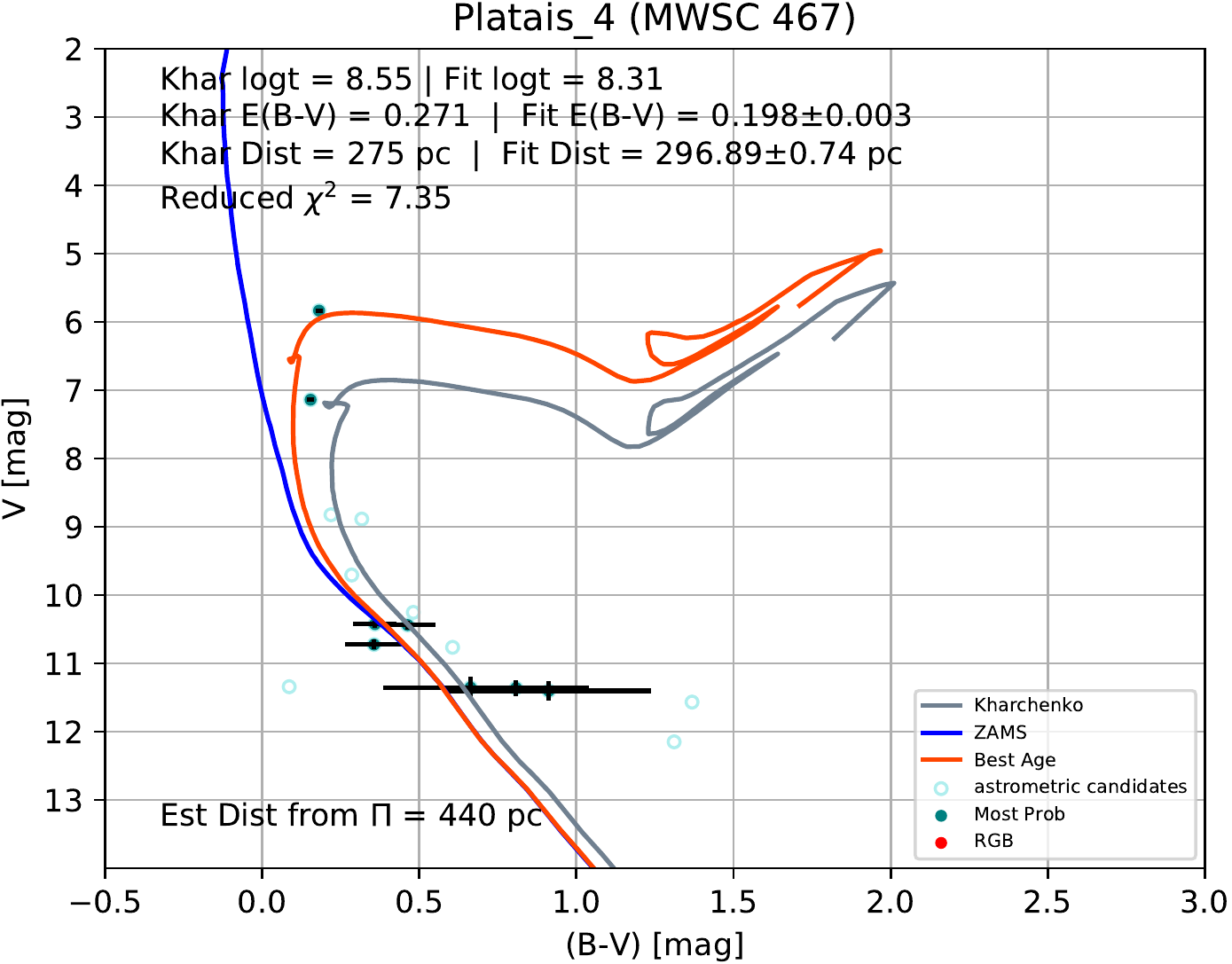}
\includegraphics[width=6cm]{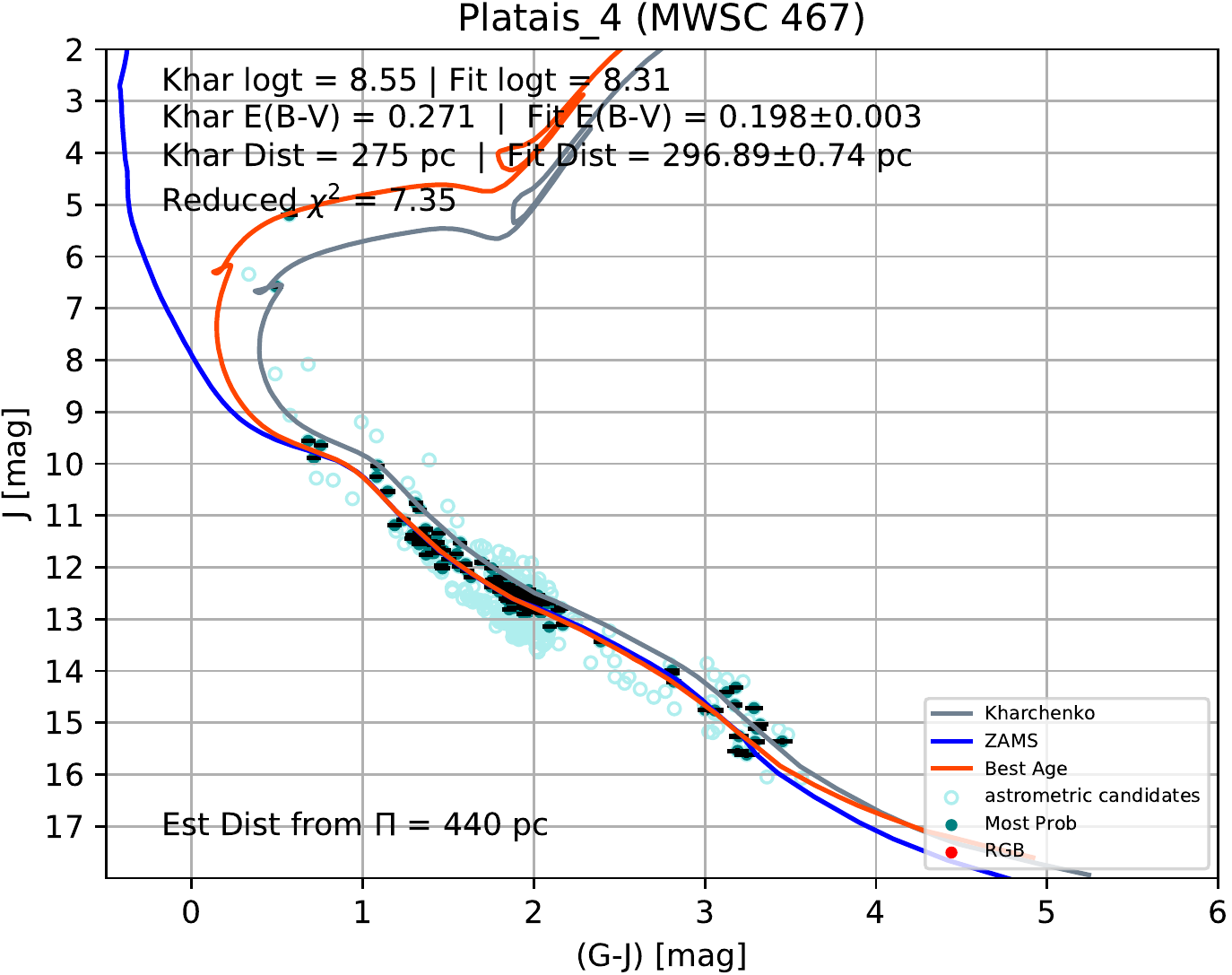}
\includegraphics[width=6cm]{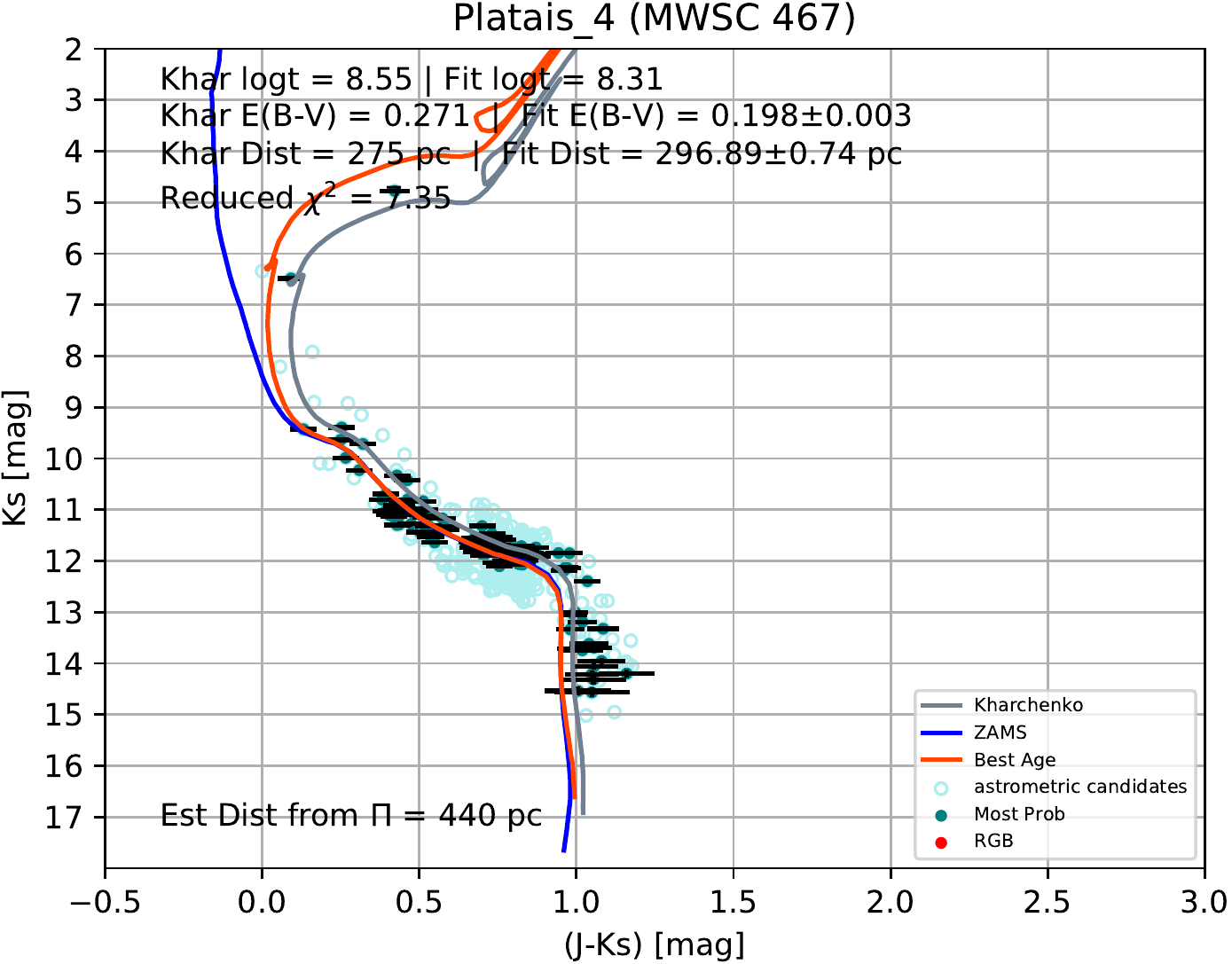}\\
\includegraphics[width=6cm]{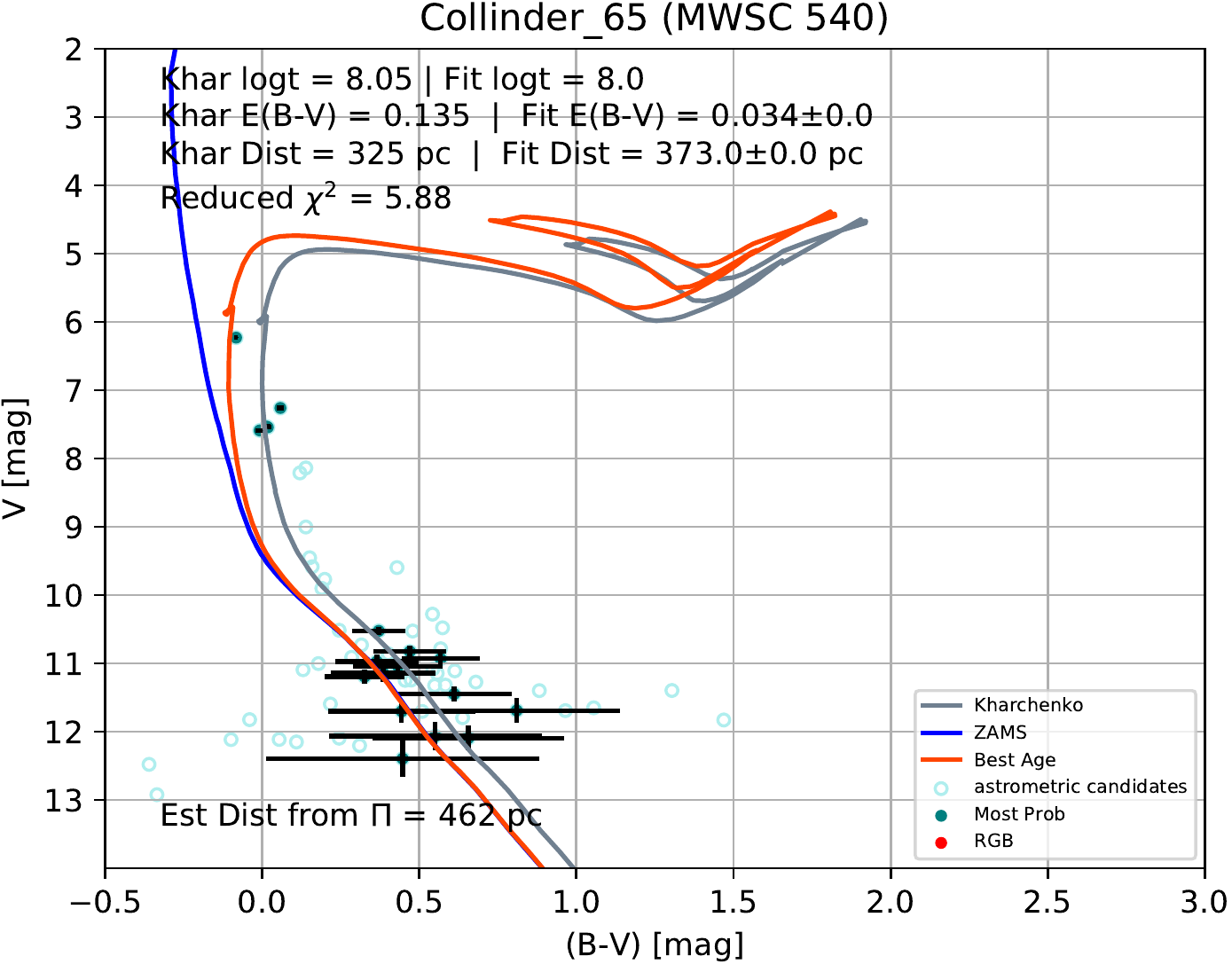}
\includegraphics[width=6cm]{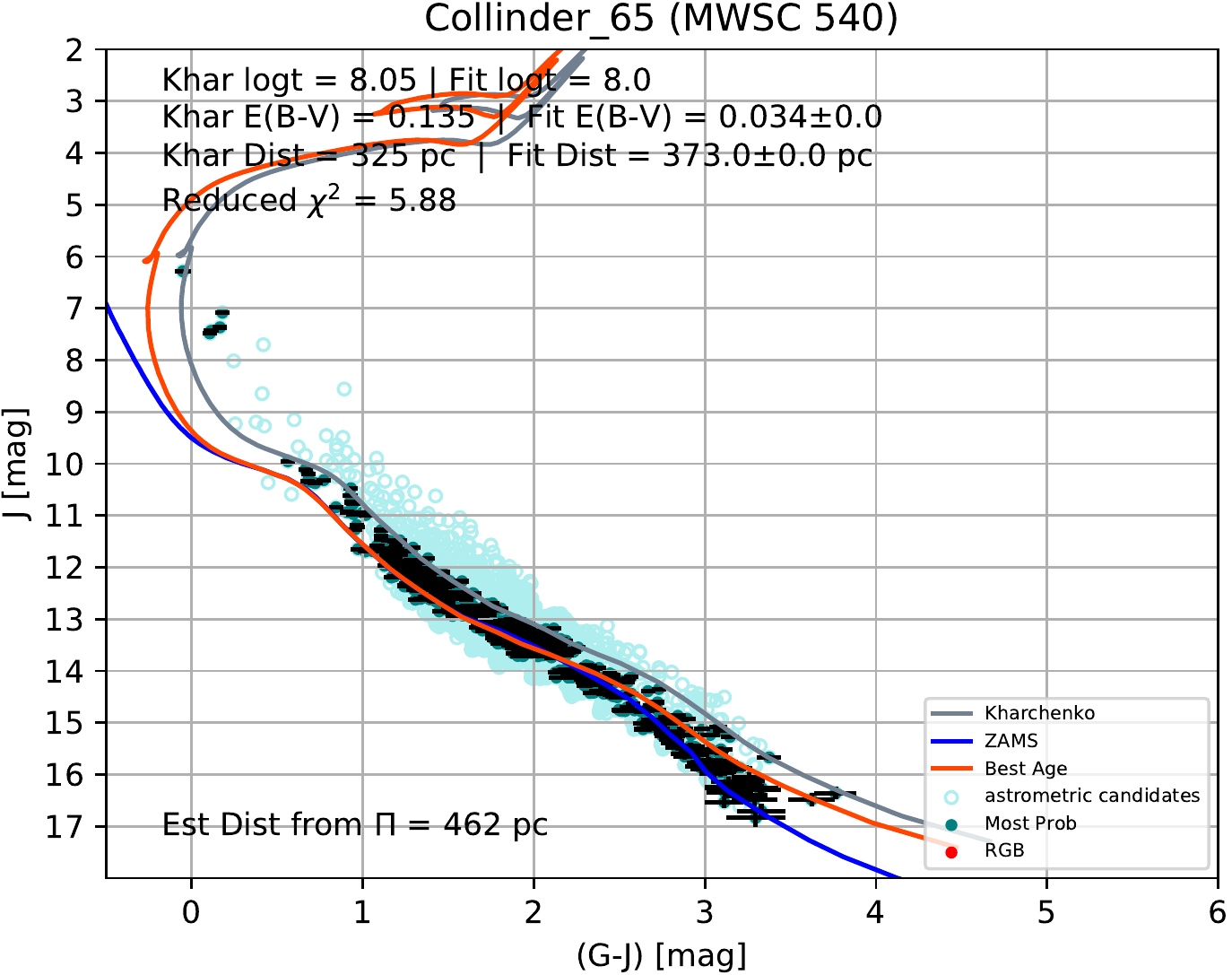}
\includegraphics[width=6cm]{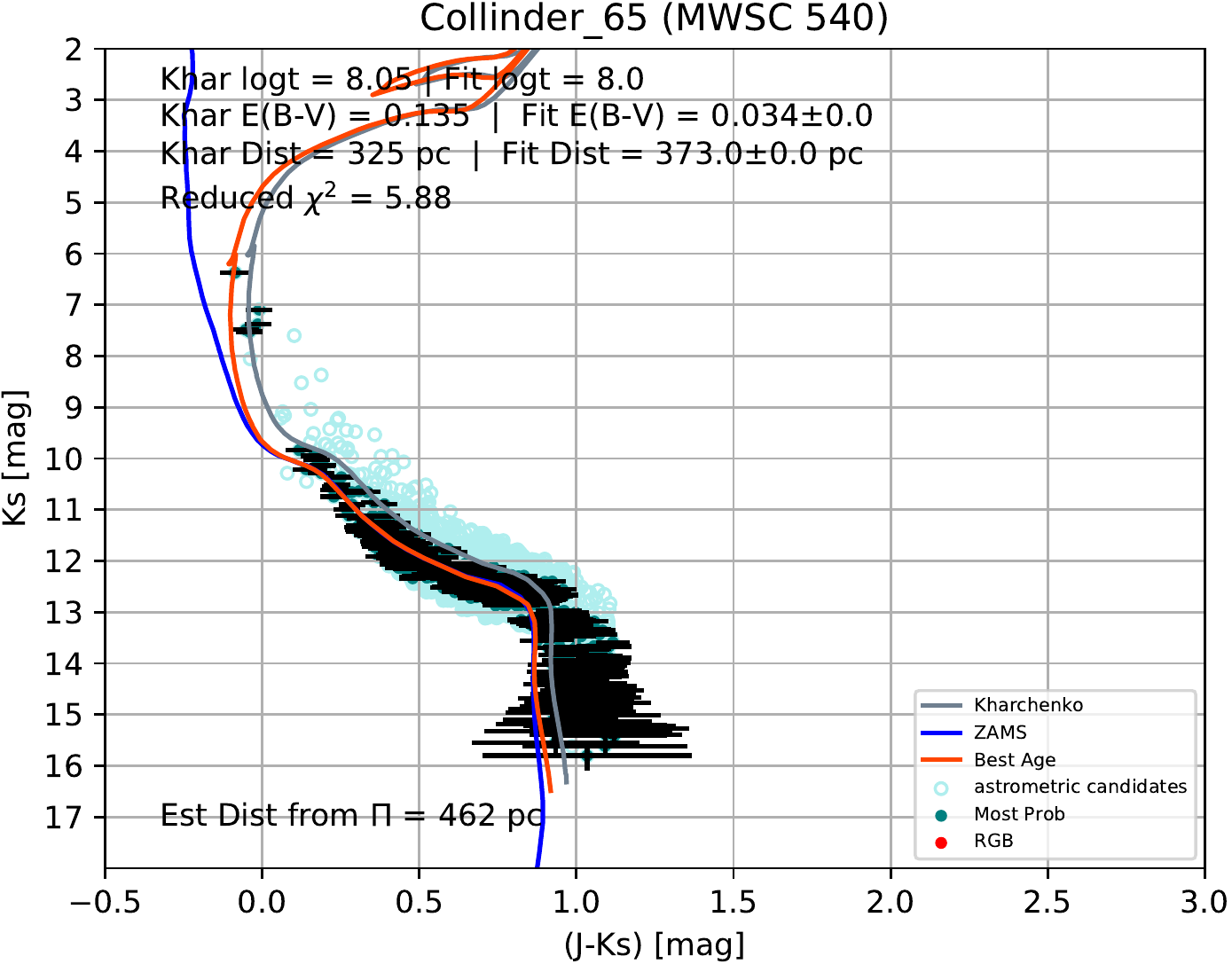}\\
\caption{Color-magnitude diagrams for clusters, from top to bottom: the Pleiades (Melotte~22), Platais~3, Platais~4, and Collinder~65. From left to right: their respective $V$ vs. $(B-V)$, $J$ vs. $(G-J)$, and $K_s$ vs. $(J-K_s)$ CMDs. The cluster members determined from the pipeline are given by teal circles with their corresponding magnitude and color error bars. The cluster astrometric candidates that were later rejected as cluster members are shown by light blue open circles. RGB stars, if any, are indicated by red circles. The red isochrone is the pipeline selected age, plotted with the fitted cluster $E(B-V)$ and $d$. This isochrone is plotted without the binary offset. The gray isochrone shows the age, $E(B-V)$, and $d$ as determined by \citet{2013A&A...558A..53K}. The blue line is the ZAMS plotted with the fitted cluster $E(B-V)$ and $d$.}
 \label{figa2}
\end{figure*}

\begin{figure*}
\centering
\includegraphics[width=6cm]{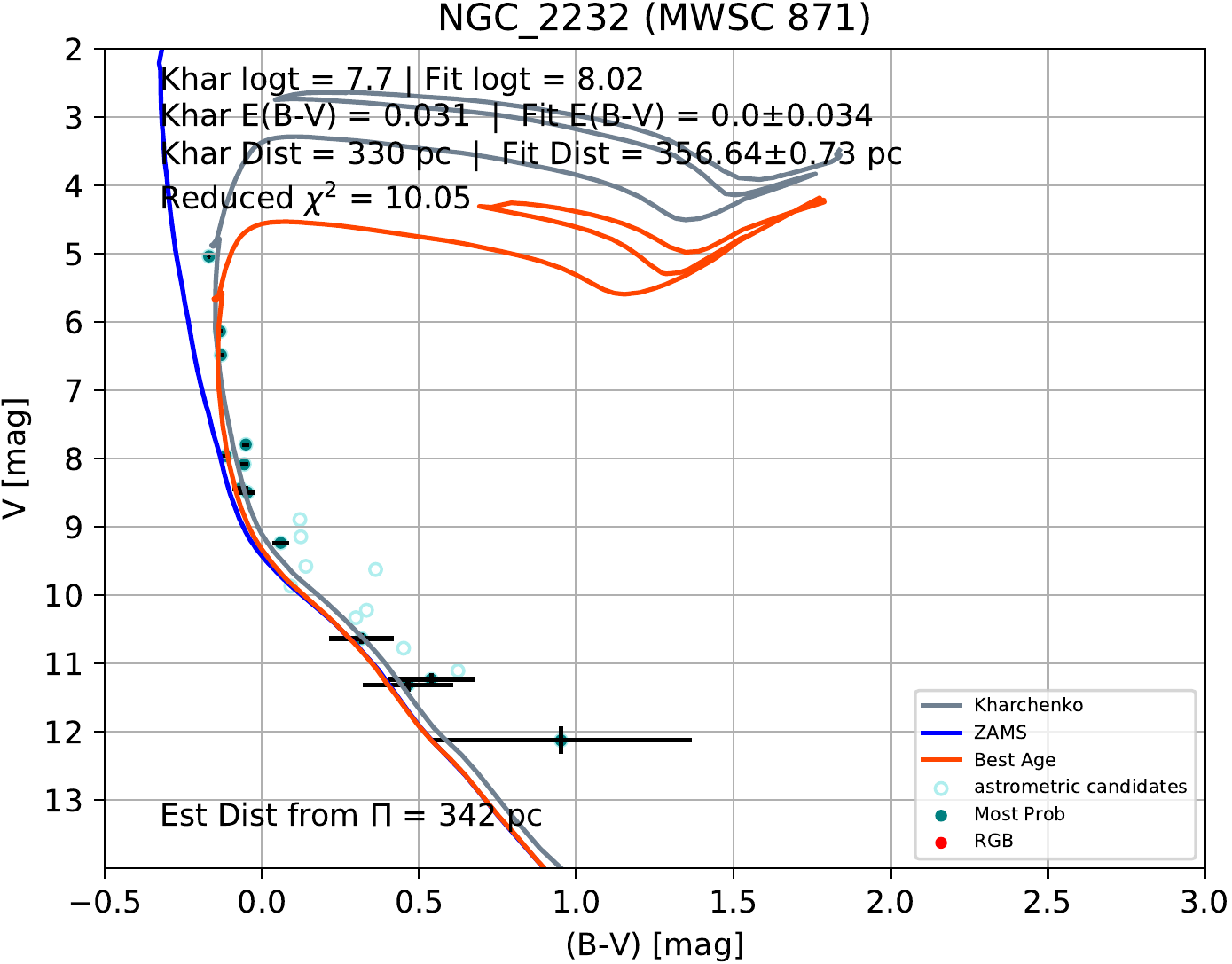}
\includegraphics[width=6cm]{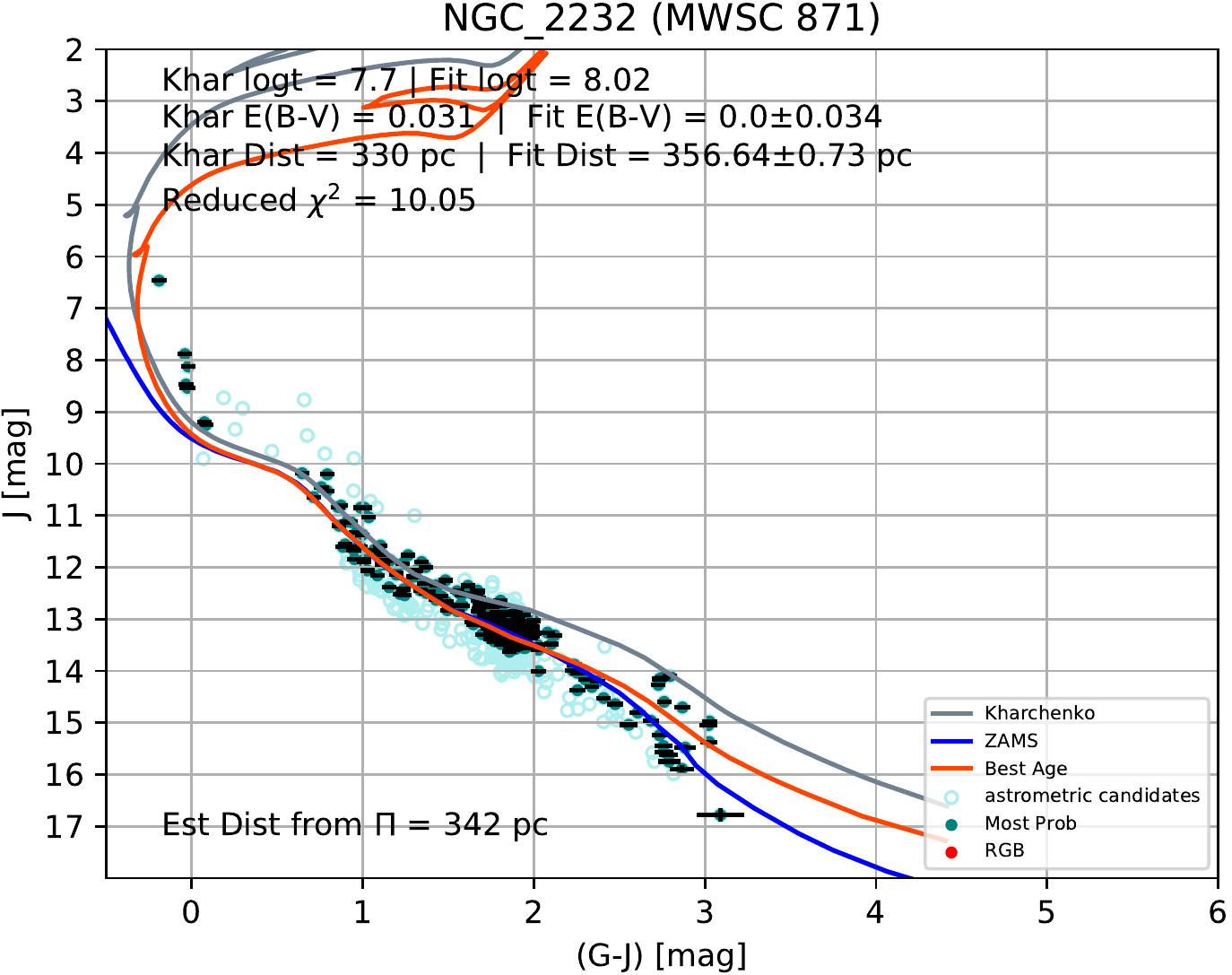}
\includegraphics[width=6cm]{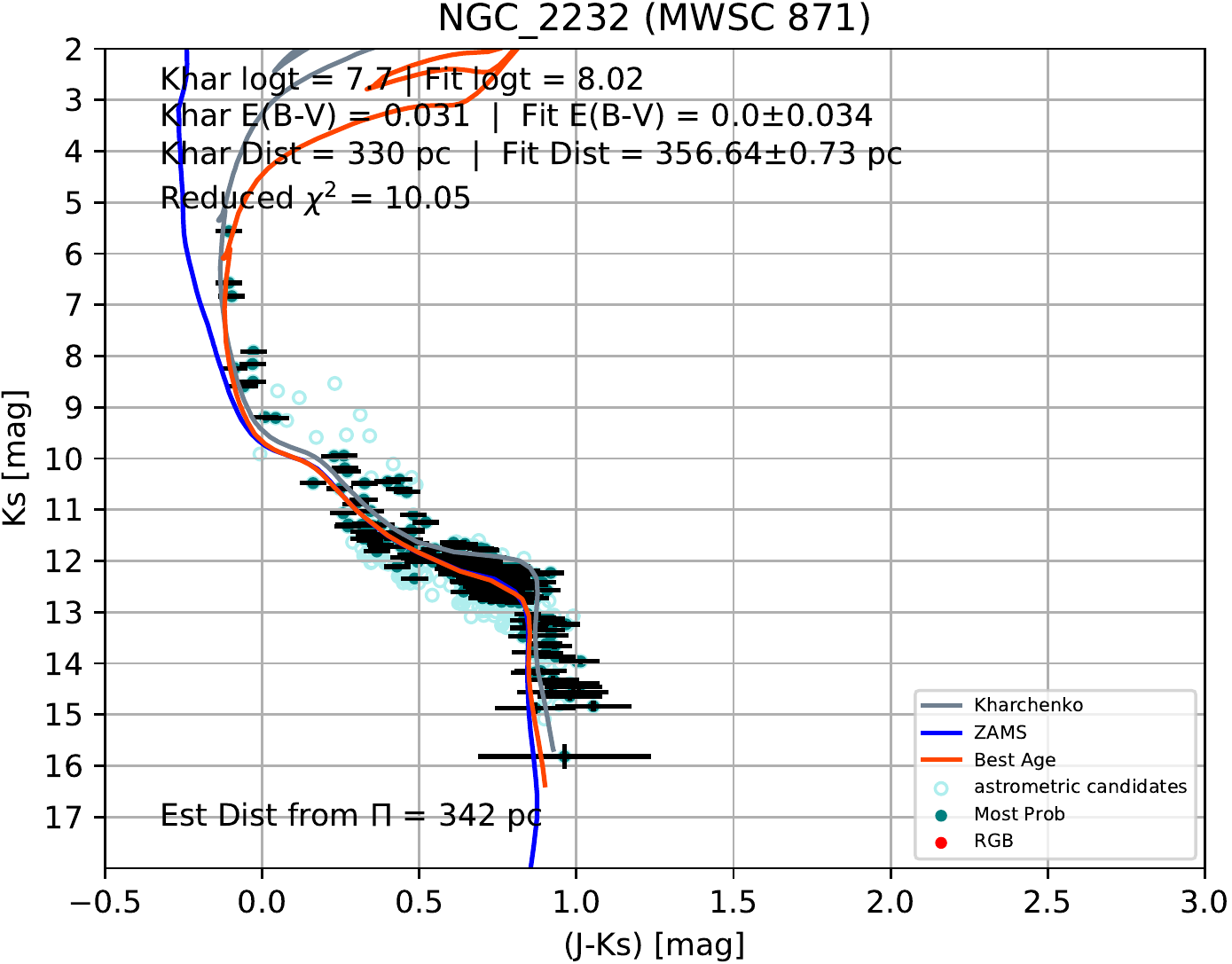}\\
\includegraphics[width=6cm]{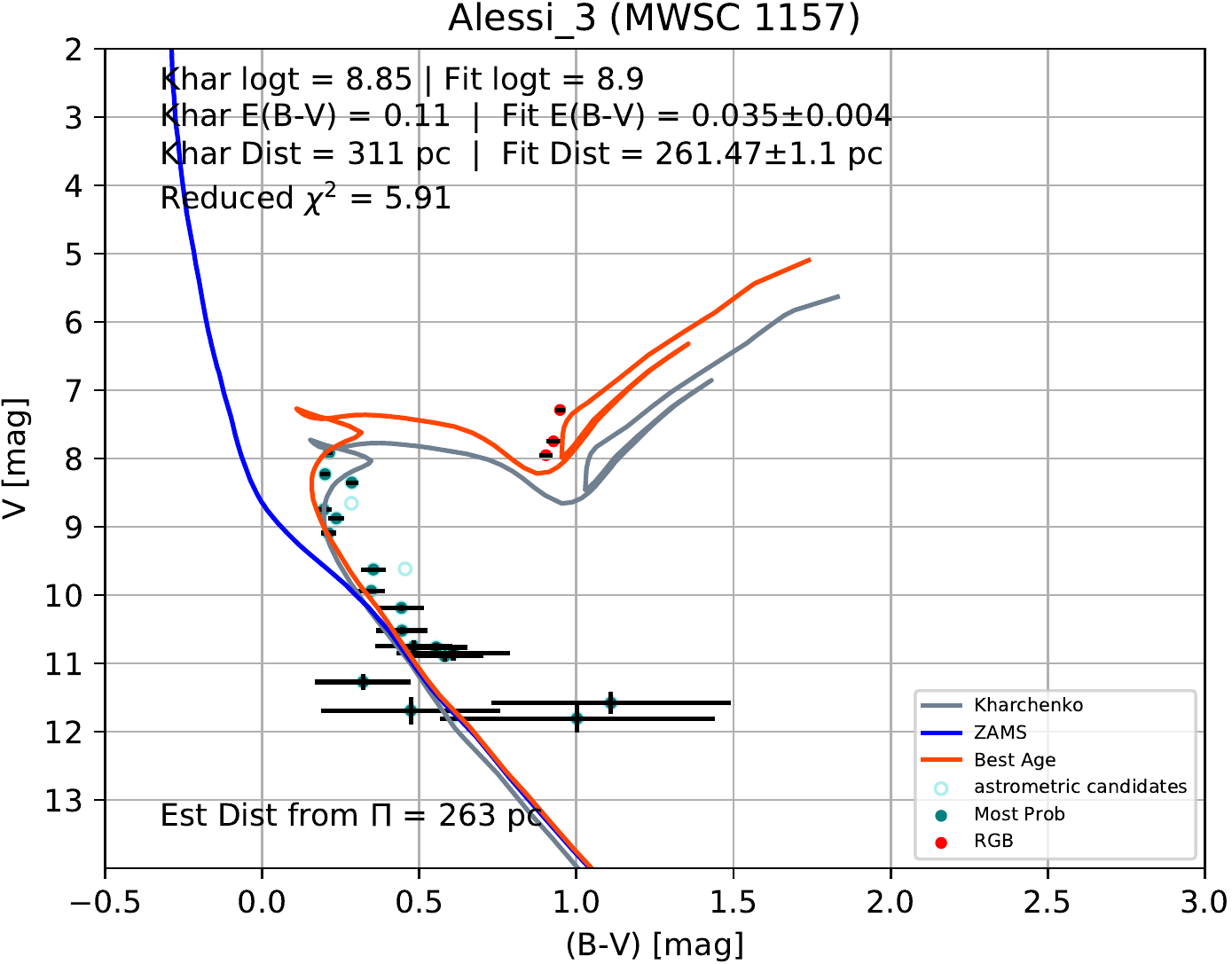}
\includegraphics[width=6cm]{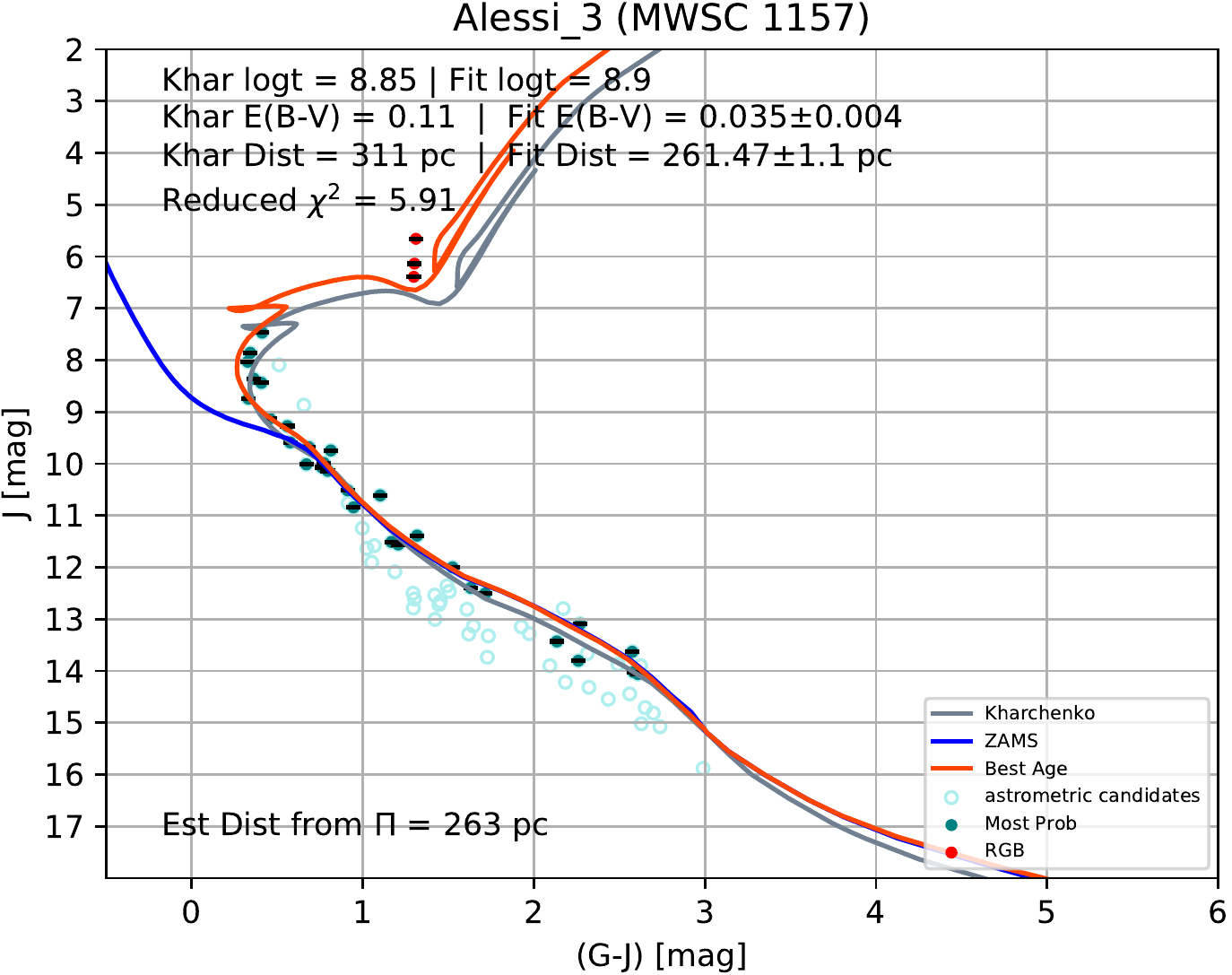}
\includegraphics[width=6cm]{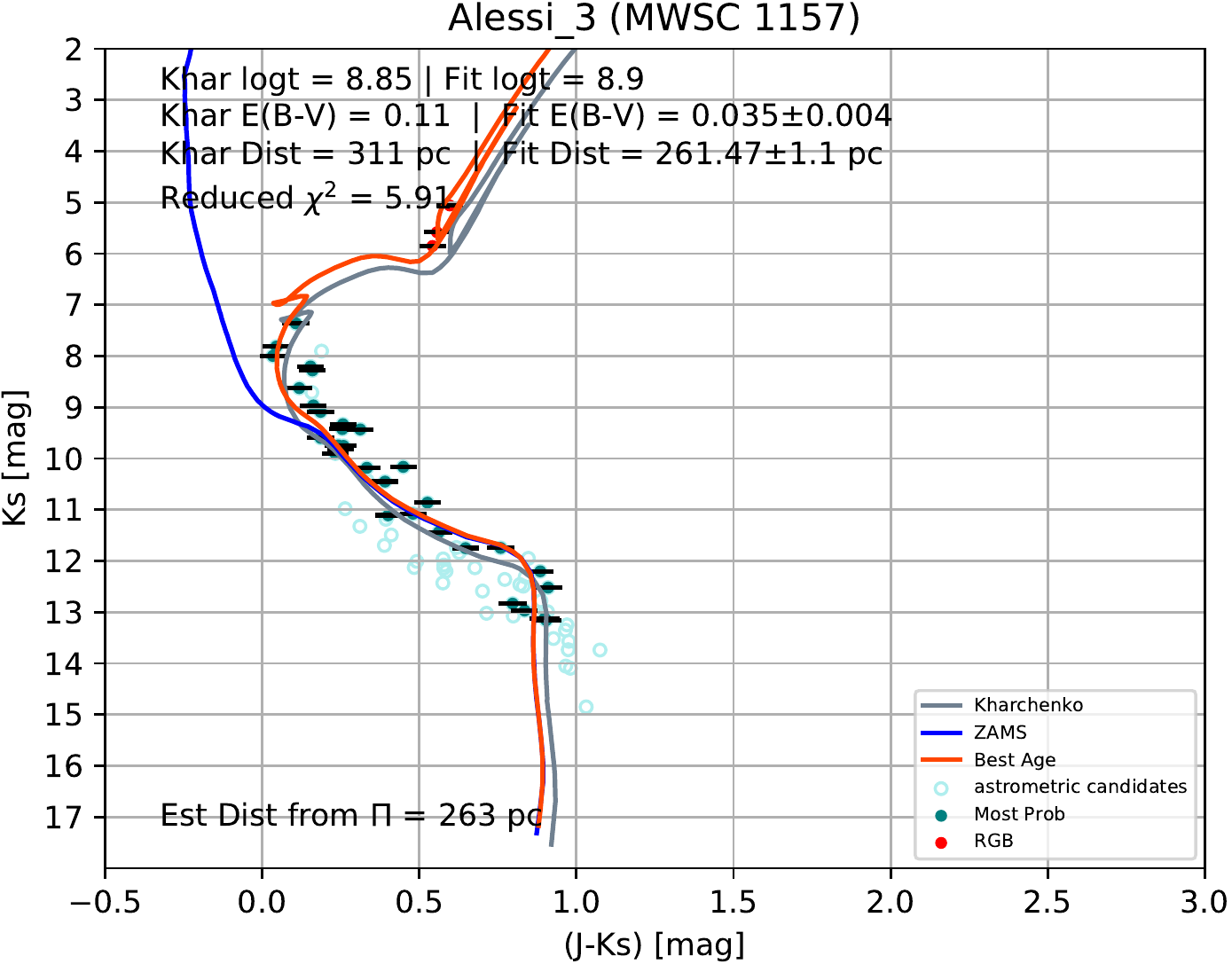}\\
\includegraphics[width=6cm]{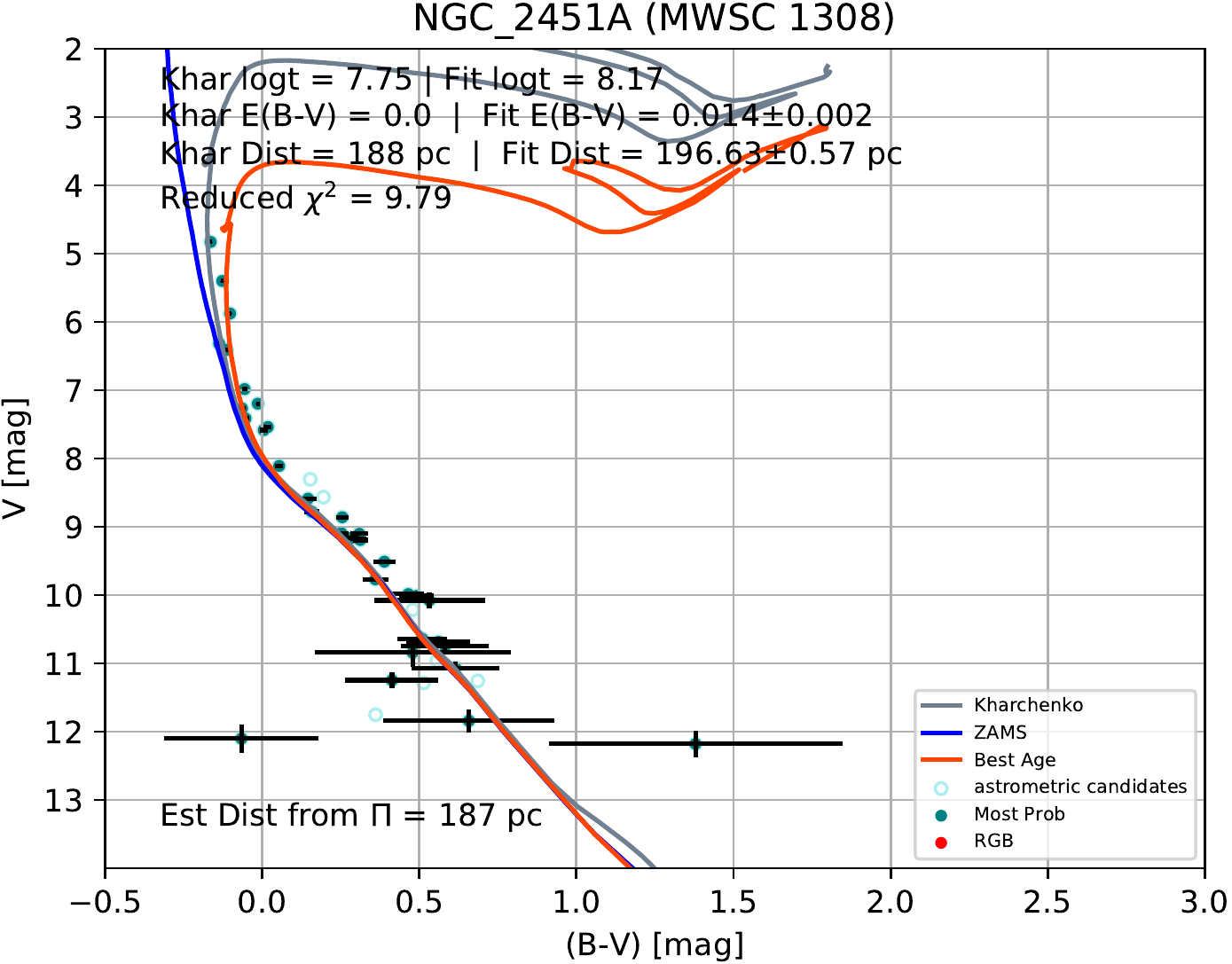}
\includegraphics[width=6cm]{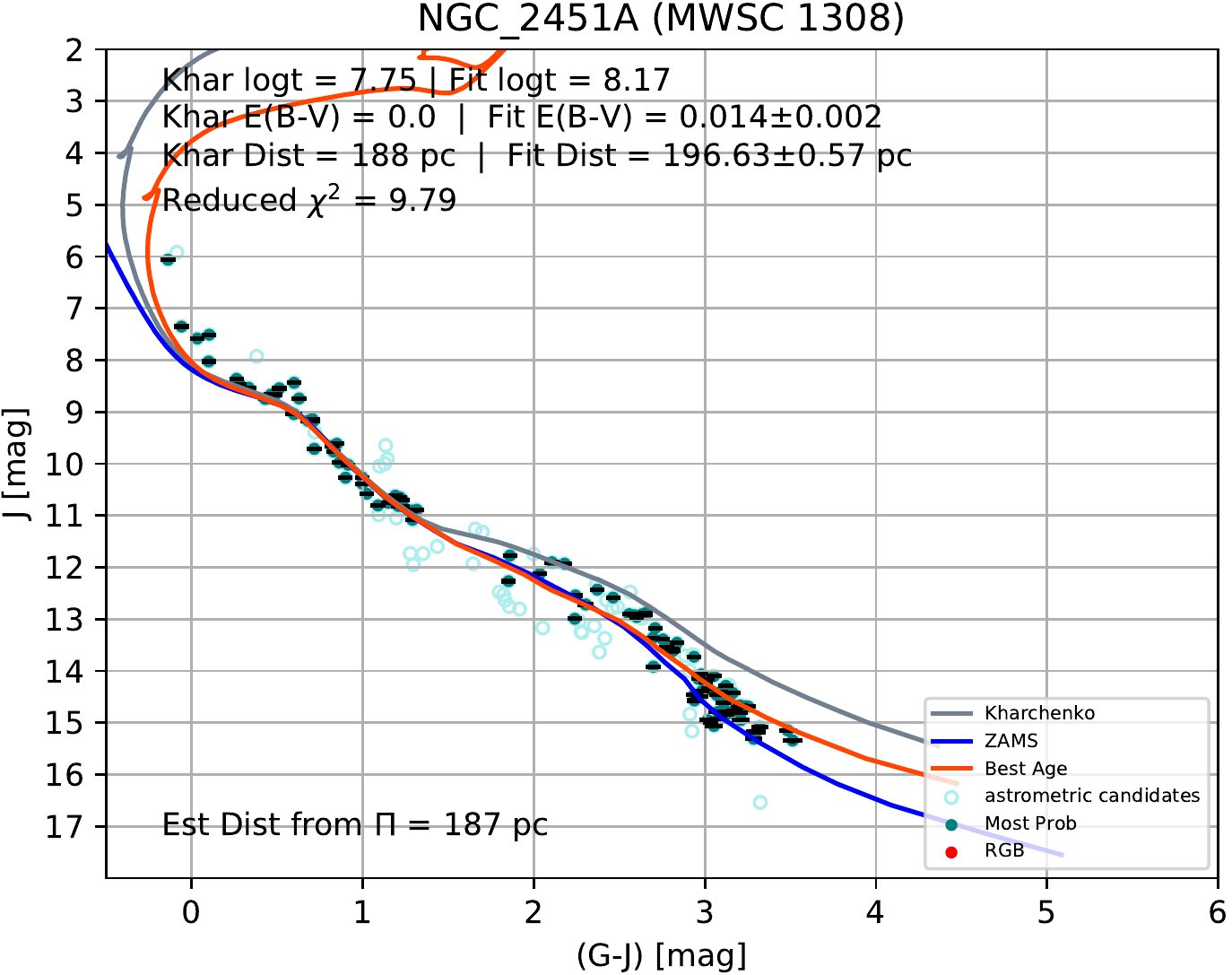}
\includegraphics[width=6cm]{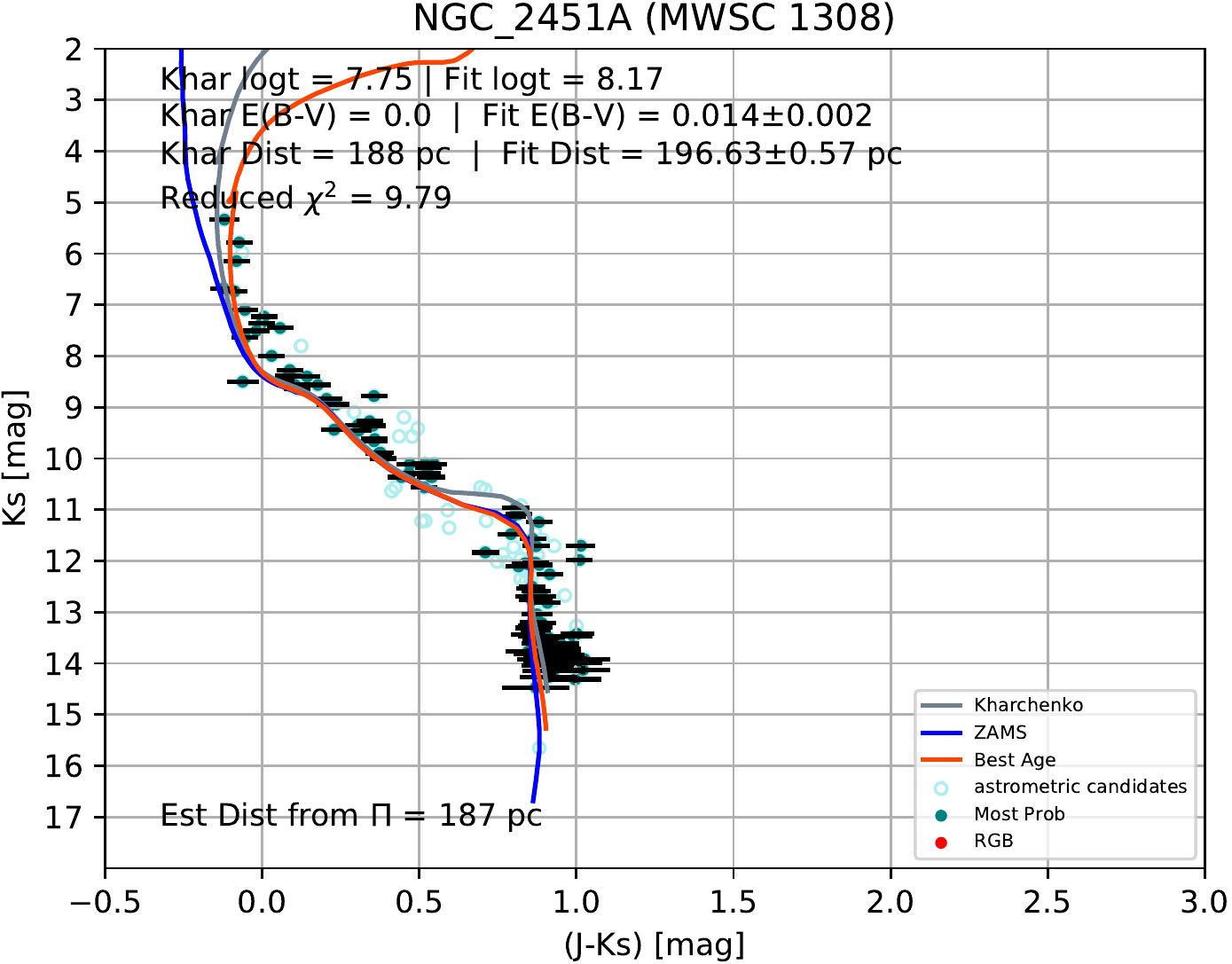}\\
\includegraphics[width=6cm]{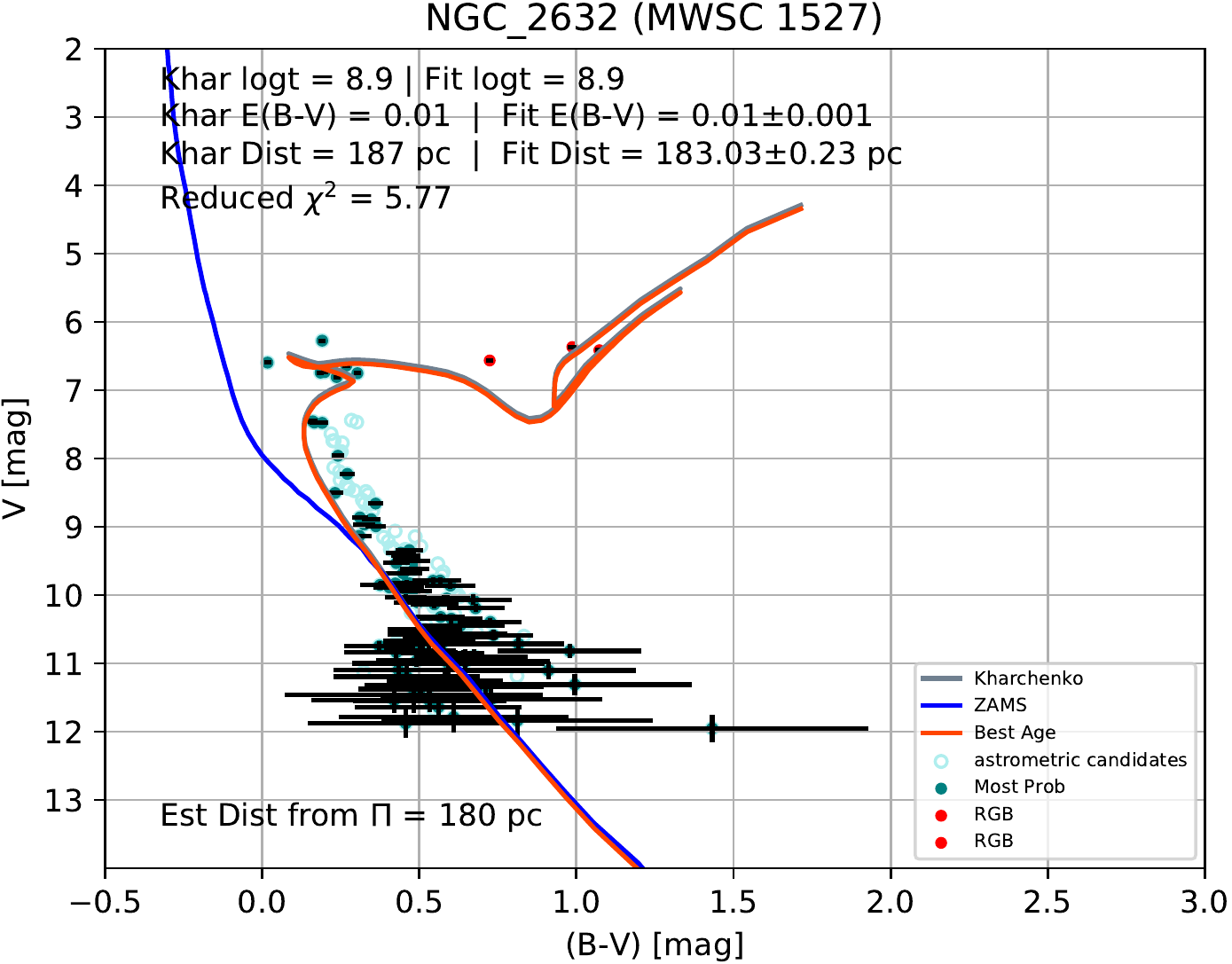}
\includegraphics[width=6cm]{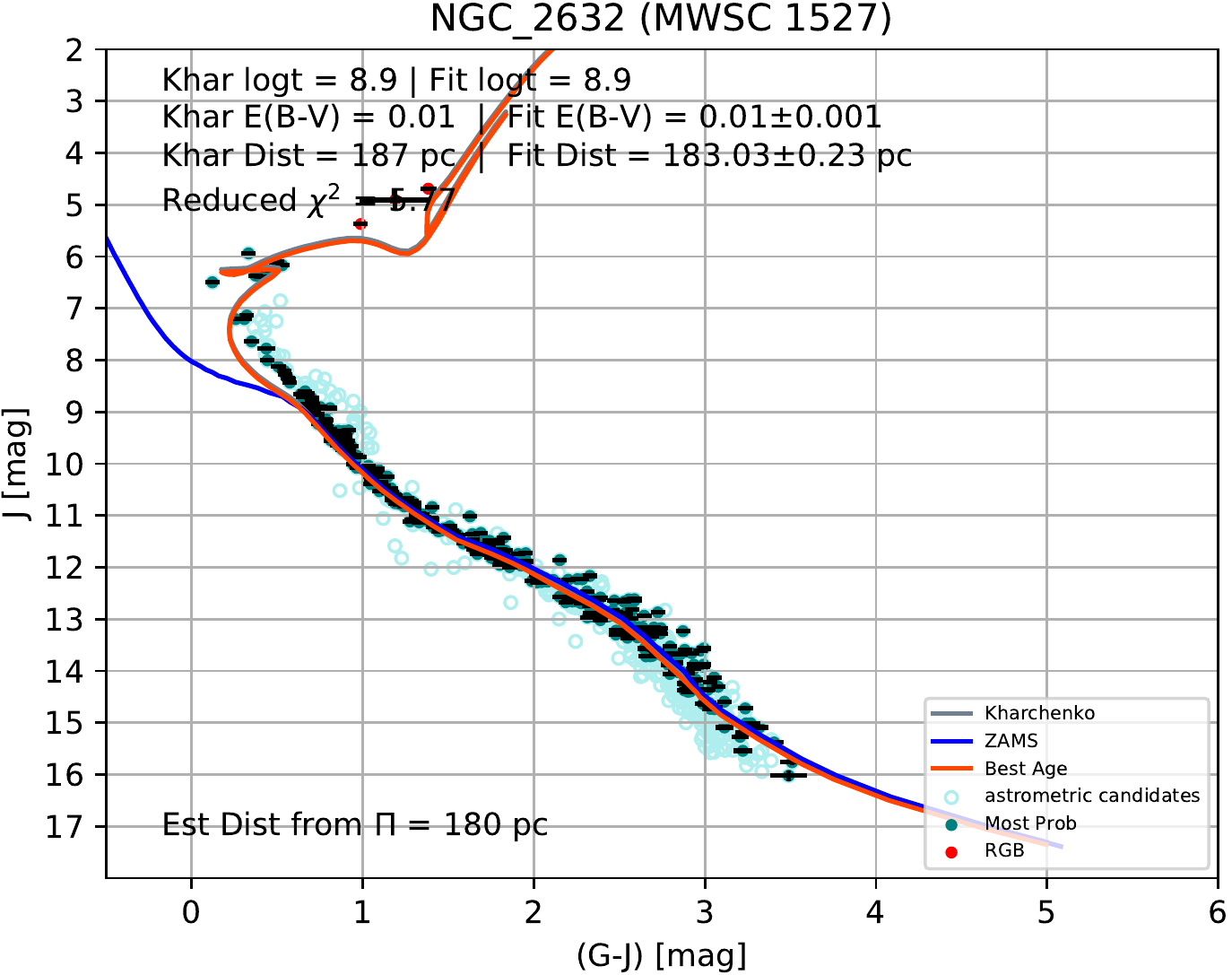}
\includegraphics[width=6cm]{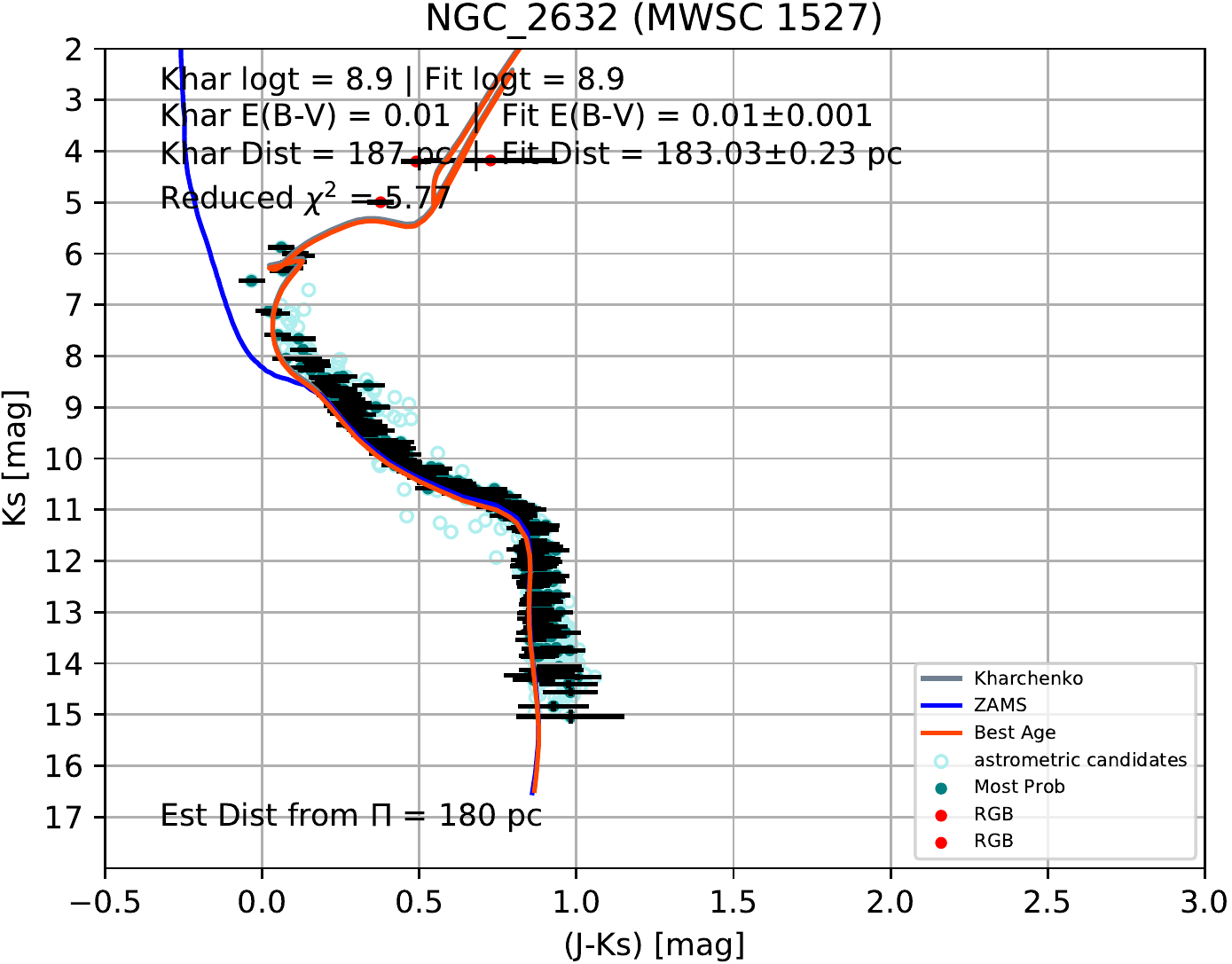}\\
\caption{Color-magnitude diagrams for clusters, from top to bottom: NGC~2232, Alessi~3, NGC~2451A, and Praesepe (NGC~2632). From left to right: their respective $V$ vs. $(B-V)$, $J$ vs. $(G-J)$, and $K_s$ vs. $(J-K_s)$ CMDs. The cluster members determined from the pipeline are given by teal circles with their corresponding magnitude and color error bars. The cluster astrometric candidates that were later rejected as cluster members are shown by light blue open circles. RGB stars, if any, are indicated by red circles. The red isochrone is the pipeline selected age, plotted with the fitted cluster $E(B-V)$ and $d$. This isochrone is plotted without the binary offset. The gray isochrone shows the age, $E(B-V)$, and $d$ as determined by \citet{2013A&A...558A..53K}. The blue line is the ZAMS plotted with the fitted cluster $E(B-V)$ and $d$.}
 \label{figa3}
\end{figure*}

\begin{figure*}
\centering
\includegraphics[width=6cm]{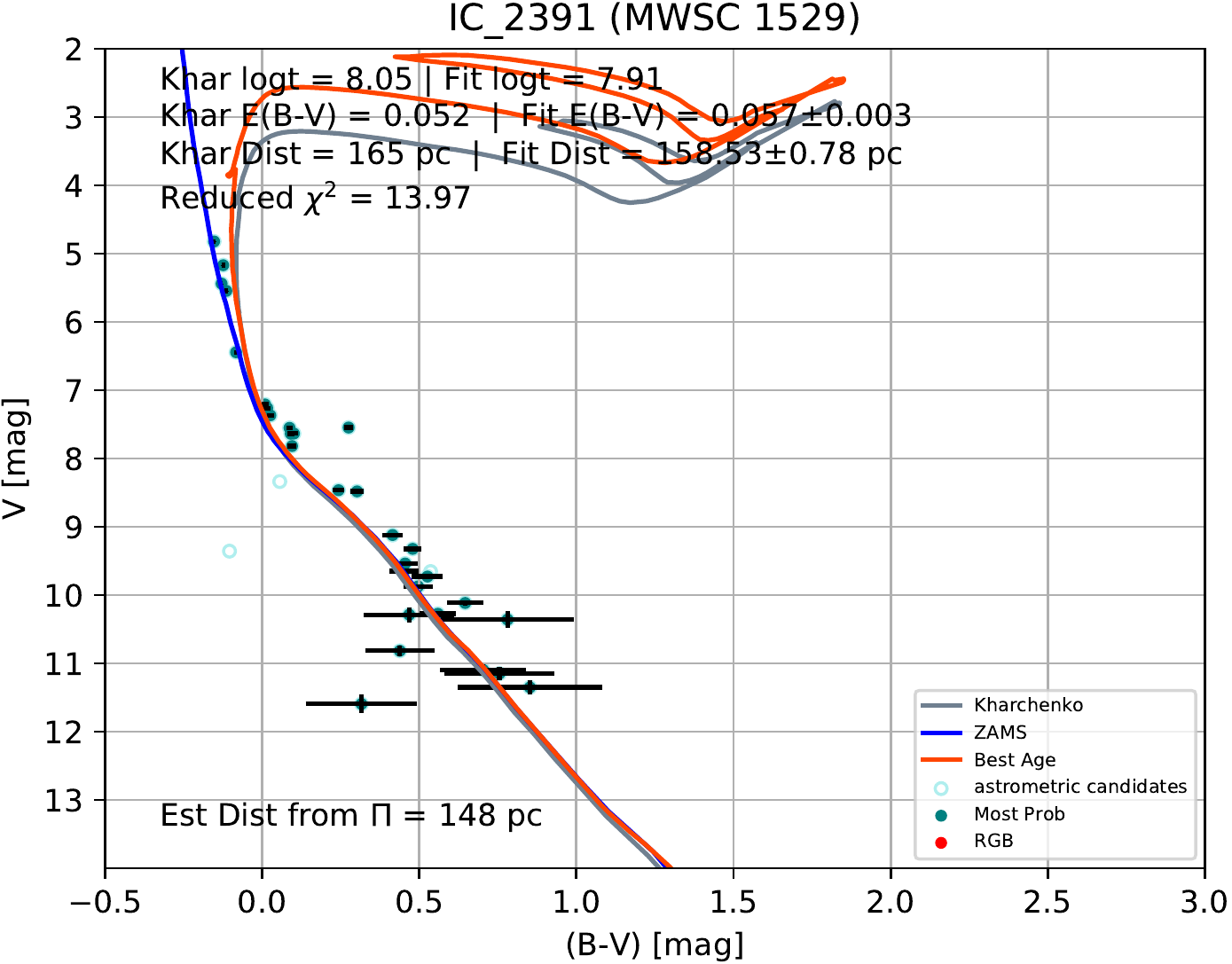}
\includegraphics[width=6cm]{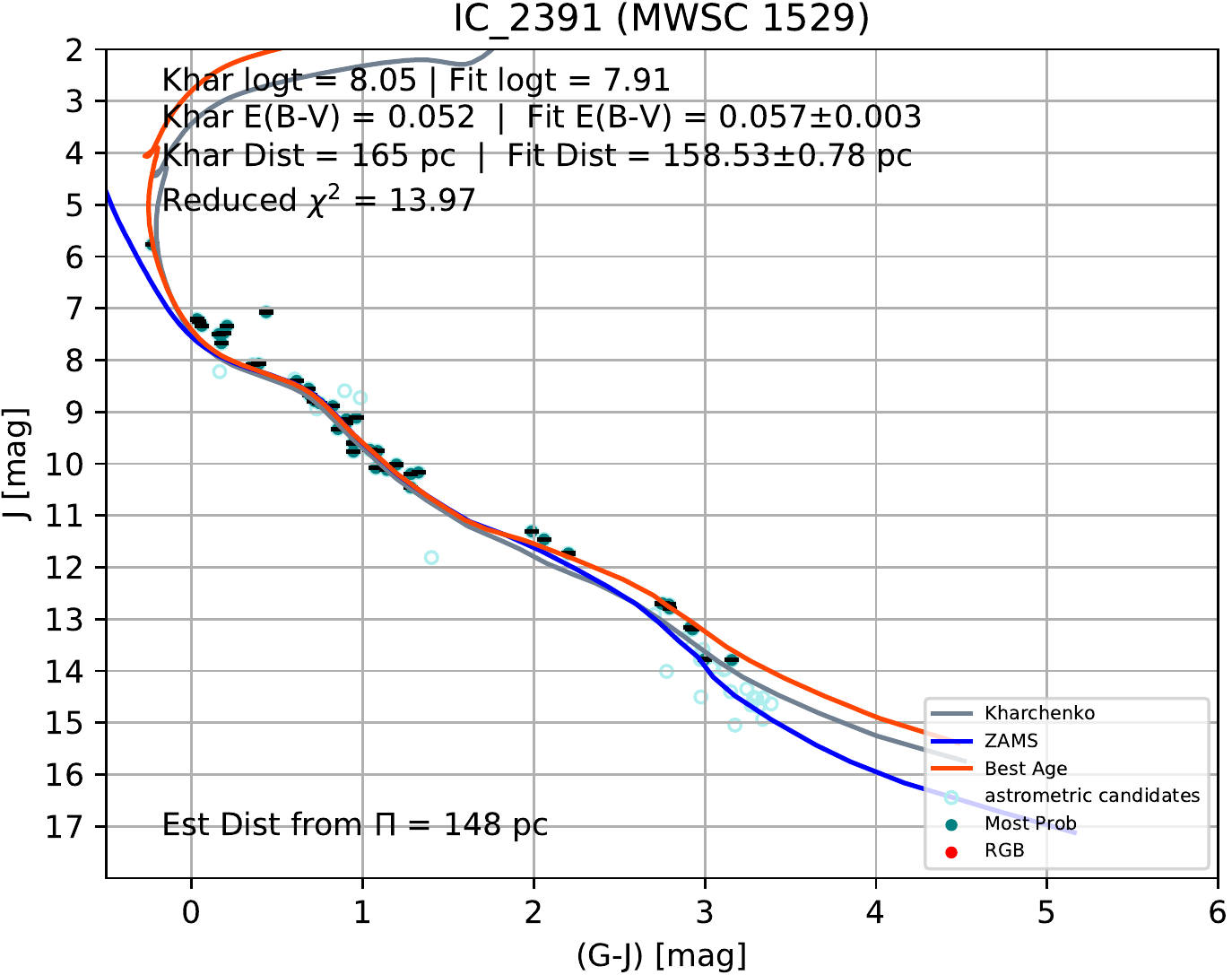}
\includegraphics[width=6cm]{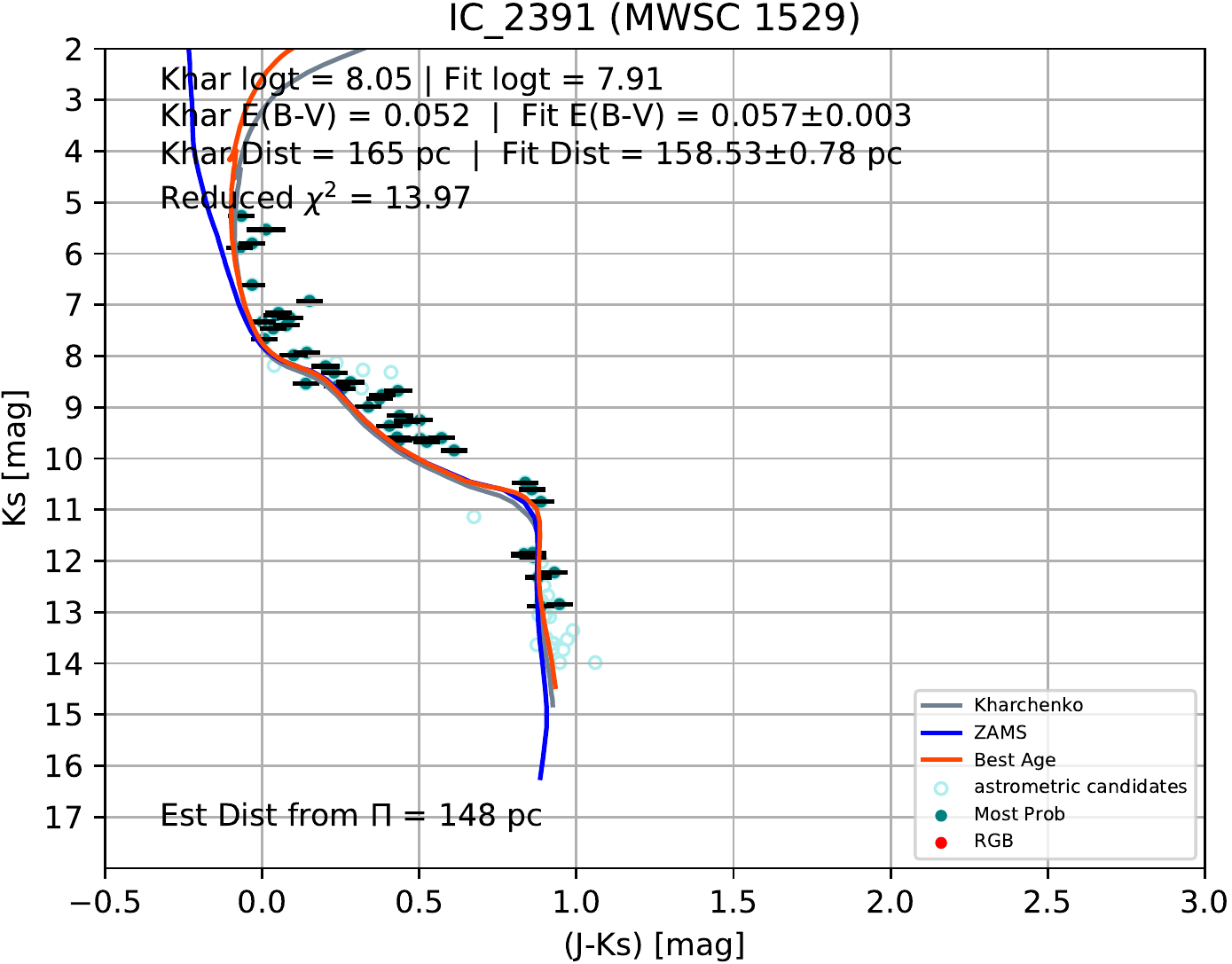}\\
\includegraphics[width=6cm]{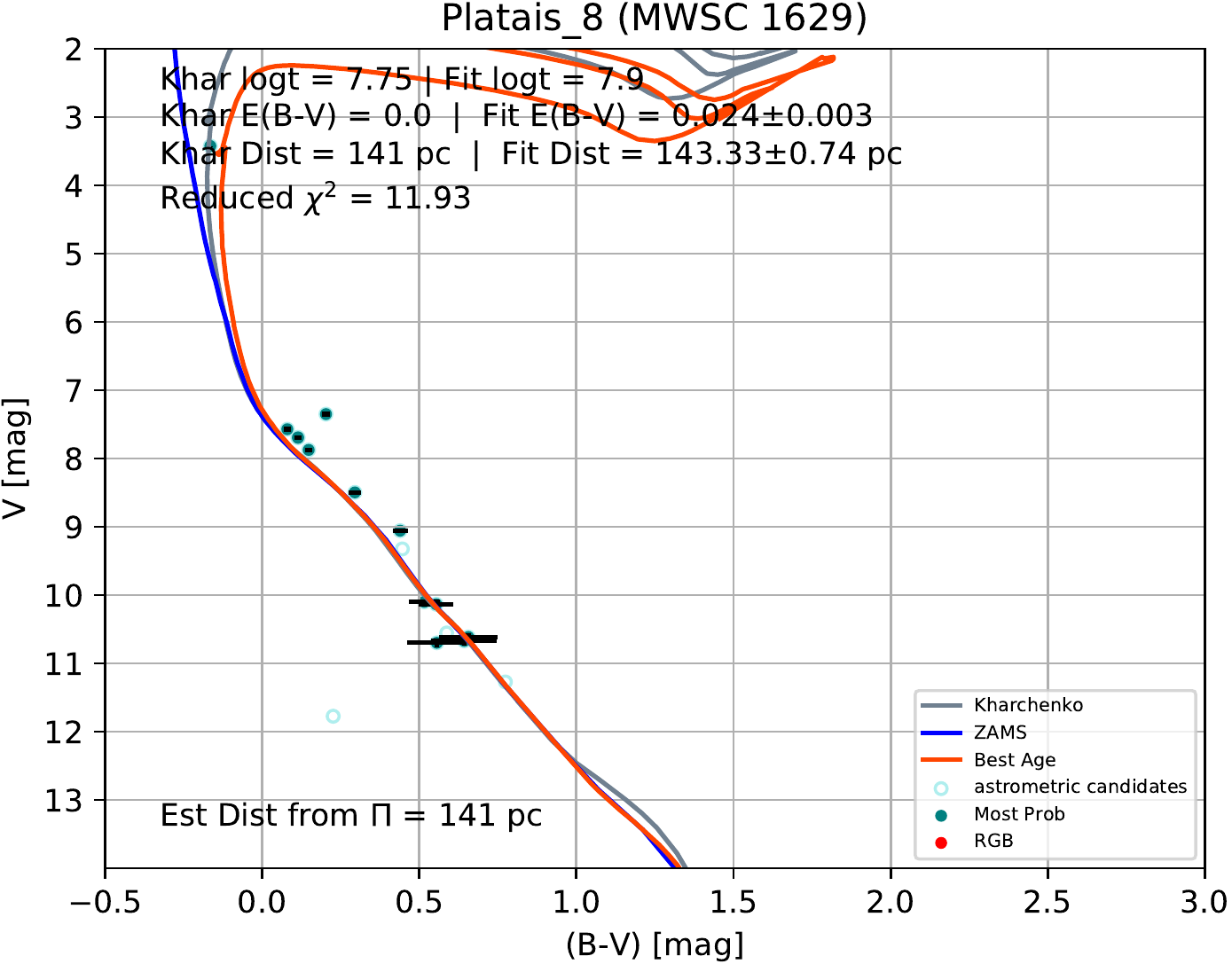}
\includegraphics[width=6cm]{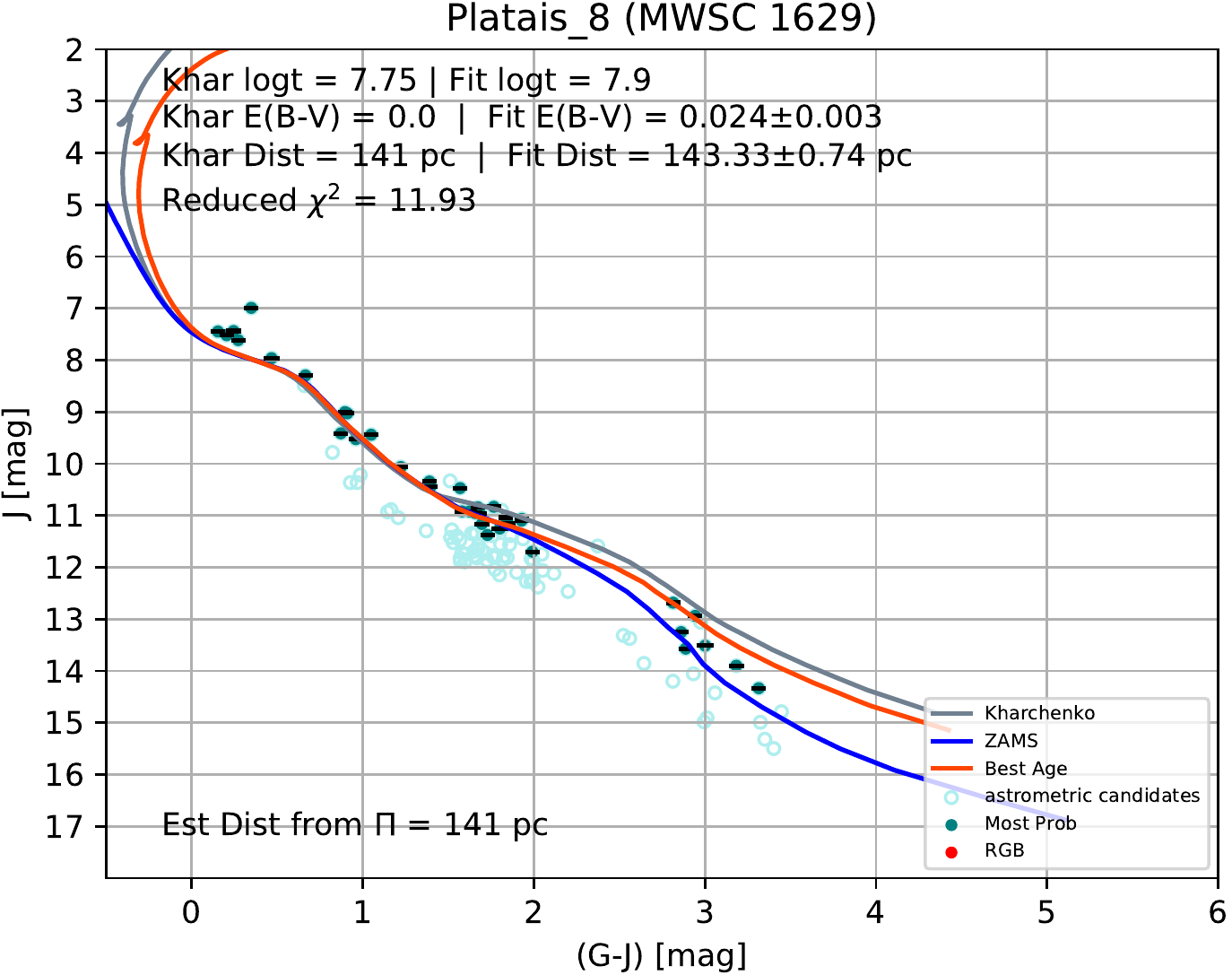}
\includegraphics[width=6cm]{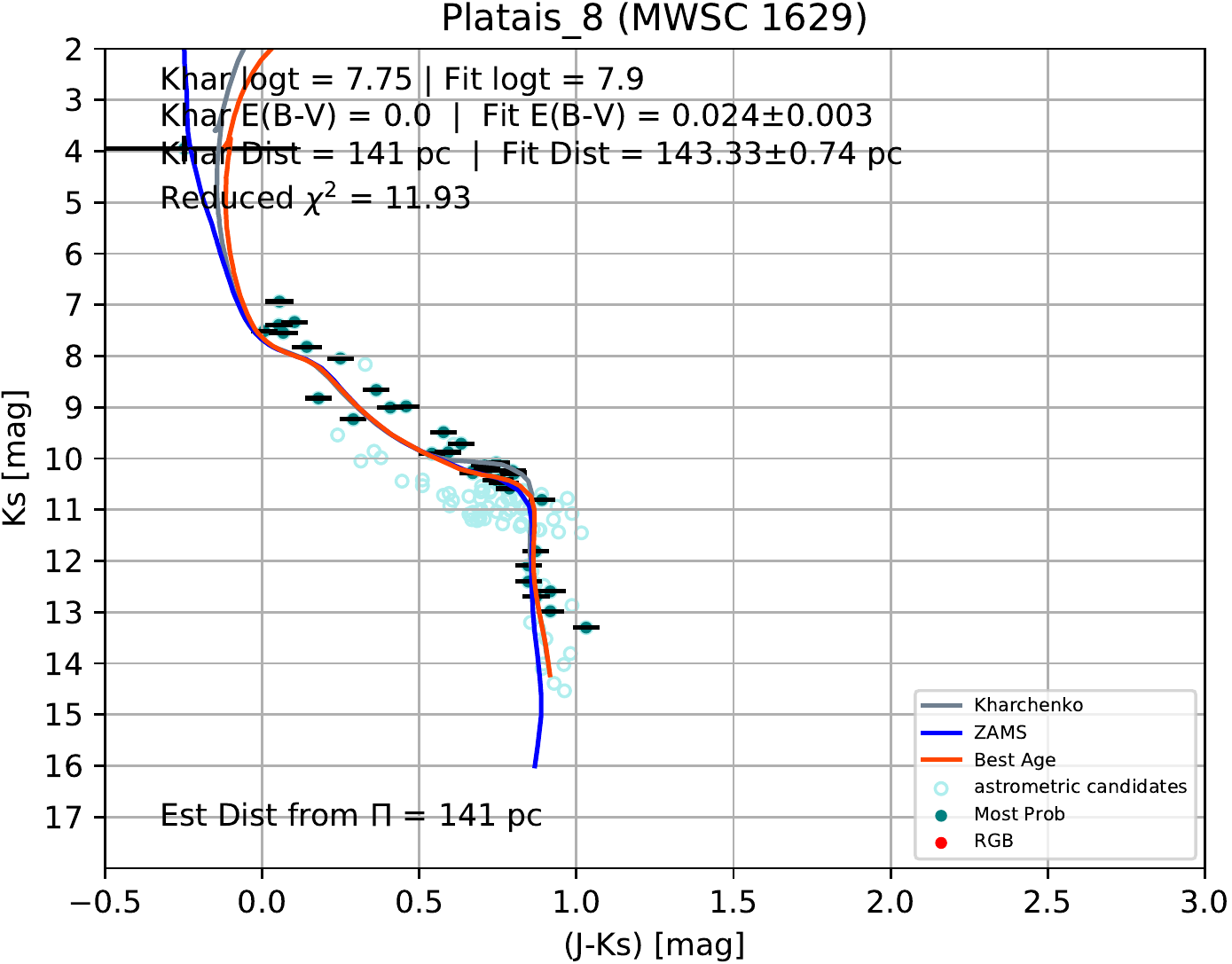}\\
\includegraphics[width=6cm]{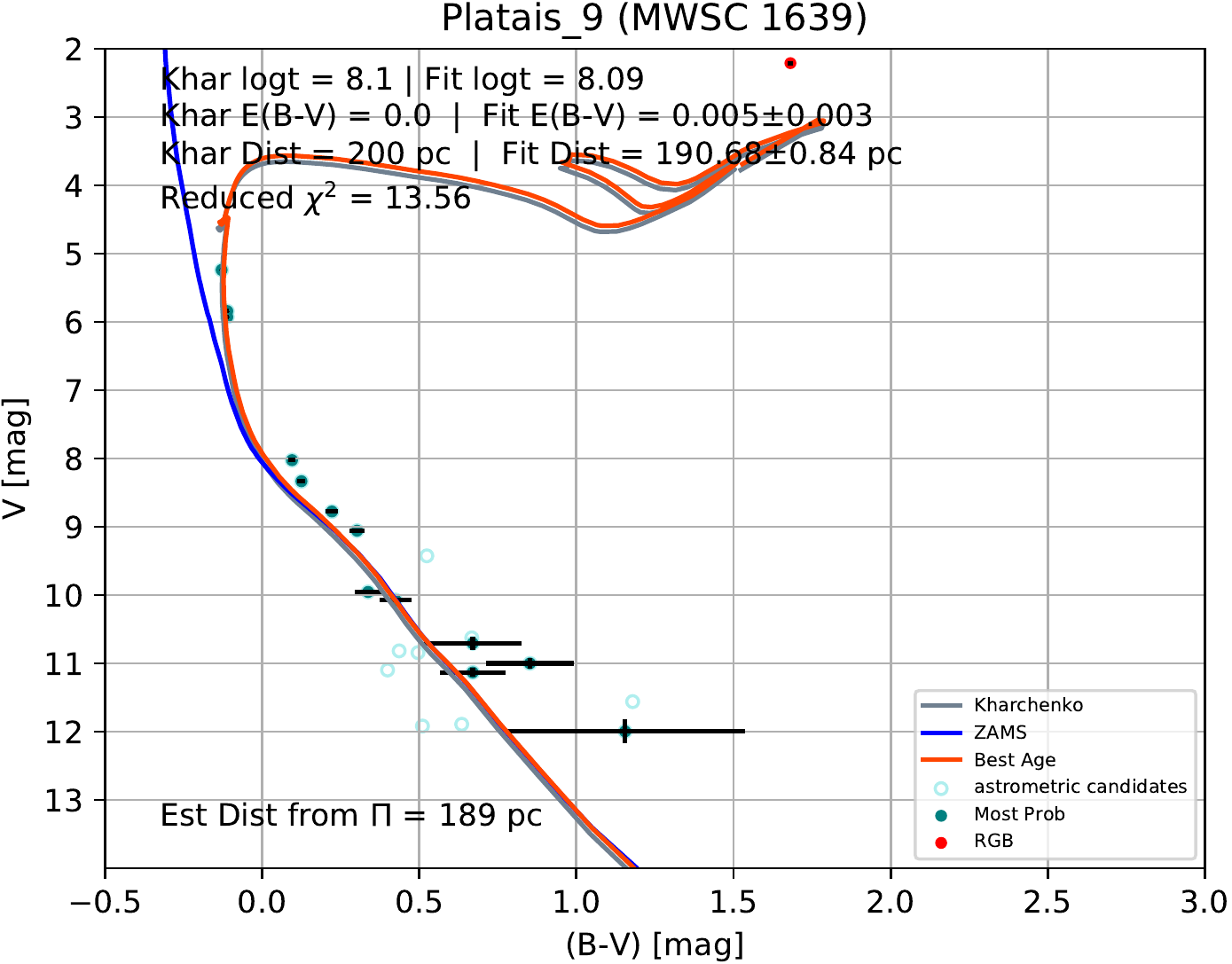}
\includegraphics[width=6cm]{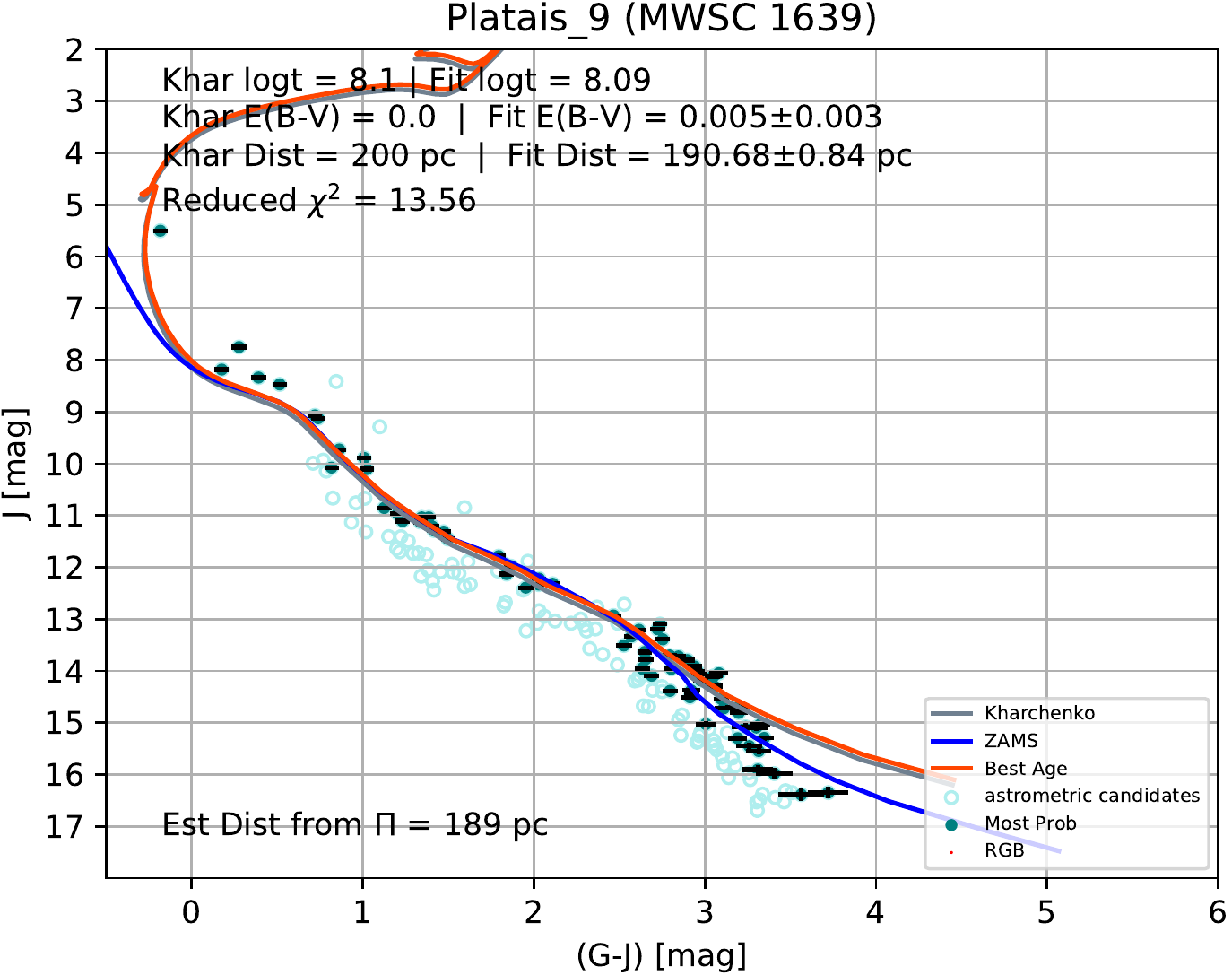}
\includegraphics[width=6cm]{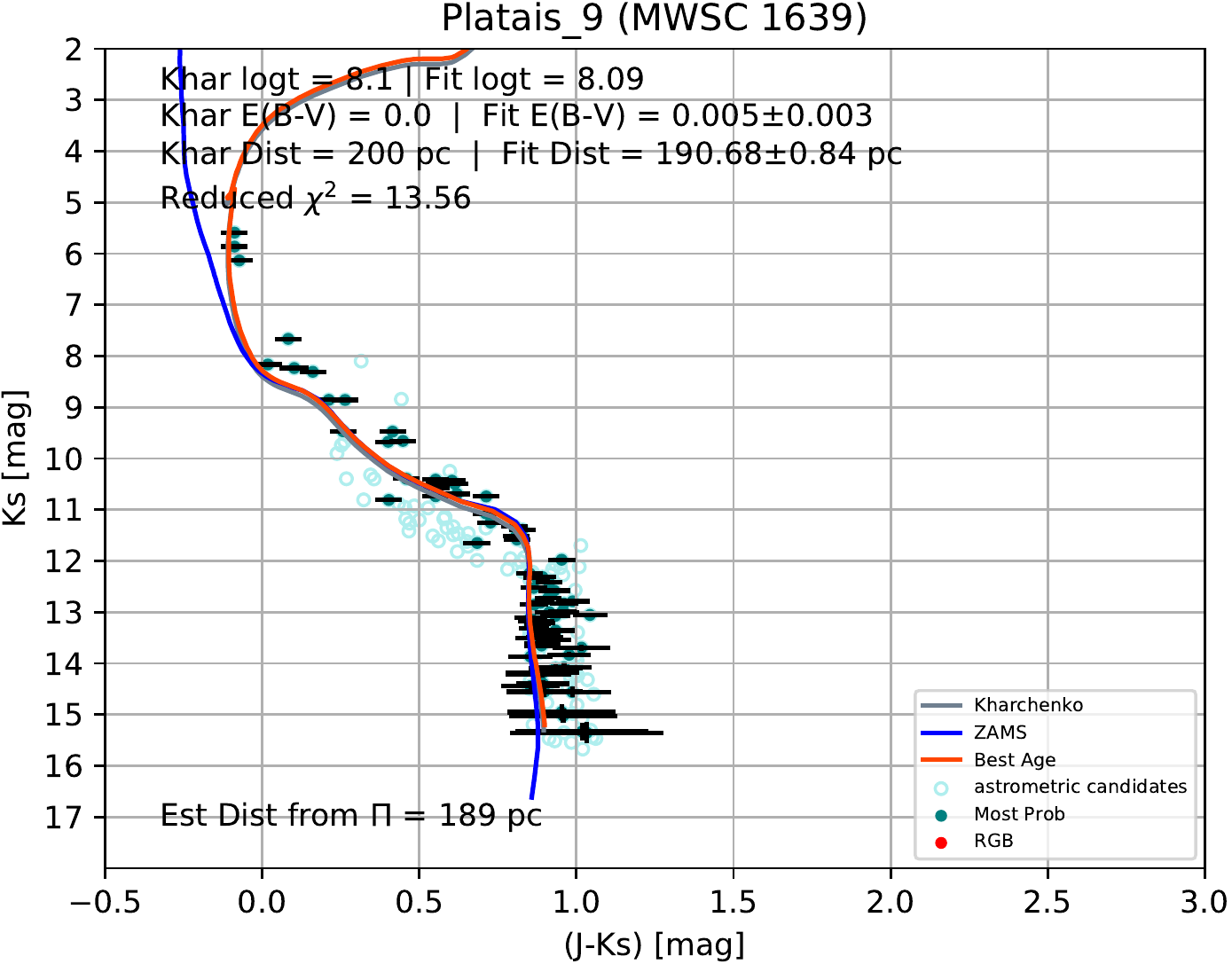}\\
\includegraphics[width=6cm]{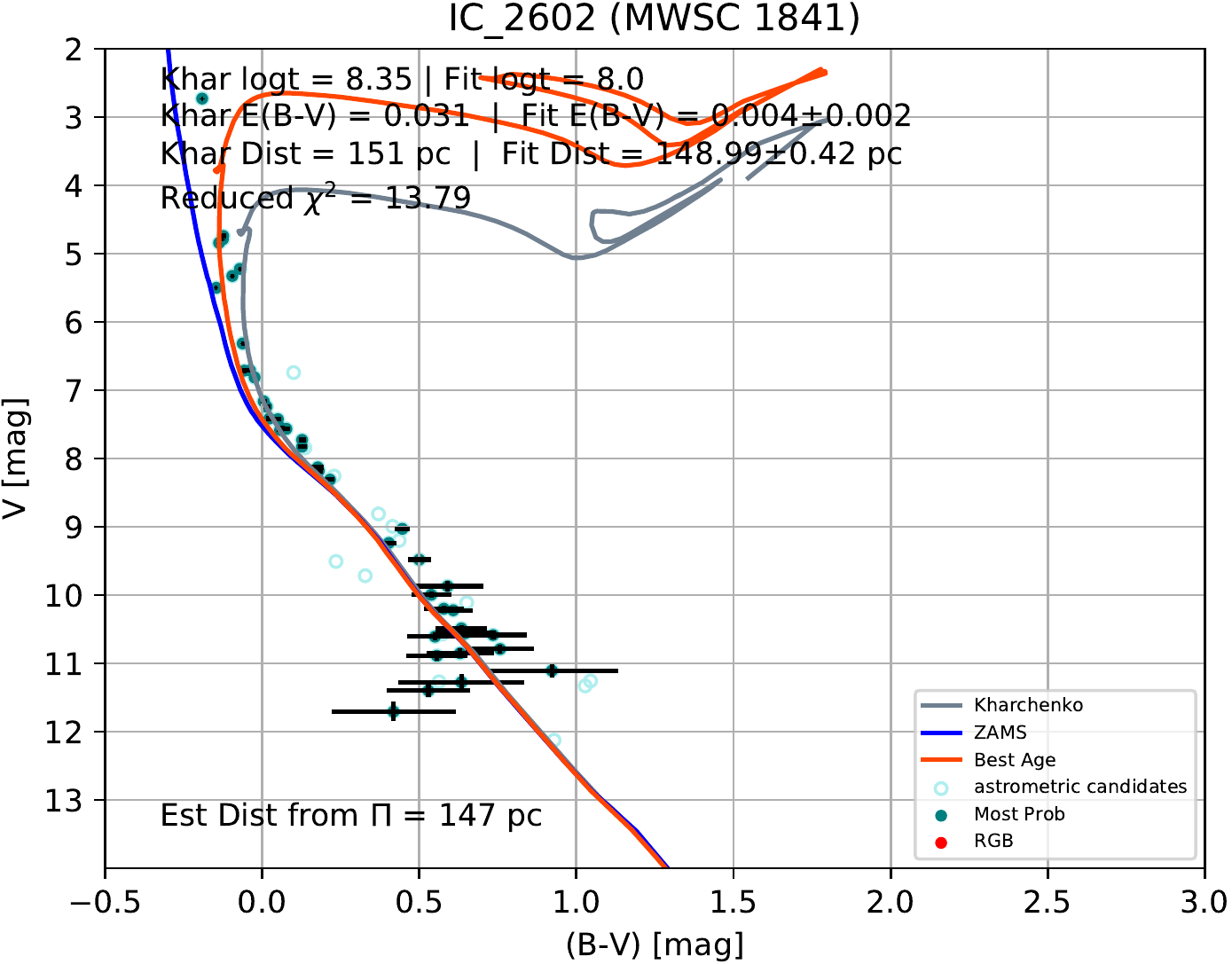}
\includegraphics[width=6cm]{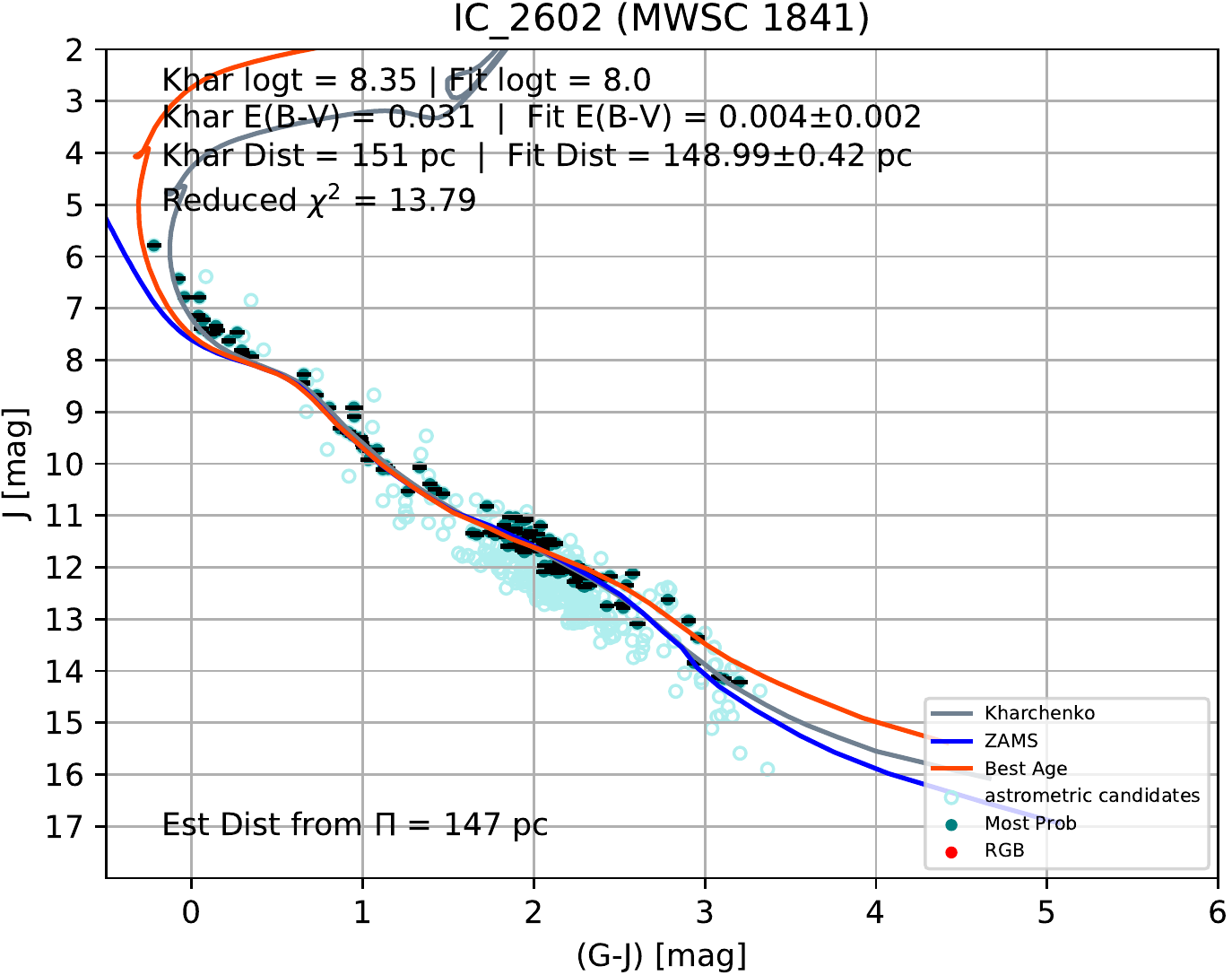}
\includegraphics[width=6cm]{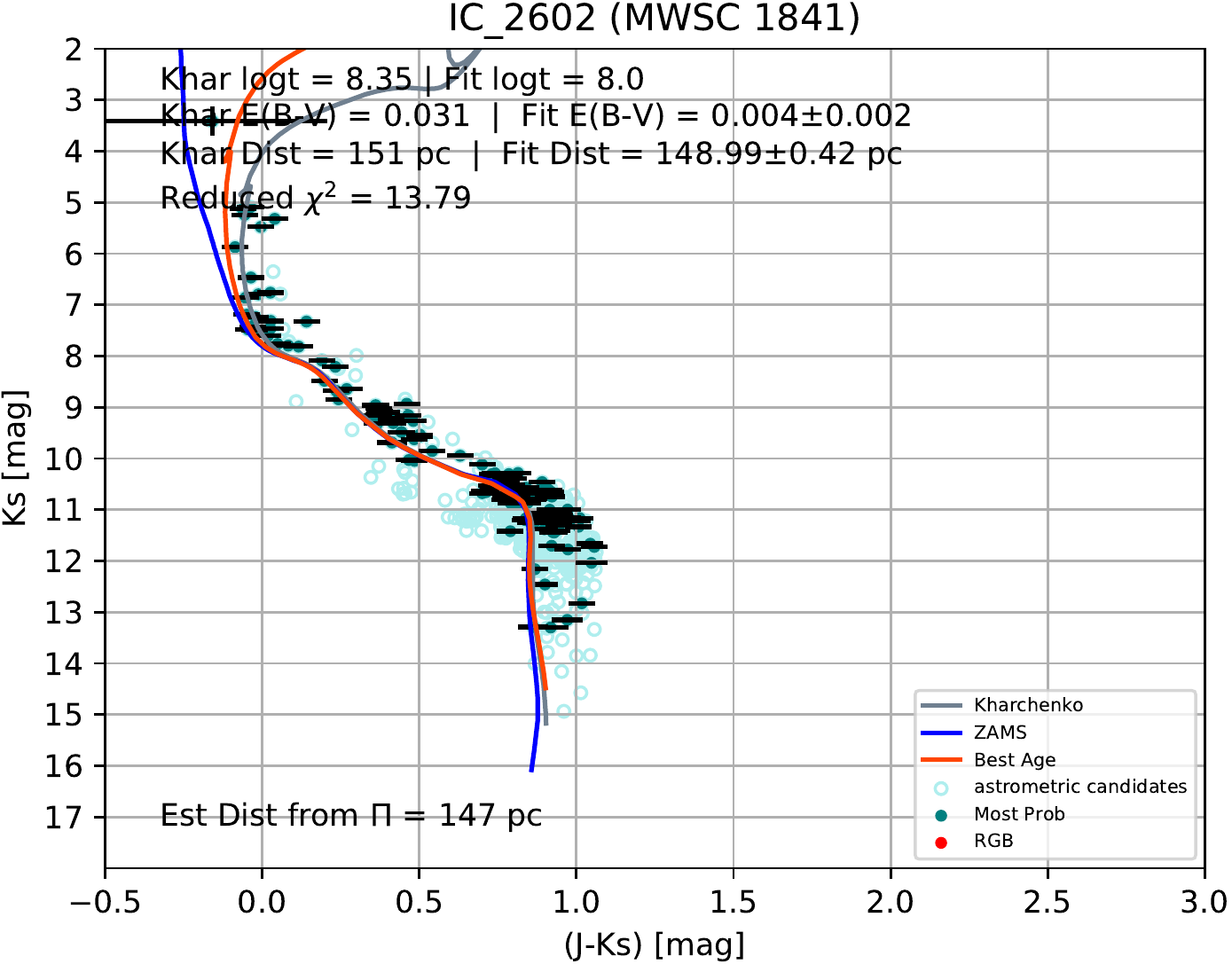}\\
\caption{Color-magnitude diagrams for clusters, from top to bottom: IC~2391, Platais~8, Platais~9, and IC~2602. From left to right: their respective $V$ vs. $(B-V)$, $J$ vs. $(G-J)$, and $K_s$ vs. $(J-K_s)$ CMDs. The cluster members determined from the pipeline are given by teal circles with their corresponding magnitude and color error bars. The cluster astrometric candidates that were later rejected as cluster members are shown by light blue open circles. RGB stars, if any, are indicated by red circles. The red isochrone is the pipeline selected age, plotted with the fitted cluster $E(B-V)$ and $d$. This isochrone is plotted without the binary offset. The gray isochrone shows the age, $E(B-V)$, and $d$ as determined by \citet{2013A&A...558A..53K}. The blue line is the ZAMS plotted with the fitted cluster $E(B-V)$ and $d$.}
 \label{figa4}
\end{figure*}

\begin{figure*}
\centering
\includegraphics[width=6cm]{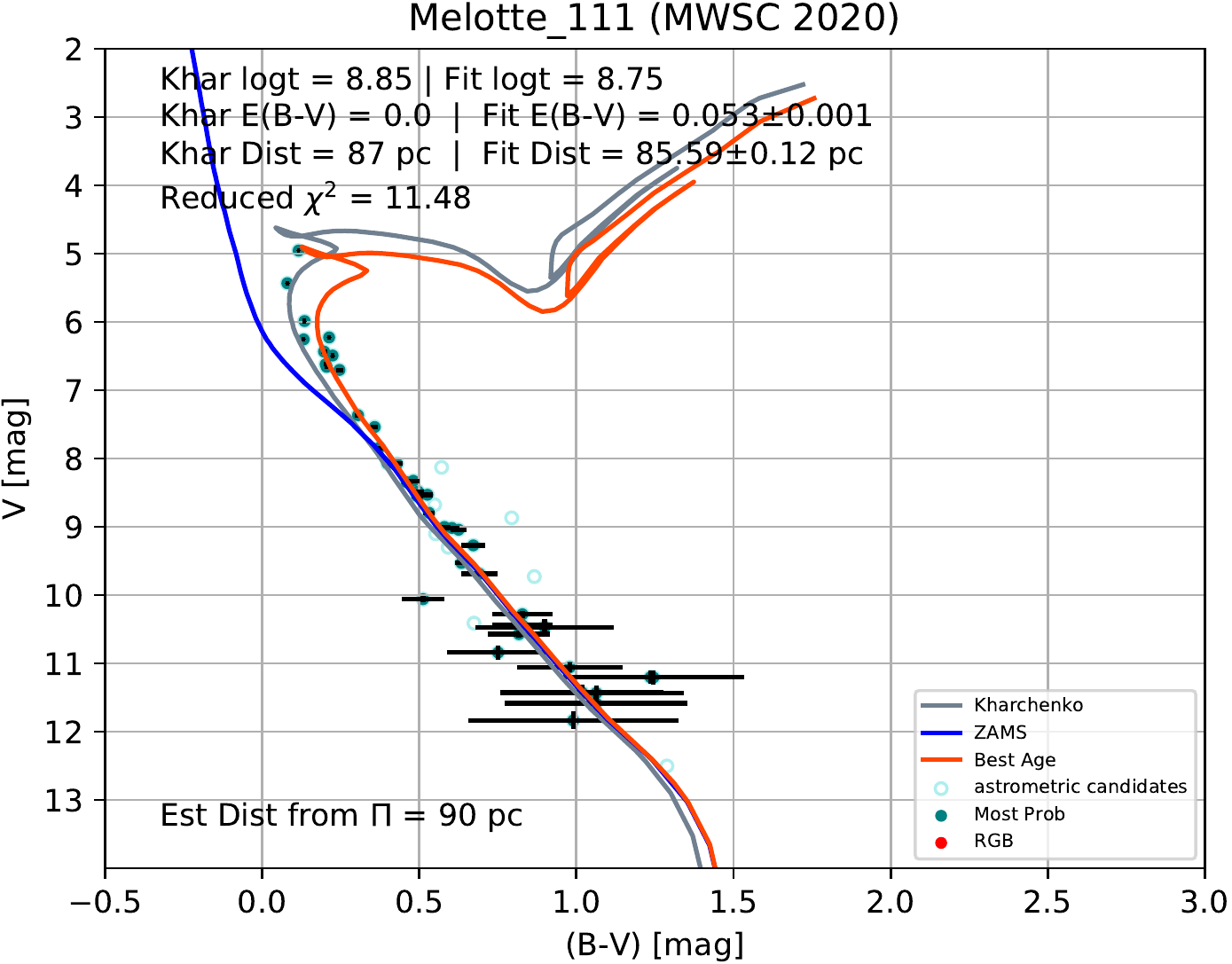}
\includegraphics[width=6cm]{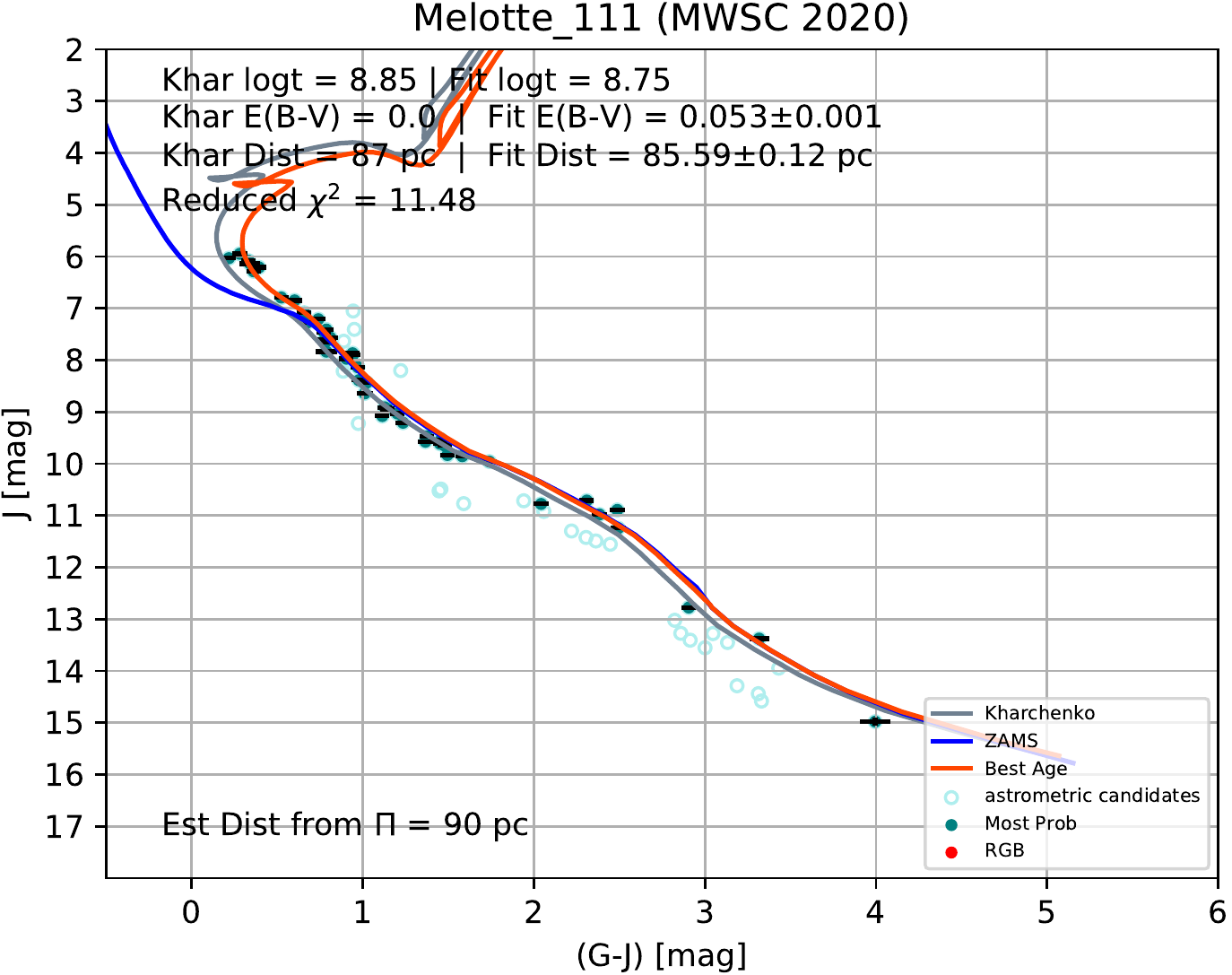}
\includegraphics[width=6cm]{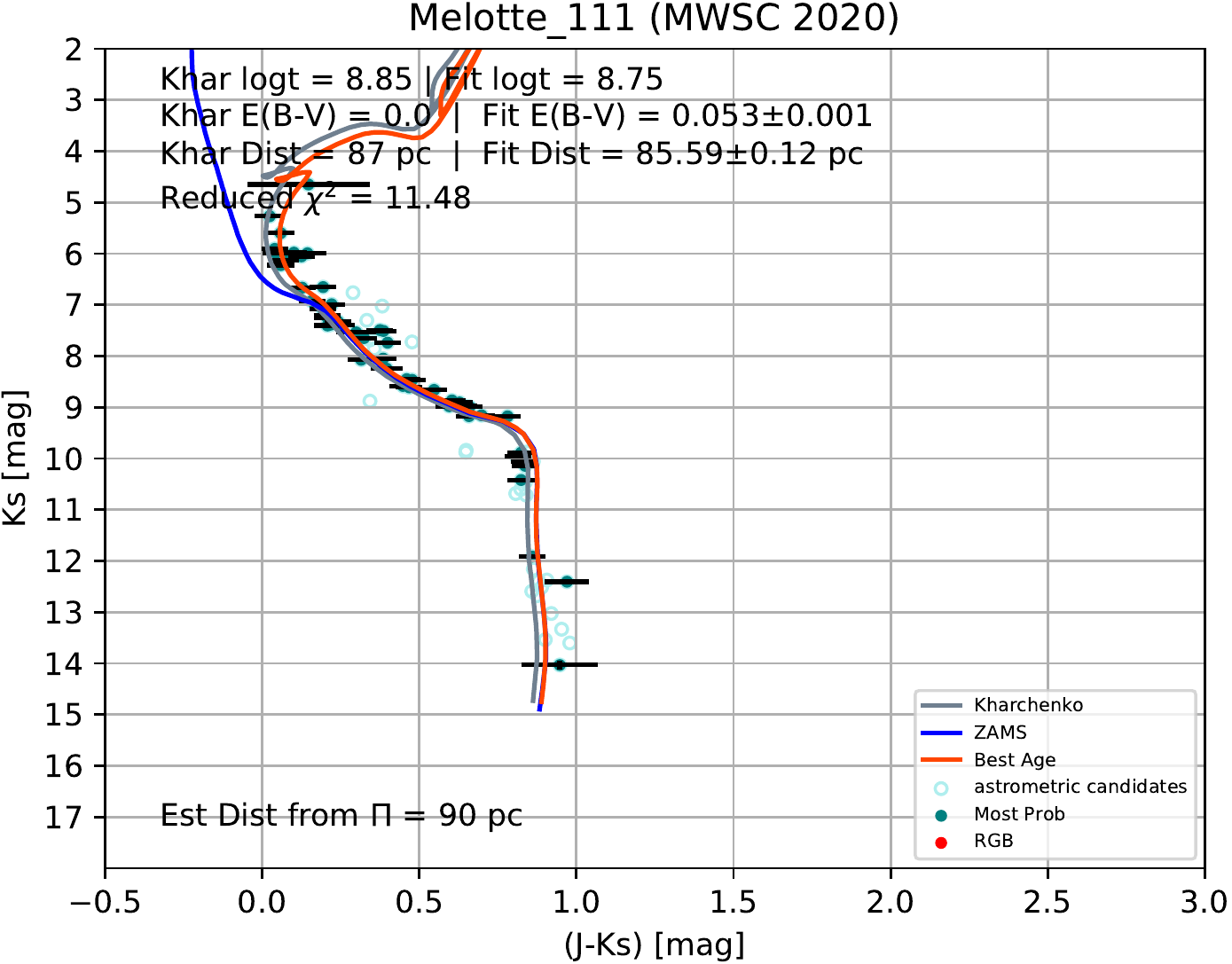}\\
\includegraphics[width=6cm]{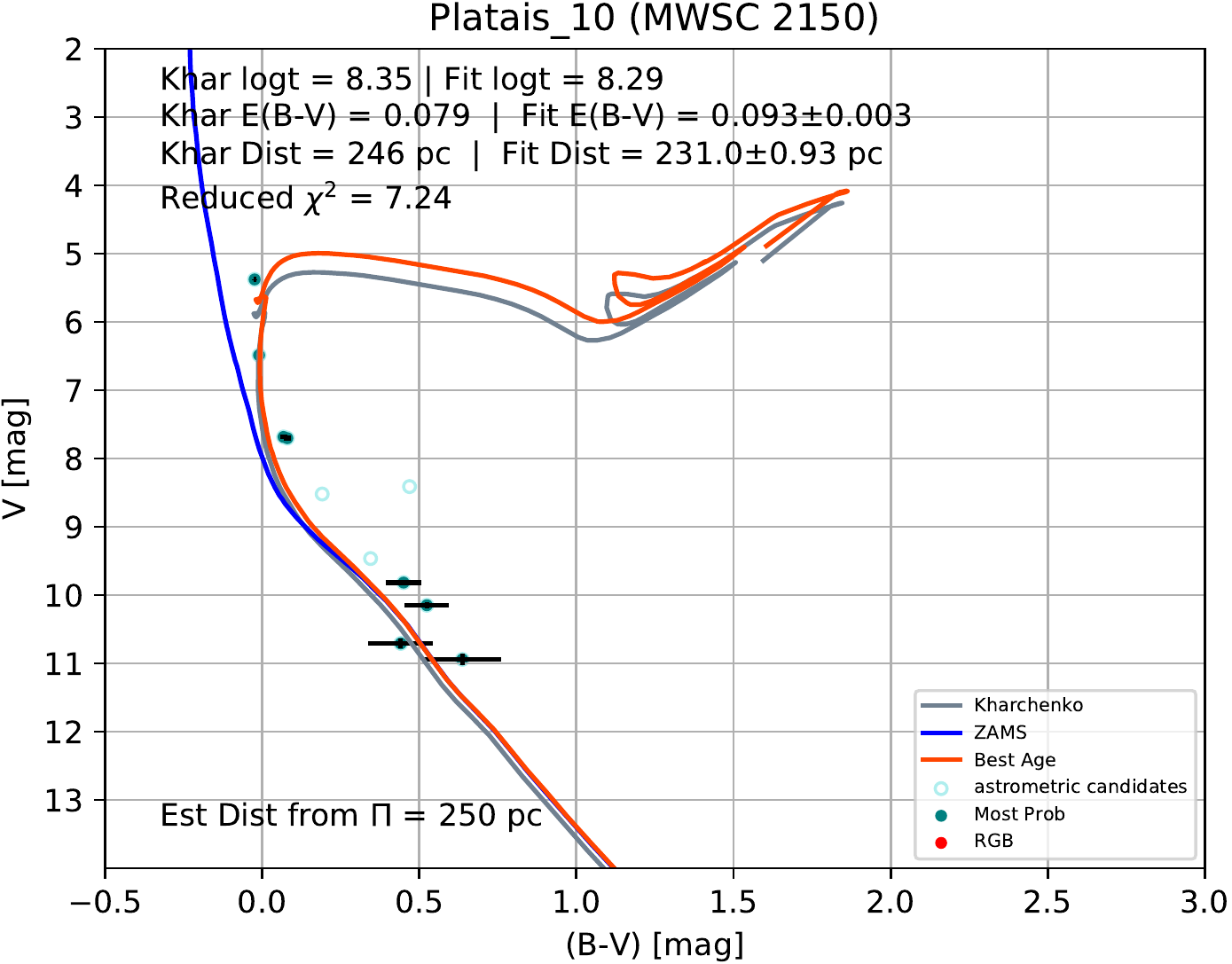}
\includegraphics[width=6cm]{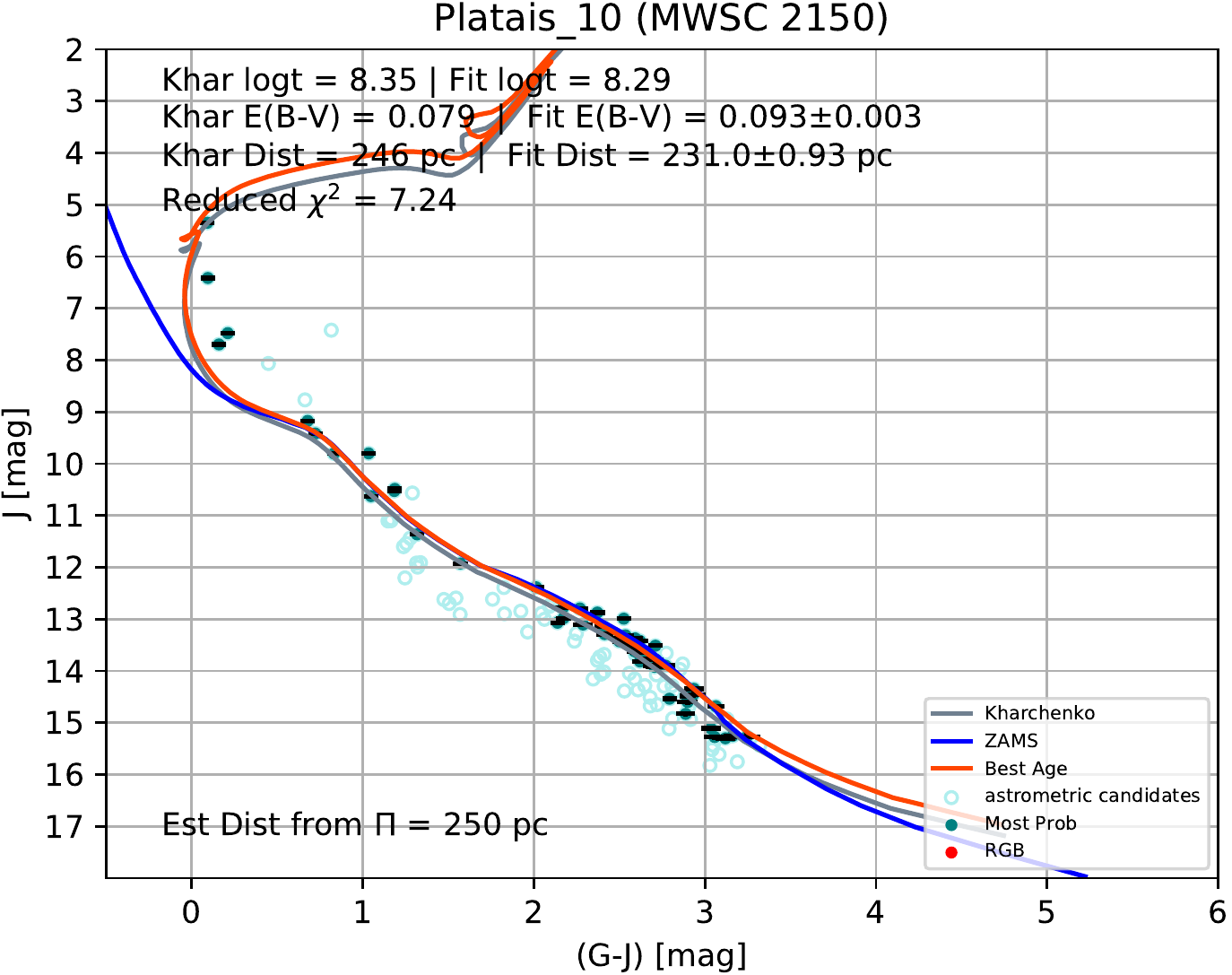}
\includegraphics[width=6cm]{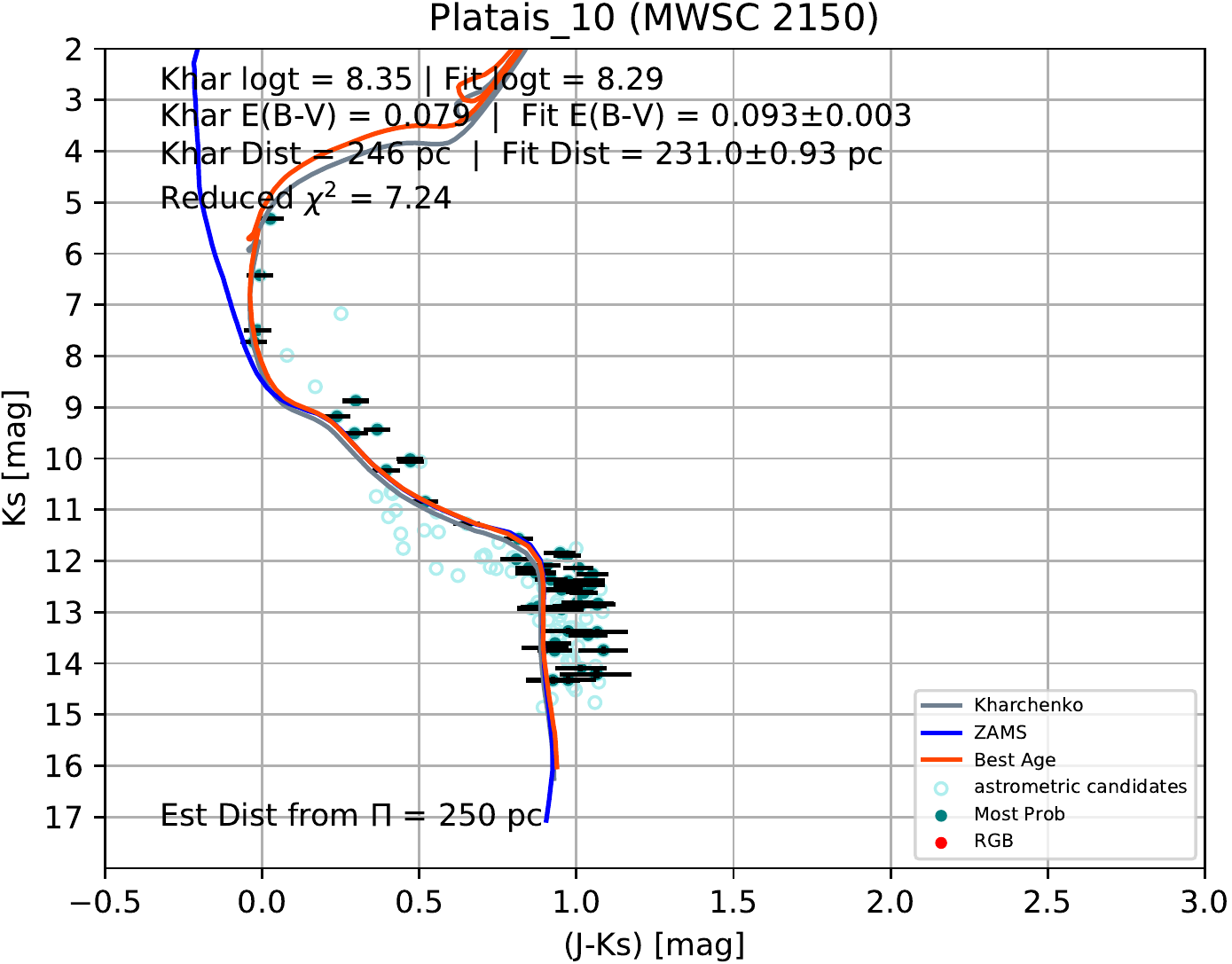}\\
\includegraphics[width=6cm]{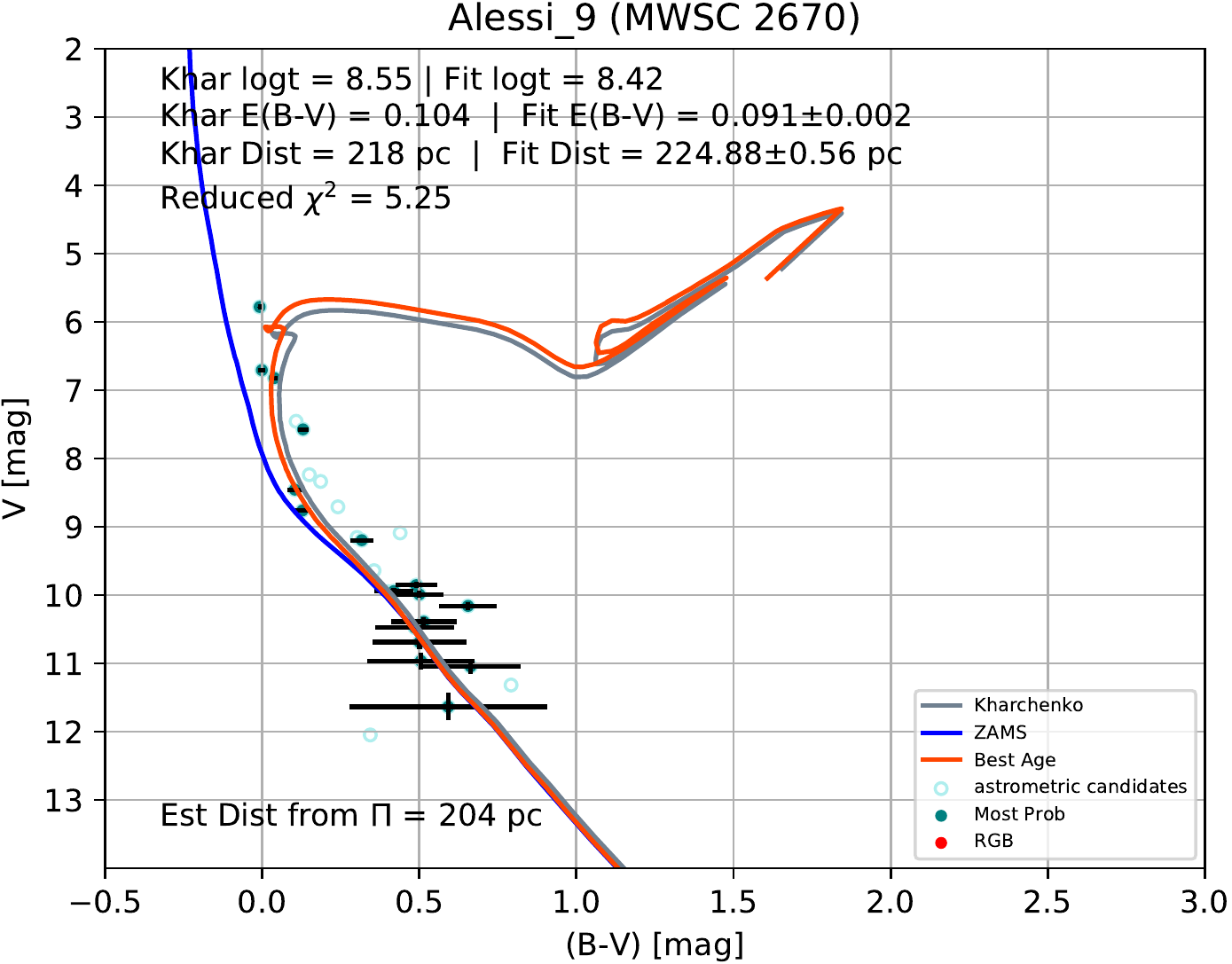}
\includegraphics[width=6cm]{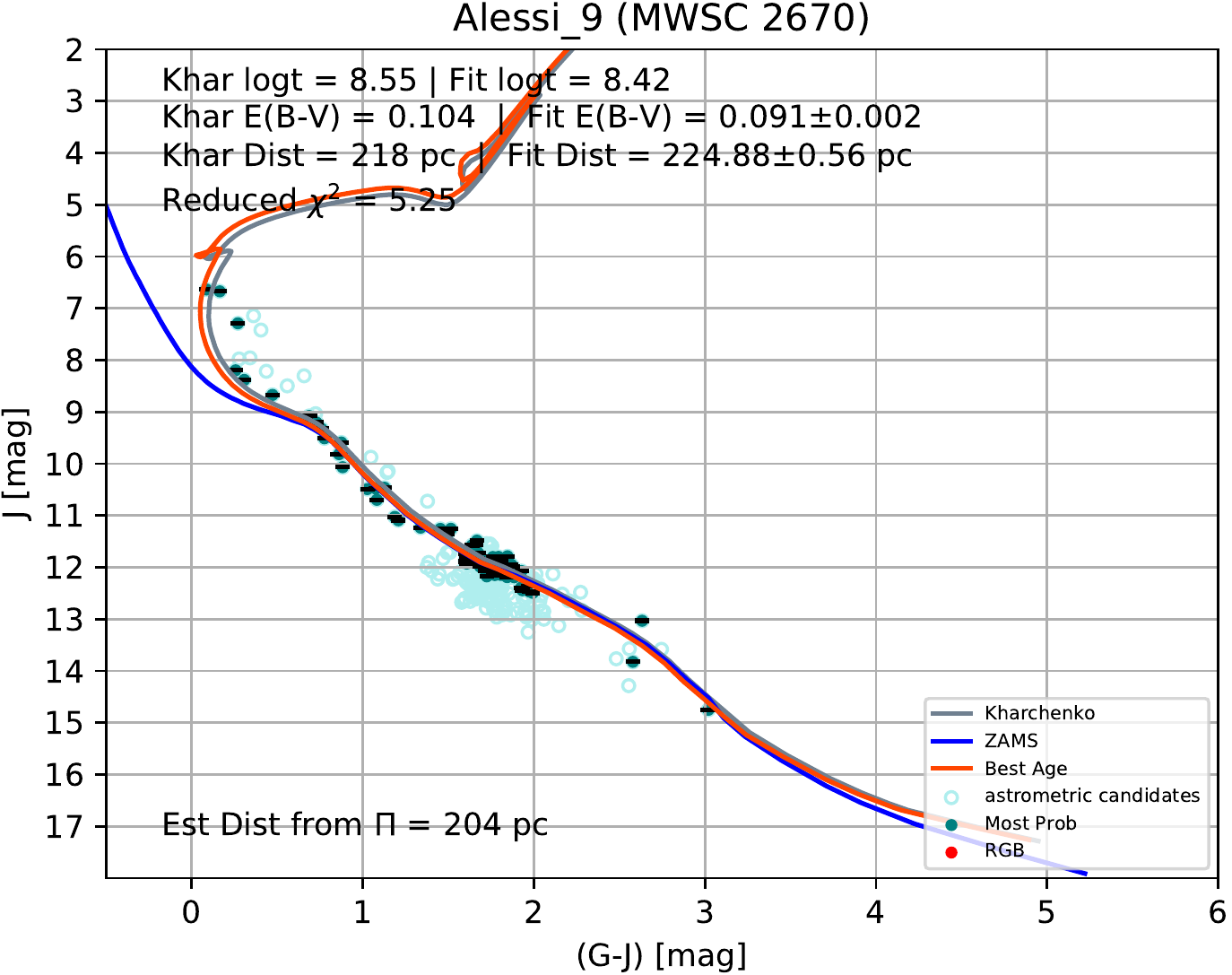}
\includegraphics[width=6cm]{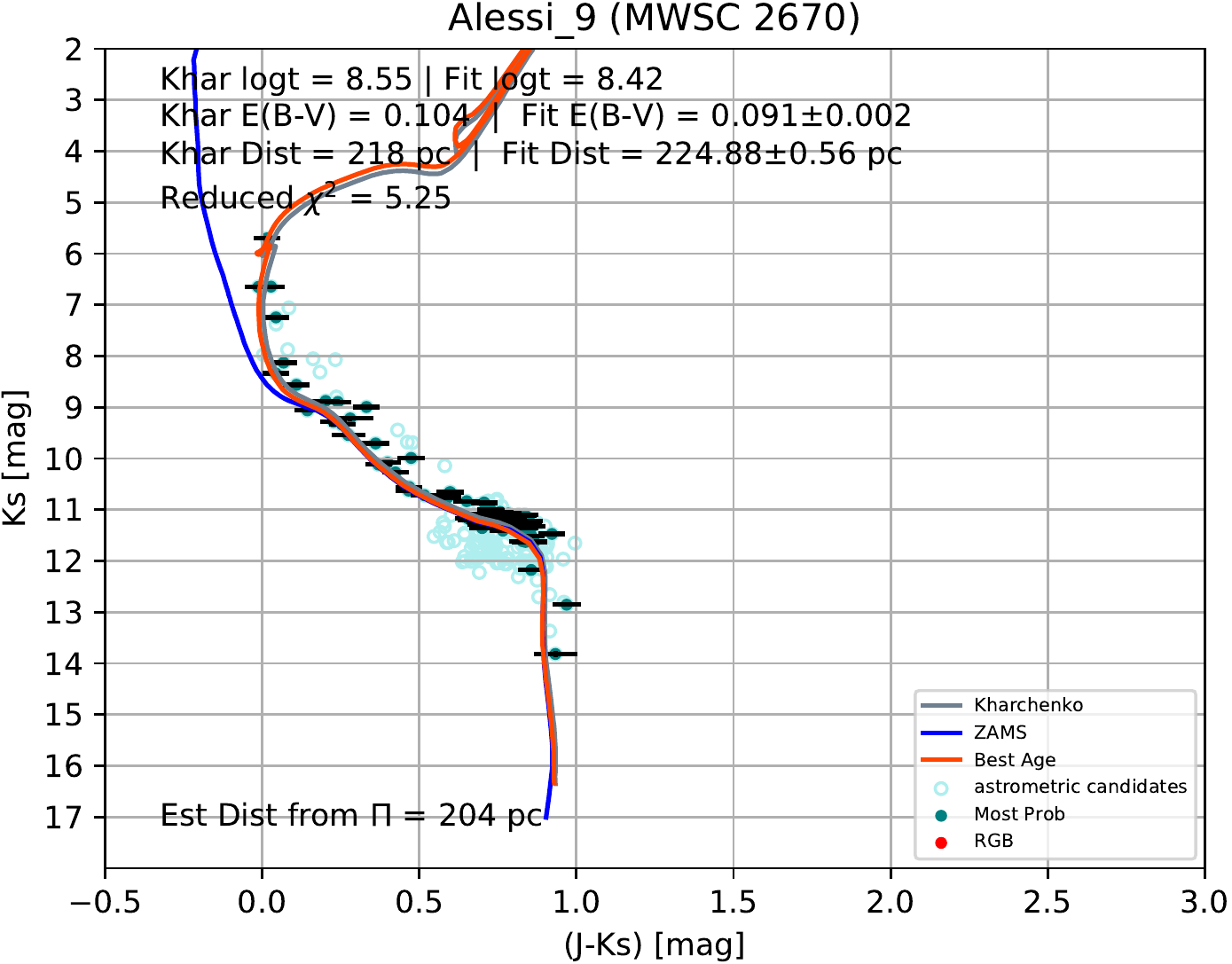}\\
\includegraphics[width=6cm]{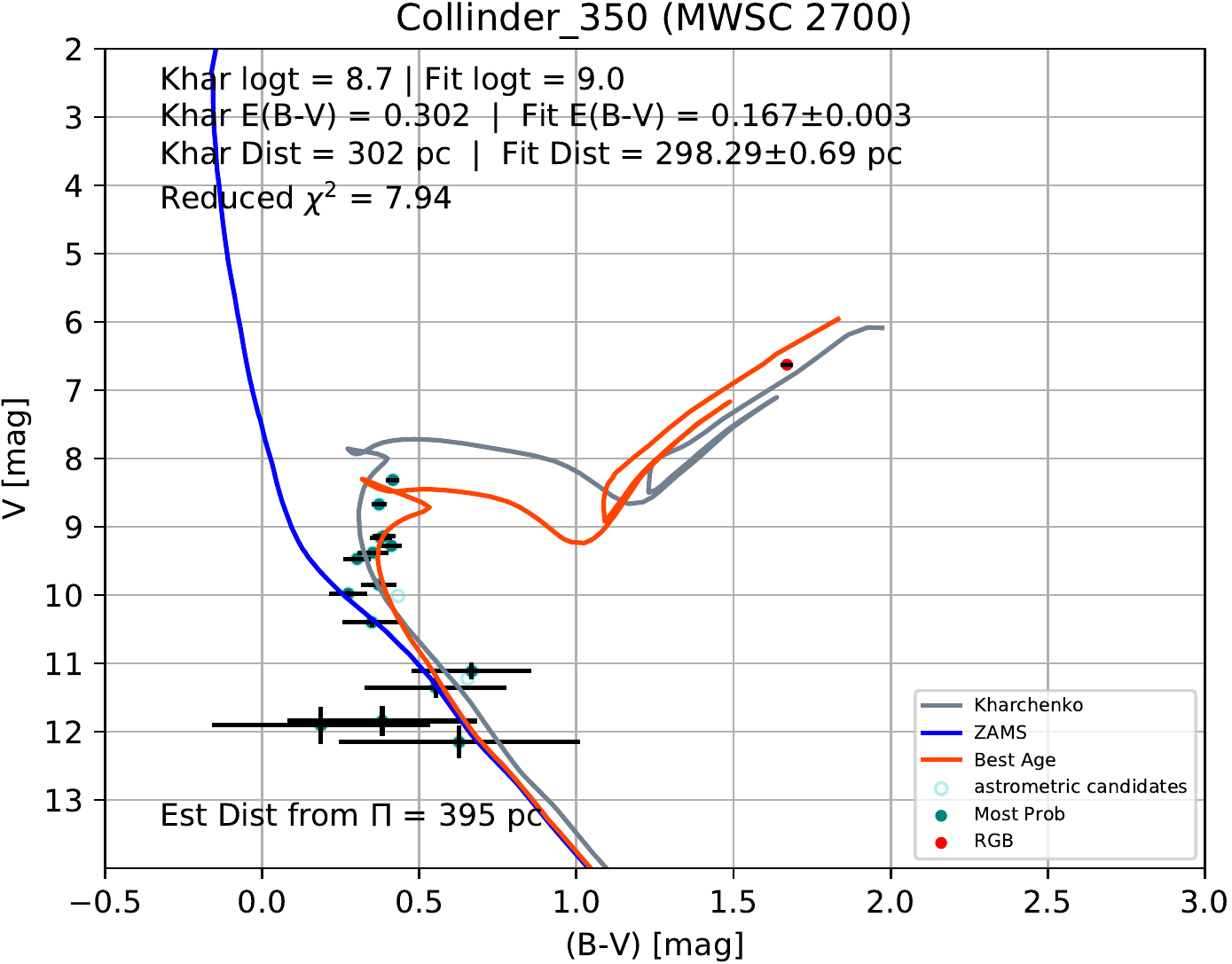}
\includegraphics[width=6cm]{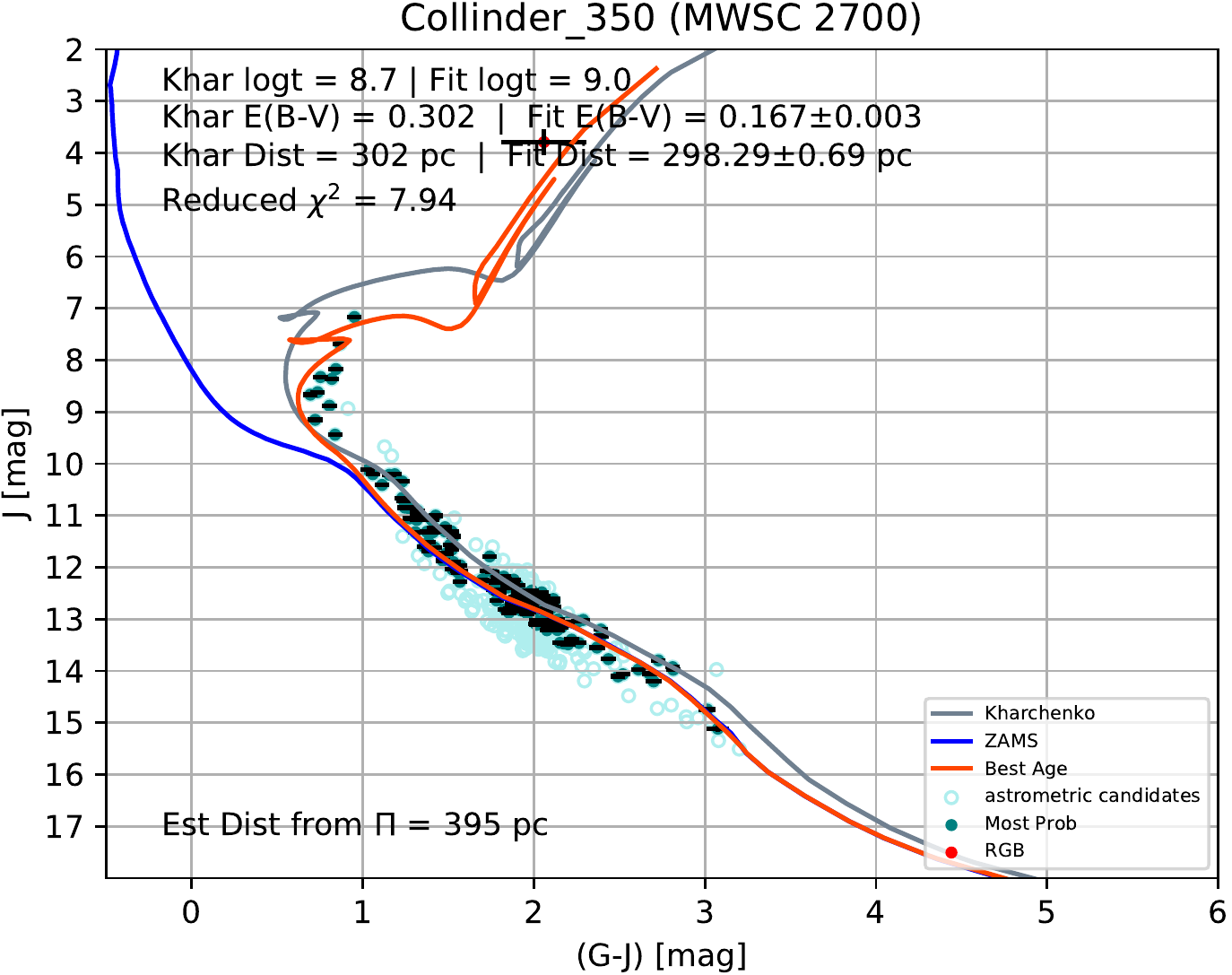}
\includegraphics[width=6cm]{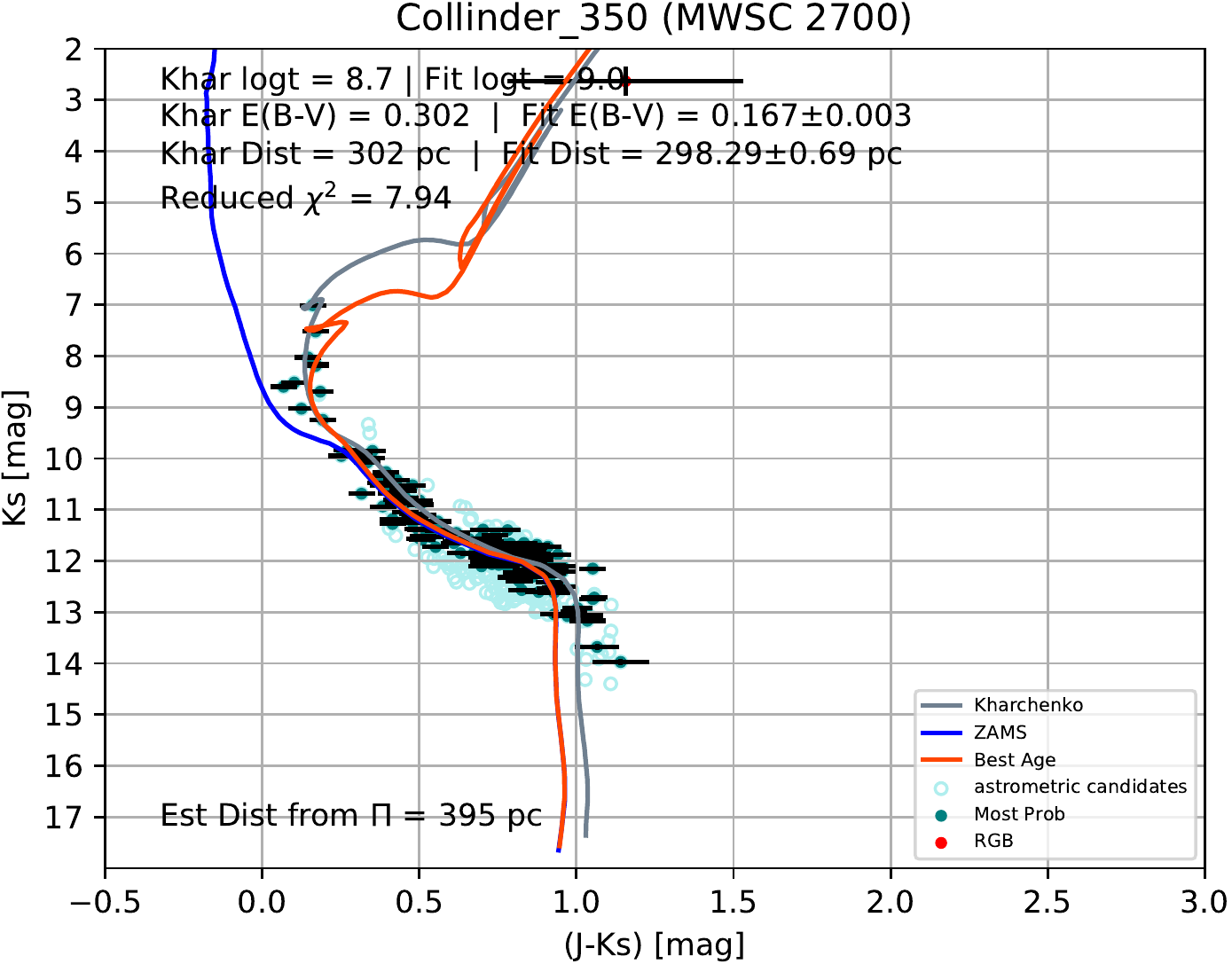}\\
\caption{Color-magnitude diagrams for clusters, from top to bottom: Coma Ber (Melotte~111), Platais~10, Alessi~9, and Collinder~350. From left to right: their respective $V$ vs. $(B-V)$, $J$ vs. $(G-J)$, and $K_s$ vs. $(J-K_s)$ CMDs. The cluster members determined from the pipeline are given by teal circles with their corresponding magnitude and color error bars. The cluster astrometric candidates that were later rejected as cluster members are shown by light blue open circles. RGB stars, if any, are indicated by red circles. The red isochrone is the pipeline selected age, plotted with the fitted cluster $E(B-V)$ and $d$. This isochrone is plotted without the binary offset. The gray isochrone shows the age, $E(B-V)$, and $d$ as determined by \citet{2013A&A...558A..53K}. The blue line is the ZAMS plotted with the fitted cluster $E(B-V)$ and $d$.}
 \label{figa5}
\end{figure*}

\begin{figure*}
\centering
\includegraphics[width=6cm]{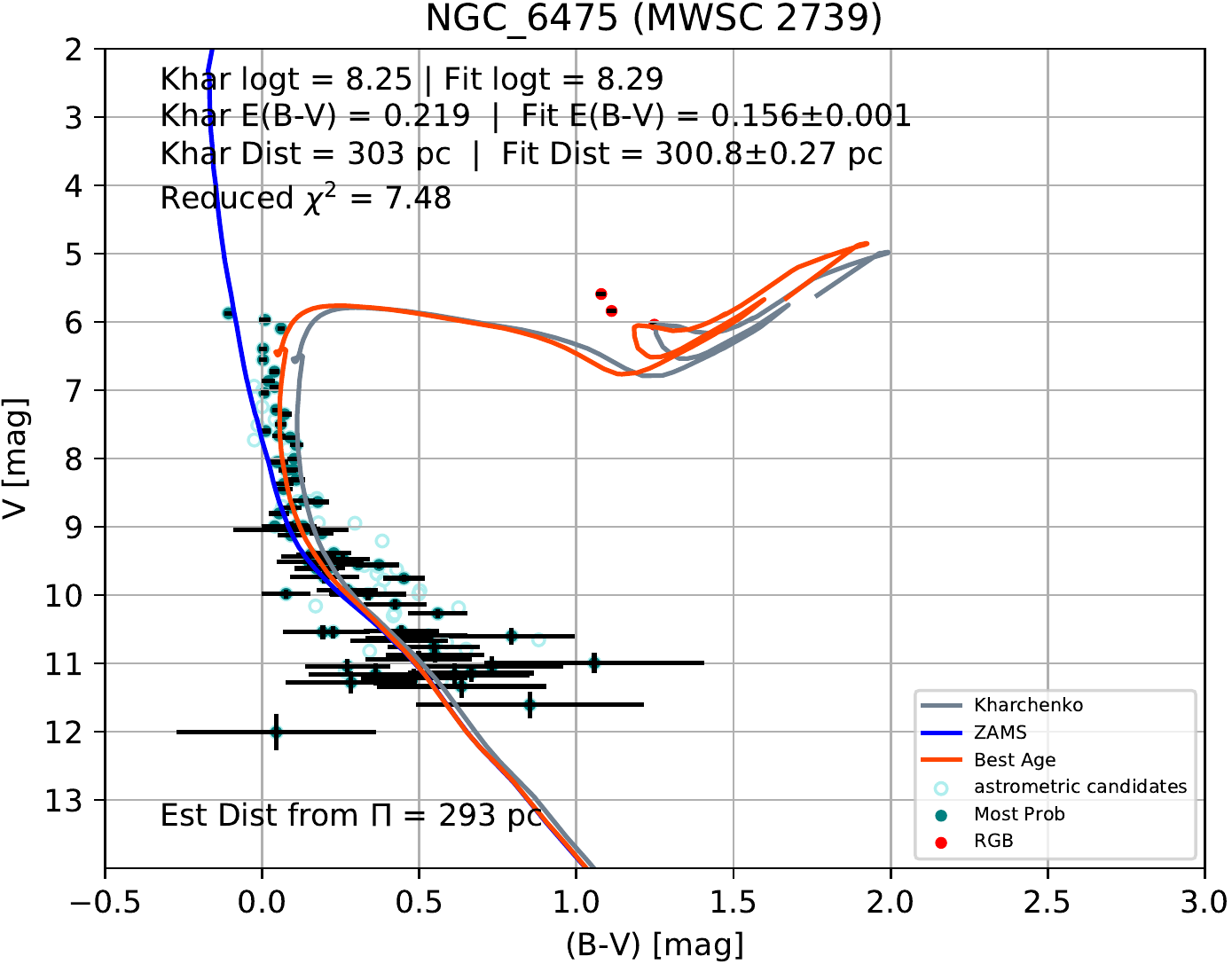}
\includegraphics[width=6cm]{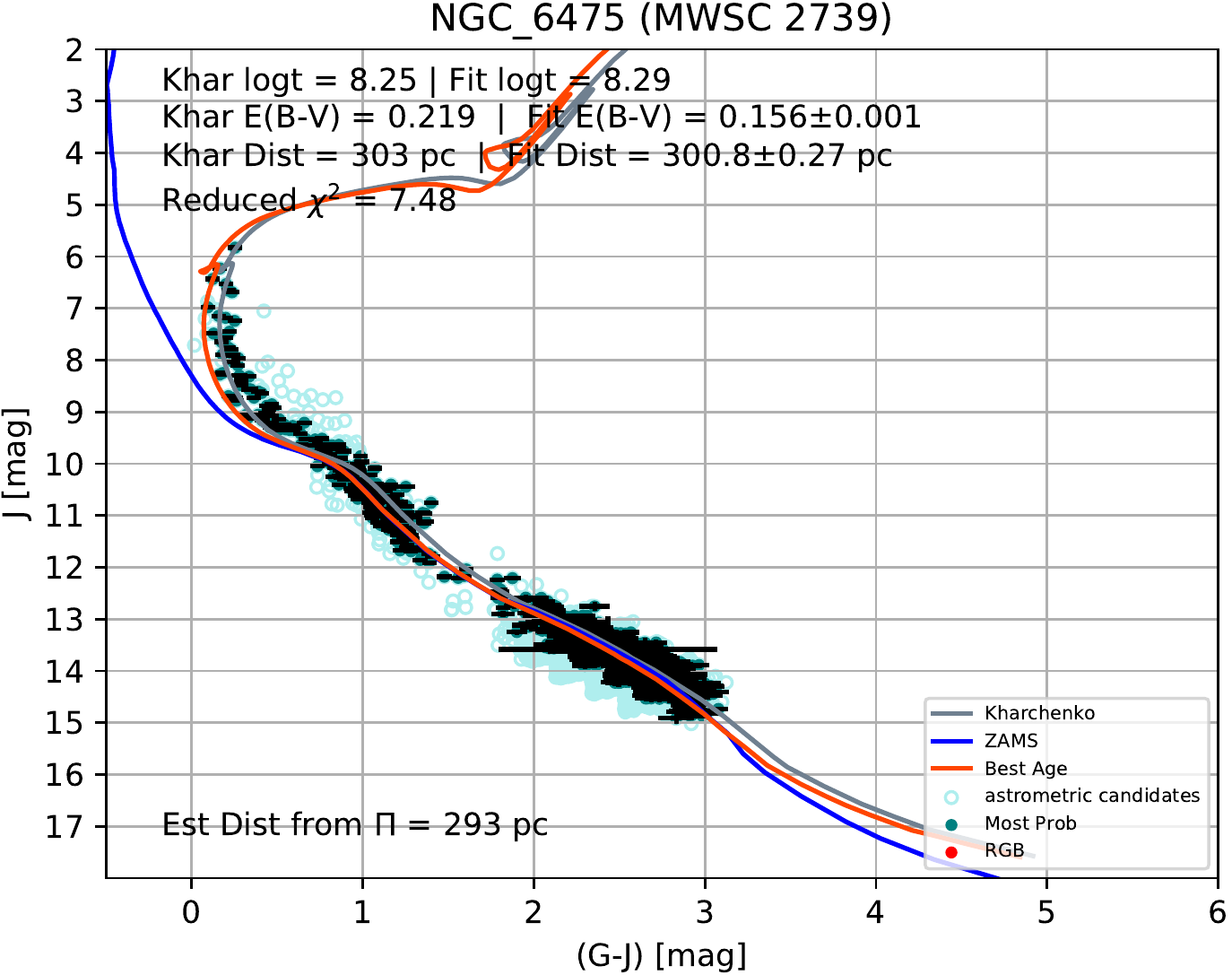}
\includegraphics[width=6cm]{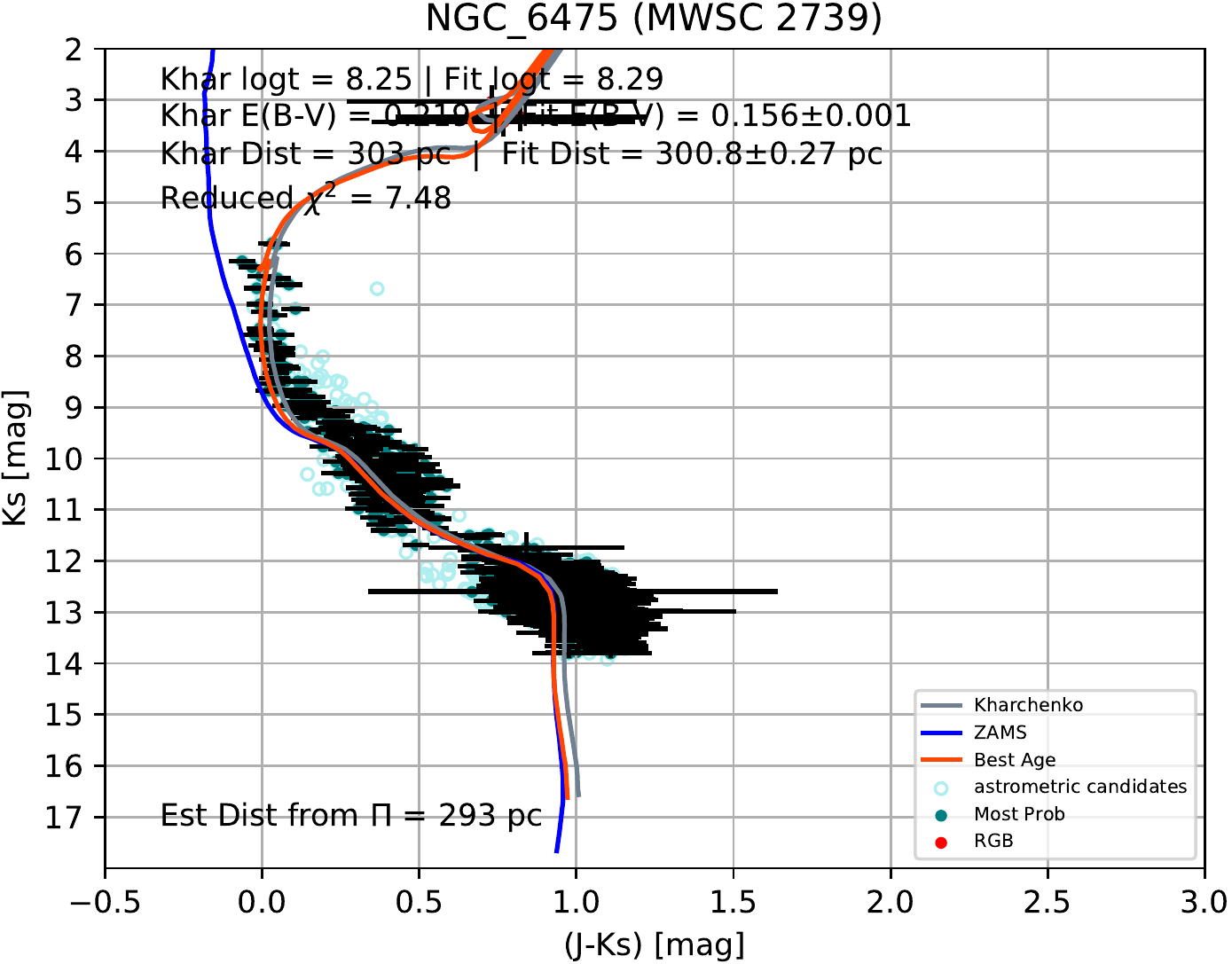}\\
\includegraphics[width=6cm]{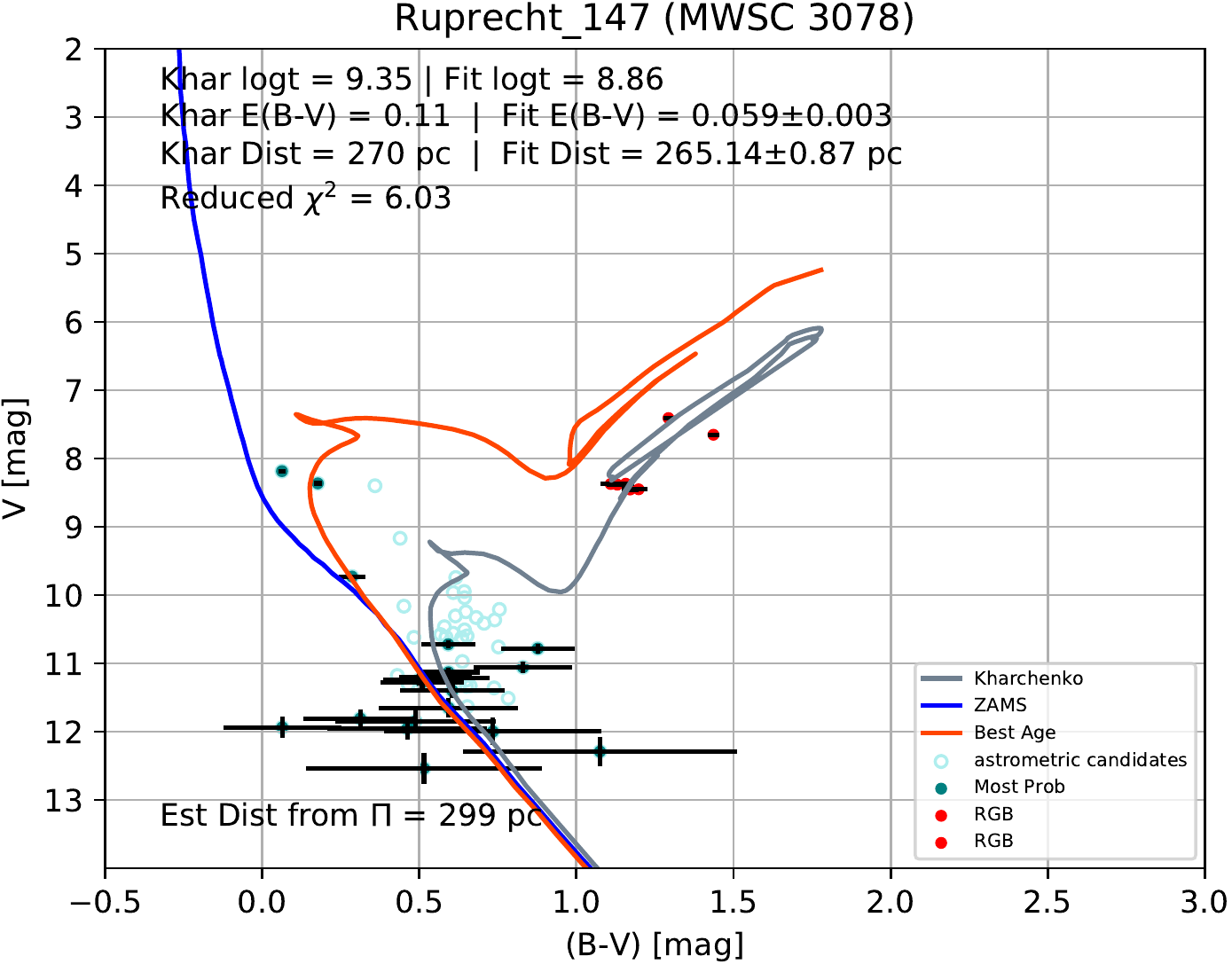}
\includegraphics[width=6cm]{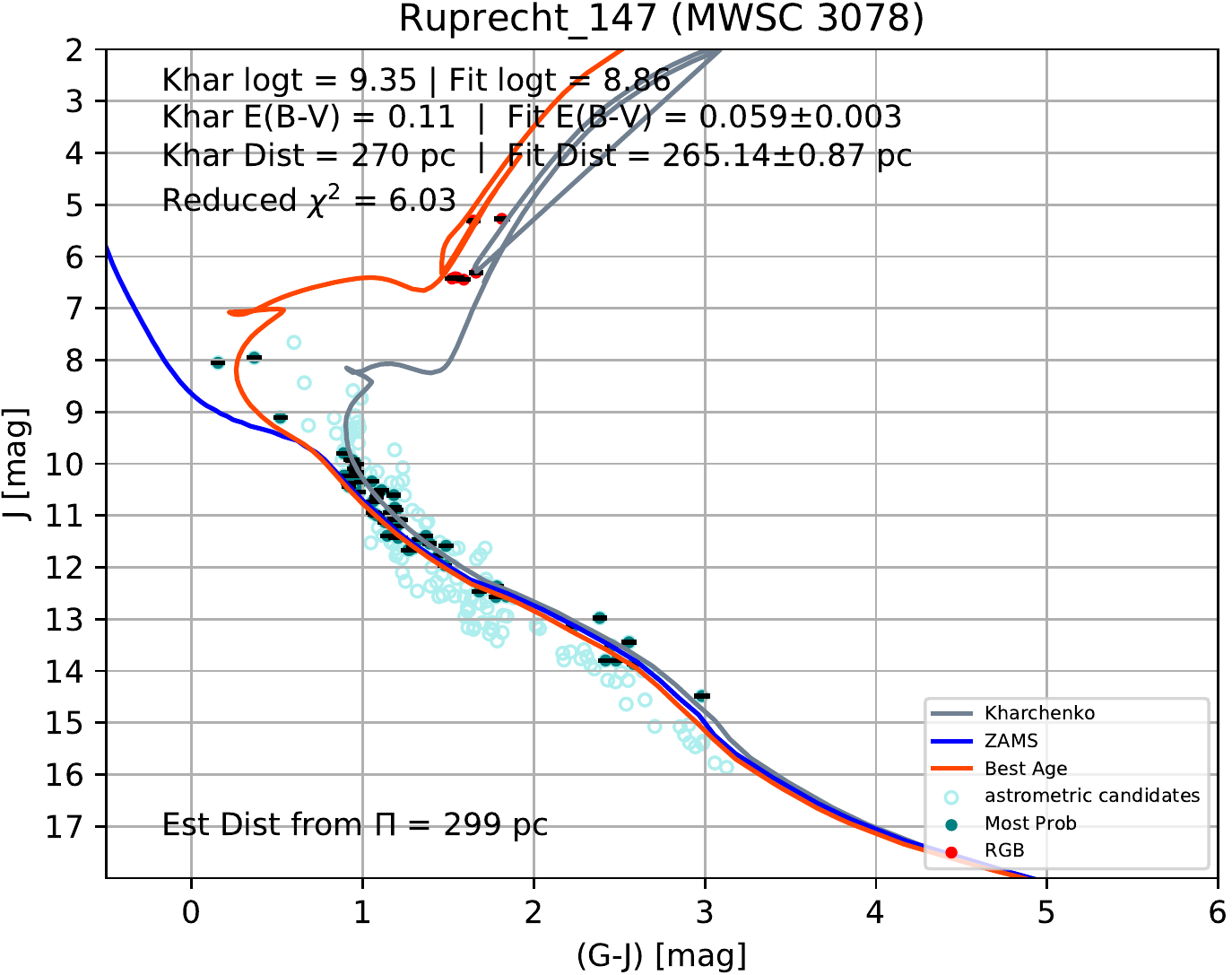}
\includegraphics[width=6cm]{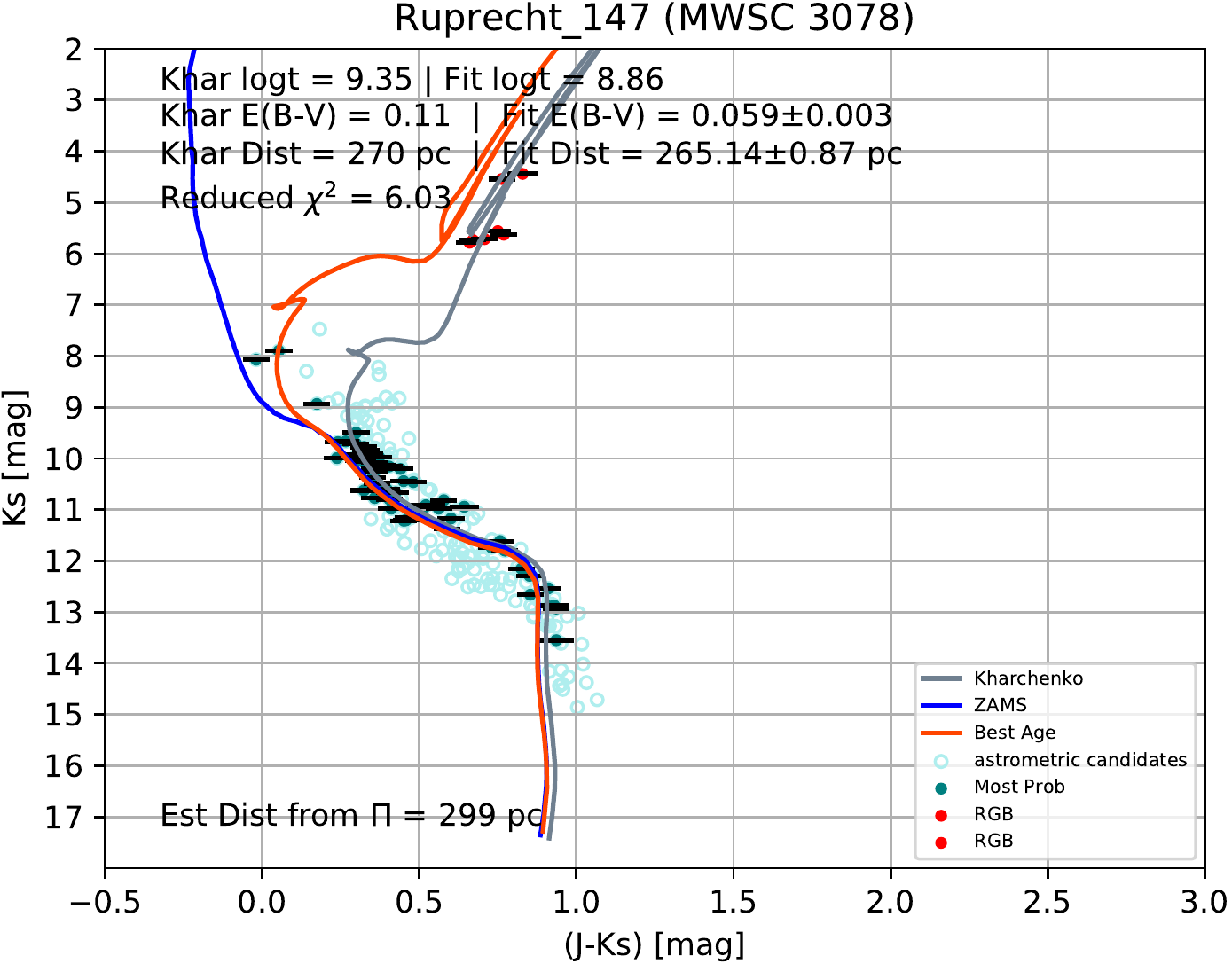}\\
\includegraphics[width=6cm]{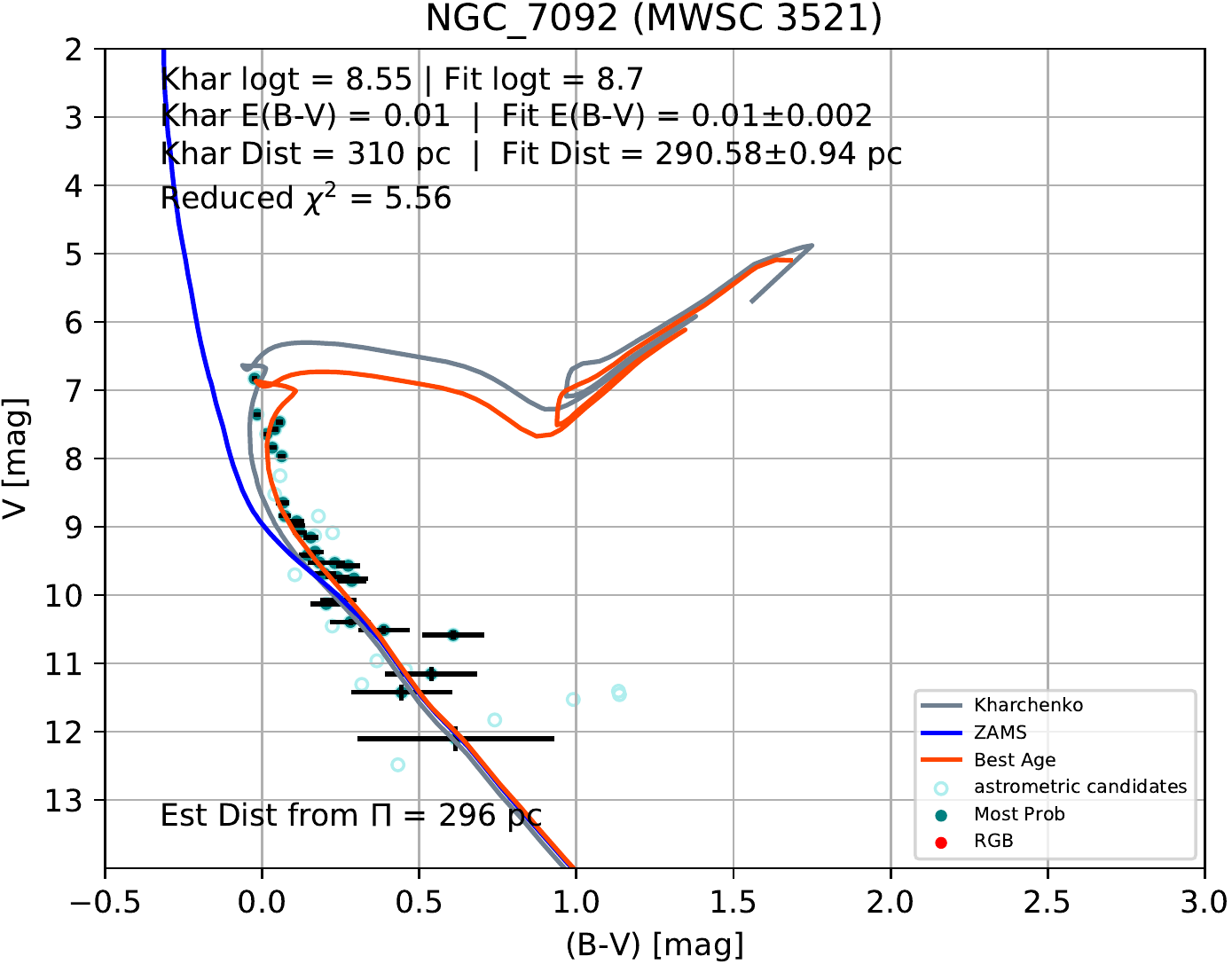}
\includegraphics[width=6cm]{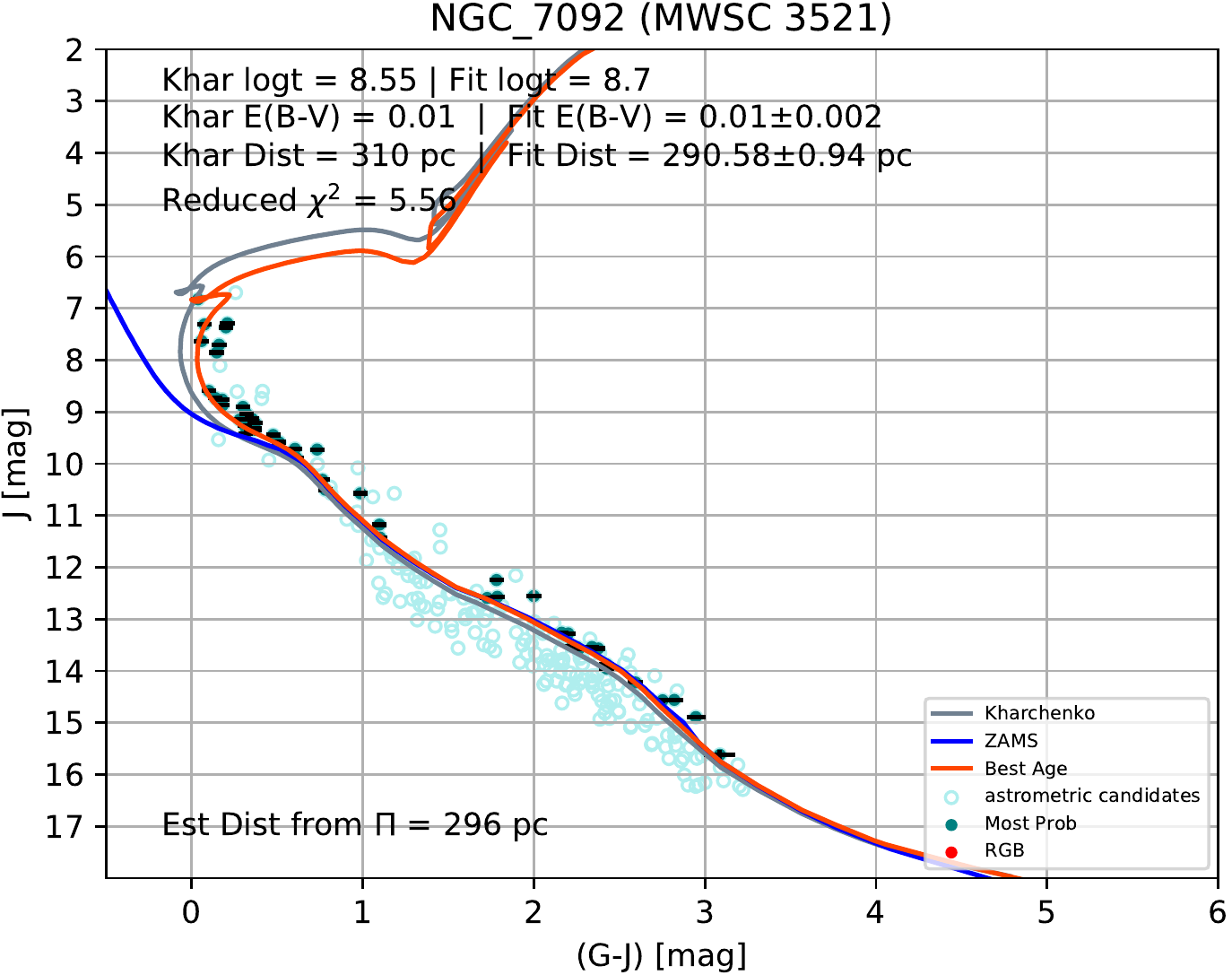}
\includegraphics[width=6cm]{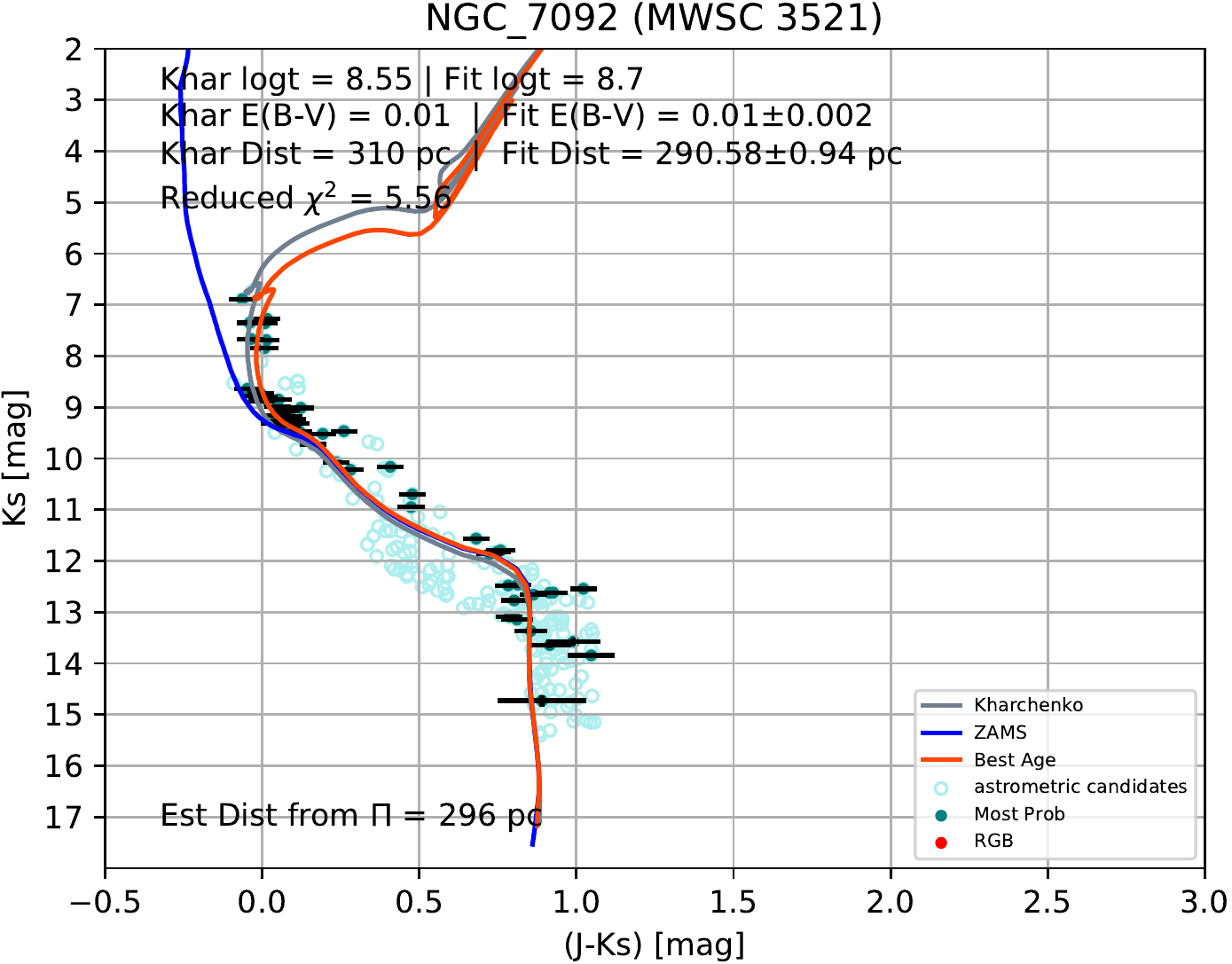}\\
\includegraphics[width=6cm]{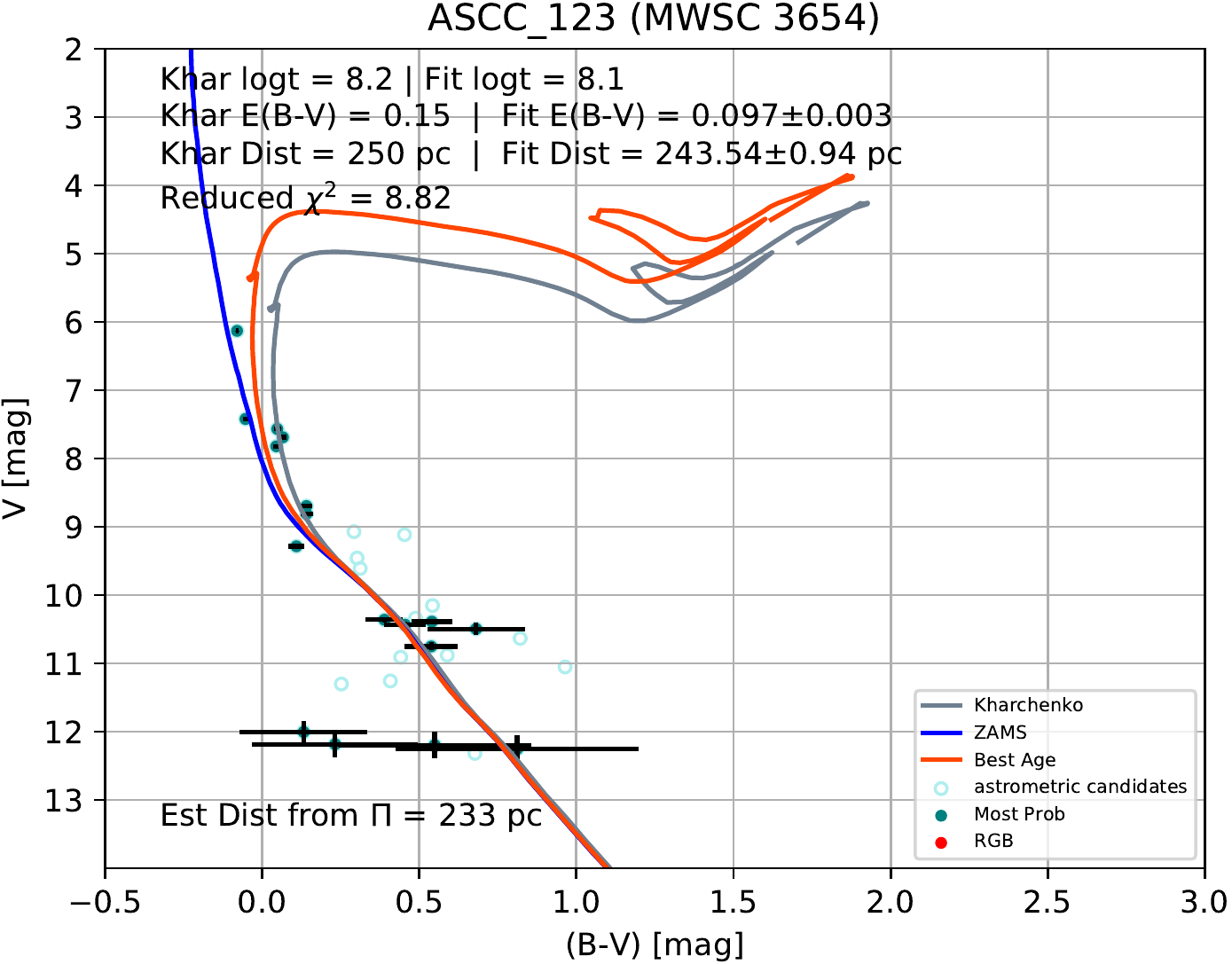}
\includegraphics[width=6cm]{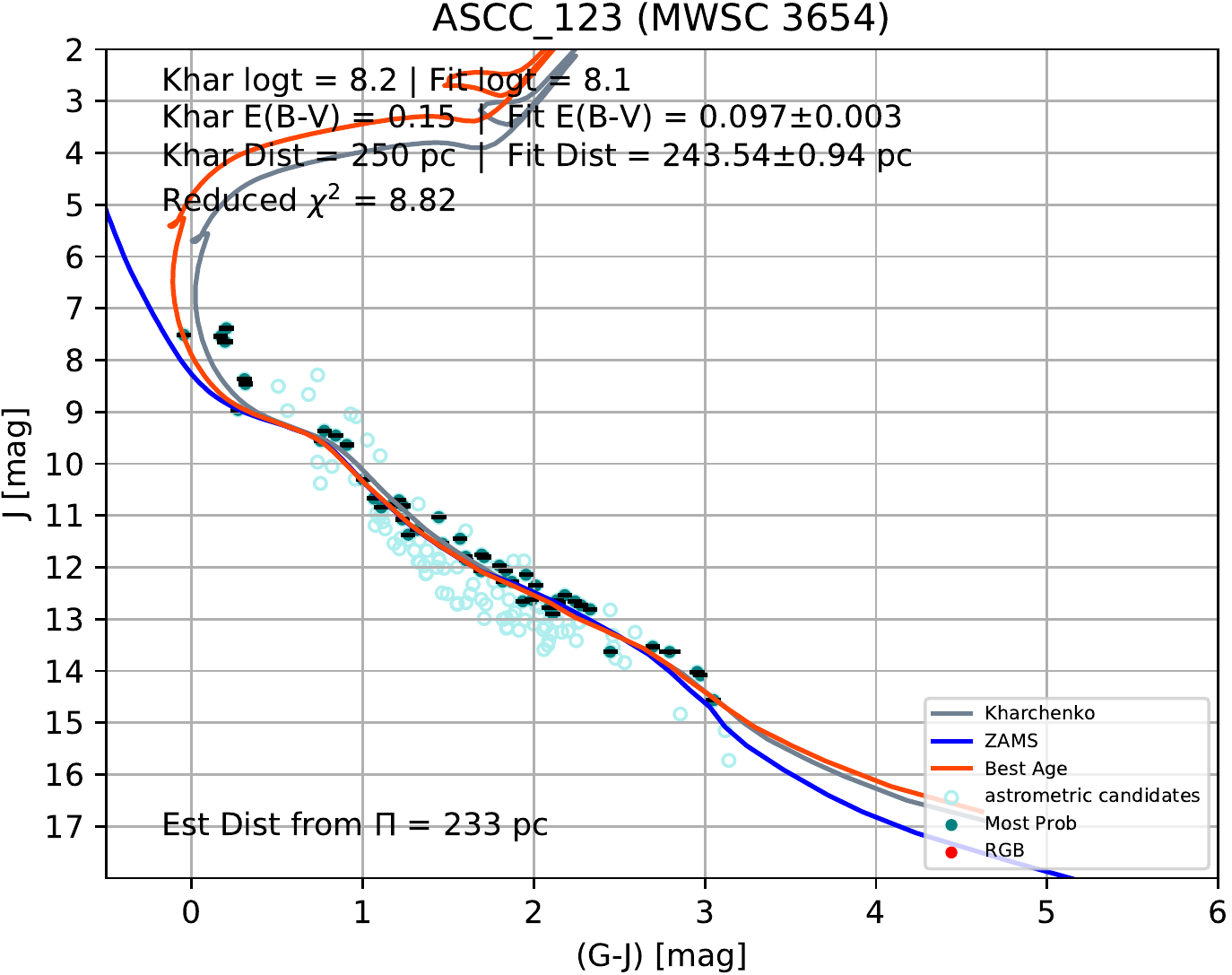}
\includegraphics[width=6cm]{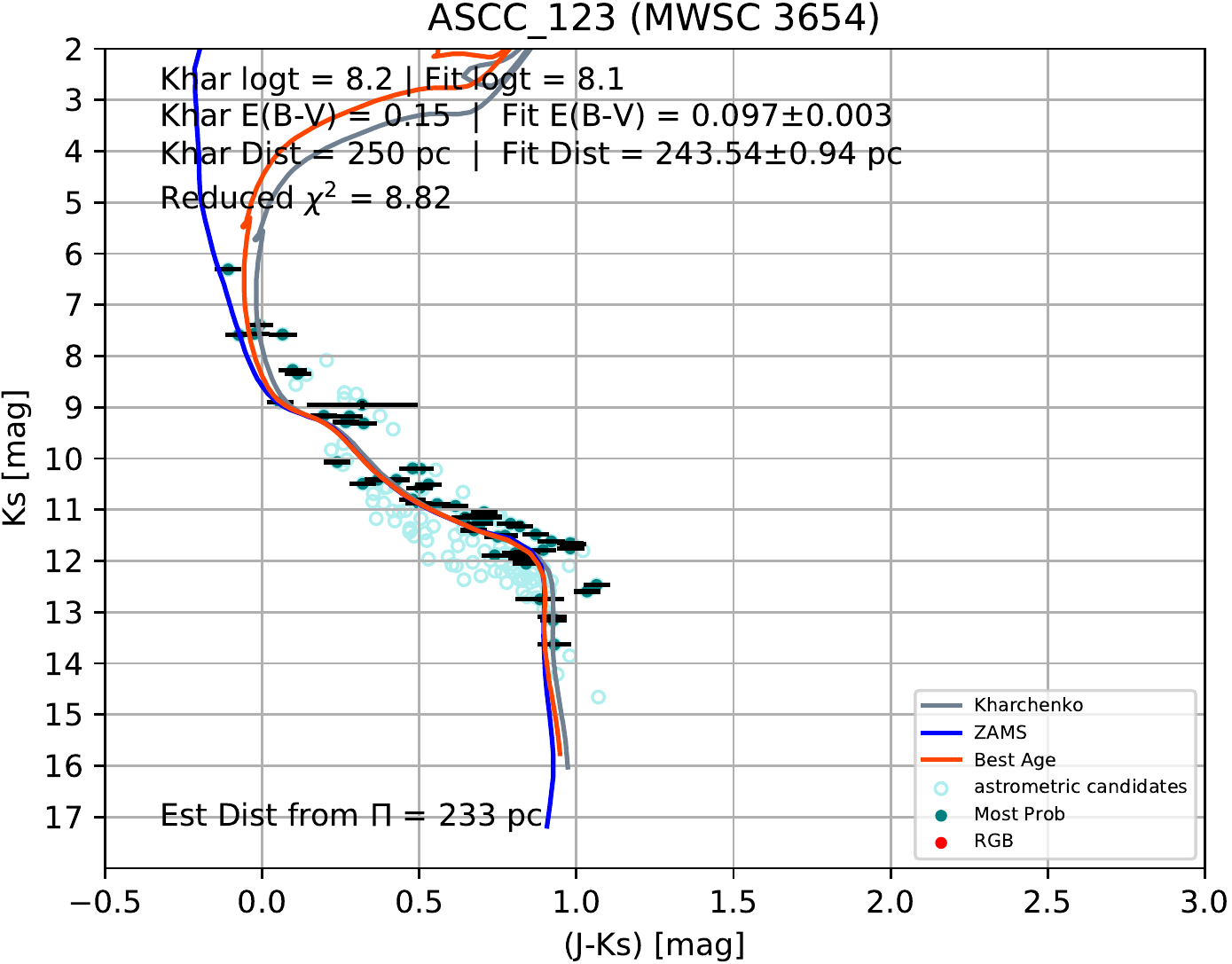}\\
\caption{Color-magnitude diagrams for clusters, from top to bottom: NGC~6475, Ruprecht~147, NGC~7092, and ASCC~123. From left to right: their respective $V$ vs. $(B-V)$, $J$ vs. $(G-J)$, and $K_s$ vs. $(J-K_s)$ CMDs. The cluster members determined from the pipeline are given by teal circles with their corresponding magnitude and color error bars. The cluster astrometric candidates that were later rejected as cluster members are shown by light blue open circles. RGB stars, if any, are indicated by red circles. The red isochrone is the pipeline selected age, plotted with the fitted cluster $E(B-V)$ and $d$. This isochrone is plotted without the binary offset. The gray isochrone shows the age, $E(B-V)$, and $d$ as determined by \citet{2013A&A...558A..53K}. The blue line is the ZAMS plotted with the fitted cluster $E(B-V)$ and $d$.}
\label{figa6}
\end{figure*}

\section{Cluster TGAS proper motion diagrams and parallax histograms}

\begin{figure*}
\centering
\includegraphics[width=7cm]{Clus7_0109_firstTGASpmcut_zoom.pdf}
\includegraphics[width=6.5cm]{Clus7_0109ParallaxHist.pdf}\\
\includegraphics[width=7cm]{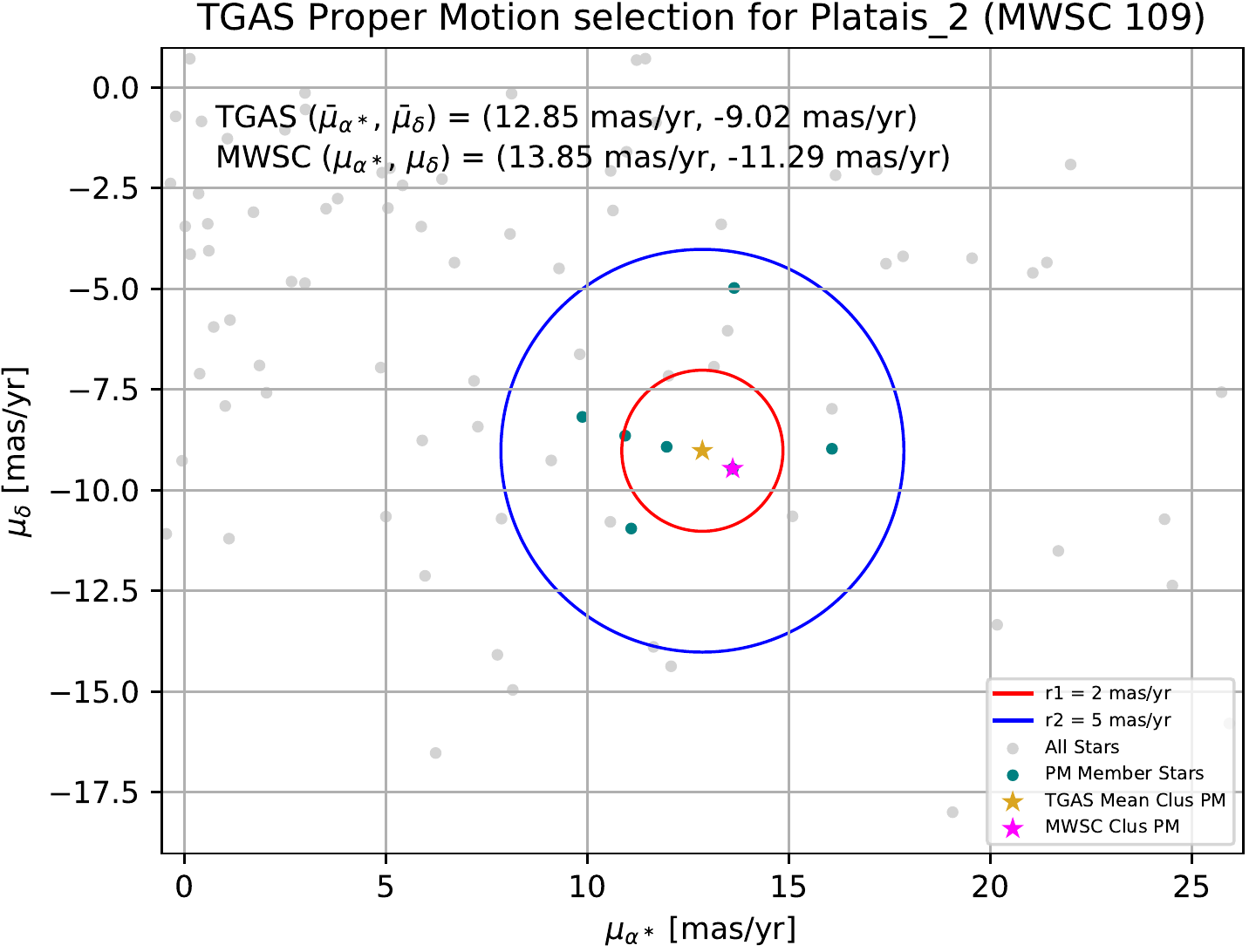}
\includegraphics[width=6.5cm]{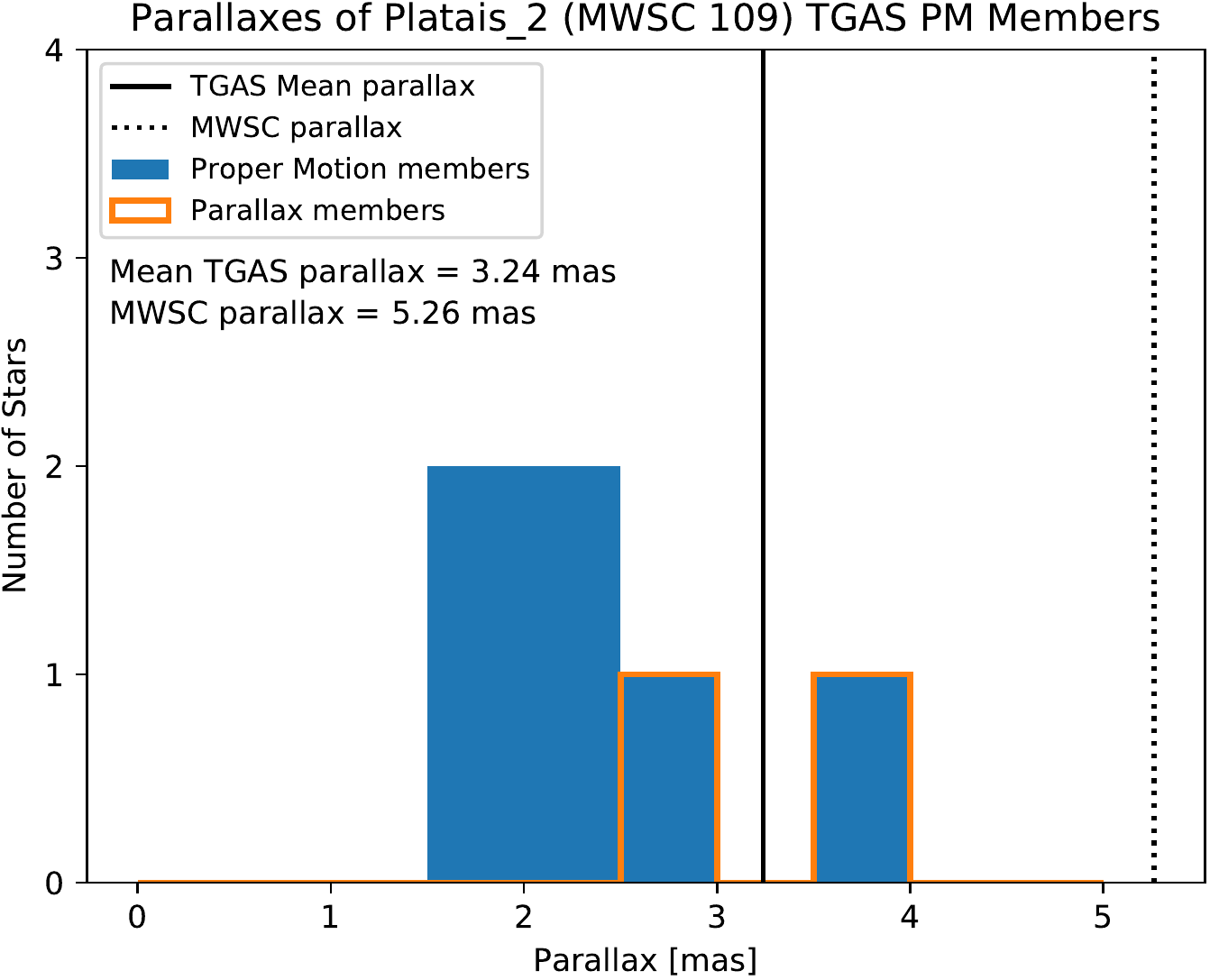}\\
\includegraphics[width=7cm]{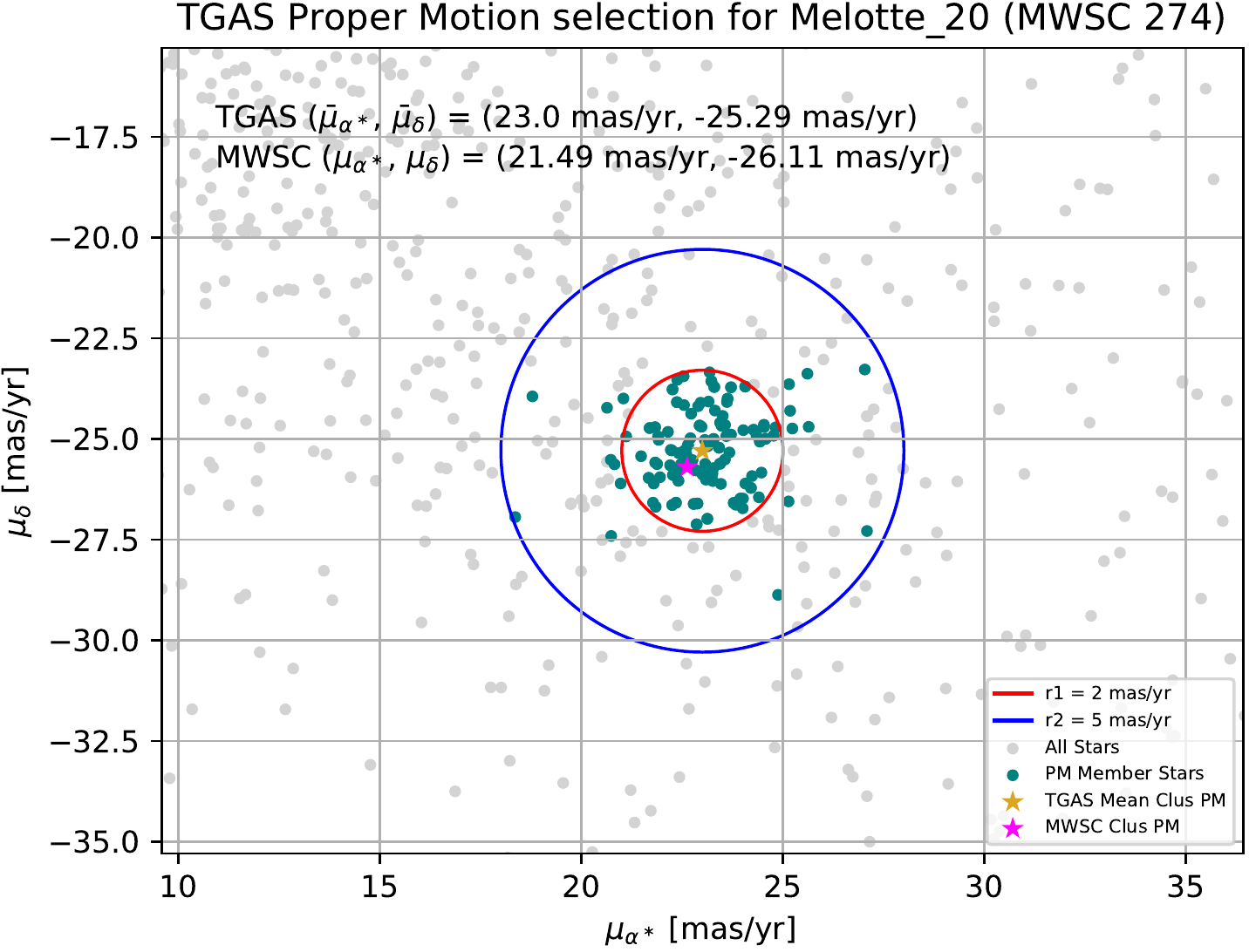}
\includegraphics[width=6.5cm]{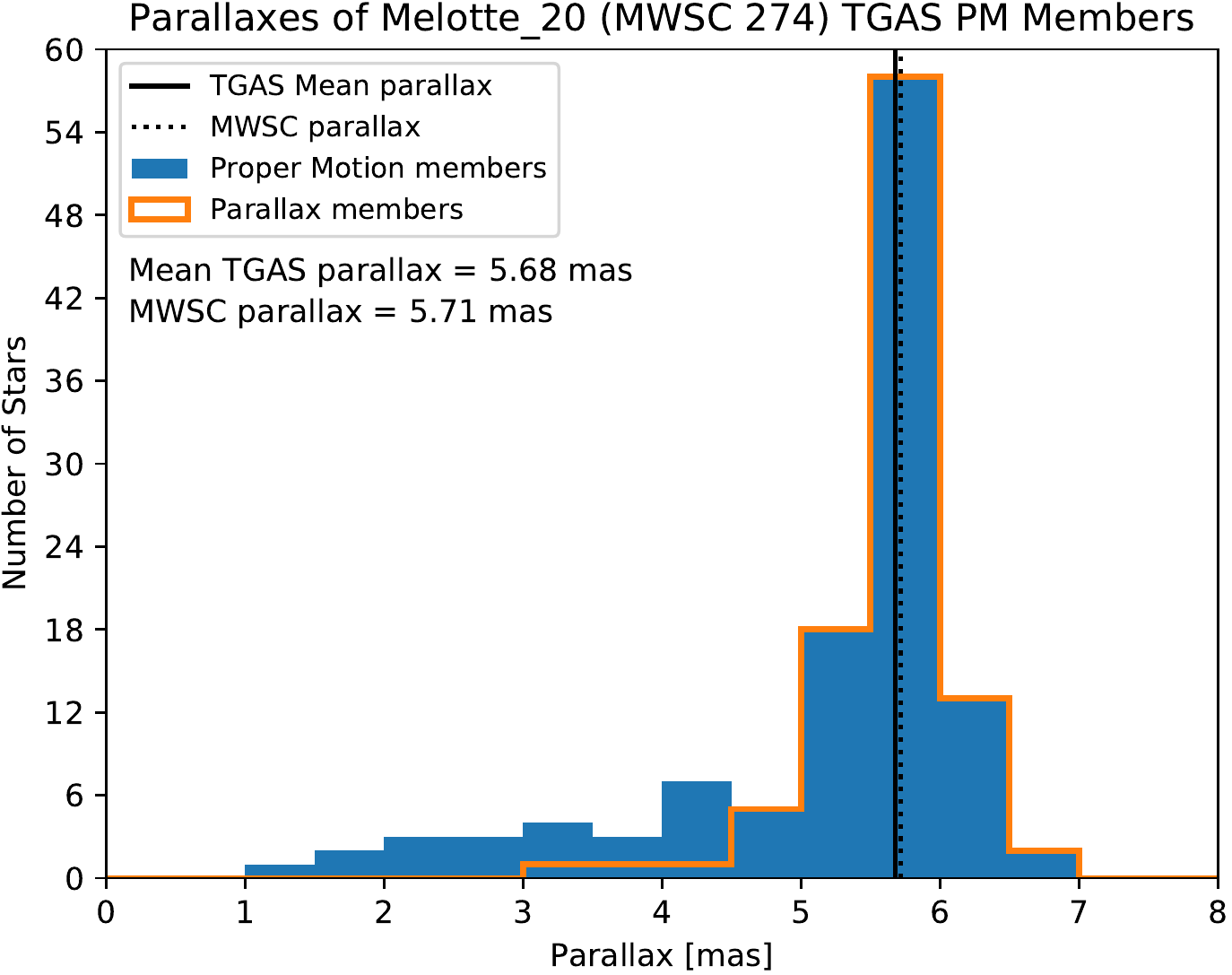}\\
\includegraphics[width=7cm]{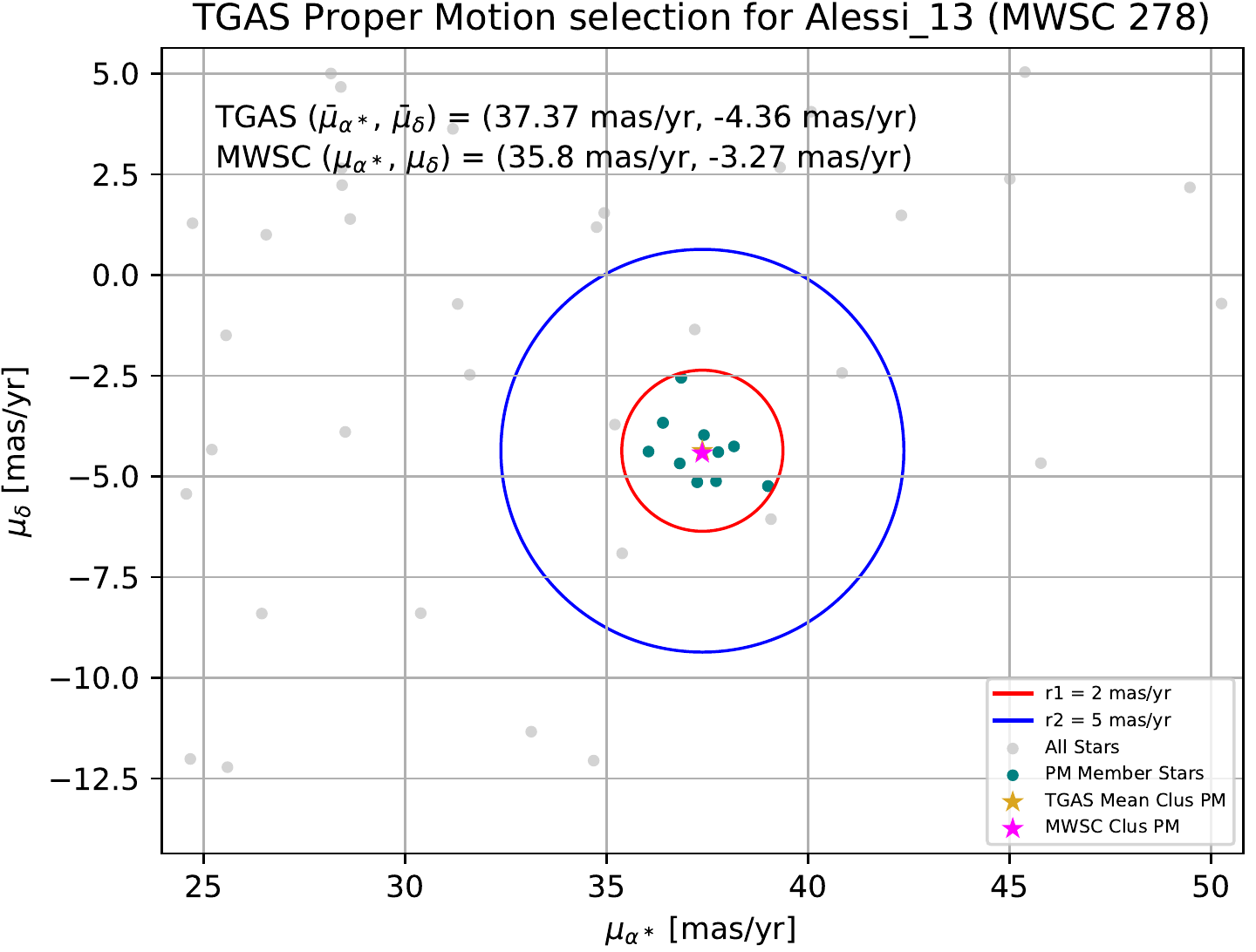}
\includegraphics[width=6.5cm]{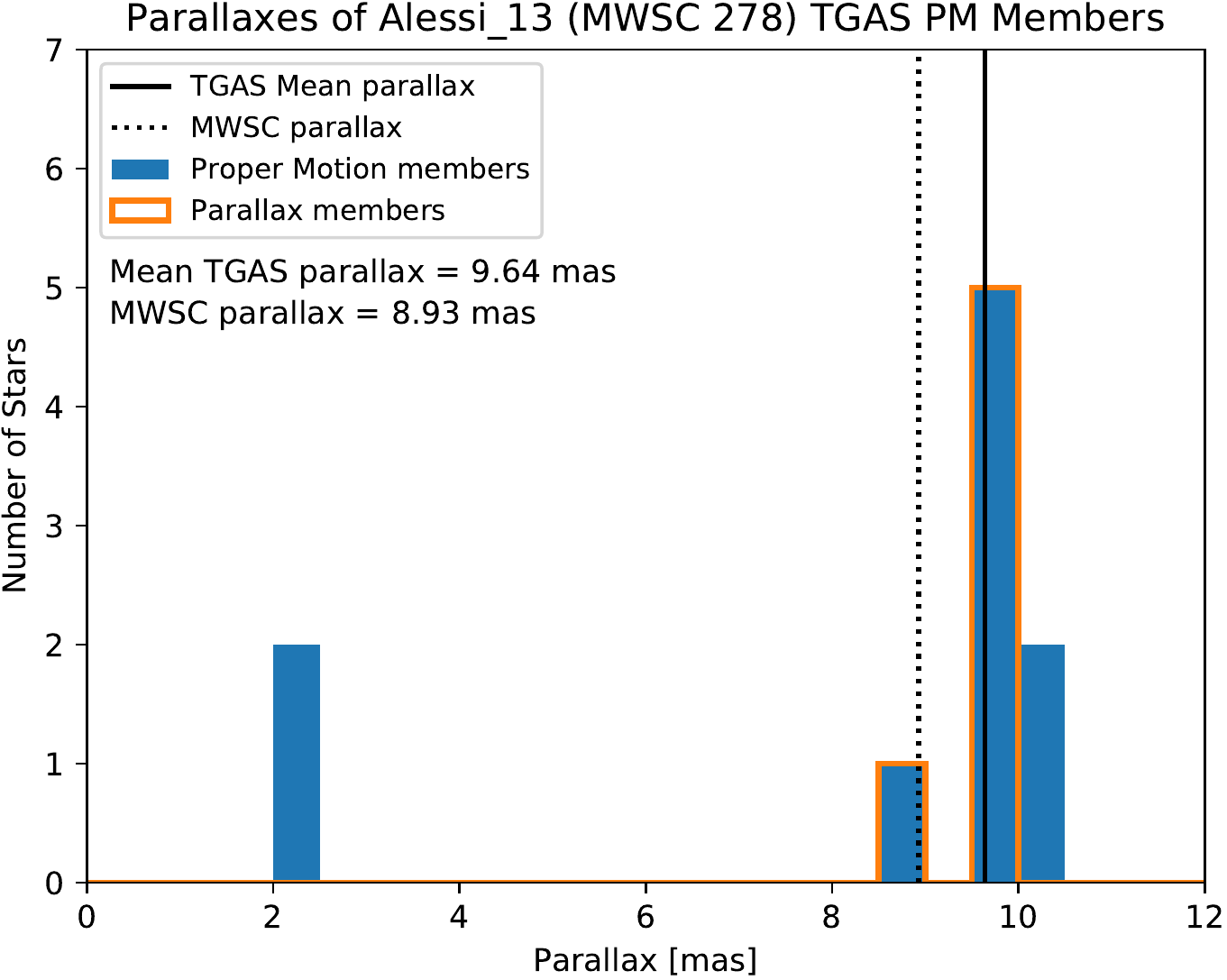}\\
\caption{TGAS proper motion (left panel) and parallax (right panel) selection diagrams for clusters, from top to bottom: Blanco~1, Platais~2, $\alpha$~Per (Melotte~20), and Alessi~13. The proper motion and parallax criteria for membership selection is described in Sect. 3.1\sybf{; the values for cluster proper motion and parallaxes in the diagrams are based on initial cluster membership. The final cluster proper motion and parallaxes are provided in Table~\ref{table:results}.} Left panel: The teal points represent the proper motion members, where all stars within the 2 \sybf{mas yr$^{-1}$} radius (red circle) of the mean cluster proper motion are selected and the stars within 5 \sybf{mas yr$^{-1}$} (blue circle) are only selected if their 3$\sigma$ errors are consistent with the mean cluster proper motion. Right panel: The orange outline illustrates the stars with 3$\sigma$ errors consistent with the mean cluster parallax; these stars are the TGAS astrometrically-selected candidates of the cluster.}
 \label{figb1}
\end{figure*}

\begin{figure*}
\centering
\includegraphics[width=7cm]{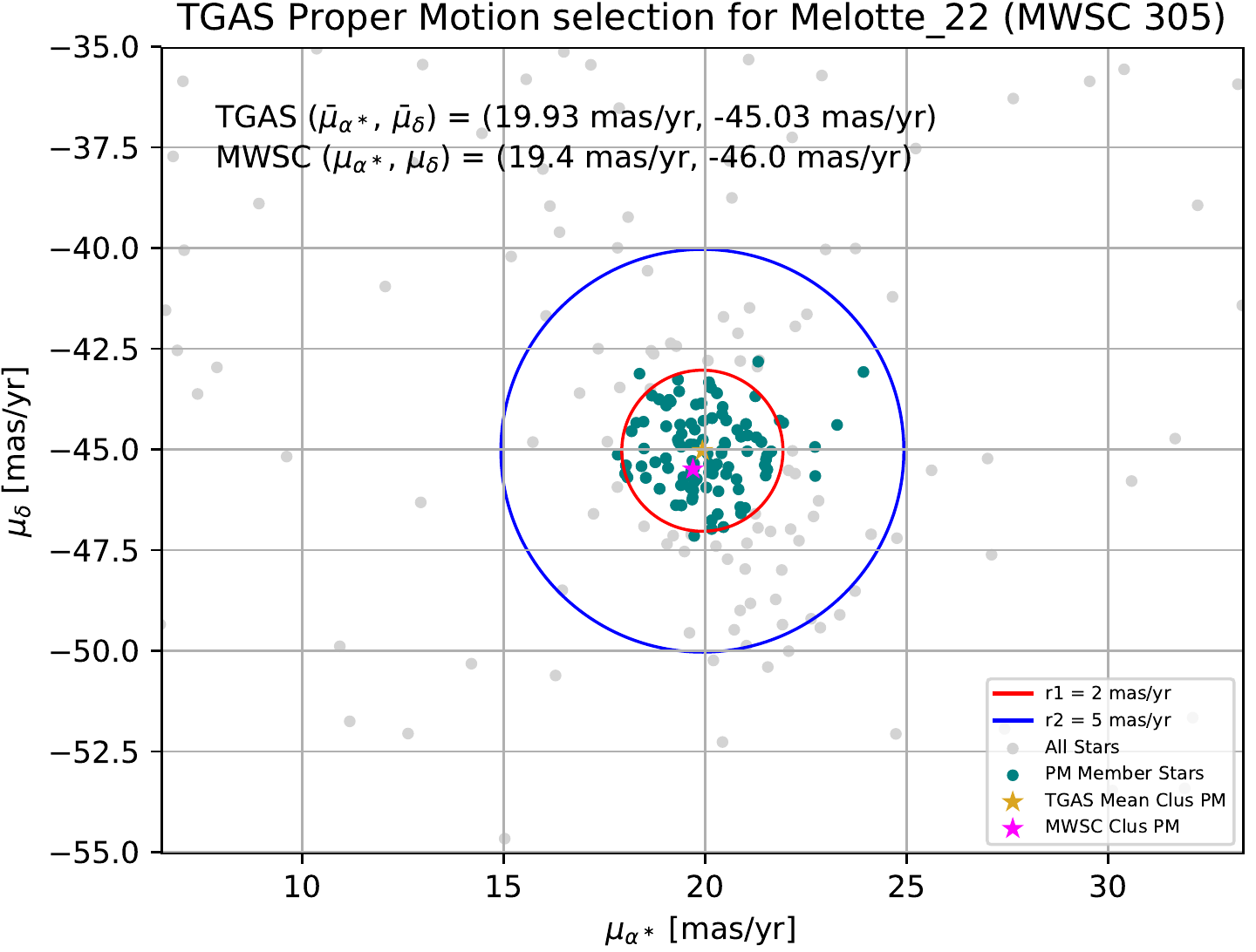}
\includegraphics[width=6.5cm]{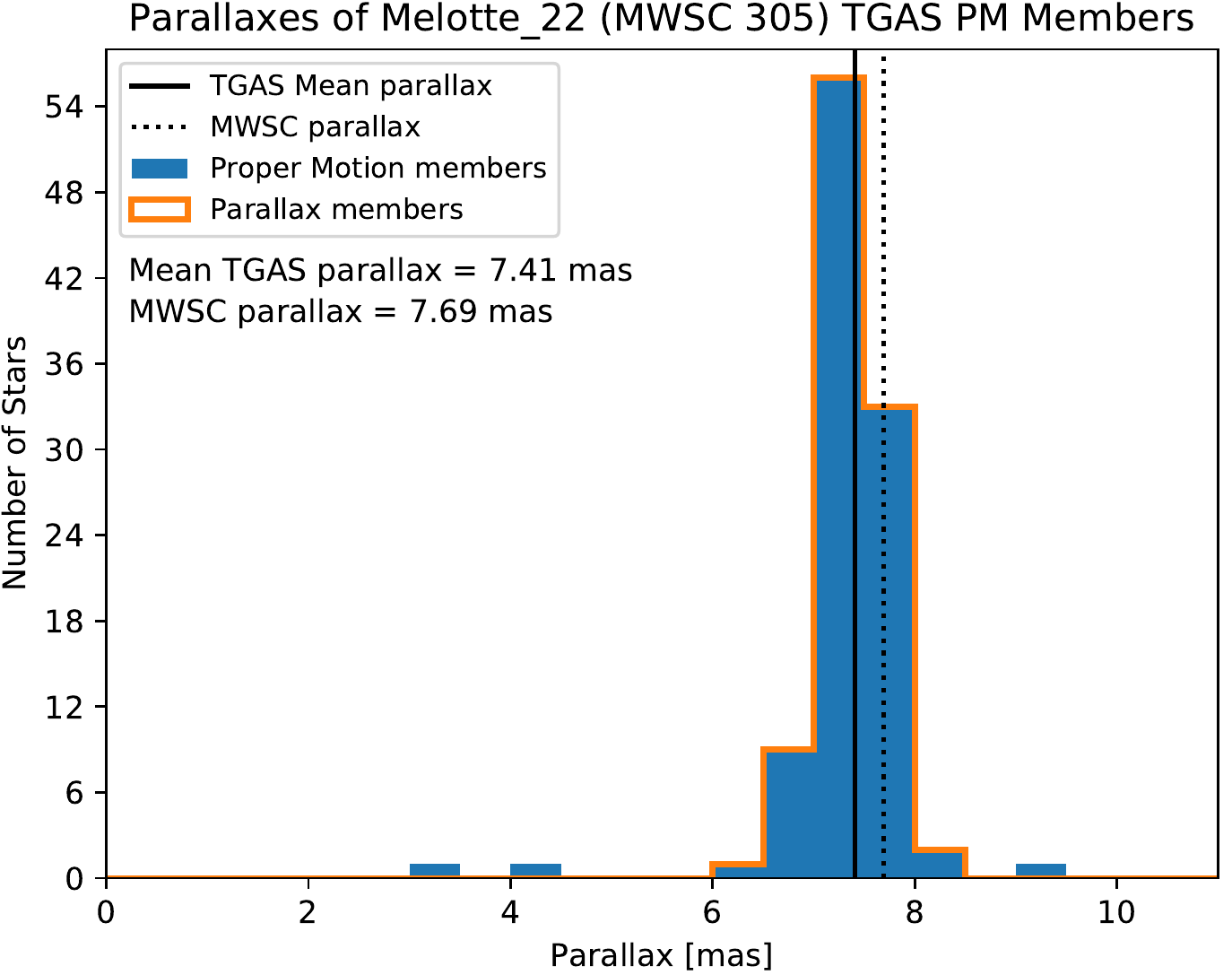}
\includegraphics[width=7cm]{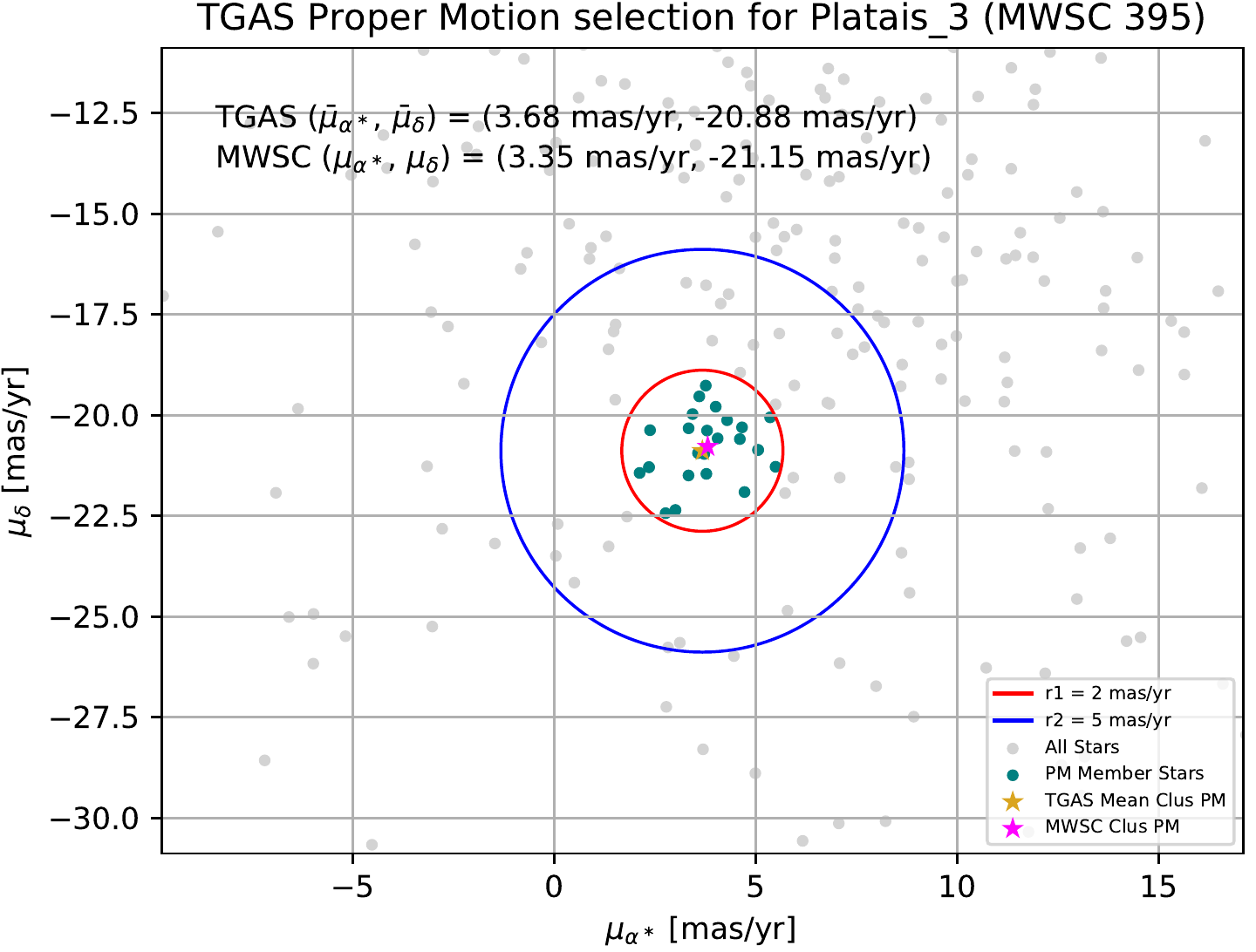}
\includegraphics[width=6.5cm]{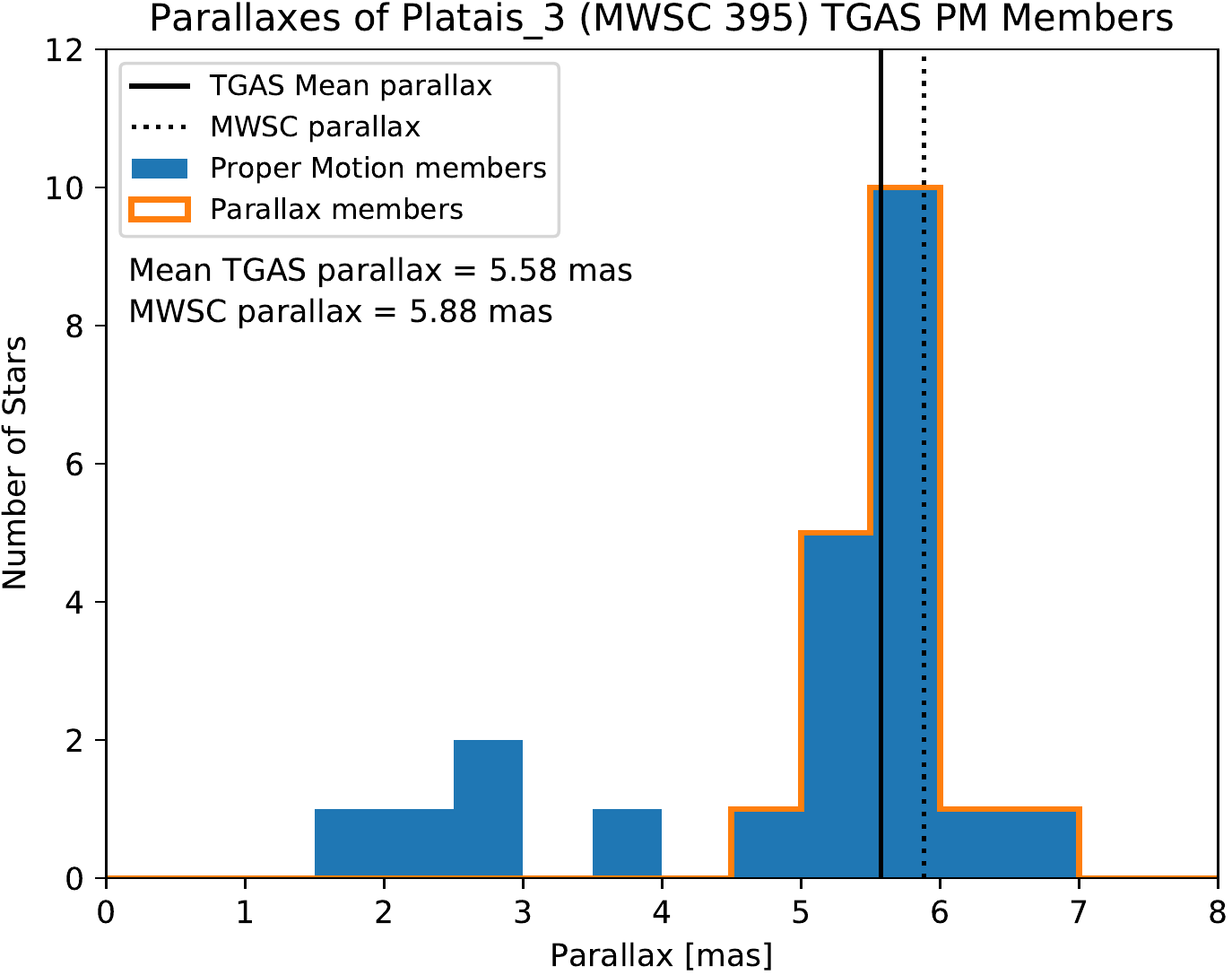}\\
\includegraphics[width=7cm]{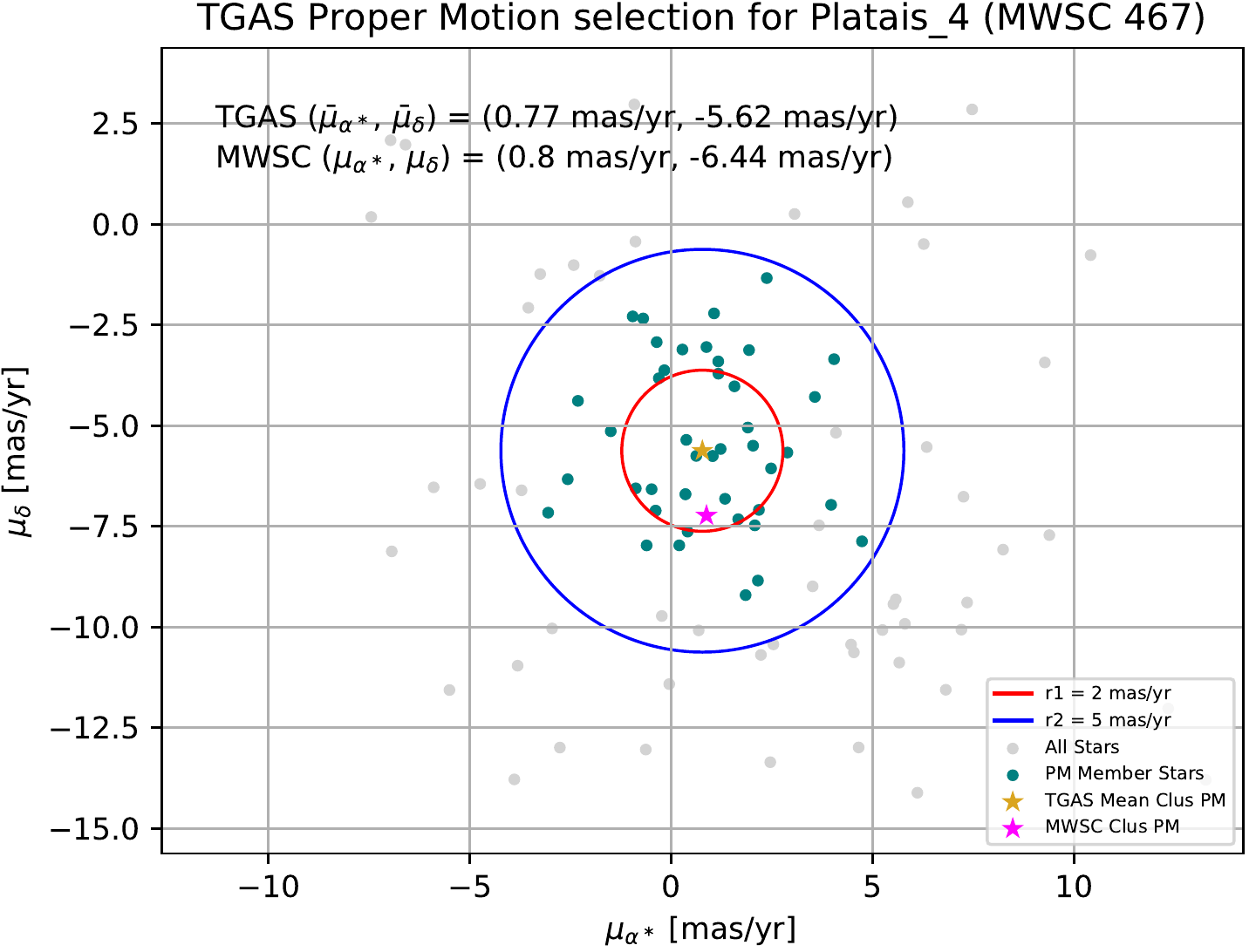}
\includegraphics[width=6.5cm]{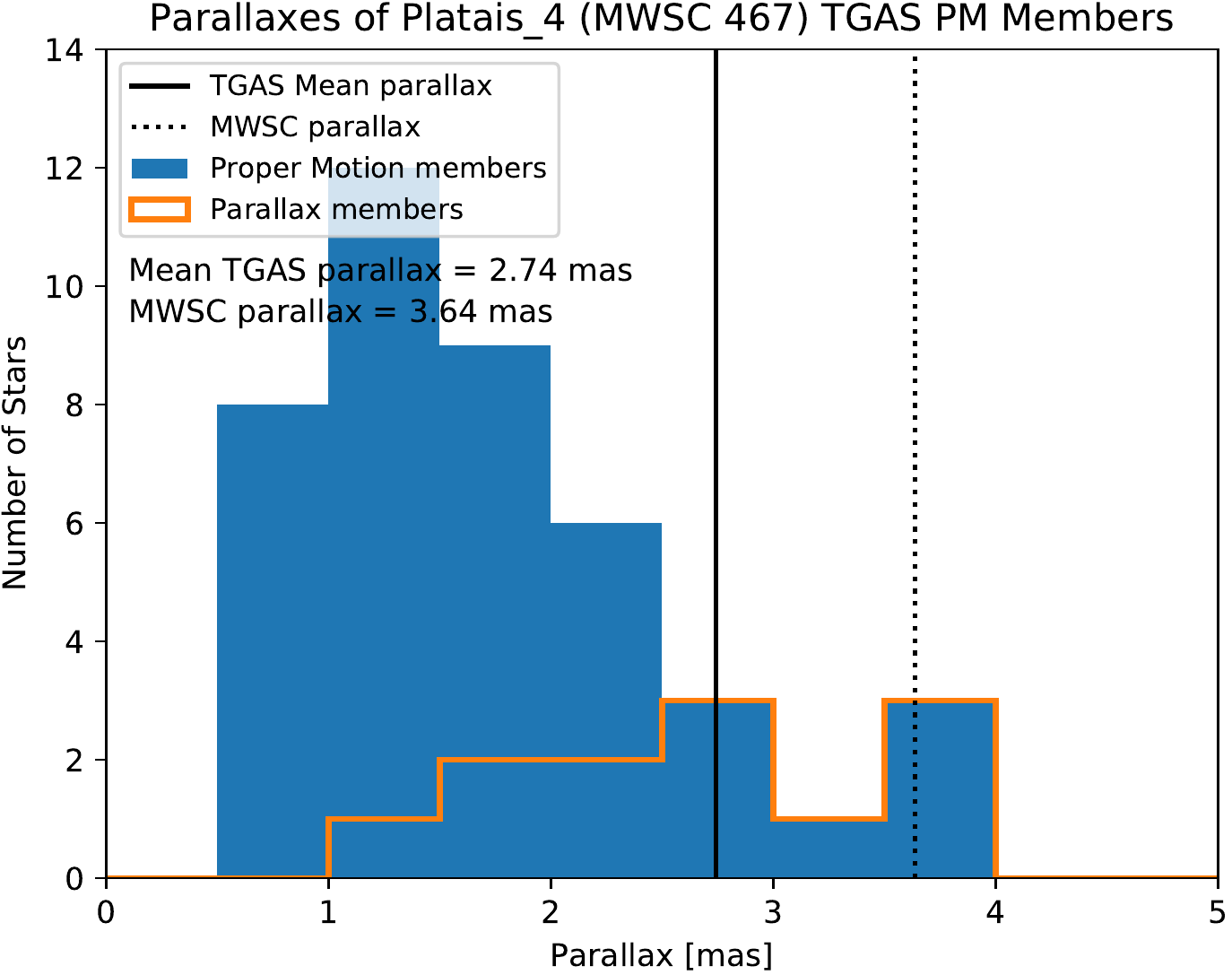}\\
\includegraphics[width=7cm]{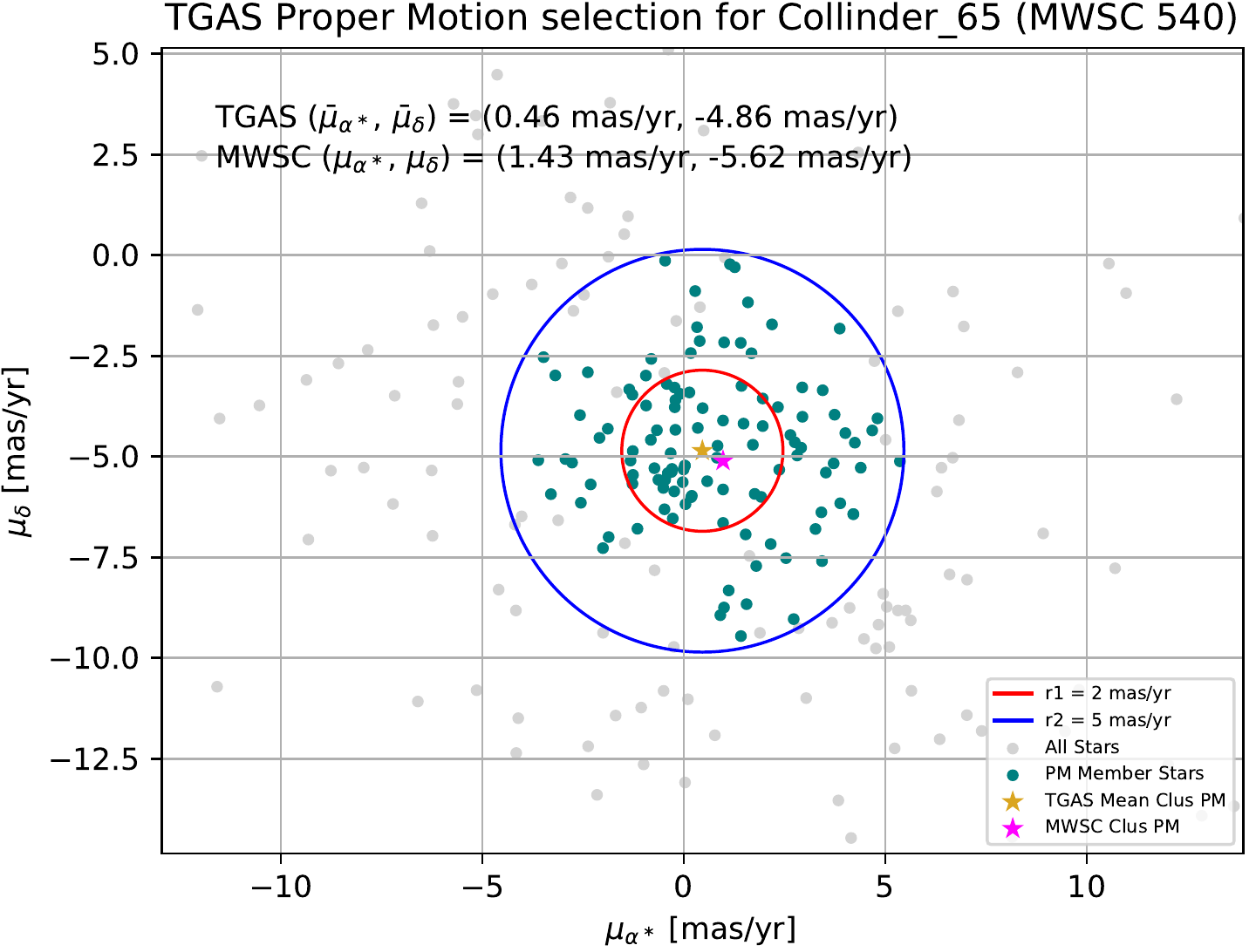}
\includegraphics[width=6.5cm]{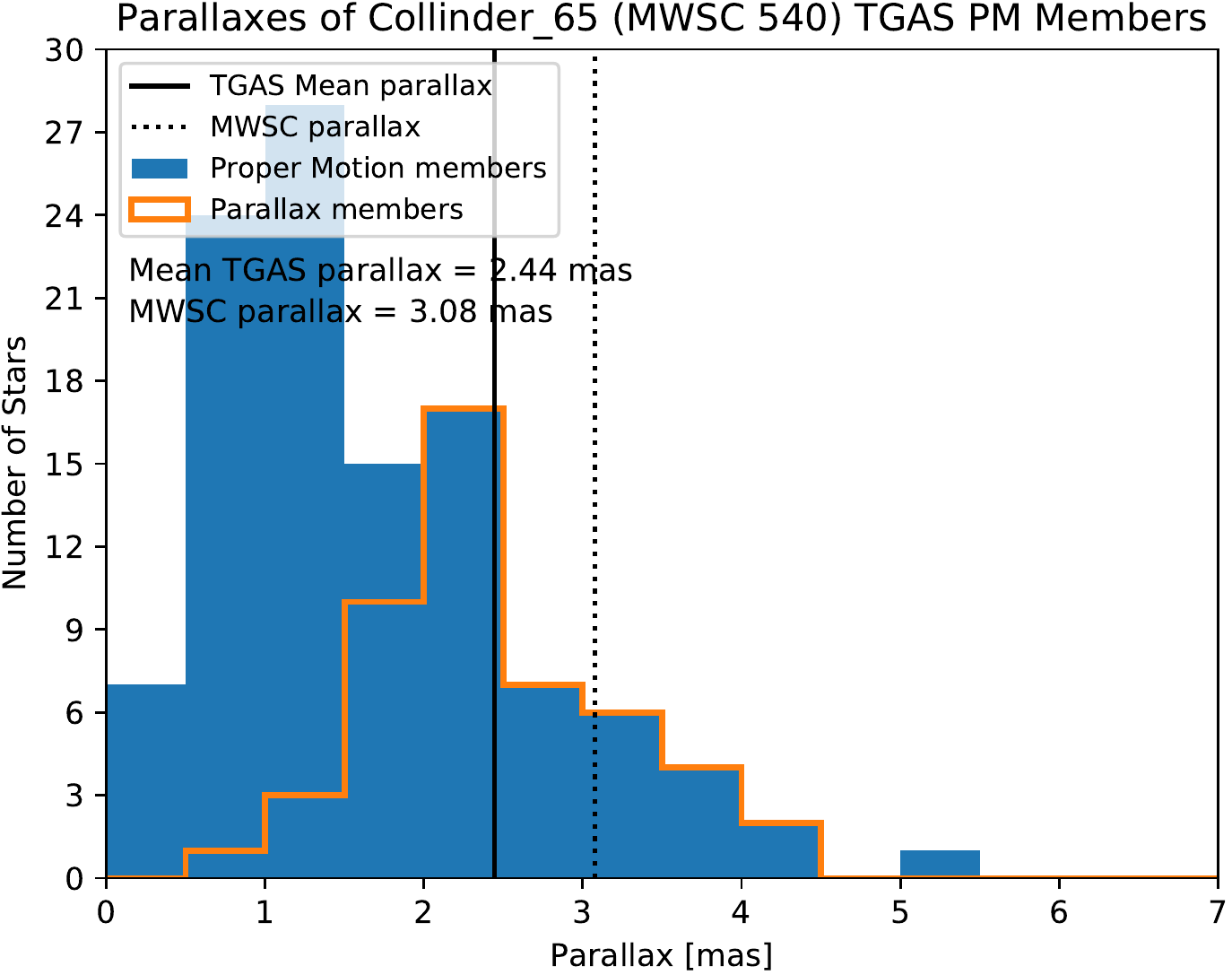}\\
\caption{TGAS proper motion (left panel) and parallax (right panel) selection diagrams for clusters, from top to bottom: the Pleiades (Melotte~22), Platais~3, Platais~4, and Collinder~65. The proper motion and parallax criteria for membership selection is described in Sect. 3.1\sybf{; the values for cluster proper motion and parallaxes in the diagrams are based on initial cluster membership. The final cluster proper motion and parallaxes are provided in Table~\ref{table:results}.} Left panel: The teal points represent the proper motion members, where all stars within the 2 \sybf{mas yr$^{-1}$} radius (red circle) of the mean cluster proper motion are selected and the stars within 5 \sybf{mas yr$^{-1}$} (blue circle) are only selected if their 3$\sigma$ errors are consistent with the mean cluster proper motion. Right panel: The orange outline illustrates the stars with 3$\sigma$ errors consistent with the mean cluster parallax; these stars are the TGAS astrometrically-selected candidates of the cluster.}
 \label{figb2}
\end{figure*}

\begin{figure*}
\centering
\includegraphics[width=7cm]{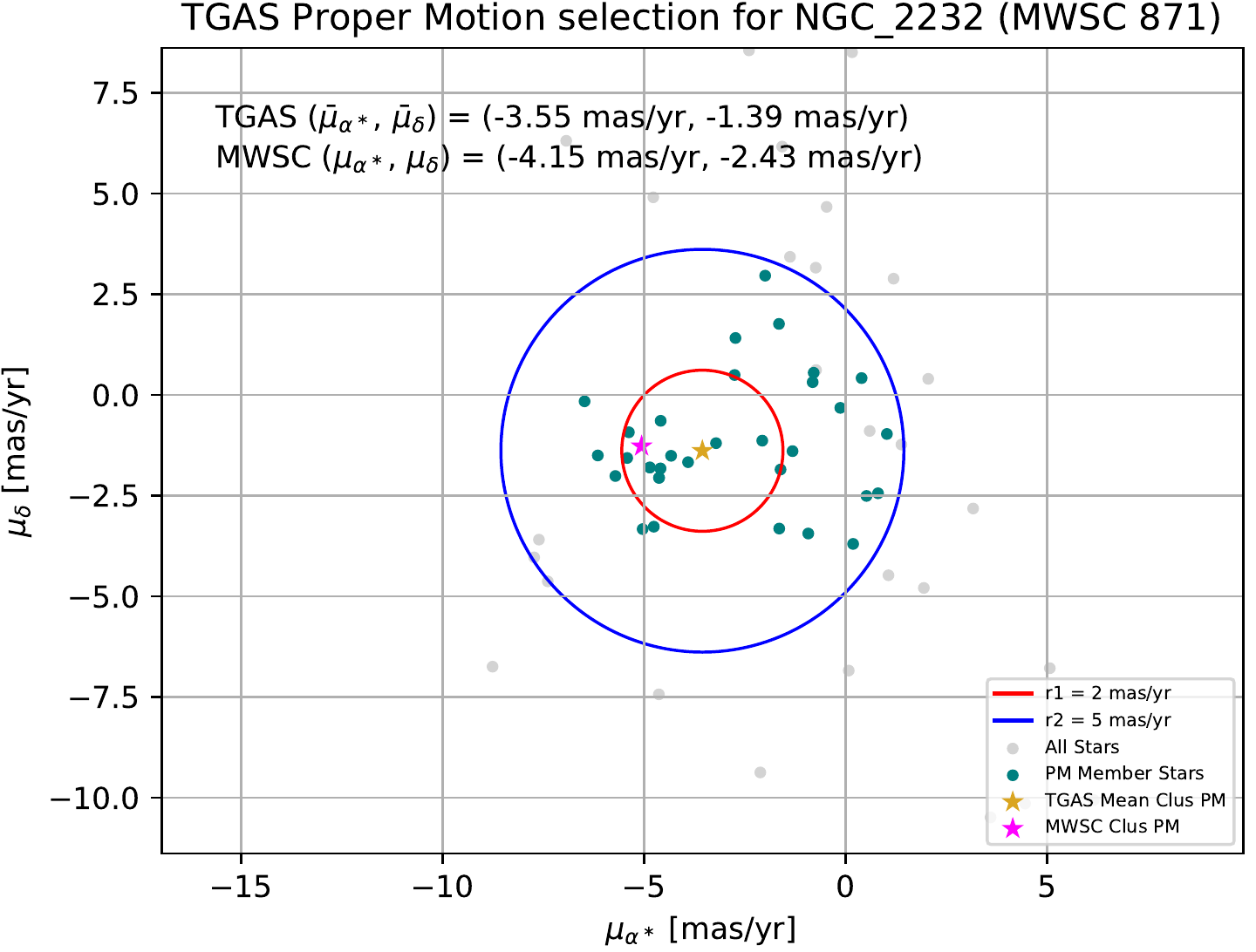}
\includegraphics[width=6.5cm]{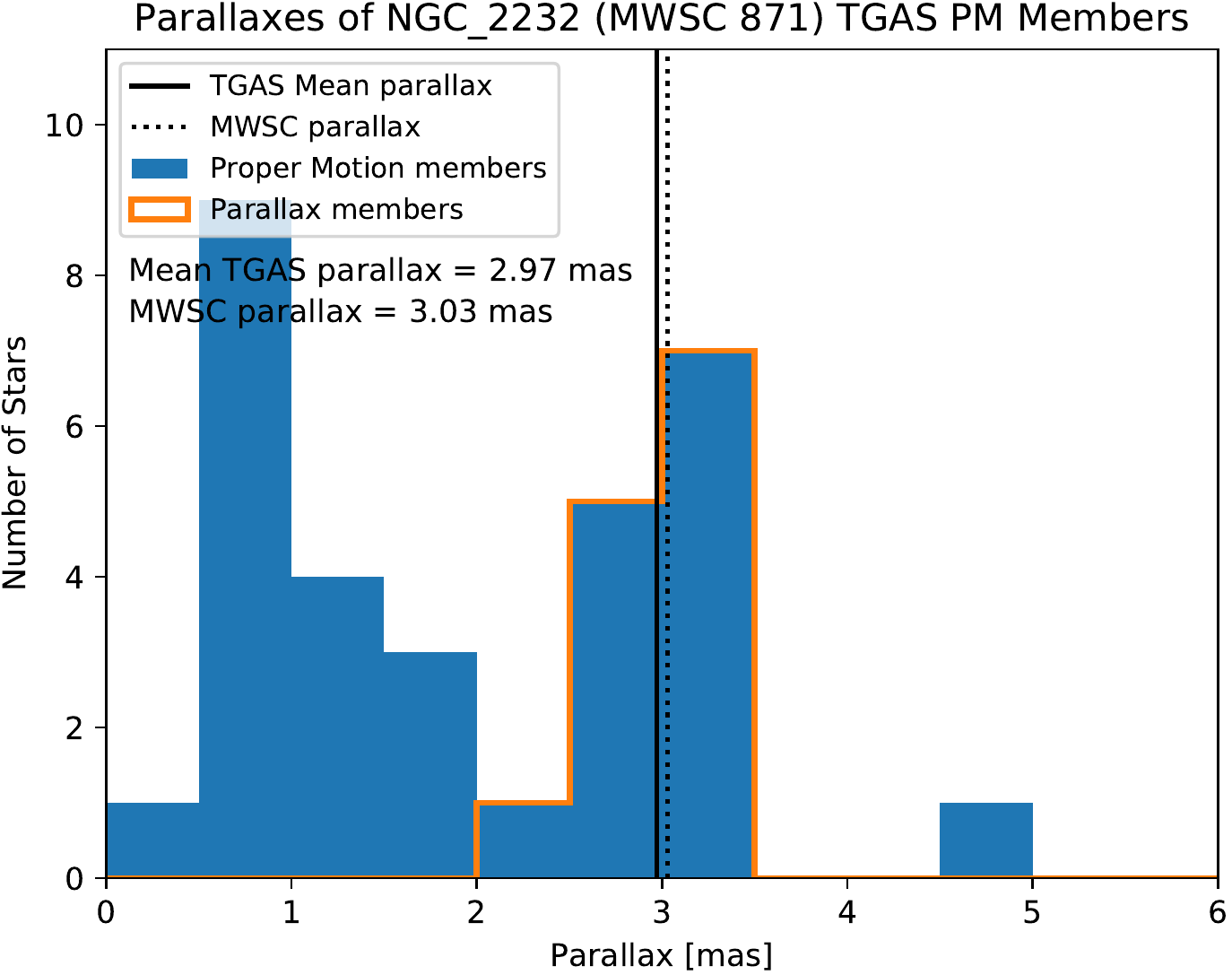}\\
\includegraphics[width=7cm]{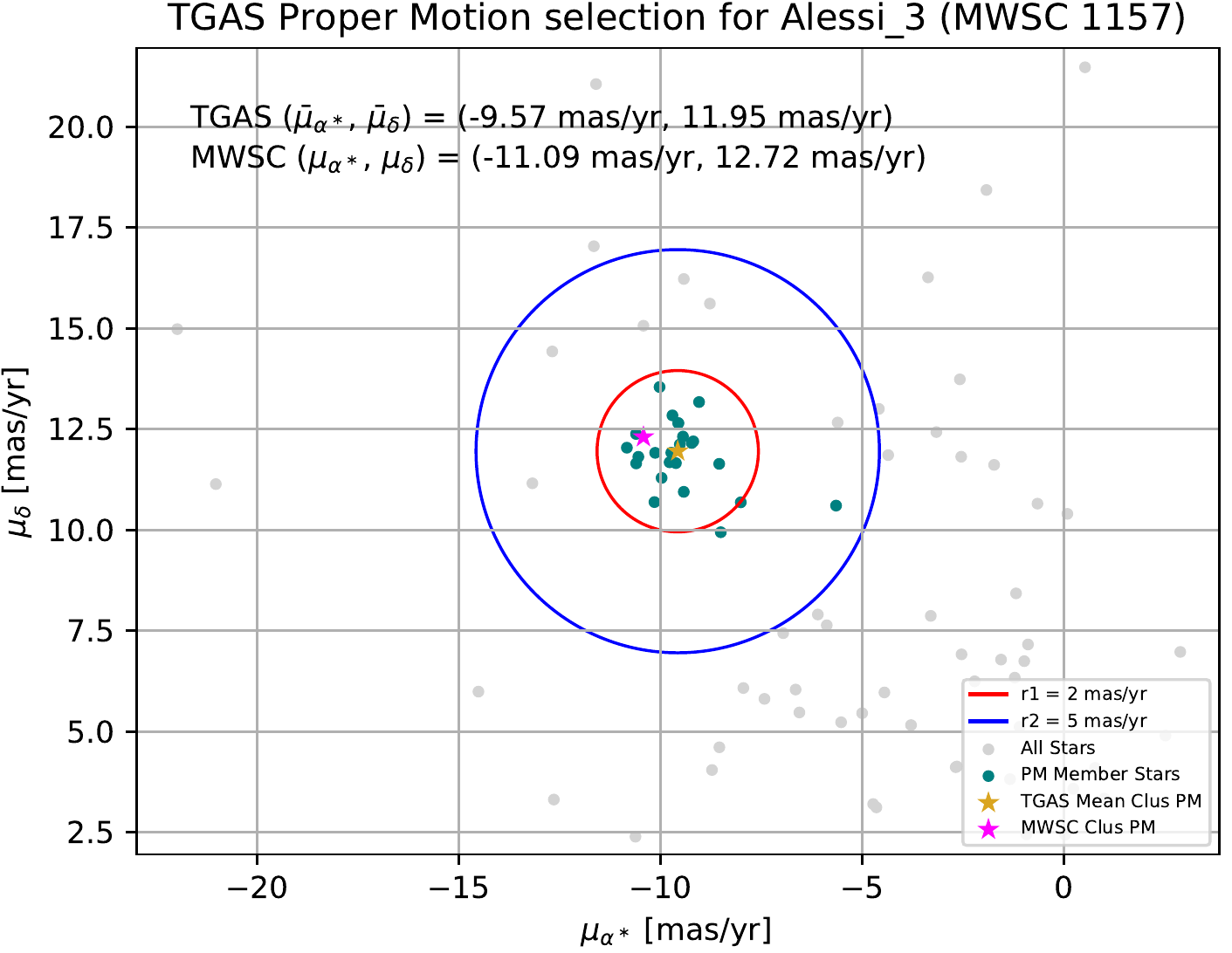}
\includegraphics[width=6.5cm]{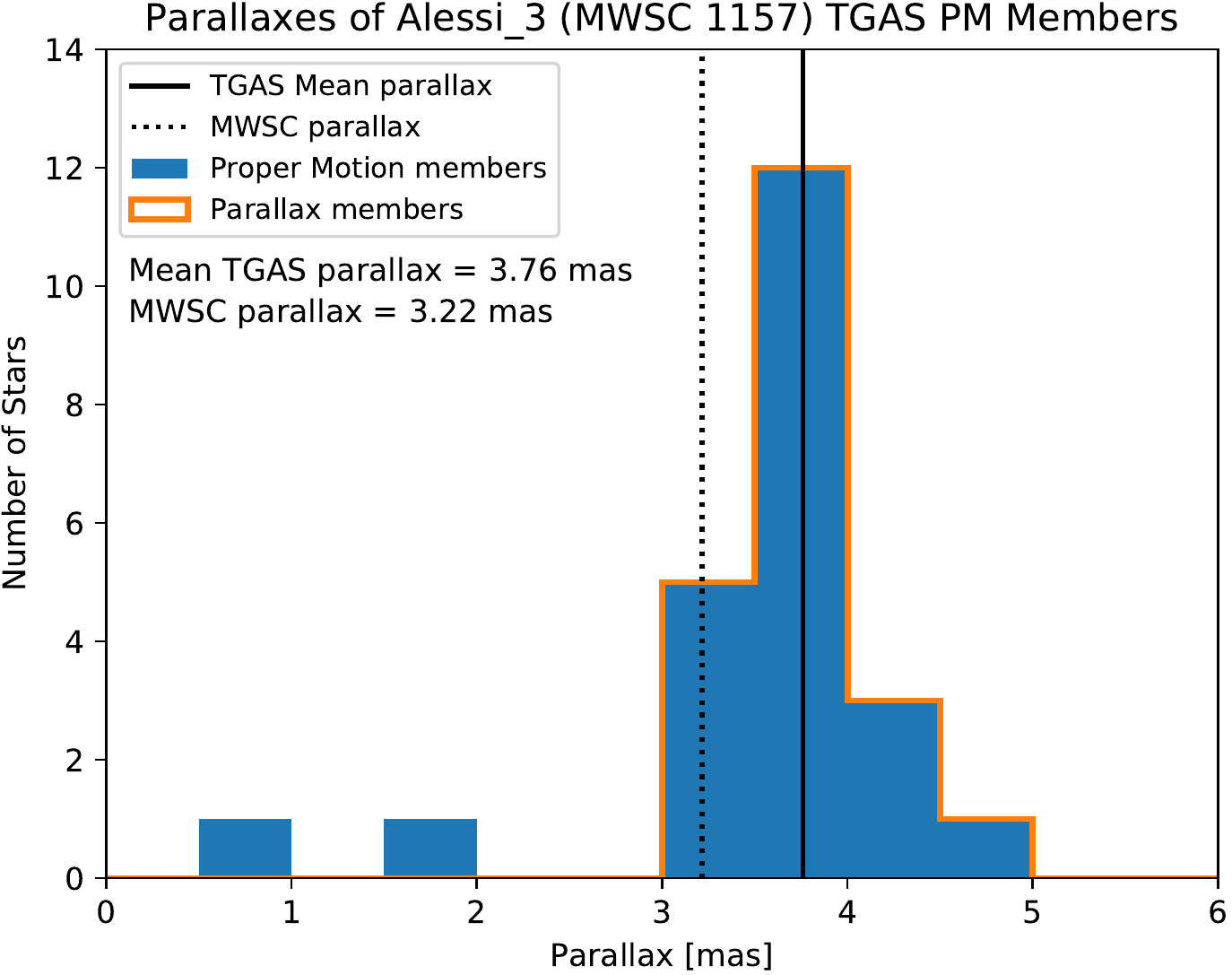}\\
\includegraphics[width=7cm]{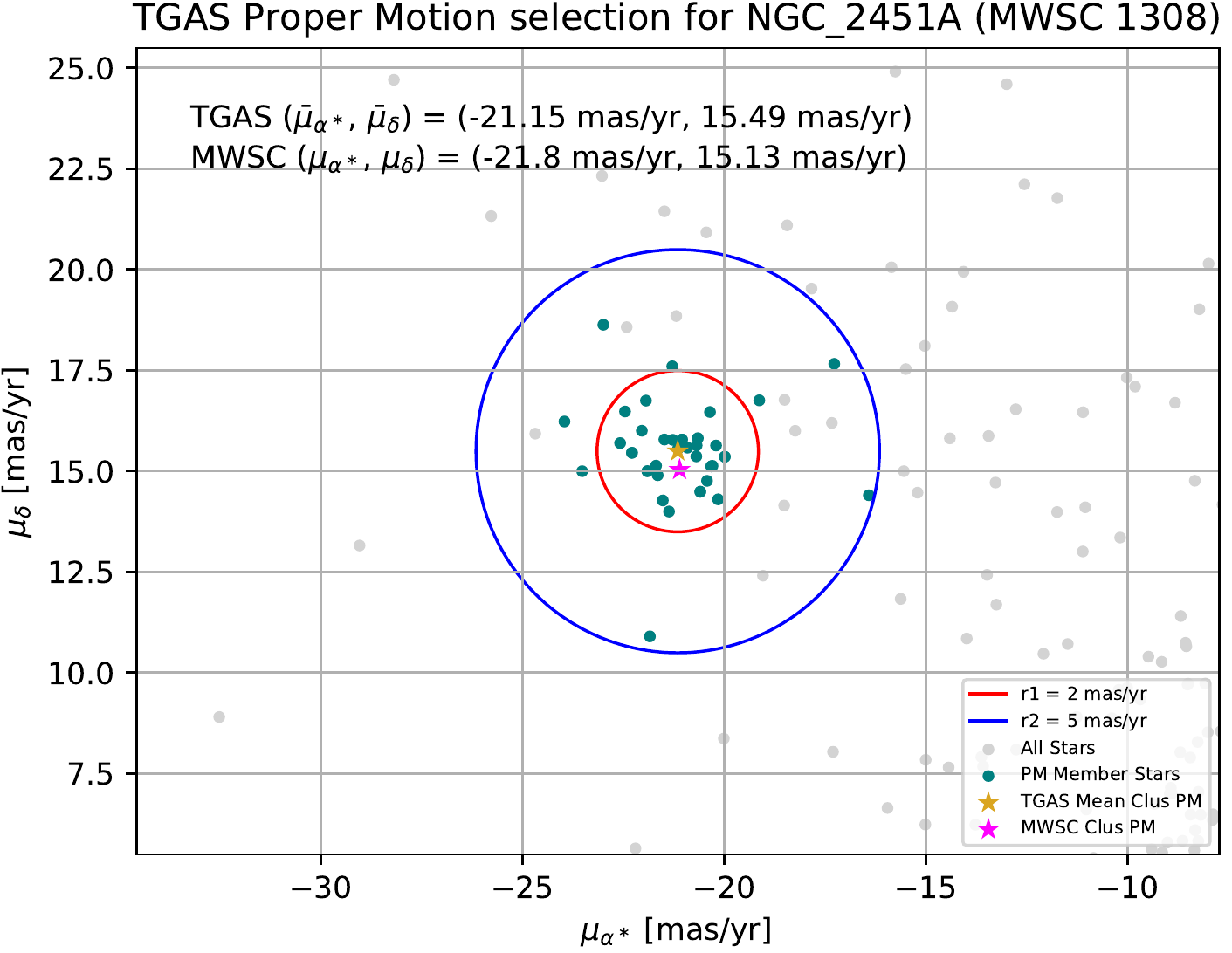}
\includegraphics[width=6.5cm]{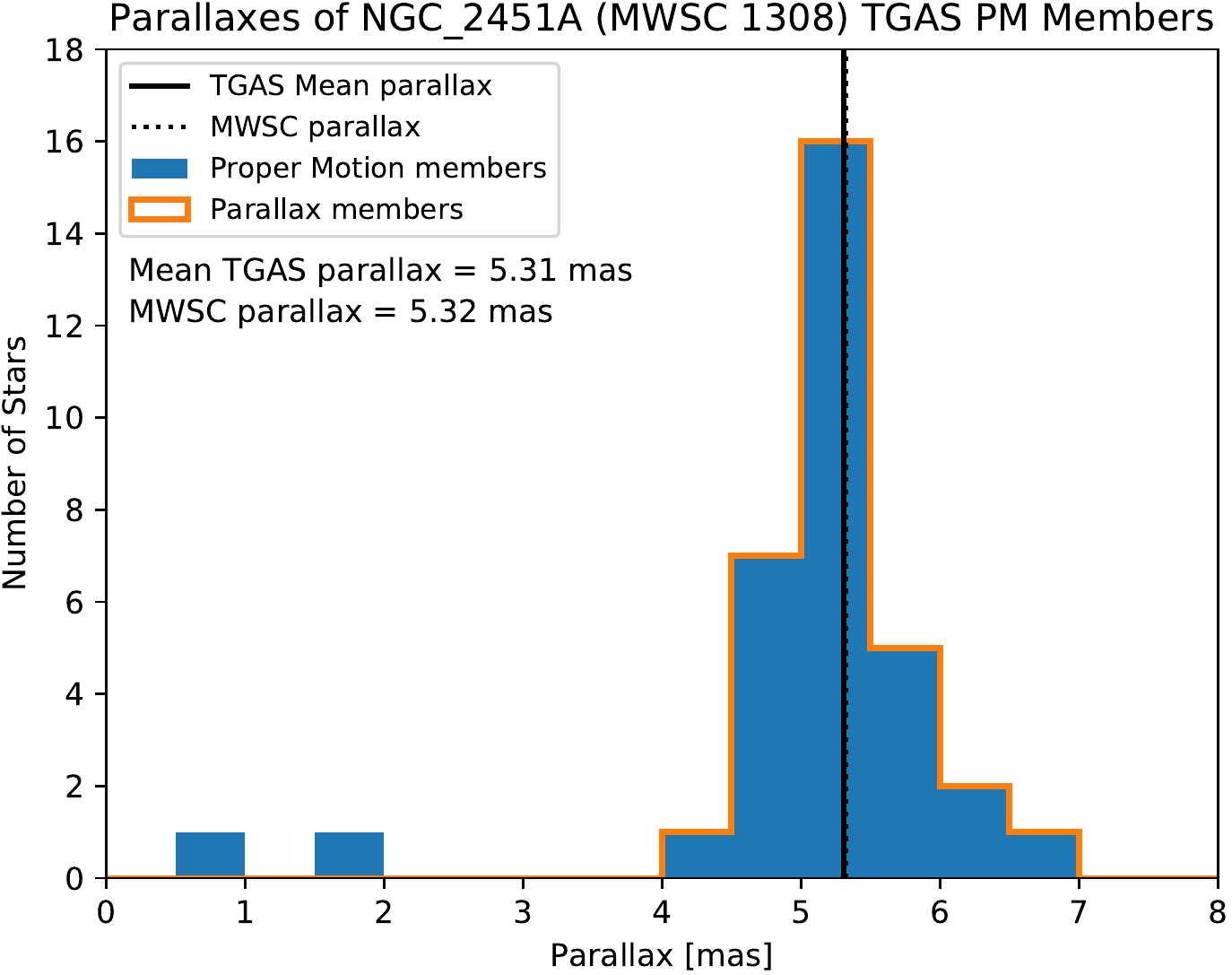}\\
\includegraphics[width=7cm]{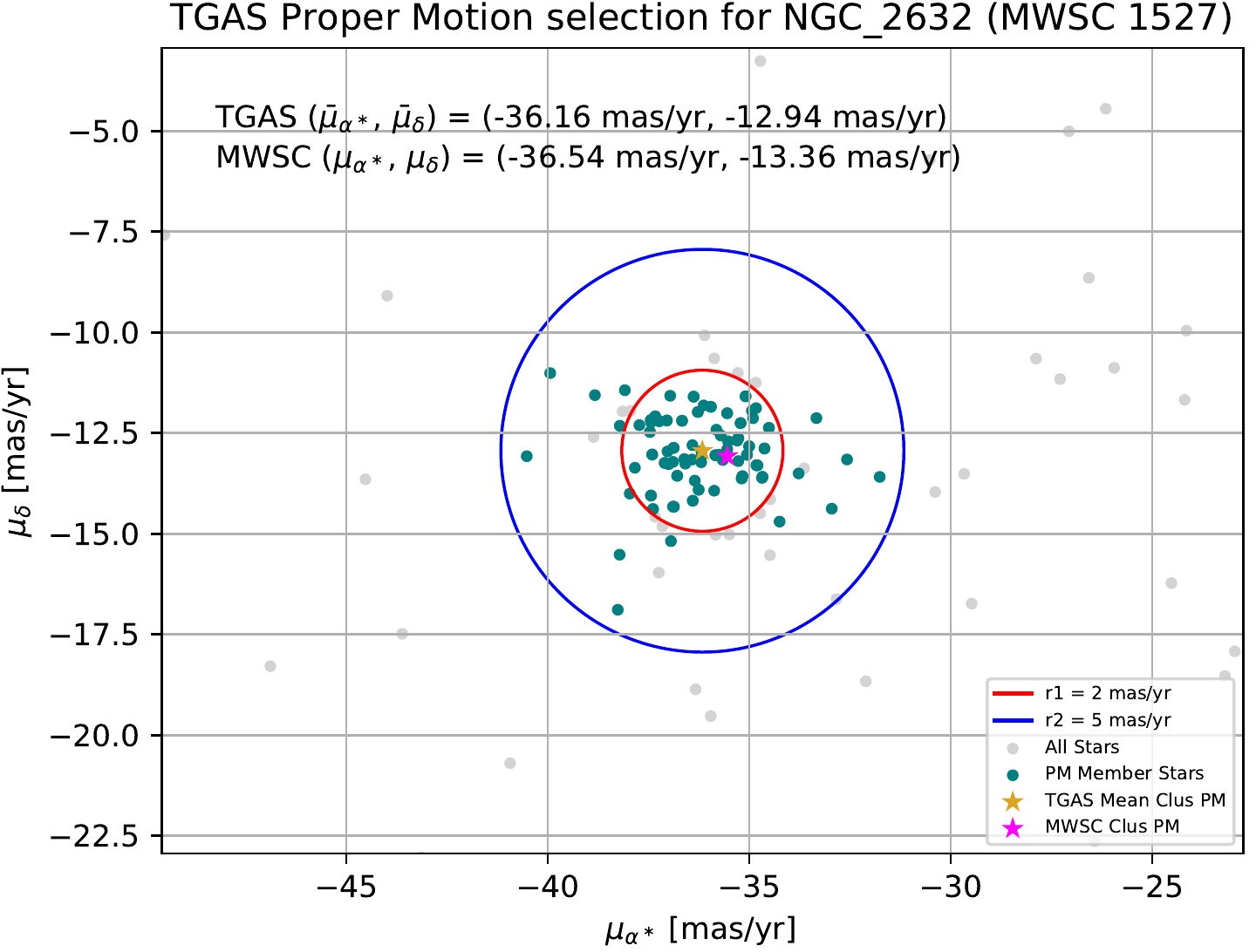}
\includegraphics[width=6.5cm]{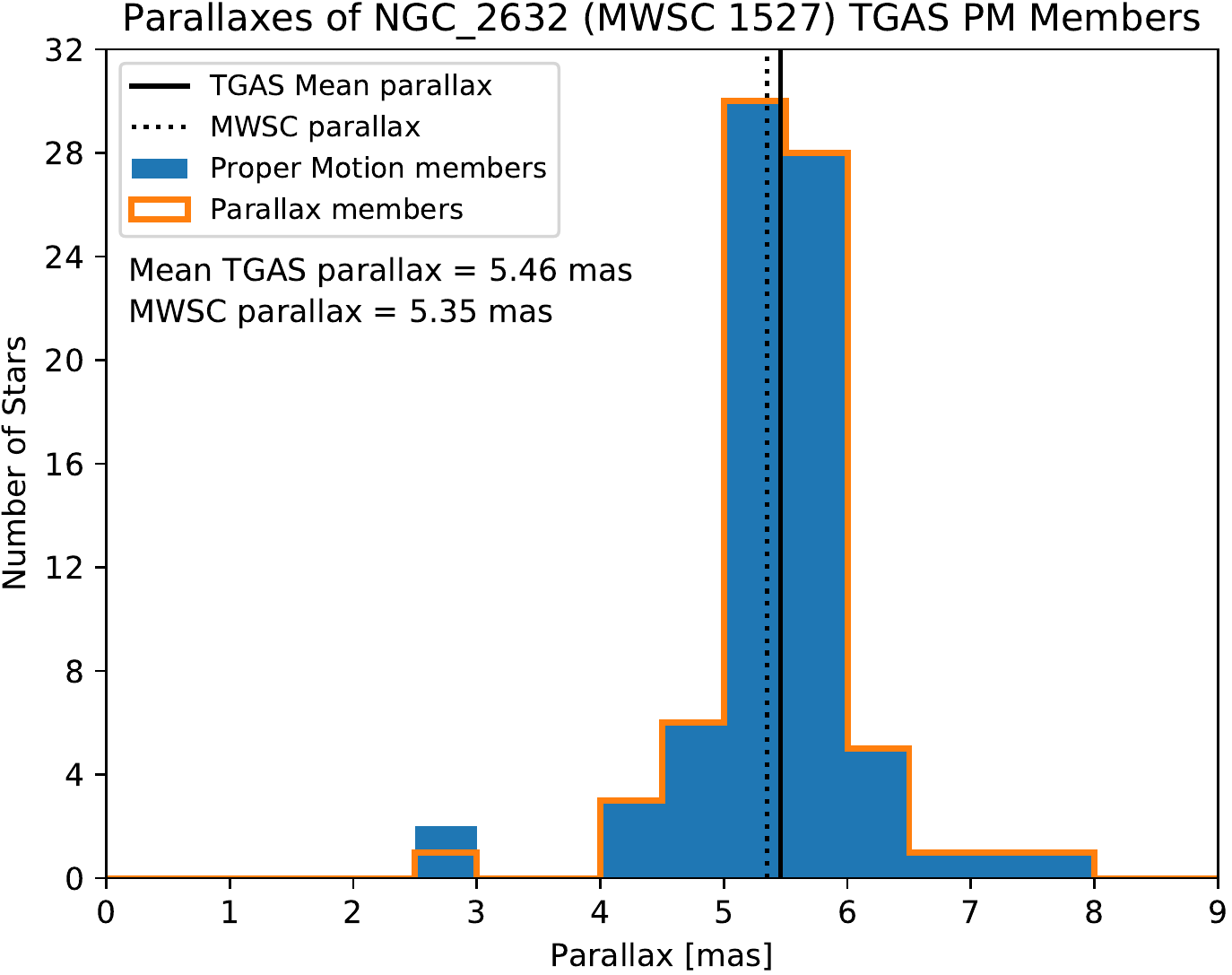}\\
\caption{TGAS proper motion (left panel) and parallax (right panel) selection diagrams for clusters, from top to bottom: NGC~2232, Alessi~3, NGC~2451A, and Praesepe (NGC~2632). The proper motion and parallax criteria for membership selection is described in Sect. 3.1\sybf{; the values for cluster proper motion and parallaxes in the diagrams are based on initial cluster membership. The final cluster proper motion and parallaxes are provided in Table~\ref{table:results}.} Left panel: The teal points represent the proper motion members, where all stars within the 2 \sybf{mas yr$^{-1}$} radius (red circle) of the mean cluster proper motion are selected and the stars within 5 \sybf{mas yr$^{-1}$} (blue circle) are only selected if their 3$\sigma$ errors are consistent with the mean cluster proper motion. Right panel: The orange outline illustrates the stars with 3$\sigma$ errors consistent with the mean cluster parallax; these stars are the TGAS astrometrically-selected candidates of the cluster.}
 \label{figb3}
\end{figure*}

\begin{figure*}
\centering
\includegraphics[width=7cm]{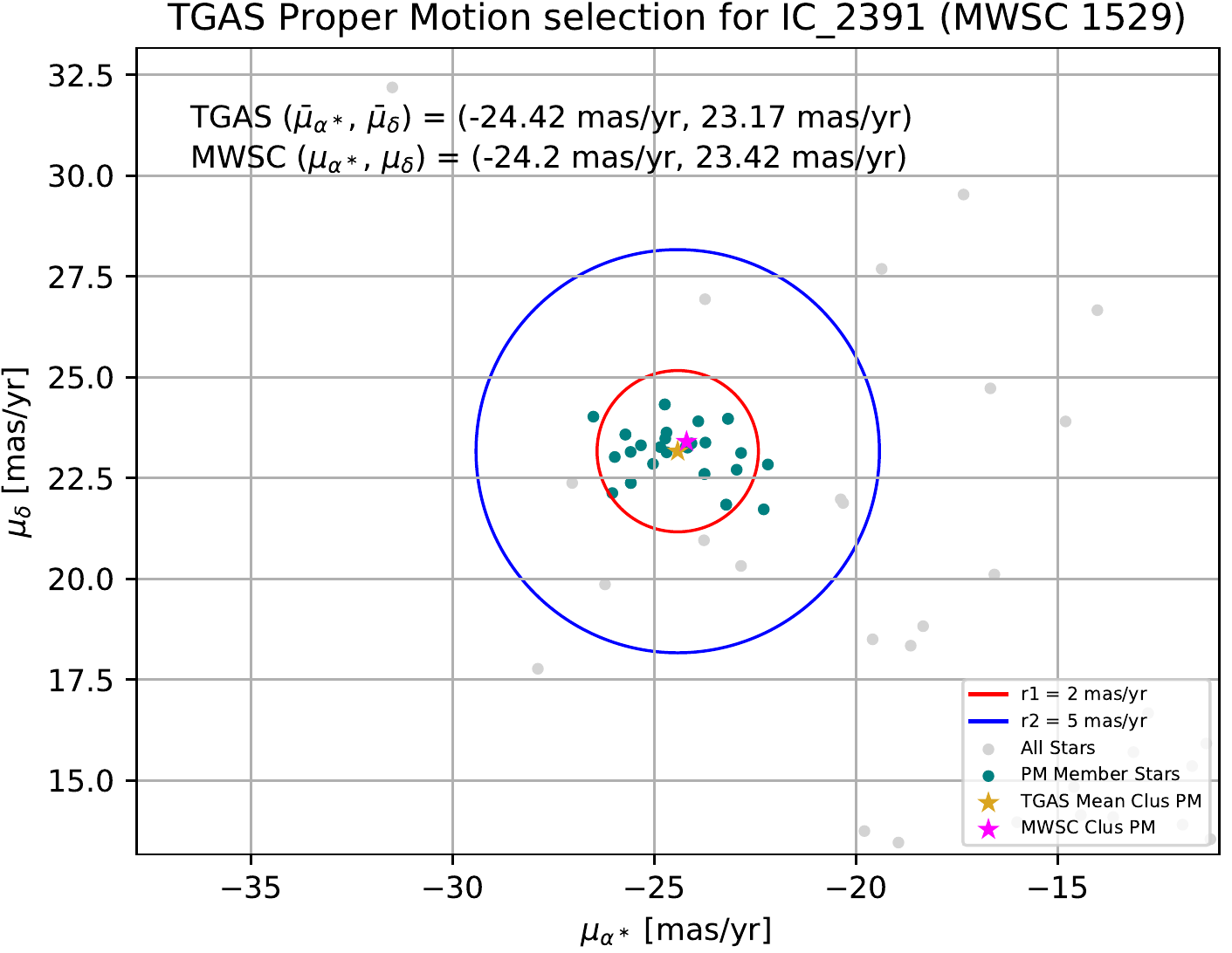}
\includegraphics[width=6.5cm]{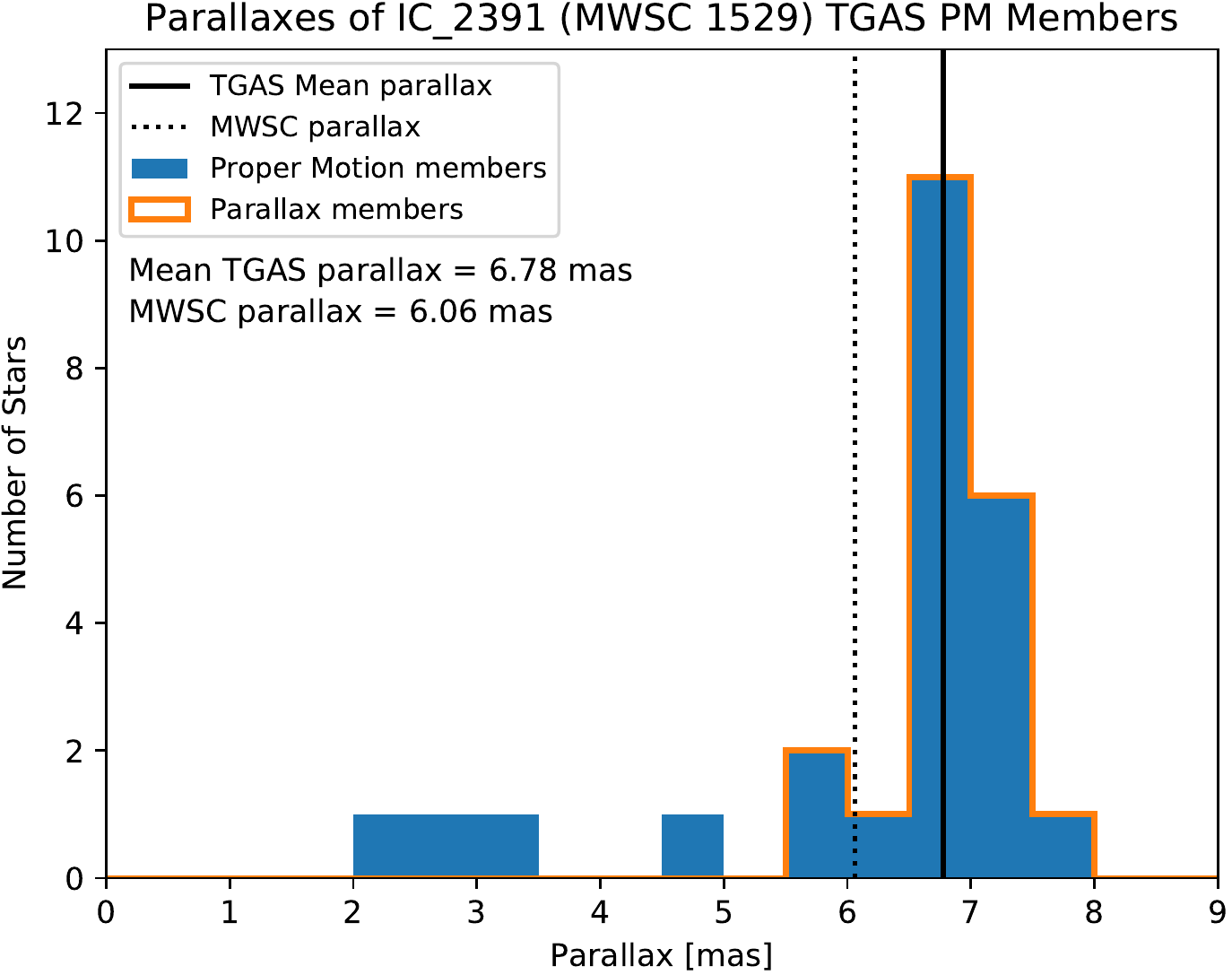}\\
\includegraphics[width=7cm]{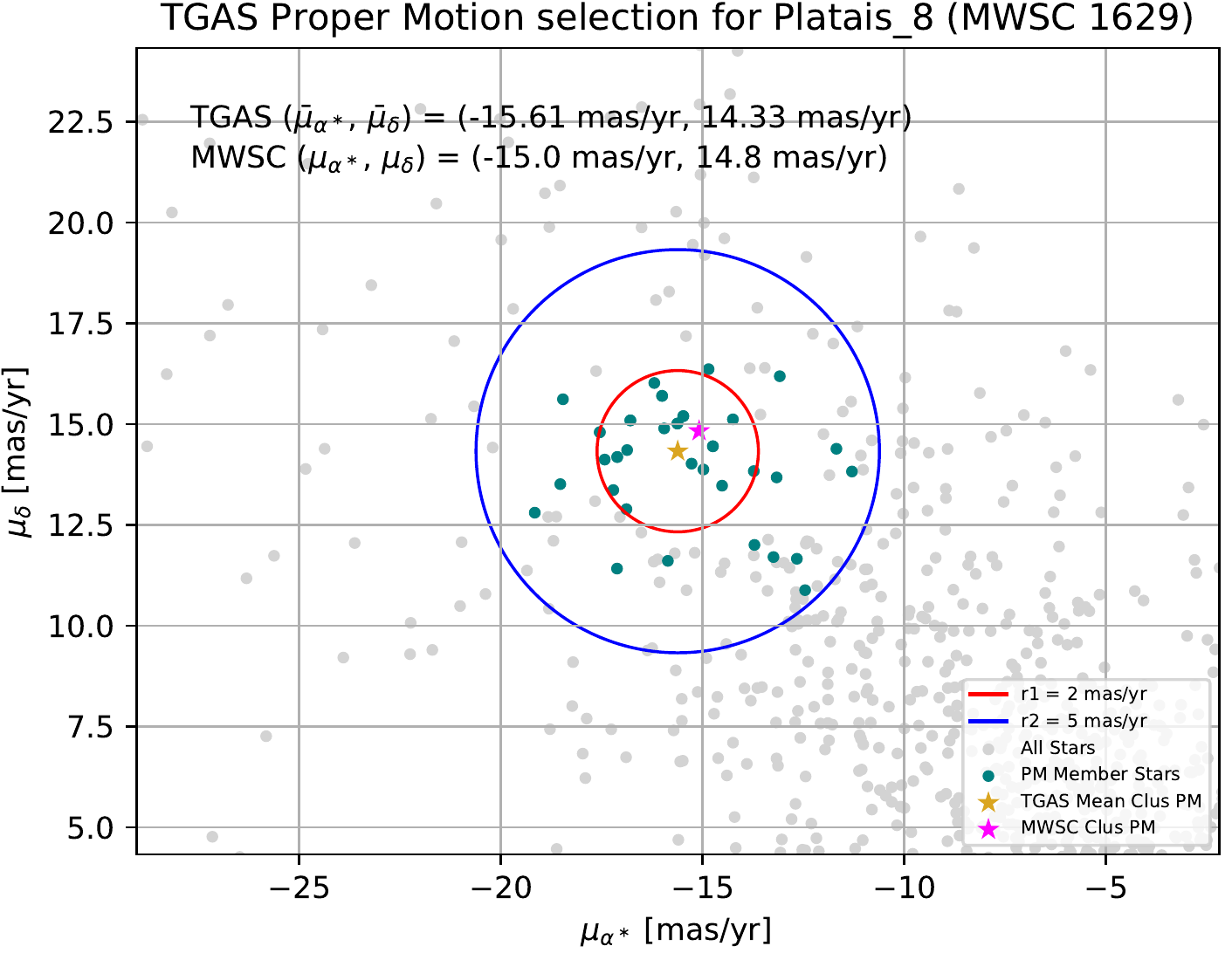}
\includegraphics[width=6.5cm]{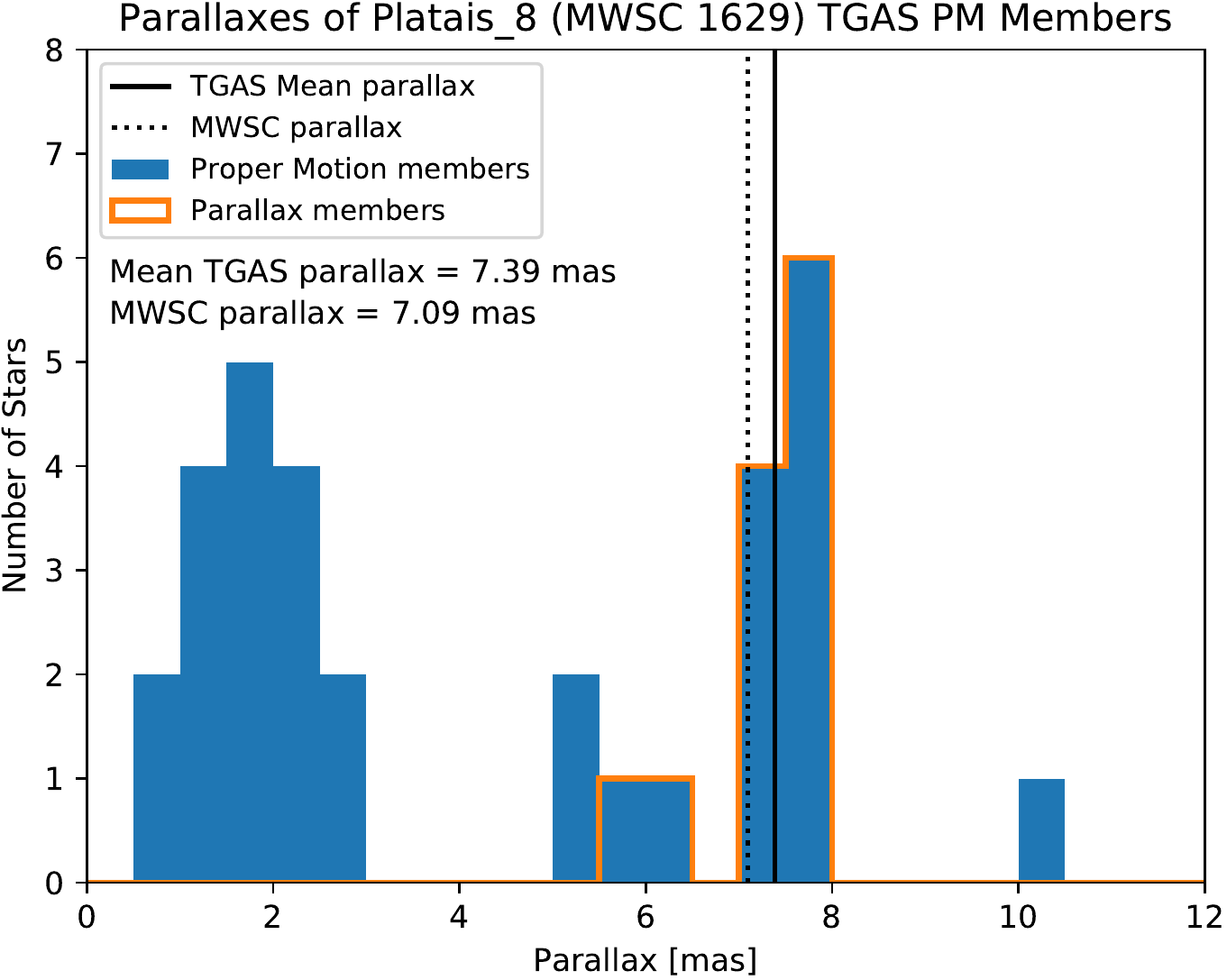}\\
\includegraphics[width=7cm]{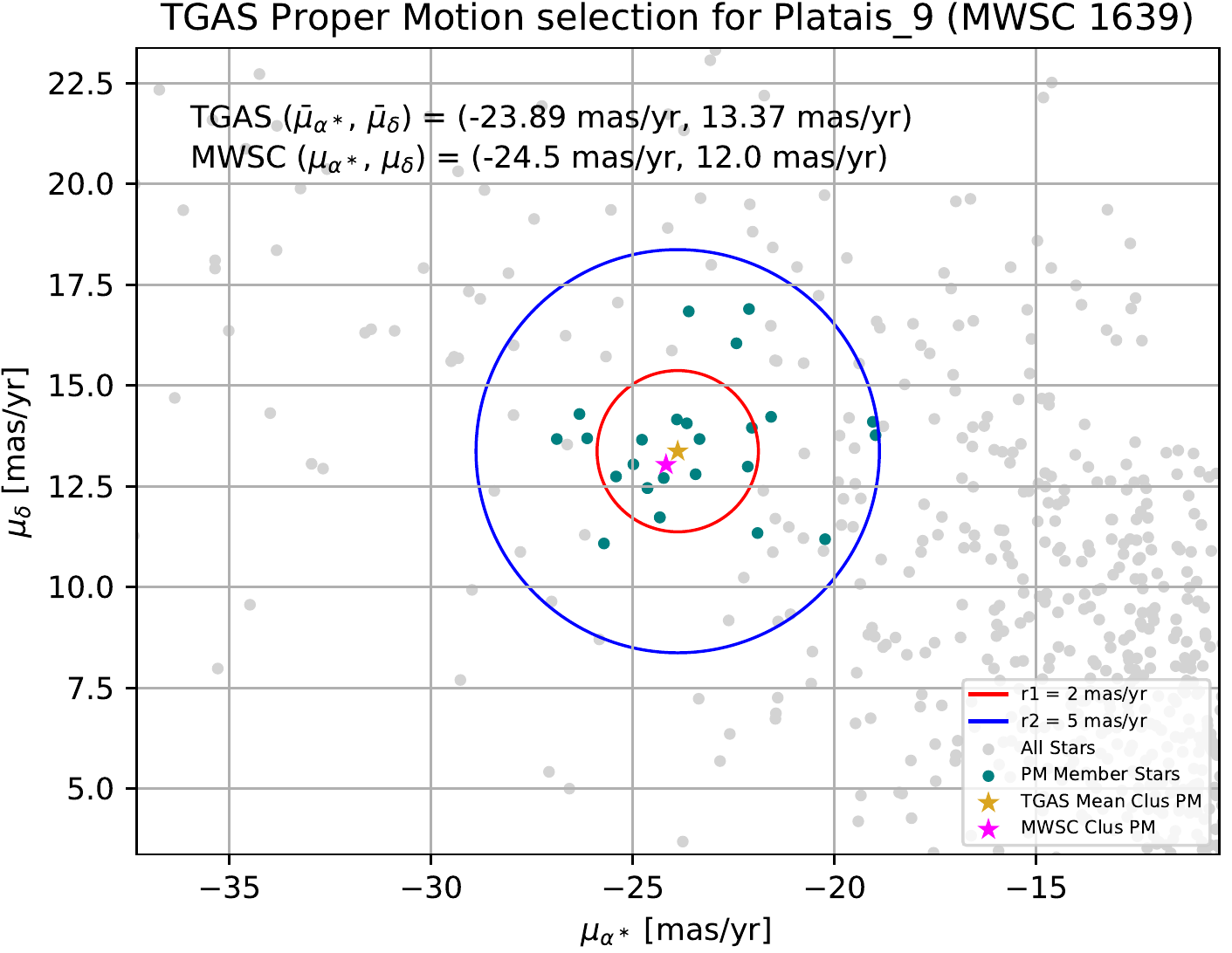}
\includegraphics[width=6.5cm]{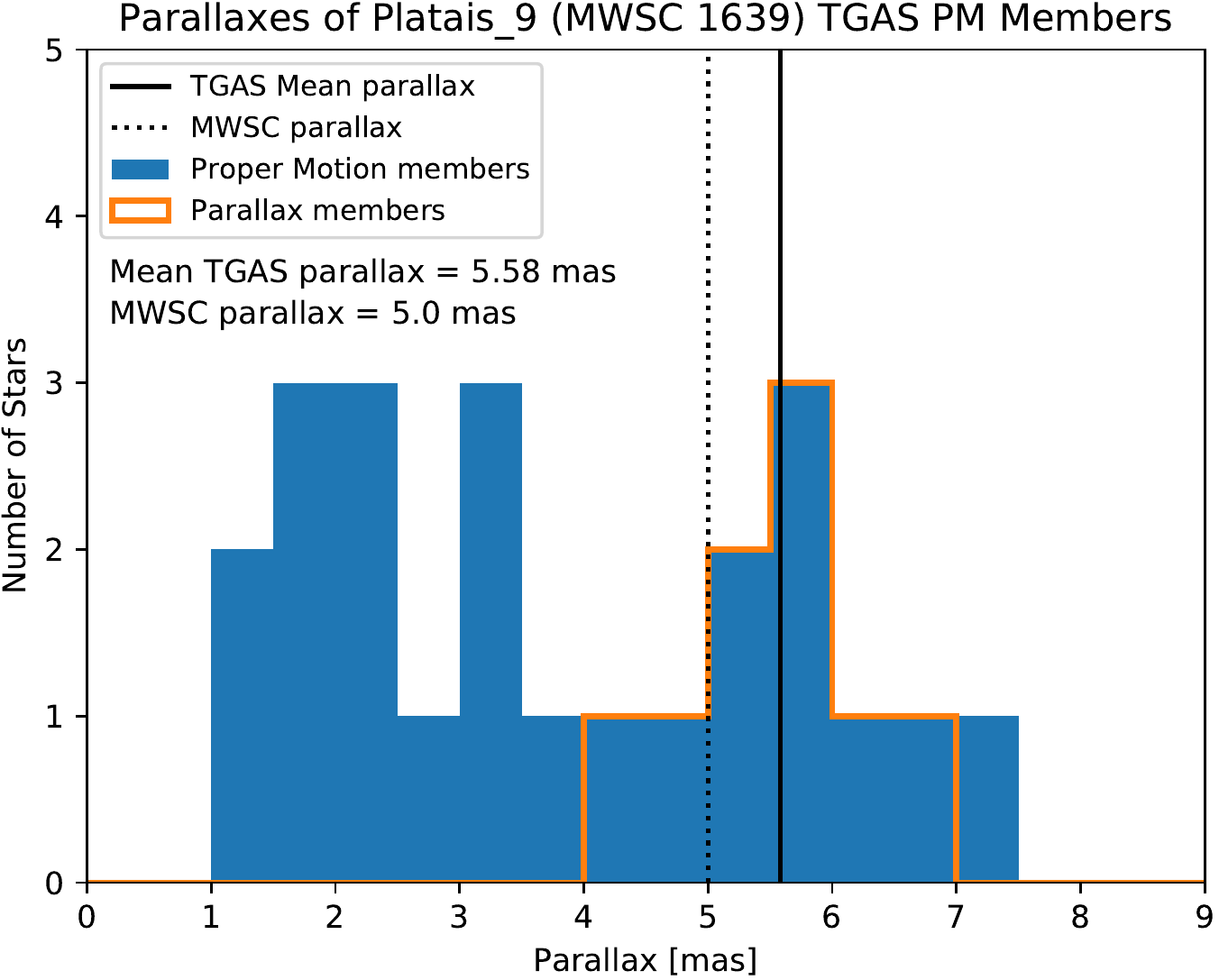}\\
\includegraphics[width=7cm]{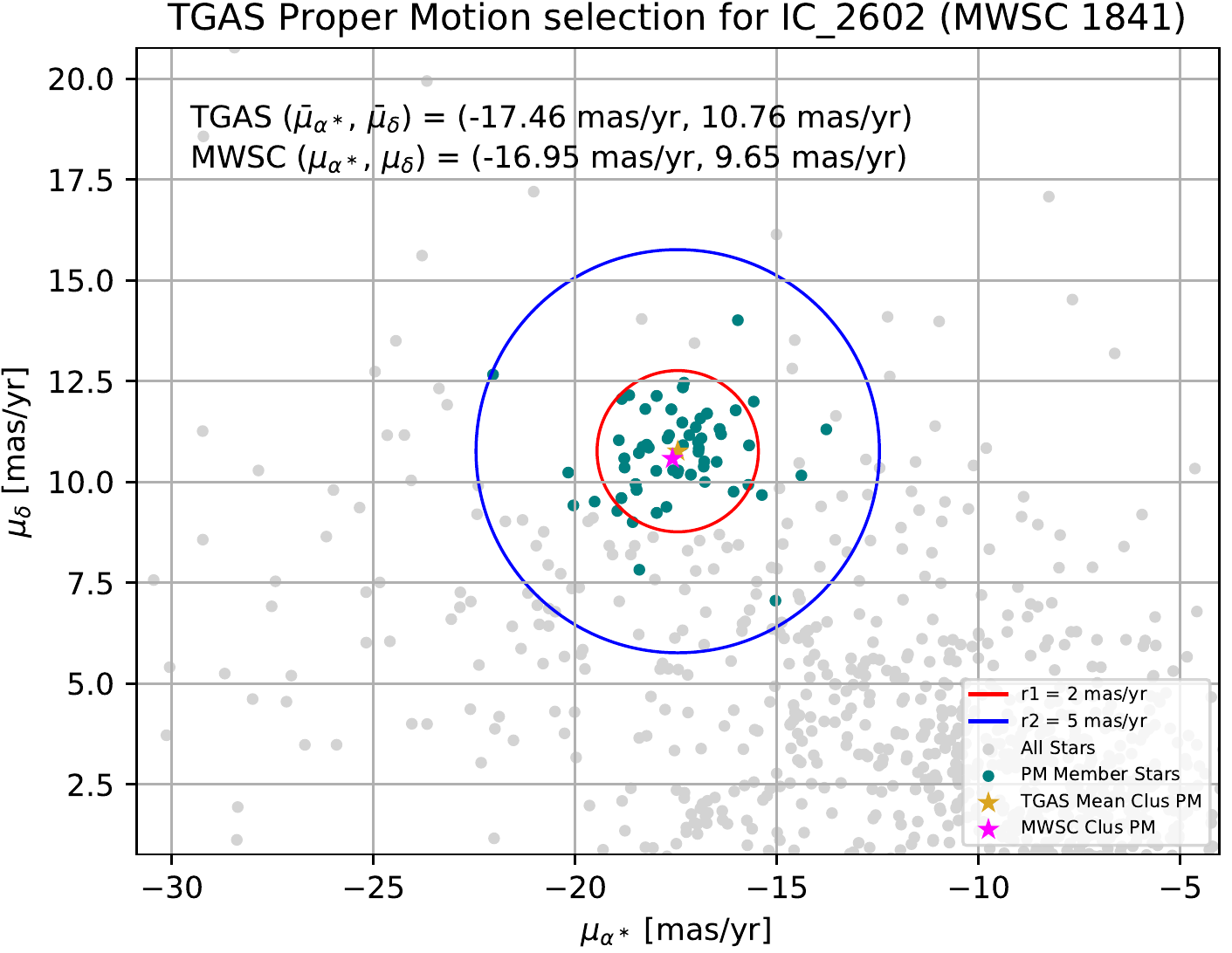}
\includegraphics[width=6.5cm]{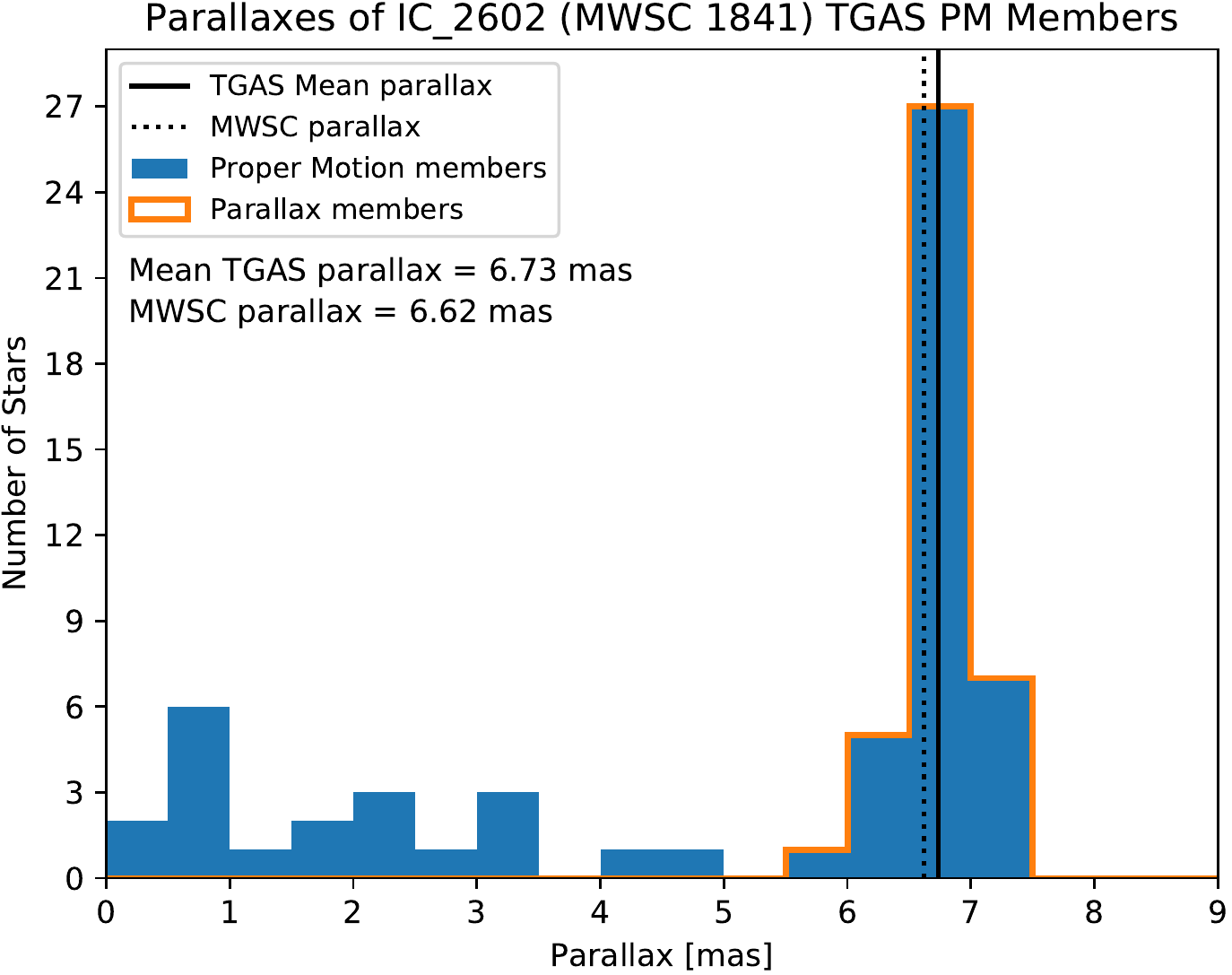}\\
\caption{TGAS proper motion (left panel) and parallax (right panel) selection diagrams for clusters, from top to bottom: IC~2391, Platais~8, Platais~9, and IC~2602. The proper motion and parallax criteria for membership selection is described in Sect. 3.1\sybf{; the values for cluster proper motion and parallaxes in the diagrams are based on initial cluster membership. The final cluster proper motion and parallaxes are provided in Table~\ref{table:results}.} Left panel: The teal points represent the proper motion members, where all stars within the 2 \sybf{mas yr$^{-1}$} radius (red circle) of the mean cluster proper motion are selected and the stars within 5 \sybf{mas yr$^{-1}$} (blue circle) are only selected if their 3$\sigma$ errors are consistent with the mean cluster proper motion. Right panel: The orange outline illustrates the stars with 3$\sigma$ errors consistent with the mean cluster parallax; these stars are the TGAS astrometrically-selected candidates of the cluster.}
 \label{figb4}
\end{figure*}

\begin{figure*}
\centering
\includegraphics[width=7cm]{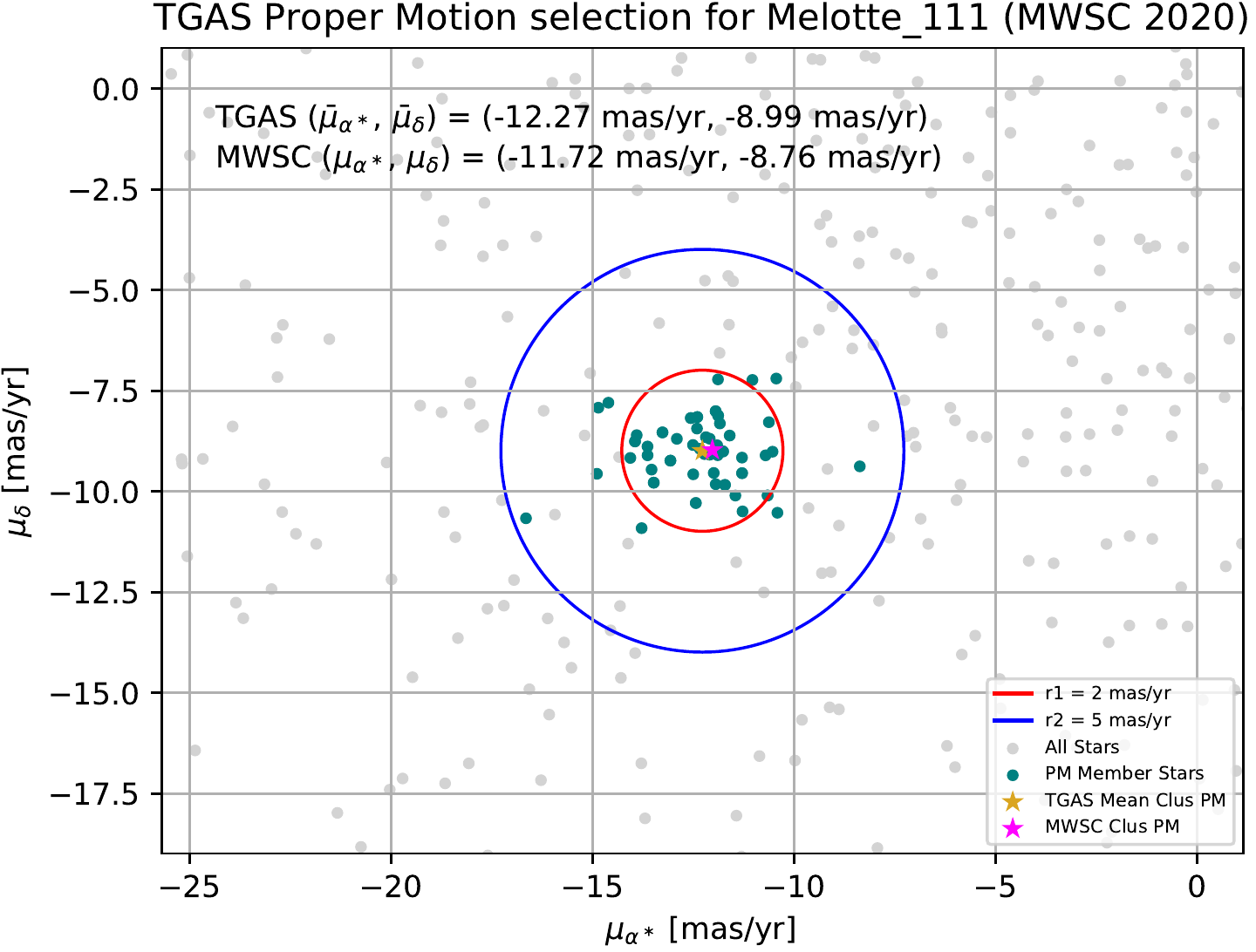}
\includegraphics[width=6.5cm]{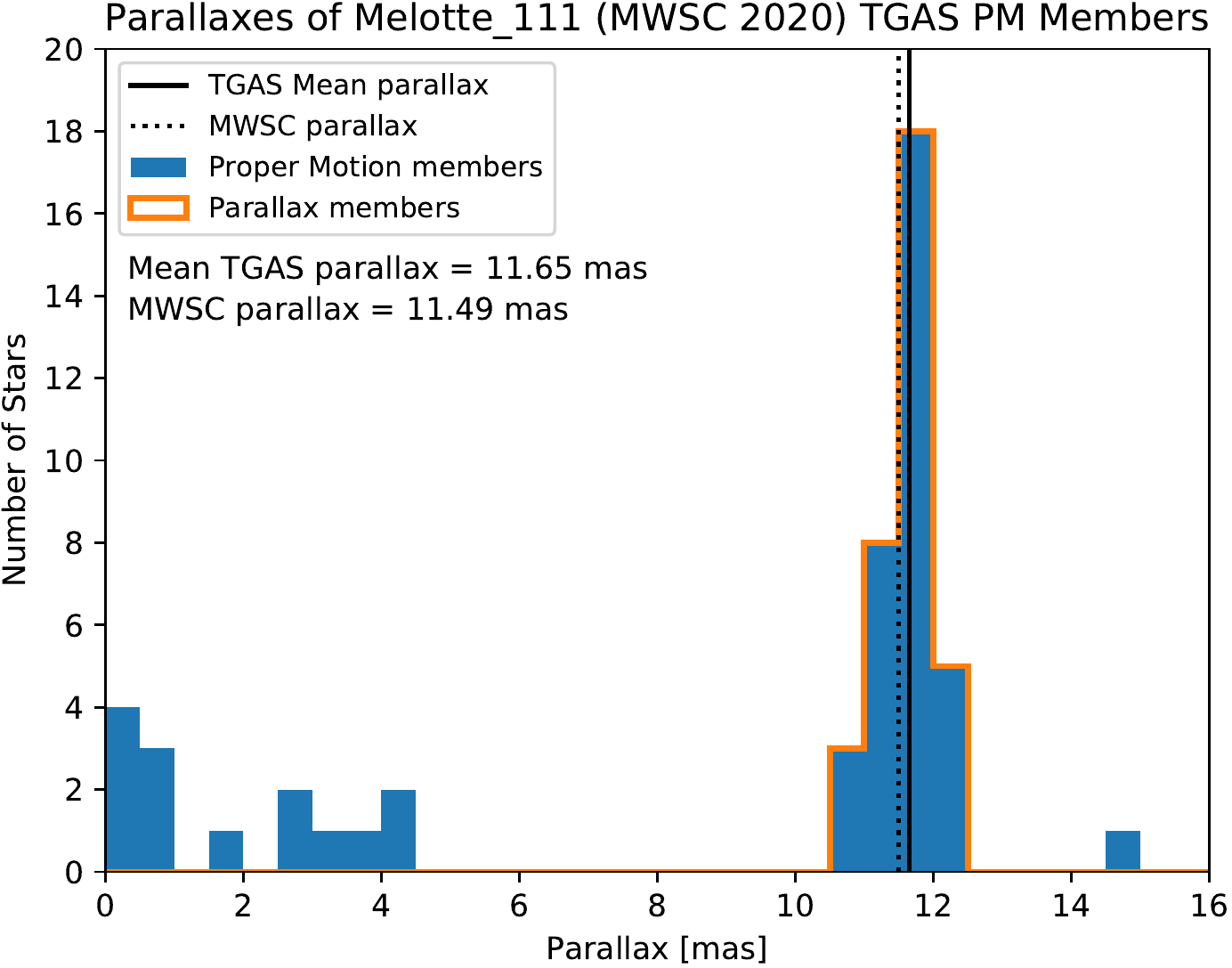}\\
\includegraphics[width=7cm]{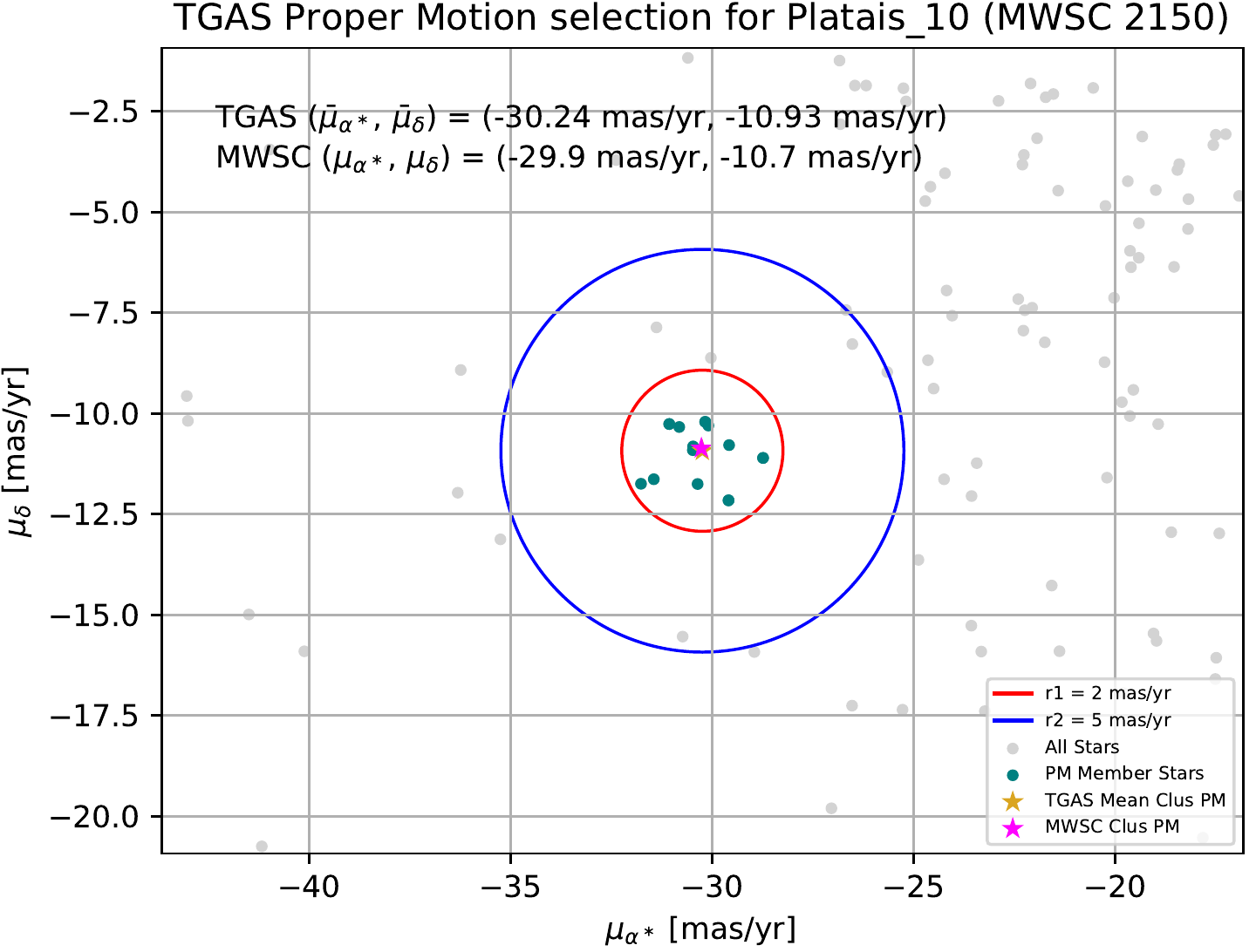}
\includegraphics[width=6.5cm]{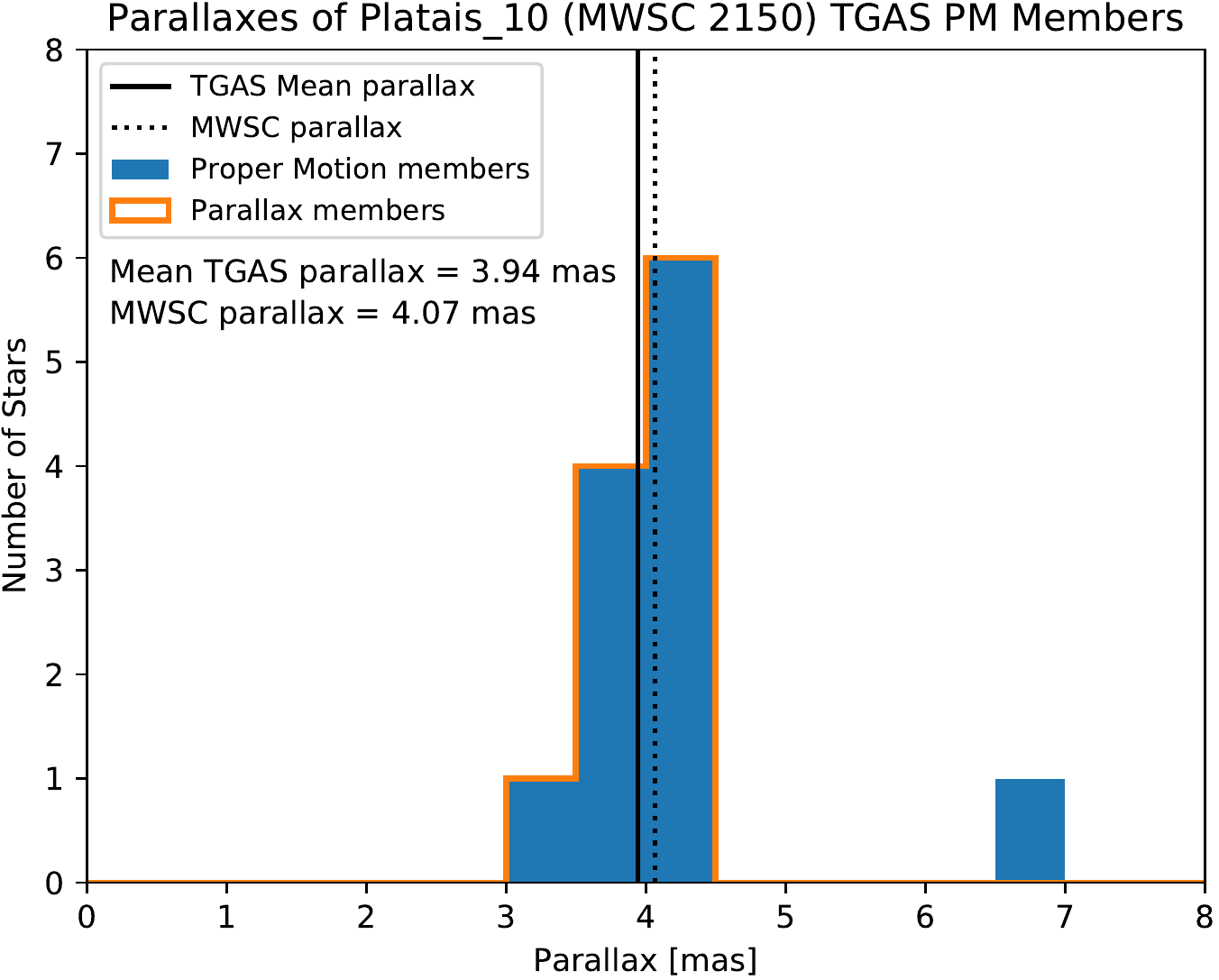}\\
\includegraphics[width=7cm]{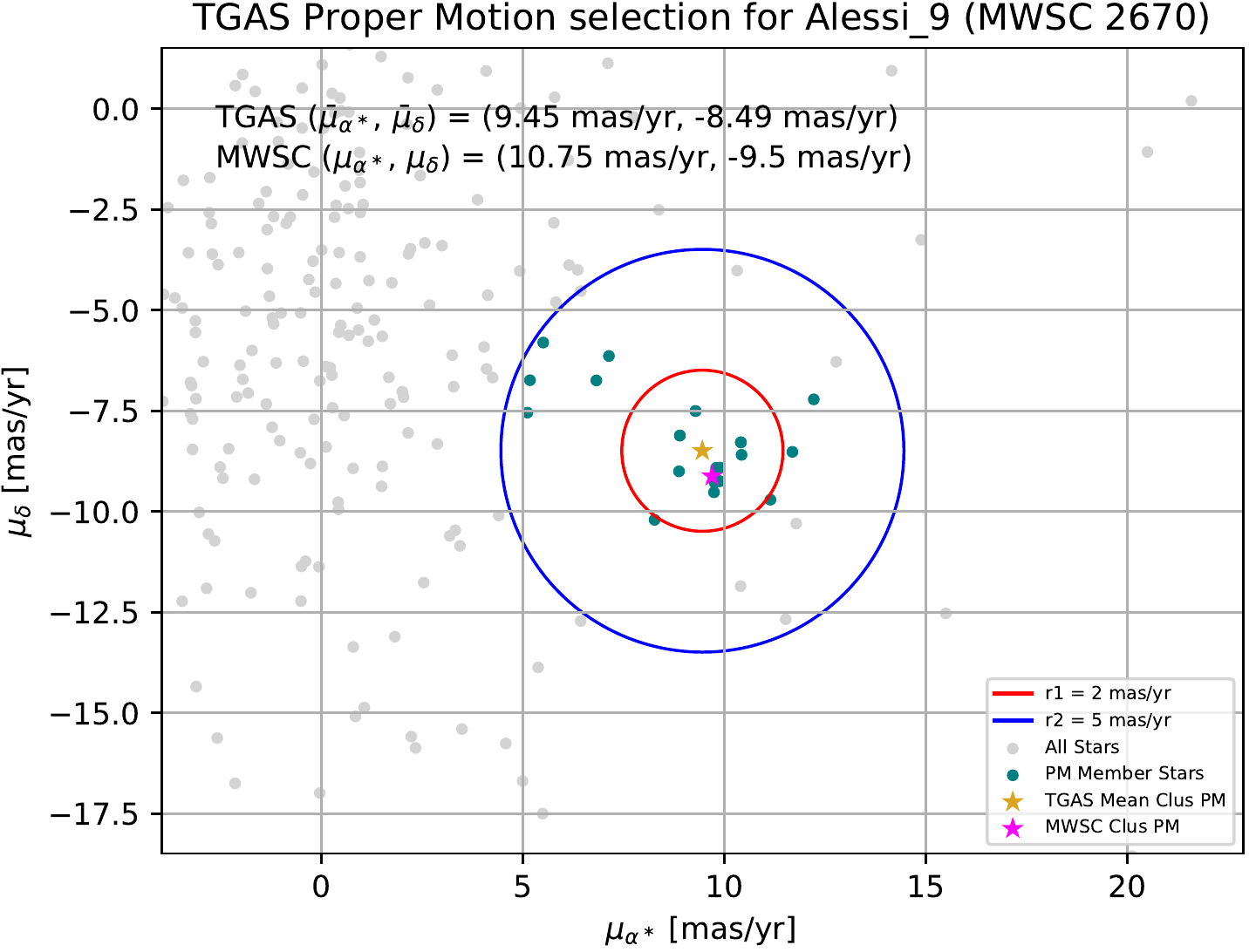}
\includegraphics[width=6.5cm]{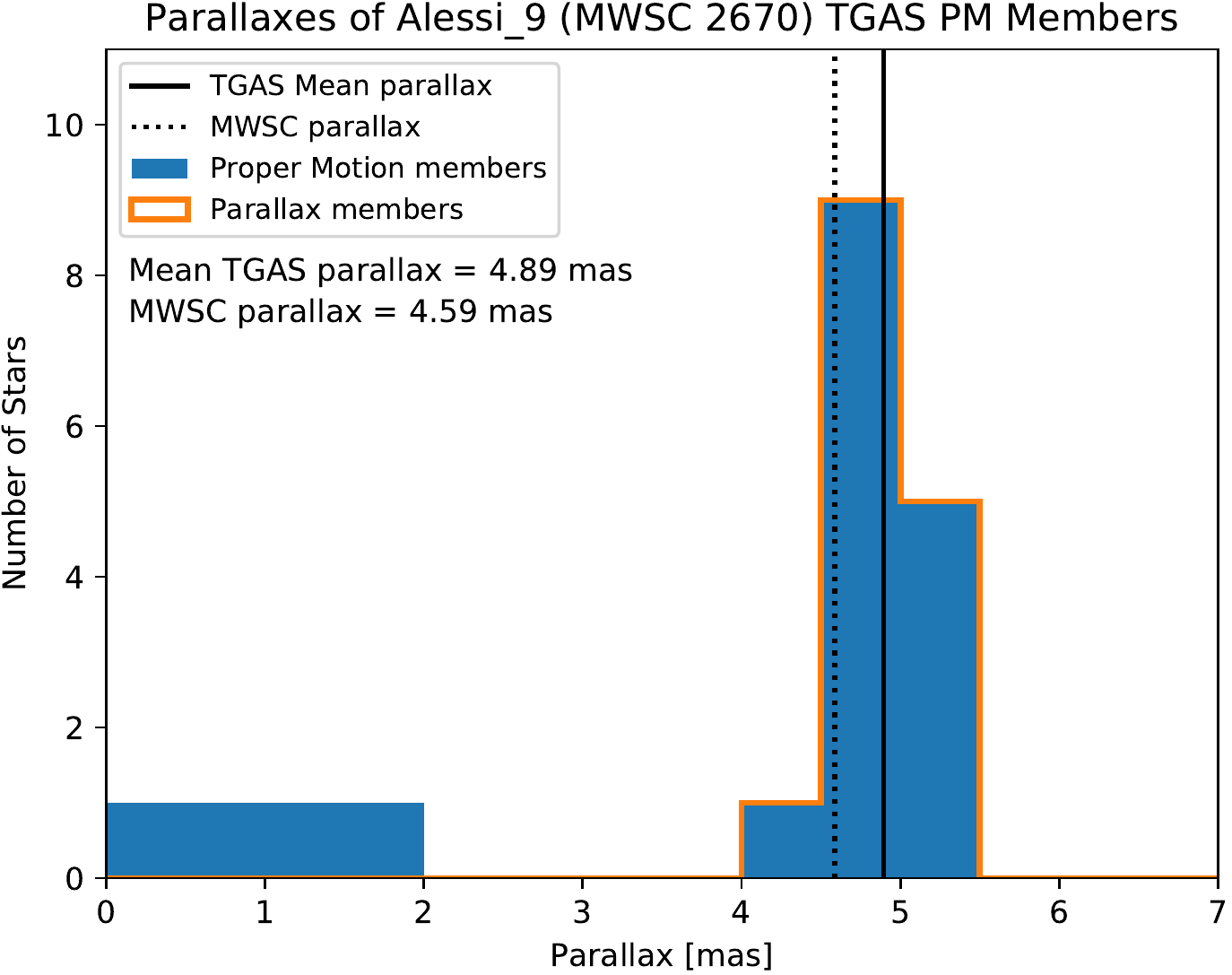}\\
\includegraphics[width=7cm]{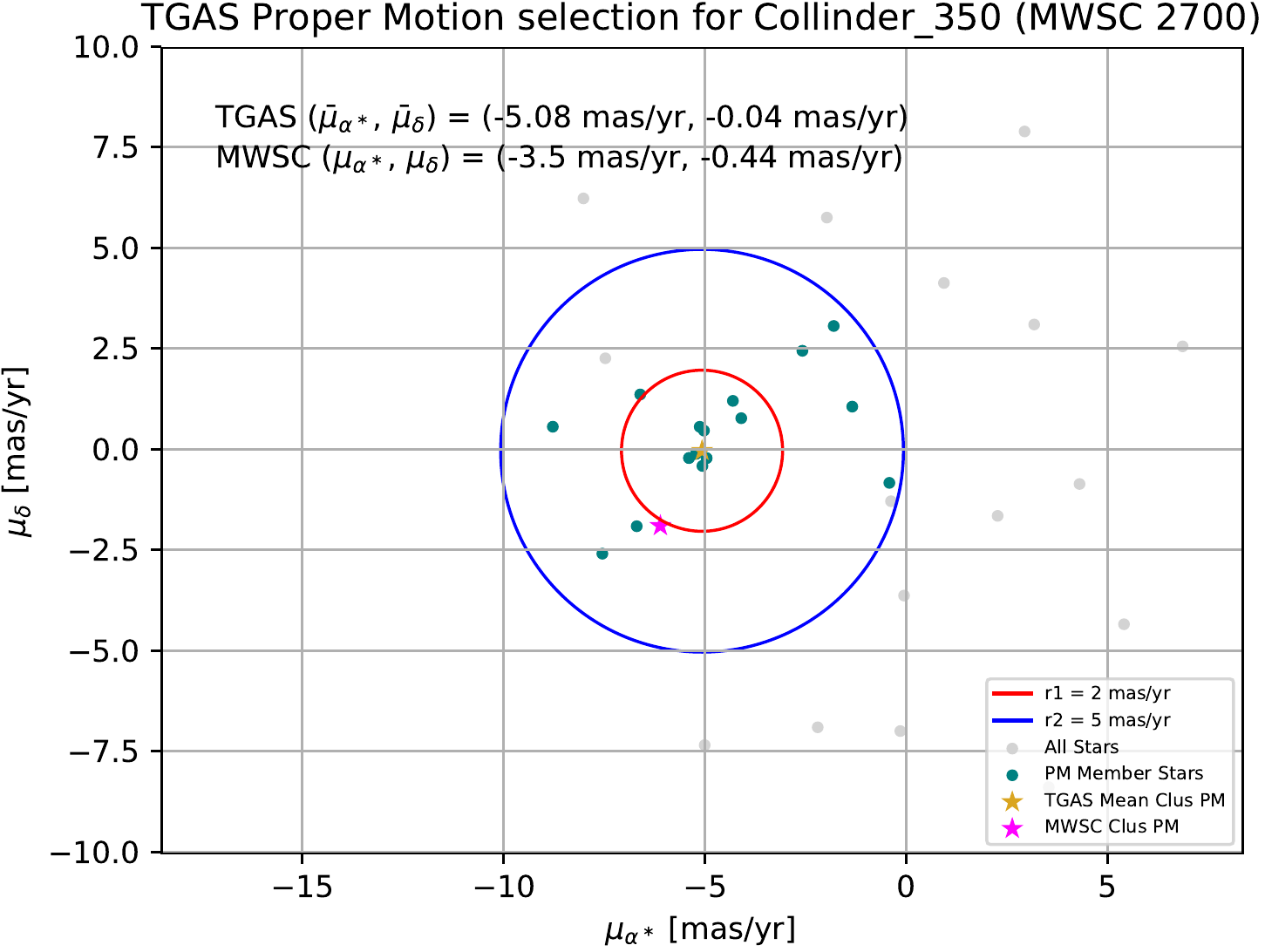}
\includegraphics[width=6.5cm]{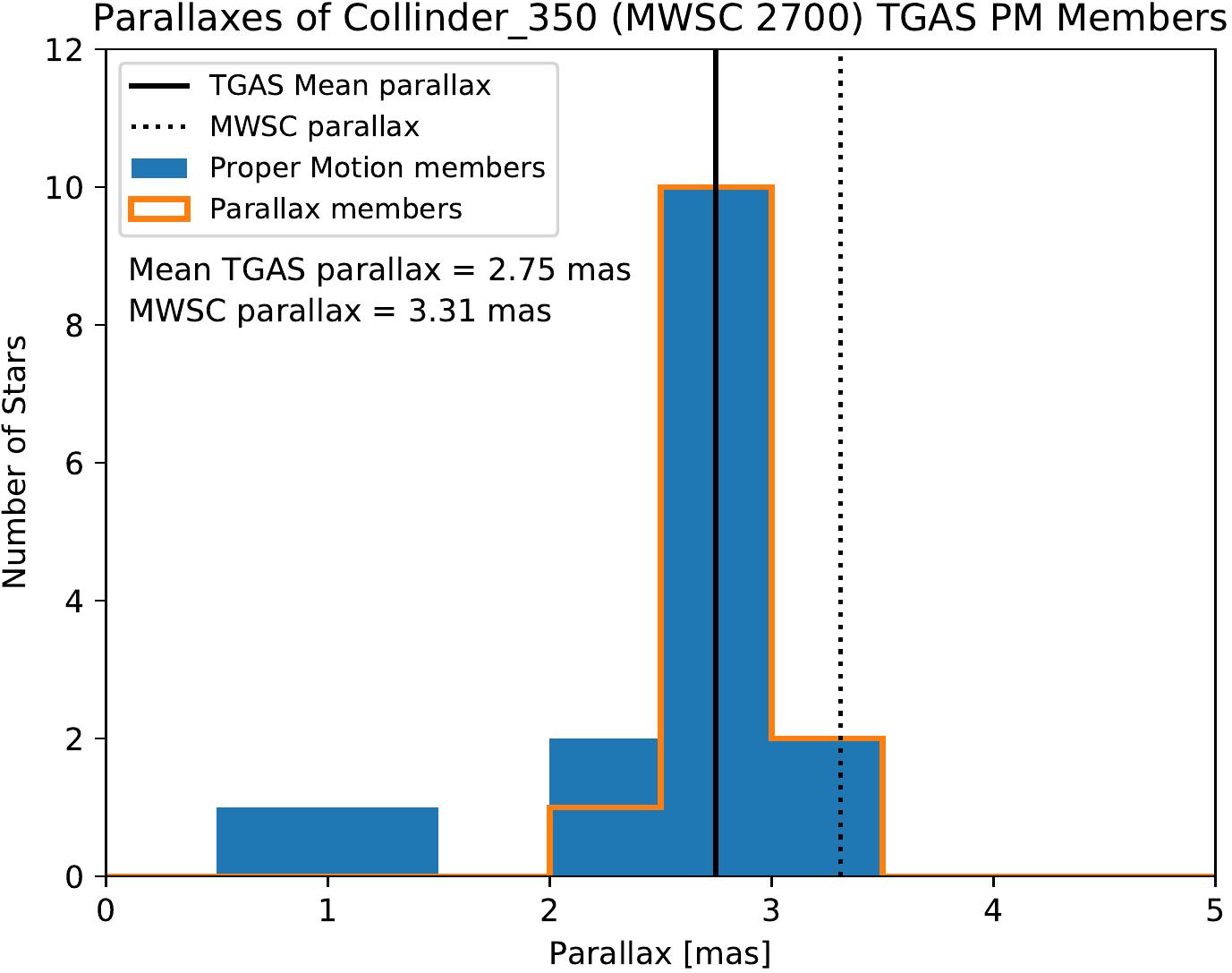}\\
\caption{TGAS proper motion (left panel) and parallax (right panel) selection diagrams for clusters, from top to bottom: Coma Ber (Melotte~111), Platais~10, Alessi~9, and Collinder~350. The proper motion and parallax criteria for membership selection is described in Sect. 3.1\sybf{; the values for cluster proper motion and parallaxes in the diagrams are based on initial cluster membership. The final cluster proper motion and parallaxes are provided in Table~\ref{table:results}.} Left panel: The teal points represent the proper motion members, where all stars within the 2 \sybf{mas yr$^{-1}$} radius (red circle) of the mean cluster proper motion are selected and the stars within 5 \sybf{mas yr$^{-1}$} (blue circle) are only selected if their 3$\sigma$ errors are consistent with the mean cluster proper motion. Right panel: The orange outline illustrates the stars with 3$\sigma$ errors consistent with the mean cluster parallax; these stars are the TGAS astrometrically-selected candidates of the cluster.}
 \label{figb5}
\end{figure*}

\begin{figure*}
\centering
\includegraphics[width=7cm]{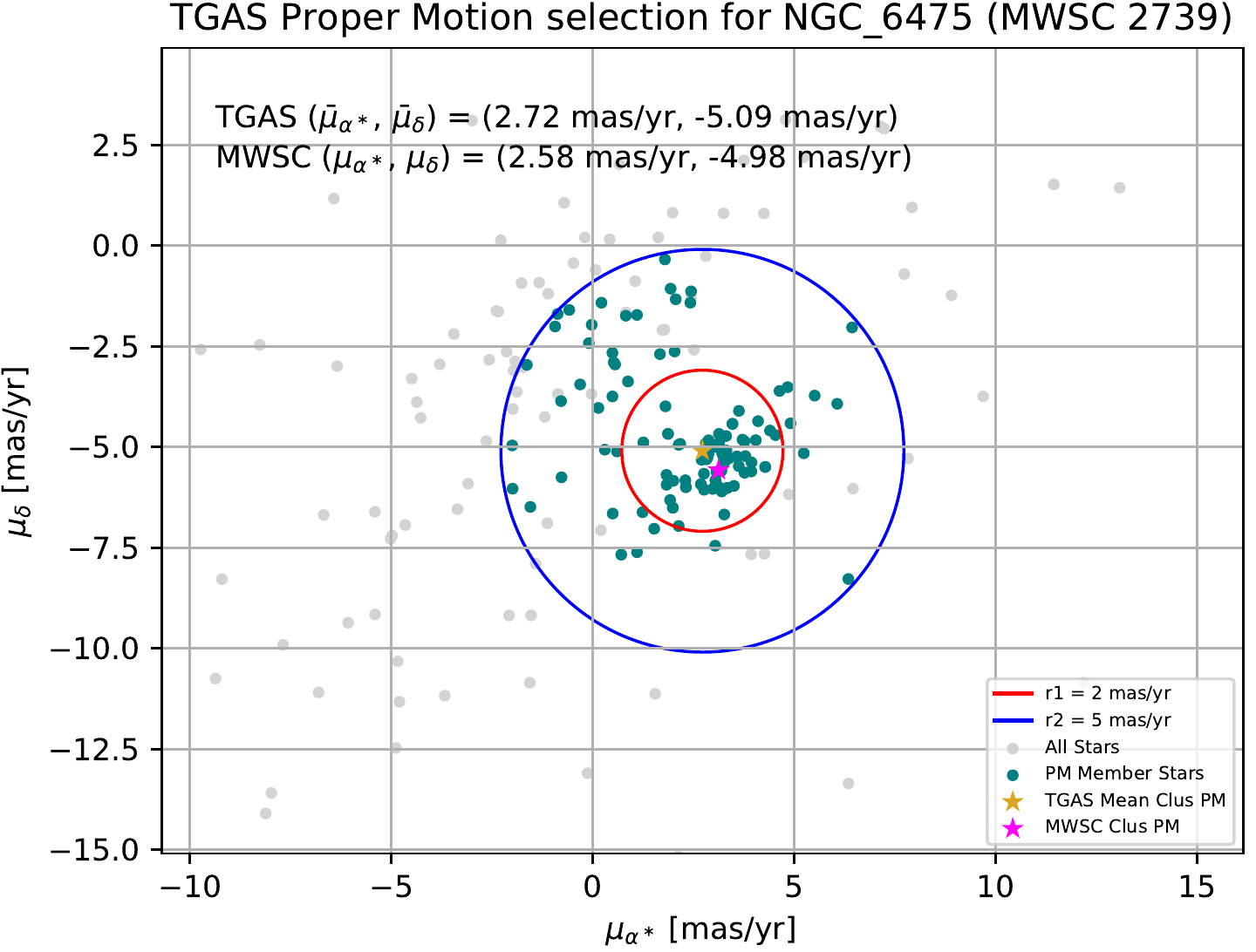}
\includegraphics[width=6.5cm]{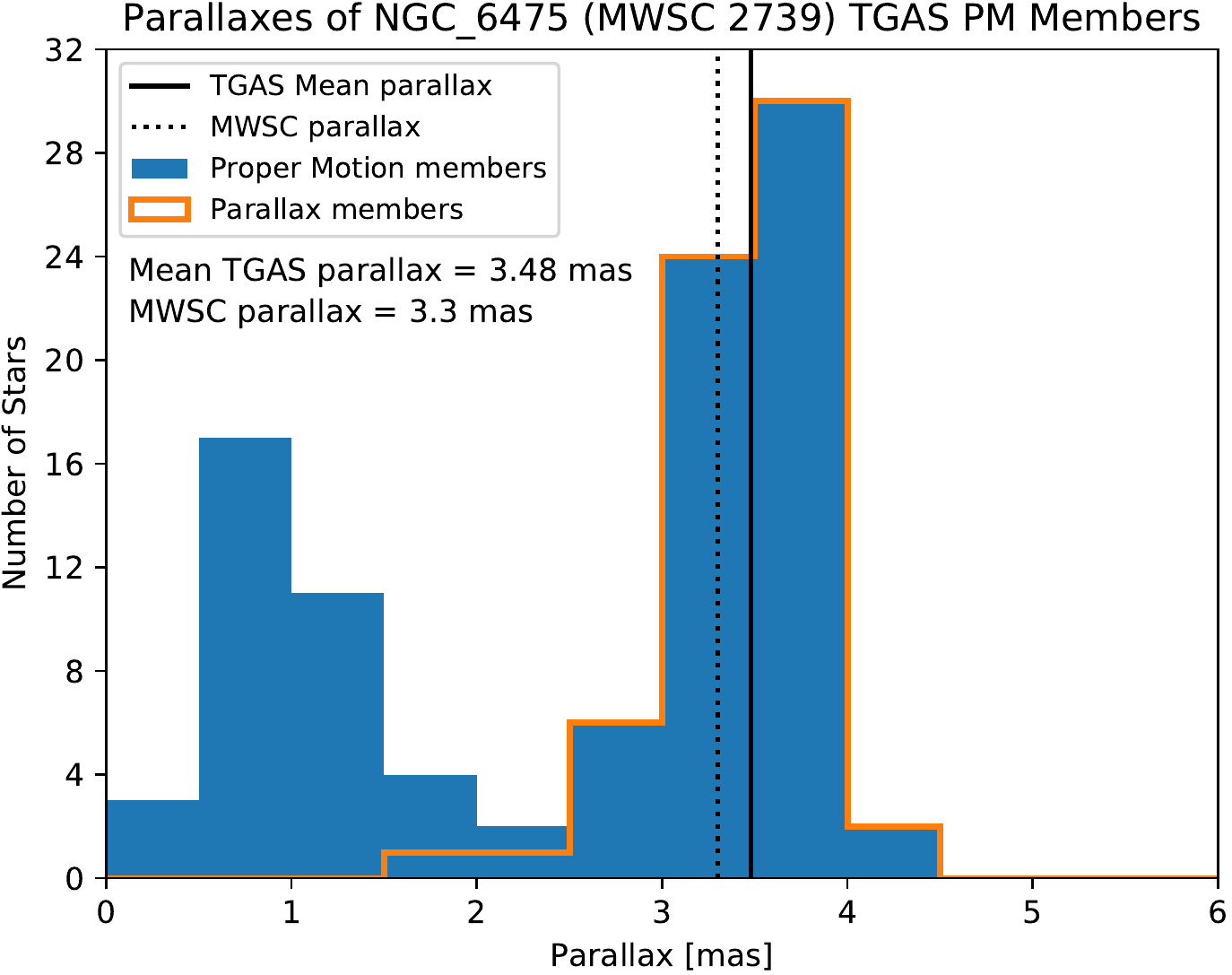}\\
\includegraphics[width=7cm]{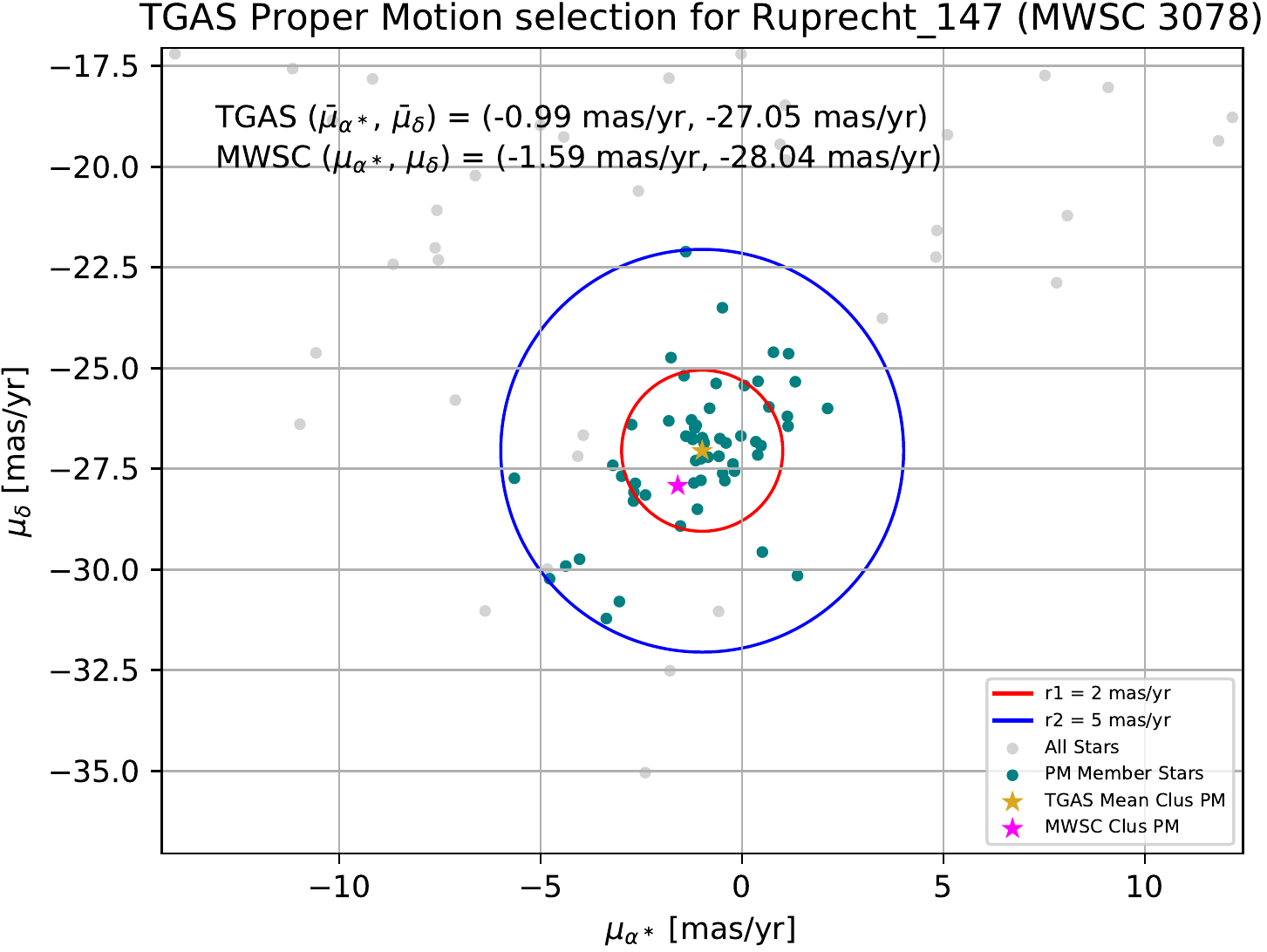}
\includegraphics[width=6.5cm]{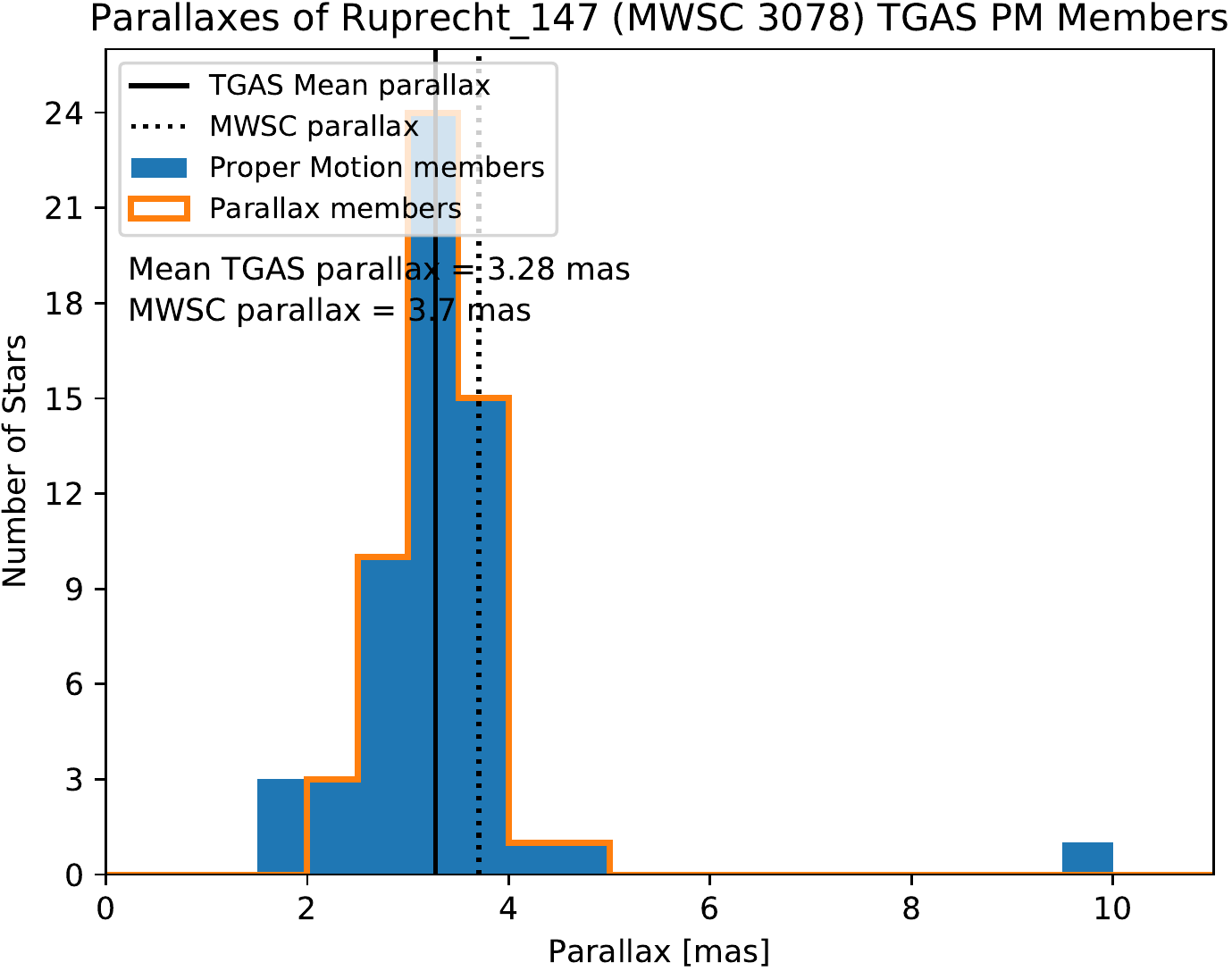}\\
\includegraphics[width=7cm]{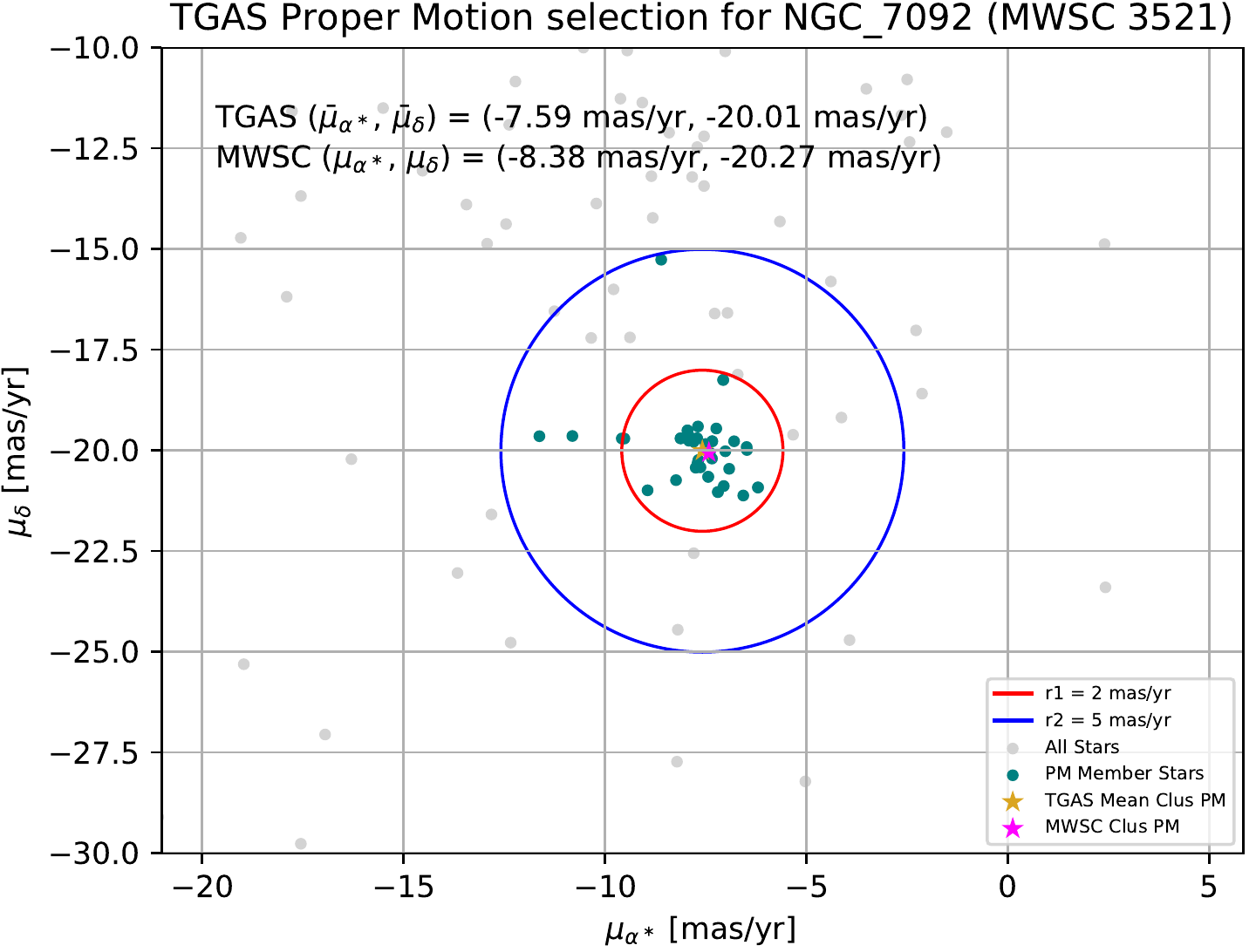}
\includegraphics[width=6.5cm]{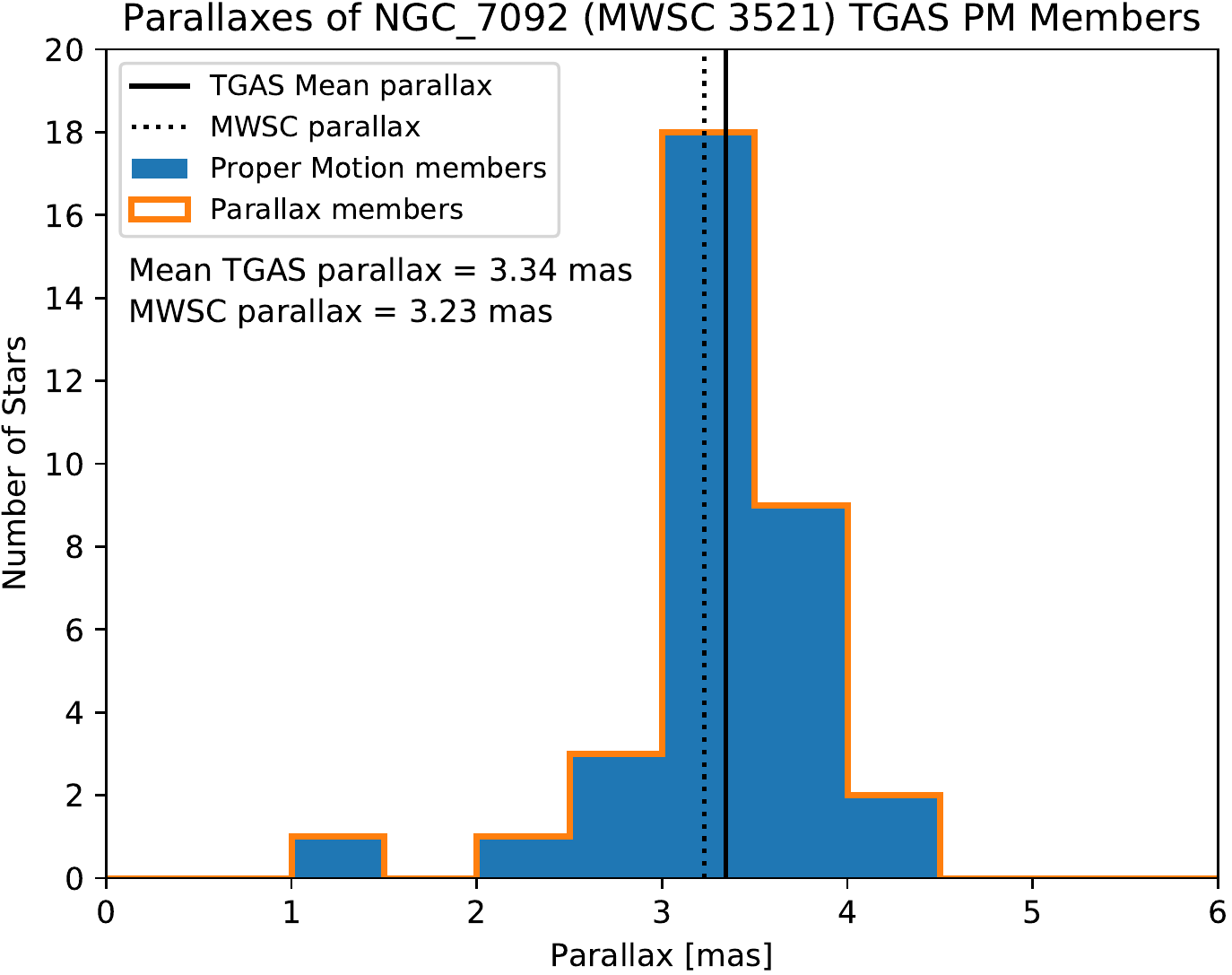}\\
\includegraphics[width=7cm]{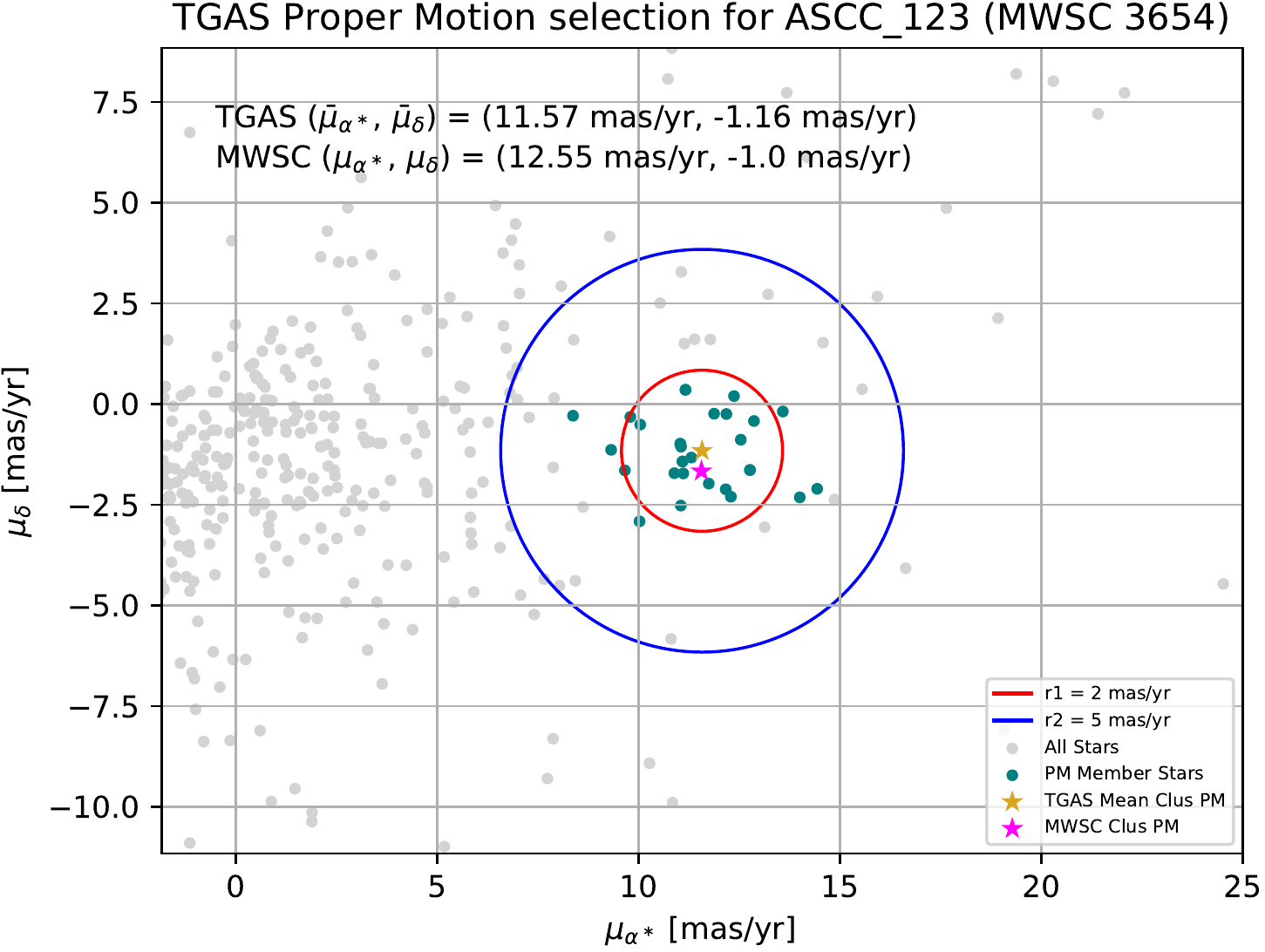}
\includegraphics[width=6.5cm]{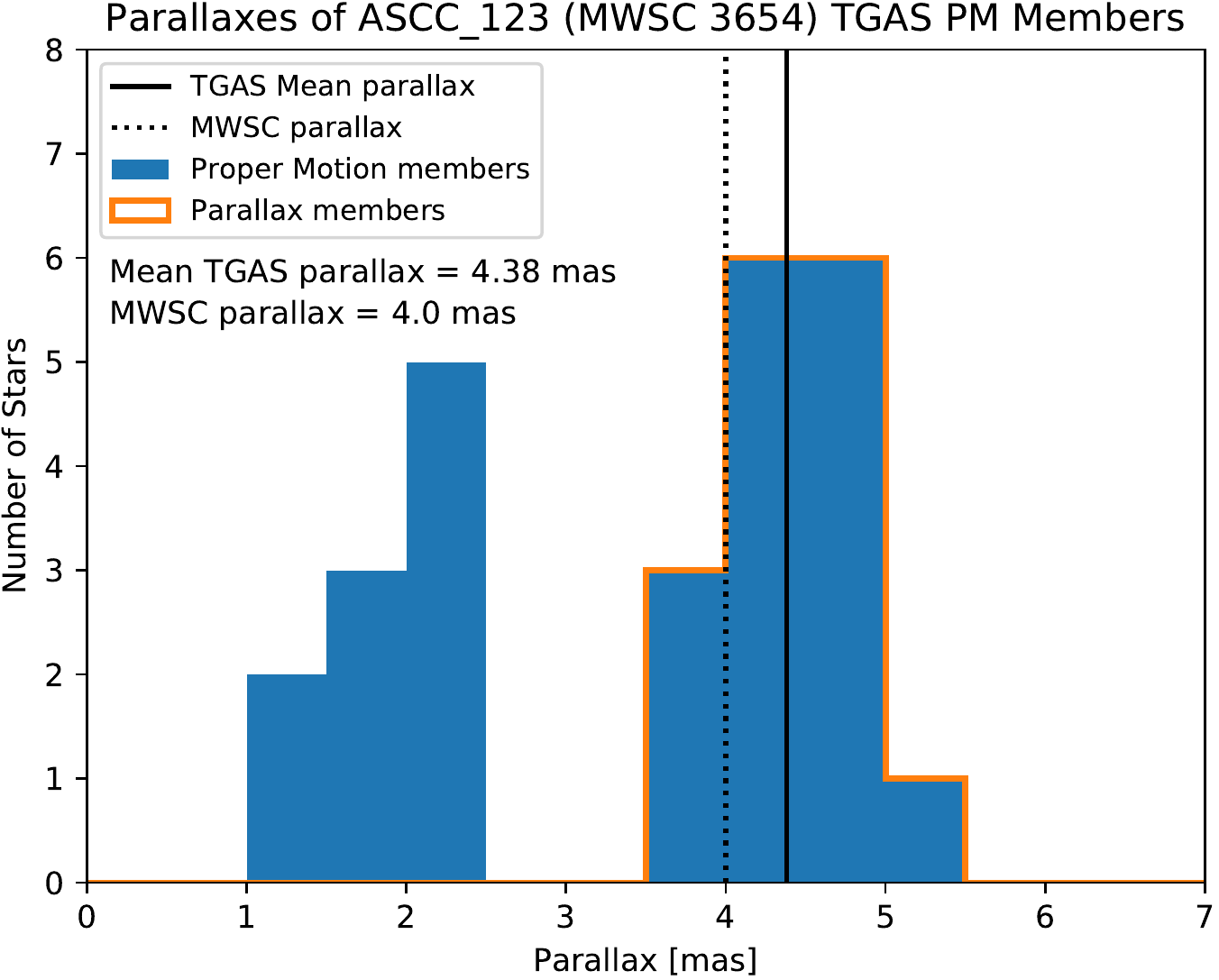}\\
\caption{TGAS proper motion (left panel) and parallax (right panel) selection diagrams for NGC~6475, Ruprecht~147, NGC~7092, and ASCC~123. The proper motion and parallax criteria for membership selection is described in Sect. 3.1\sybf{; the values for cluster proper motion and parallaxes in the diagrams are based on initial cluster membership. The final cluster proper motion and parallaxes are provided in Table~\ref{table:results}.} Left panel: The teal points represent the proper motion members, where all stars within the 2 \sybf{mas yr$^{-1}$} radius (red circle) of the mean cluster proper motion are selected and the stars within 5 \sybf{mas yr$^{-1}$} (blue circle) are only selected if their 3$\sigma$ errors are consistent with the mean cluster proper motion. Right panel: The orange outline illustrates the stars with 3$\sigma$ errors consistent with the mean cluster parallax; these stars are the TGAS astrometrically-selected candidates of the cluster.}
 \label{figb6}
\end{figure*}

\end{appendix}

\end{document}